\documentclass[a4paper,12pt,twoside]{report}
\pdfoutput=1
\usepackage[english]{babel}

\usepackage{PhD} 
\usepackage[centertags]{amsmath}
\usepackage{graphicx}
\usepackage{epigraph,amsfonts,amssymb,amsthm,newlfont,textcomp,array,dsfont,feynmf,microtype,mathtools}
\usepackage{lscape}
\usepackage{rotating}
\usepackage{epstopdf}
\usepackage[sort&compress]{natbib}
\usepackage{color,tabularx,booktabs,lscape,afterpage,longtable,dcolumn,truncate}
\usepackage[font={footnotesize}]{caption}
\usepackage[perpage,marginal]{footmisc}
\usepackage[charter,greekuppercase=italicized]{mathdesign}
\usepackage{wrapfig,floatflt,hvfloat,underscore}
\usepackage[shortlabels]{enumitem}
\usepackage[autoplay]{animate}
\usepackage{upgreek}                        
\usepackage[nottoc]{tocbibind}              
\usepackage{tikz}                           
\usepackage{tablefootnote,scrextend}        

\usepackage[colorlinks,plainpages=false,linkcolor=black,urlcolor=blue,citecolor=black,pdfpagemode=UseNone,pdfstartview=FitBH,
            pdftitle={},
            pdfsubject={INS}, pdfauthor={Pavlo Y. Portnichenko}, pdfcreator={Pavlo Y. Portnichenko},
            pdfkeywords={}]{hyperref}

\usepackage{cleveref}                       

\makeatletter
\def\lambdabar{\protect\@lambdabar}
\def\@lambdabar{%
\relax
\bgroup
\def\@tempa{\hbox{\raise.7\ht0
\hbox to0pt{\kern.25\wd0\vrule width.5\wd0
height.5pt depth.1pt\hss}\box0}}%
\mathchoice{\setbox0\hbox{$\displaystyle\lambda$}\@tempa}%
{\setbox0\hbox{$\textstyle\lambda$}\@tempa}%
{\setbox0\hbox{$\scriptstyle\lambda$}\@tempa}%
{\setbox0\hbox{$\scriptscriptstyle\lambda$}\@tempa}%
\egroup
}
\makeatother

\DeclareMathAlphabet{\pazocal}{OMS}{zplm}{m}{n}

\DeclareRobustCommand{\subsup}[3]{{%
  \mathpalette\makesubsup{{#1}{#2}{#3}}%
}}

\makeatletter
\providecommand{\@firstofthree}[3]{#1}
\providecommand{\@secondofthree}[3]{#2}
\providecommand{\@thirdofthree}[3]{#3}
\newcommand{\makesubsup}[2]{%
  \sbox\z@{$\m@th#1{}_{\@secondofthree#2}$}%
  \sbox\tw@{$\m@th#1{}^{\@thirdofthree#2}$}%
  \dimen@=\wd\z@
  \ifdim\wd\tw@>\wd\z@ \dimen@=\wd\tw@\fi
  { _{\mathmakebox[\dimen@][r]{\@secondofthree#2}}%
    ^{\mathmakebox[\dimen@][r]{\@thirdofthree#2}}%
      \mkern0mu \@firstofthree#2}%
}
\makeatother

\makeatletter
\newcommand{\bibstyle@mycitestyle}{\bibpunct[, ]{[}{]}{,}{n}{,}{,\hspace{-2.1pt}}%
\gdef\bibnumfmt##1{[##1]}}
\makeatother
\newcommand{\fint}{\text{\raisebox{4.5pt}{\rotatebox{-60}{\footnotesize|}}}\kern-12pt\int}
\renewcommand{\iint}{\int\kern-7pt\int}
\renewcommand{\iiint}{\int\kern-7pt\int\kern-7pt\int}

\begin{document} \phd
\nolistoffigures
\nolistoftables 
\permissionfalse
\graphicspath{{Images/}} \sloppy
\newcolumntype{.}{D{.}{.}{-1}} 
\title{Magnetic Dynamics in Heavy-Fermion Systems with Multipolar Ordering Studied by Neutron Scattering}

\author{Pavlo Y. Portnichenko}

\email{\href{mailto:portnichenko@gmail.com}{portnichenko@gmail.com}}

\copyrightyear{2013--2018} \submitdate{18 May 2018}

\examiner{Prof. Dr. Hans-Henning Klau{\ss}}
\supervisor{Jun.-Prof. Dr. Dmytro Inosov}
\firstreader{Prof. Dr. Carsten Timm}
\secondreader{DR Emeritus Dr. Jean-Michel Mignot}

\university{Fakult\"{a}t Physik  \\der Technischen Universit\"{a}t Dresden}

\address{Dresden, Germany}

\dept{Fakult\"{a}t Physik, TU Dresden}
\beforepreface

\setlength{\epigraphwidth}{0.618\textwidth}
\renewcommand{\textflush}{flushepinormal}

%
%

\prefacesection{Introduction} \fancyhead[LO]{Introduction} \thispagestyle{fancy}

\setlength{\epigraphwidth}{0.5\textwidth}
\epigraph{I don't know anything, but I do know that everything is interesting if you go into it deeply enough.}{\vspace{0.3em} Richard Feynman}

Despite more than half a century of studies in heavy-fermion compounds, a full understanding of the various possible magnetic ordering phenomena is still far from complete. Some heavy-fermion materials show so-called hidden-order phases, which are invisible to conventional diffraction techniques. The multipolar moments of the $f$ electrons in their specific crystal field environment play a decisive role in the formation of these phases. Such hidden-order phases have been observed in a variety of compounds containing $4f$ and $5f$ elements, like URu$_2$Si$_2$, NpO$_2$, YbRu$_2$Ge$_2$, and CeB$_6$. The competition or coexistence of multipolar ordering with more conventional magnetic order parameters, such as ferro- or antiferromagnetism, gives rise to complex magnetic-field--temperature phase diagrams in these compounds that provide a rich playground for experimental and theoretical investigations.

Because the multipolar ordering is often coupled to another magnetically ordered state, the instances where orbital ordering appears as a separate solitary phase are rare. The most well-studied member of this family of compounds is the pure CeB$_6$ material, which is considered a textbook example of a system with the so-called magnetically hidden order, typically associated with an antiferromagnetic arrangement of magnetic quadrupole moments. In this compound an antiferroquadrupolar (AFQ) ordered phase is found in a narrow range between  $T_{\text{N}}=2.3$\,K and $T_{\text{Q}}=3.2$\,K. Application of a magnetic field  stabilizes the AFQ phase over a broader temperature range, up to the fields of $H_{c}\sim80$\,T, when a transition to the paramagnetic phase occurs.

Close inspection of the phase diagram of Ce$_{3}$Pd$_{20}$Si$_{6}$ shows remarkably similar behavior to CeB$_6$, but with reduced temperature and field scales, with  $T_{\text{N}}=0.2$\,K, $T_{\text{Q}}=0.45$\,K and $H_{c}\sim10$\,T, thus being in a range of fields and temperatures accessible to modern neutron scattering techniques. Structurally this material is far more complex than CeB$_6$, as it consists of two interpenetrating sublattices of cerium. One of them possesses a simple-cubic structure like in CeB$_6$, whereas the other one  has a geometrically frustrated face-centered-cubic structure. This unprecedented coexistence in the same material of two inequivalent Kondo lattices with different symmetry was predicted to exhibit two drastically different Kondo temperatures due to the competitive Kondo-screening effects.

Here I examine the pure and La-doped $f$\!-electron system CeB$_6$, as well as  Ce$_{3}$Pd$_{20}$Si$_{6}$, by means of neutron scattering. The magnetic field dependence of  the zone center exciton mode in CeB$_6$ was studied in fields up to 14.5\,T and compared with the existing theory. A long standing question of the applicability of the phenomenological temperature dependence $\Gamma=k_{\text B}T_{\text K}+A \sqrt{T}$, where $T_{\text K}$ is the Kondo temperature, for two limiting cases, the Kondo lattice and Kondo impurity, has been investigated on a set of samples diluted with lanthanum. In Ce$_{3}$Pd$_{20}$Si$_{6}$, the order parameter of the AFQ phase was uncovered, and its field dependence was studied  in  fields up to 9\,T.

\textbf{Structure of this thesis}. In the \textit{first chapter}, I give a brief introduction to neutron scattering, which covers the two most important aspects: the theory and experimental details of this technique. There are several important reasons which allowed neutrons to become an ideal probe of condensed matter physics. The absence of an electric charge allows them to penetrate deep inside the material, and the presence of one-half spin is responsible for the interaction with the unpaired electron spin, thus probing magnetic properties. However, the use of neutron scattering was strongly deterred  by the low flux neutron sources, which existed at the very beginning. Therefore, in the experimental part I give a brief summary of possible mechanisms for neutron production  and detection, as well as some examples  of available  neutron sources. A detailed explanation of the most important instruments used in neutron scattering, the triple-axis spectrometer and time-of-flight spectrometer, is also given.

The \textit{second chapter} mainly addresses the  physics of heavy-fermion systems. They are usually intermetallic compounds where partially filled orbitals on each atom, considered as localized magnetic moments, are hybridized with the conduction electrons, forming heavy fermion bands. Properties of these systems are mainly governed by the competition between the Kondo screening and the Ruderman-Kittel-Kasuya-Yosida coupling mechanism via the conduction electrons. An important part of physics in the $f$\!-electron system is the strong coupling between the spin and orbital degrees of freedom.  This leads to the presence of new eigenstates, that may be described in  terms of the multipole degrees of freedom. It turned out that the interaction between different multipoles is responsible for the appearance of the ordered phases in various materials. One well known example is CeB$_6$, where the interaction between quadrupole moments leads to the appearance of the so called ``magnetically hidden'' order. The reason for this order to be called ``hidden'' comes from the fact that quadrupole ordering is the ordering of the charge degrees of freedom, and because the neutron possesses no charge, it is invisible to neutron scattering. In magnetic materials, time dependent spin fluctuations can form collective excitations, known as magnons. At the first glance, all magnons may appear to be the same, however, as will be shown, they exhibit significantly different behavior upon applying a magnetic field. Thus the external magnetic field can help to uncover the underlying mechanism of the spin interactions in matter.

The \textit{third} and \textit{fourth chapters} are focused on the main results of the inelastic neutron scattering (INS) experiments on Ce$_{\text{1}-x}$La$_x$B$_\text{6}$ and Ce$_{3}$Pd$_{20}$Si$_{6}$ respectively. Until now, no systematic investigations of the spin dynamics under the influence of magnetic field have been reported for these materials, despite the fact that the already published  articles explicitly mentioned that ``more accurate INS experiments as function of field and temperature than presently available are desirable''. Therefore my research focused on the evolution of the recently discovered  ferromagnetic resonance in CeB$_6$ upon application of the magnetic field. Using both triple-axis and time-of-flight spectrometers, a systematic investigation of the zone-center  excitations was done, as well as changes of the resonance energy  associated with the presence of anisotropy with respect to the field direction. The obtained results were directly compared with existing theories, which predict the evolution of excitation energies, and a reasonably good agreement was found.

It was found that in many  heavy-fermion metals, the quasielastic line width follows phenomenological temperature dependence $\Gamma=k_{\text B}T_{\text K}+A \sqrt{T}$, where $T_{\text K}$ is the Kondo temperature. There have been no systematic studies of the dependence of the Kondo temperature and the parameter $A$ on doping, therefore availability of the La-doped series of Ce$_{\text{1}-x}$La$_x$B$_\text{6}$ single crystals gave me an opportunity to address these long-standing open questions.  An unambiguous confirmation of the $\sqrt{T}$ law scaling was found in all available samples, and the value of the neutron-deduced Kondo temperature was found to be in a good agreement with the published values determined from resistivity measurements.
 \enlargethispage{10pt}

In the \textit{fourth chapter} in contrast to previous results on  polycrystalline samples, which showed presence of crystal electric field excitations that correspond to different sublattices, my measurements on a single crystal of Ce$_{3}$Pd$_{20}$Si$_{6}$ excluded the presence of the second lower-energy crystal field line. Instead, within the paramagnetic state, a signal that can be well described by a quasielastic Lorentzian line shape centered at the $(1\,1\,1)$ wave vector was found. According to its $\mathbf{Q}$-space symmetry, the signal is associated with the  simple-cubic sublattice and corresponds to the zone corner  for the simple-cubic sublattice. Upon warming, the observed signal exhibits gradual broadening of the quasielastic signal linewidth in accordance with the expected $\sqrt{T}$ law.

\clearpage
In a structurally simpler  CeB$_6$ sample, the corner of the Brillouin zone corresponds to the propagation vector of the antiferroquadrupolar phase. By analogy with cerium hexaboride we expect that at a $\mathbf{Q}$-point, where a maximum of the quasielastic intensity can be found, an analog of the antiferroquadrupolar Bragg peak should exist. Application of the magnetic field along either the $[1\,1\,0]$ or the $[0\,0\,1]$ direction induces a broad diffuse peak, in addition to the sharp structural Bragg reflection. This magnetic peak contribution reaches its maximal intensity in the middle of the antiferroquadrupolar phase  and then starts to decrease, thus repeating behavior of the antiferroquadrupolar Bragg peak intensity in CeB$_6$. In addition, for magnetic fields applied along the $[0\,0\,1]$ direction, the diffuse magnetic peaks in Ce$_{3}$Pd$_{20}$Si$_{6}$ become incommensurate, suggesting a more complex modulated structure of the underlying multipolar order that can be continuously tuned by a magnetic field.

Another technical chapter is presented in the \textit{Appendix A}. In my personal opinion this part is extremely important. Because of non-ideal sample environment, as well as due to technical nuances of neutron scattering, various unexpected signals which in some cases mimic intrinsic scattering features from the crystal, can  be easily confused with a real signal. Therefore, in this section I describe in detail various spurions  which I had to deal with during the experiments, as well as some instruction on  ways to avoid them or take them into account.

\cleardoublepage

\smallskip

\afterpreface

\fancyhead[LO]{\truncate[...]{0.95\textwidth}{\nouppercase\rightmark}}
\fancyhead[RE]{\truncate[...]{0.95\textwidth}{\hfill\nouppercase\leftmark}}


\chapter{Neutron scattering}\label{Chap:Neutronscattering}

\section{Introduction}\label{Sec:Introduction}
Since ancient times, humanity has been interested in the structure of matter. The ancient Greek philosophers Leucippus and Democritus in the 5th century BC suggested that all matter is built from the ungenerated and indivisible smallest particles called atoms and void~\cite{Taylor99}. Although later this conception for some time was superseded by   Aristotle's doctrine of four ``simple bodies'' \cite{Lloyd68}, thanks to the atomism supporters at the turn of the $16^{\text{th}}$\,--\,$17^{\text{th}}$ centuries, molecular-atomic theory was developed later in the 19$^{\text{th}}$ century. Discovery of significant chemical laws, in particular the law of conservation of mass together with the law of definite proportions and the law of multiple proportions played a major role in the establishment of atomic theory \cite{Dalton08}.

Observation  of the electron indicated that the atom consists of smaller particles and is divisible. The suggested plum pudding model, in which a cloud of positive charge, equal to the size of the atom, contains small, negatively charged ``corpuscles'', was the first attempt to describe the atom \cite{Thomson04}. Later this model was contradicted with the Rutherford model, based on the experimental results of $\alpha$ and $\beta$ particles scattered by a thin gold foil. A small number of $\alpha$ particles were deflected  by very large angles, indicating that the scattering center has a small size and a substantial electric charge, and the mass of the atom is concentrated in it. This heavy, positively charged center core is surrounded by negatively charged electrons, compensating each other so that atoms are neutrally charged. Electrons were thought to be flying around the center core like planets around the sun \cite{Rutherford11}. However the planetary model had a significant shortcoming, since it  was unable to explain the stability of the atom. According to classical electrodynamics, electrons orbiting the nucleus experience acceleration and emit electromagnetic waves. As a result of continuous energy loss by radiation, electrons should spiral toward the nucleus and as a result fall on it. Observed emission and absorption spectral lines of atoms cannot be explained by the planetary model either. The following progress in understanding principles of quantum mechanics led to the emergence of the Bohr model, complemented later by Sommerfeld. Introduction of certain orbits, where electrons can orbit infinitely without losing energy by radiation \cite{Bohr13,Bohr13a,Bohr13b}, together with the  quantization of the $z$-component of the angular momentum that granted the existence of elliptical orbits with the same energy, thus allowing energy levels degeneracy \cite{Sommerfeld31}, gave good description of the hydrogen atom and explained the previously observed Zeeman effect \cite{Zeeman97a,Zeeman97b,Zeeman97}. Because Bohr's theory utilized laws of classical mechanics combined with the quantum rule, it was not sufficiently consistent and general, and therefore later was replaced by modern quantum mechanics.

Introduction of the concept of the atomic orbital, as a solution of the Schr\"{o}dinger equation, was the next step in attempts to obtain an absolute understanding of the atom \cite{Schroedinger26}. The electron was imagined as a negatively charged cloud, whose density is proportional to the square of the wave function at the corresponding point of the atom \cite{Born55}. The term orbital in particular emphasizes the fact, that the state of an electron in an atom is fundamentally different from the classical motion and is described by the laws of quantum mechanics. Each orbital is described by quantum numbers: $n$\,--\,principal, $l$\,--\,azimuthal, and $m$\,--\,magnetic  quantum numbers. The principal quantum number is used to characterize orbital energy, therefore larger values of $n$  correspond to higher potential energy. Azimuthal and magnetic quantum numbers are used to describe the shape of the orbital and the spatial arrangement of the atomic orbitals, respectively.

While understanding of the atomic orbitals rose to new levels, progress in the physics of atomic nucleus lagged behind. As far back as in the beginning of the $20^{\text{th}}$  century, it was known that substances having the same chemical behavior but different atomic weight can exist \cite{StroemholmSvedberg09,StroemholmSvedberg09a}. The empirical rule, which described the transmutation of elements during radioactive decay, known nowadays as the radioactive displacement law, was also discovered \cite{Fajans13,Soddy13}. Thus, despite the fact that the neutron was discovered later, existence of isotopes was confirmed.

Experiments with irradiation of certain light elements like Li, Be, B, F, Mg, and Al by $\alpha$-particles, obtained from a radioactive polonium source, revealed  a new type of  unusually penetrating radiation \cite{BotheBecker30,BeckerBothe32}. Erroneously, the discovered emission was first thought to have $\gamma$-nature, as it was not influenced by an electric field. Later on, a similar experiment was done by James Chadwick, wherein he irradiated paraffin with $\alpha$-particles and made a conclusion that this radiation is likely to be  uncharged particles with the same mass as the proton \cite{Chadwick32,Chadwick33,Chadwick32a}. Several years later, in 1935, he won the Nobel Prize in Physics ``for the discovery of the neutron'' \cite{ChadwickNobelPrize}. With his discovery, the proton--neutron model of the nucleus gained impetus and was quickly accepted.

For the purpose of research,  neutrons described in this thesis are interesting due to their unique nature as probing particles. We will discuss neutron properties in detail later in sec.\,\ref{Sec:Neutronproperties}, while now we will briefly go through the history and formation of  neutron scattering as a versatile tool for probing magnetism. Development of high-flux sources, to allow scattering experiments as we know them now, took a while. However, the first attempts of neutron diffraction experiments that followed just after the discovery of the neutron  using a radioisotope Rn-Be source, successfully demonstrated Bragg scattering, thereby confirming wave-particle duality of the neutron \cite{DunningPegram35,MitchellPowers36}. Further development was directly linked to the progress in nuclear reactor construction.

\begin{wrapfigure}[13]{r}{0.35\textwidth}\vspace{-1em}
\includegraphics[width=0.35\textwidth]{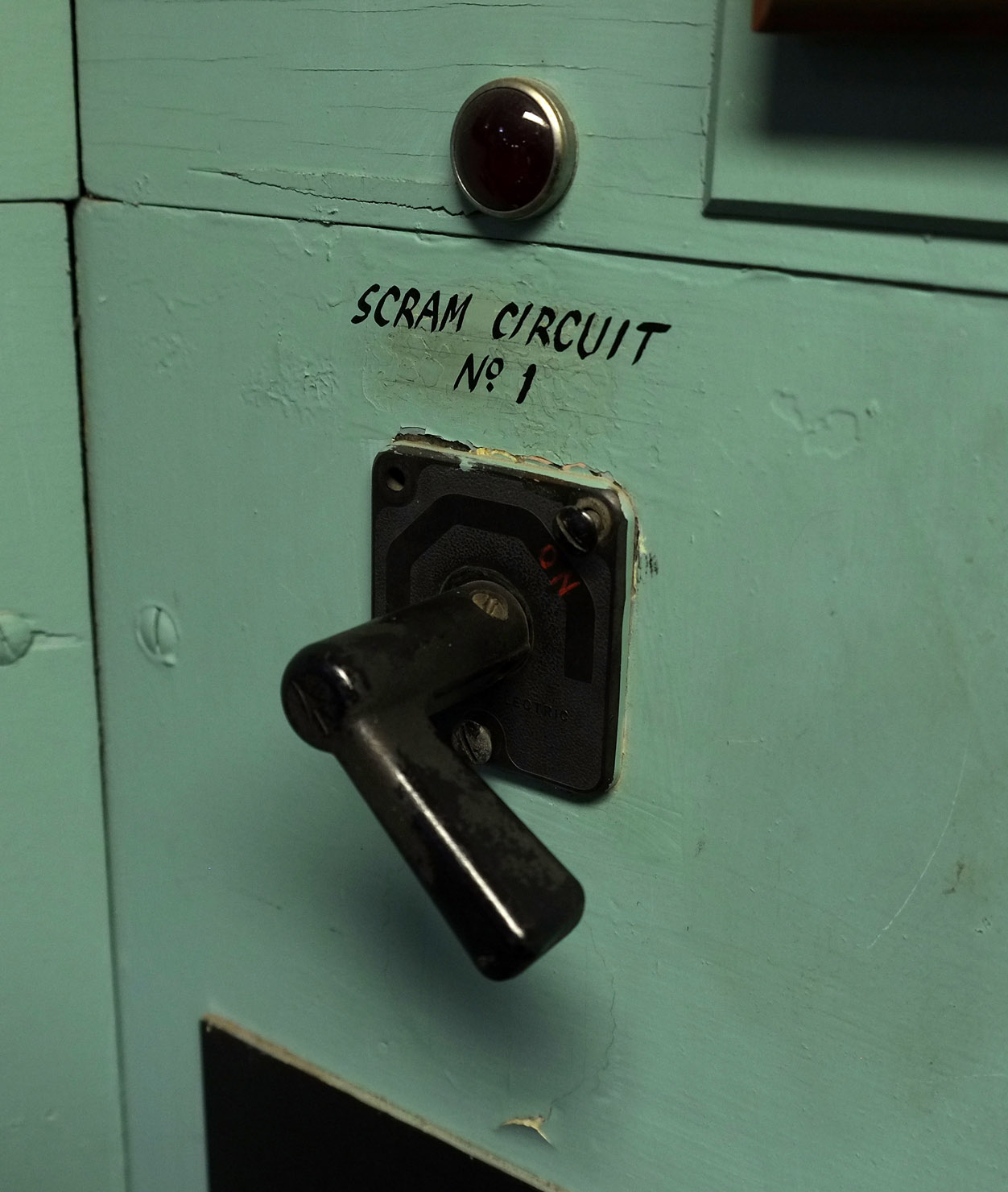}
\caption{SCRAM button in the control room of the X-10 Graphite Reactor at Oak Ridge National Laboratory.}
\label{Fig:SCRAM}\vspace{-4pt}
\end{wrapfigure}

The first nuclear reactor Chicago Pile-1 was built in 1942 at the University of Chicago under the supervision of Enrico Fermi \cite{Fermi52}. Although this reactor is irrelevant to the neutron scattering experiments, as it was used to test the feasibility of a controlled self-sustaining nuclear chain reaction, it would be dishonest not to mention it. In addition, one curious fact is related to this ``reactor'': in case of emergency, control rods made of cadmium (a good neutron absorber), suspended on a rope, were supposed to be dropped into the pile and stop the reaction. As there was a chance that this would fail to happen automatically, a man with an axe to cut the rope in case of unforeseen circumstances was on duty. This person was the \textit{safety control rod axe man}, or acronym SCRAM, and since then SCRAM is used as a name for an emergency shutdown of a nuclear reactor (fig.\,\ref{Fig:SCRAM}).

Later in 1943, at Oak Ridge National Laboratory, the second graphite reactor X-10 was constructed. This reactor was designed mostly to produce weapons-grade plutonium from natural uranium, but in addition was used to obtain the first neutron diffraction patterns on gypsum and rocksalt crystals in 1944 by E. Wollan and C. Shull \cite{MasonGawne13}. The first two-axis diffractometer used for these experiments actually was a remodeled x-ray diffractometer, while the first specially designed spectrometer for neutron-scattering experiments  was built later in 1949 \cite{Shull95}. Another significant result obtained  due to the presence of the neutron's magnetic moment was the observation of antiferromagnetism in MnO, in which magnetic reflections were observed at the forbidden positions of the structural unit cell \cite{ShullSmart49}. In order to measure inelastically scattered neutrons, it was necessary to add an additional axis with the analyzer crystal, which resulted in the triple-axis spectrometer, first created in 1961 by B. Brockhouse at the Chalk River NRX reactor in Canada \cite{Brockhouse61,BrockhouseWoods64}. Years later, in 1994, the Nobel Prize in Physics was awarded to B. Brockhouse  and C. Shull ``for pioneering contributions to the development of neutron scattering techniques for studies of condensed matter''  \cite{BrockhouseShullNobelPrize}.

Neutron scattering has been developing for more than half a century, and now it has become an important tool of modern materials science. Significant improvement of the reactor neutron sources as well as development of neutron spallation sources was essential to obtain  present-day knowledge in condensed matter. Rapid development of the neutron scattering technique led to the rise of innovative instruments, like time-of-flight, backscattering, and spin-echo spectrometers, which greatly increased the applicability of the neutron scattering method. In this chapter I would like to focus mainly on two important techniques used in my research: triple-axis and time-of-flight spectroscopy, as most of the results in my thesis were obtained using these methods. Besides, I will point out various types of spurious signals that can occur in such experiments.

\vspace{-7.5pt}
\section{Neutron Properties and its Interaction with Matter}\label{Sec:Neutronproperties}

Due to unique properties of the neutron it became a highly effective probe of condensed matter physics. The most important properties of the neutron, vital for its use as a probe, are the absence  of electric charge and presence of spin $\frac{1}{2}$. Because the neutron is a chargeless particle, it experiences no  Coulomb interaction and can penetrate deep inside the material, scattering only on the nuclei due to the strong interaction. In addition, the neutron's magnetic moment can interact with the unpaired electron spins via the dipole-dipole interaction, becoming an indispensable probe of magnetic properties. Depending on whether a neutron is free or bound in an atomic nuclei, it can be unstable with  a mean lifetime of  $\sim$15  min. \cite{Nakamura10,WietfeldtGreene11} or stable, respectively. The neutron with a mass $m_{\text{n}}=1.0087{\text u}$ is slightly heavier than a proton and has a magnetic moment $\mu_{\text{n}}=-1.913\mu_{\text N}$, where $\mu_{\text N}=e\hbar/2m_{\text p}$ is the nuclear magneton.

\vspace{-7.5pt}\enlargethispage{8pt}
\subsection{Neutron absorption }\label{SubSec:NeutronAbsorption}

Because of these properties, neutrons show significantly different results after interaction with the sample as compared to photons or electrons. First of all, the result of interaction with matter can be divided into two different cases: absorption and scattering. In case of  the neutron's absorption by nuclei, they form a compound nucleus in an excited state, which after some time decays into the ground state or undergoes fission. During the decay process, nuclei can relax into the ground state or to another excited state with lower energy.  In particular, natural cobalt  $^{59}\text{Co}$, after being irradiated with neutrons, transforms into unstable $^{60}\text{Co}$ with a half-life of 5.2713 years, which undergoes $\beta^{-}$ decay into Ni:
$^{60}_{27}\text{Co}\,\longrightarrow \,^{60}_{28}\text{Ni}\!+\!e^{-}\!+\!\overline{\nu}_{e}$
\cite{AudiBersillon03}. A similar process can be observed for gold
$^{198}_{79}\text{Au}\,\longrightarrow \,^{198}_{80}\text{Hg}\!+\!e^{-}\!+\!\overline{\nu}_{e}$
\cite{Lieser01} or, for instance, for platinum
$^{197}_{78}\text{Pt}\,\longrightarrow \,^{197}_{79}\text{Au}\!+\!e^{-}\!+\!\overline{\nu}_{e}$
\cite{AudiBersillon03}, which certainly would interest ancient seekers of \textit{lapis philosophorum}. Another result is observed in case of nuclear fission, in particular important for neutron scattering,  boron
$\subsup{\text{B}}{5}{10}\,+^{1}_{0}\!\text{n}\,\longrightarrow \,^{4}_{2}\text{He}\!+\!^{7}_{3}\text{Li}\!+\!2.78\,\text{MeV}$, lithium
$^{6}_{3}\text{Li}\,+^{1}_{0}\!\text{n}\,\longrightarrow \,^{4}_{2}\text{He}\!+\!^{3}_{1}\text{H}\!+\!4.78\,\text{MeV}$ or helium
$^{3}_{2}\text{He}\,+^{1}_{0}\!\text{n}\,\longrightarrow \,^{1}_{1}\text{H}\!+\!^{3}_{1}\text{H}\!+\!765\,\text{keV}$
\cite{Tsoulfanidis95}. These reactions are used in neutron detection due to their huge cross sections and will be  described in details later in sec.\,(\ref{SubSec:NeutronDetector}).

\begin{figure}[t]\vspace{-10pt}
\includegraphics[width=\textwidth]{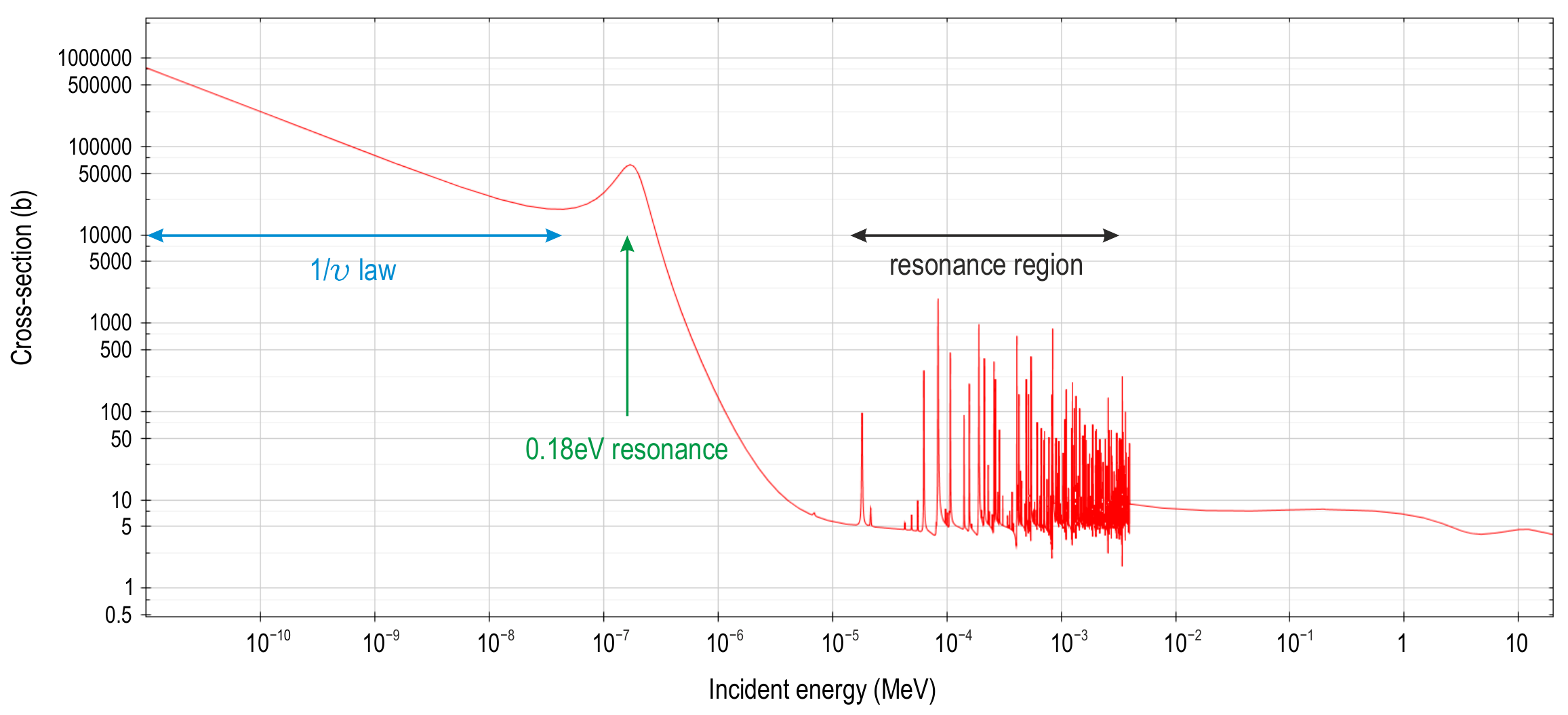}\vspace{-10pt}
\caption{Total neutron cross-section of $^{113}\text{Cd}$. Figure source obtained from the Joint Evaluated Fission and Fusion File (JEFF) library using JANIS \cite{SopperaBossant14}.}
\label{Fig:CdCrossSection}\vspace{-10pt}
\end{figure}

Of particular interest is the absorption of neutrons by cadmium. This material is commonly used as neutron absorber due to its very high neutron absorption cross-section. Naturally occurring Cd consists of various isotopes, however we will pay our attention to $^{113}\text{Cd}$.  After being irradiated with neutrons, it is converted into resultant nucleus $^{114}\text{Cd}$ in an excited state, which leads to the emission of $\gamma$-rays during relaxation: $^{113}_{48}\text{Cd}\,+^{1}_{0}\!\text{n}\,\longrightarrow \,^{114}_{48}\text{Cd}^\star\longrightarrow \,^{114}_{48}\text{Cd}\,+\gamma$ \cite{RusevJandel13}. Absorption cross-sections strongly vary with neutron energy, and as shown in fig.\,\ref{Fig:CdCrossSection}, $^{113}\text{Cd}$ is especially good as shielding against thermal neutrons with energies below 1\,meV, however it is several orders of magnitude worse in absorbing fast neutrons. The cross-section decreases with increasing incident neutron energy as a general rule, which is valid as well for $^{10}\text{B}$, making these two materials particulary useful for thermal-neutron shielding.

The absorption cross-section for neutrons was for the first time explained in 1936  by Breit and Wigner \cite{BreitWigner36} in an attempt to understand accumulated experimental results, like cross-section decrease with increasing velocity of neutrons \cite{TillmanMoon35, FrischPlaczek36}, and aspired to explain large  cross-sections as a resonance effect \cite{Bethe35}. According to the Breit-Wigner formula, neutron absorption cross-section for a specific element is
\vspace{-0pt}\begin{equation}
\sigma_{n,\gamma}(E)=\pi\lambdabar^{2}\frac{\Gamma_{n}\Gamma_{\gamma}}{(E-E_{\text{R}})^2+(\Gamma/2)^2},
\label{eq:BreitWignerFormula}
\vspace{-0pt}\end{equation}
where $\lambdabar=\frac{\lambda}{2\pi}=\frac{\hbar}{mv}$, and $E_{\text{R}}$ is the resonance energy. As there is always a certain probability that the excited state will decay, $\Gamma$ can be related to the finite lifetime as $\Gamma=\hbar/\tau$. Decay occurs via various processes, like $\gamma$-emission, re-emission as a neutron, emission of charged particles and others, thus the total width $\Gamma$ can be written as a sum of different partial widths: $\Gamma=\Gamma_{\gamma}+\Gamma_{n}+\Gamma_{\alpha}+...$. Generally speaking radiation width changes from resonance to resonance differently for $\Gamma_{\gamma}$ and $\Gamma_{n}$. While the radiation width value changes slightly,  neutron width  varies quite strongly from resonance to resonance and  can be introduced through the relation:

\vspace{-15pt}\begin{equation}
\Gamma_{n}=\Gamma_{n0}\surd{E}.
\label{eq:NeutronWidth}
\vspace{-0pt}\end{equation}
When $E=E_{\text{R}}$ the  cross section has a maximum and can be written as
\vspace{-0pt}\begin{equation}
\sigma_{n,\gamma}^{0}=4\pi\lambdabar^{2}_{\text{R}}\frac{\Gamma_{n}\Gamma_{\gamma}}{\Gamma^2}.
\label{eq:ResonanceCrossSection}
\vspace{-0pt}\end{equation}
Another limiting case is when $E\!\rightarrow\!0$. When $E_{\text{R}}\gg\Gamma$, eq.\,\ref{eq:BreitWignerFormula} transform into
\vspace{-0pt}\begin{equation}
\sigma_{n,\gamma}(E)=\pi\lambdabar^{2}\frac{\Gamma_{n}\Gamma_{\gamma}}{E_{\text{R}}^2}.
\label{eq:E0CrossSection}
\vspace{-0pt}\end{equation}
Taking into account eq.\,\ref{eq:NeutronWidth} and  relation  between the wavelength and the neutron energy $E=8.17\times10^{-18}/\lambda^2$, eq.\,\ref{eq:E0CrossSection} can be written in the following form:
\vspace{-0pt}\begin{align}
\sigma_{n,\gamma}(E) &= \pi\lambdabar^{2}\frac{\Gamma_{n0}\Gamma_{\gamma}}{E_{\text{R}}^2}\surd{E}
\nonumber \
\\&=\pi\lambdabar^{2}_{\text{R}}\frac{\Gamma_{n0}\Gamma_{\gamma}}{E_{\text{R}}}\frac{1}{\surd{E}},
\label{eq:1divideV}
\vspace{-0pt}\end{align}
indicating that the absorption of neutrons is proportional to $1/\surd{E}$, or $1/v$, therefore explaining the previously observed cross-section decrease with increasing velocity of neutrons. The ${1}/{v}$ law has a simple explanation: neutrons with lower energy spend more time in the vicinity of the nucleus and thus have a higher probability of being  captured by the nucleus \cite{BeckurtsWirtz64}.

The described capture cross-section is shown in fig.\,\ref{Fig:CdCrossSection} for the $^{113}\text{Cd}$ isotope. At low energies below $10^{-2}$\,eV, highlighted with the blue arrow in fig.\,\ref{Fig:CdCrossSection}, the cross-section follows the ${1}/{v}$ law. Unlike many other materials, in Cd the first resonance emerges at a relatively small energy of ~0.18\,eV (green arrow), immediately after which the cross-section rapidly falls down by several orders of magnitude, until it reaches the resonance region around 100\,eV \cite{Molnar04}.

\subsection{Nuclear and magnetic scattering}\label{SubSec:NuclearAndMagneticScattering}

We have already discussed two possible outcomes from the interaction of nuclei with a neutron, absorbtion and fission, and now let us consider interaction through the nuclear force. In this case the differential cross section can be written as:
\vspace{-0pt}\begin{equation}
\frac{\text{d}\sigma}{\text{d}\Omega}=b^{2},
\label{eq:DifferentialCrossSection}
\vspace{-0pt}\end{equation}
where \textit{b} is a property only of the nucleus that defines the scattering length. It strongly depends not only on the atomic number but also on the isotope, and in comparison with the scattering length for  X-rays which monotonically increases as a function of $Z$ (atomic-mass numbers), changes in an irregular fashion. This important property makes neutrons sensitive to both light and heavy atoms \cite{SkoeldPrice86,LucatortoParr13}. In addition, the scattering length \textit{b} also varies with the nuclear spin. When a neutron interacts with the nuclei,  the formed ``compound nucleus'' can take two different values for neutrons with spin $+1/2$ and $-1/2$ \cite{Kostorz79}.

Now let us consider matter which consists of different isotopes, thus the scattering length $b_{i}$ varies from one nucleus to another. In this case the differential cross section averaged over all the system can be written as:
\vspace{-0pt}\begin{equation}
\frac{\text{d}^2\sigma}{\text{d}\Omega\text{d}E}=\frac{k_\text{f}}{k_\text{i}}\frac{1}{2\pi\hbar}{\sum_{jj^\prime}\overline{b_{j^\prime}b_{j}}}\int\langle {j^\prime},{j}\rangle\text{exp}(-\text{i}\omega t)\text{d}t
\label{eq:DifferentialCrossSectionForCohAndIncoh}
\vspace{-0pt}\end{equation}
where $\langle {j^\prime},{j}\rangle=\langle\text{exp}\{-\text{i}\mathbf{Q}\cdot\mathbf{R}_{j^\prime}(0)\}\text{exp}\{\text{i}\mathbf{Q}\cdot\mathbf{R}_{j}(t)\} \rangle$, and $\mathbf{Q} $ is known as the scattering vector (see \ref{eq:ScatteringVector}). Taking into account  the assumption that there is no correlation between $b$ values for different nuclei, equation\,\ref{eq:DifferentialCrossSectionForCohAndIncoh} can be rewritten in the following form:
\vspace{-0pt}\begin{align}
\frac{\text{d}^2\sigma}{\text{d}\Omega\text{d}E} &= \frac{k_\text{f}}{k_\text{i}}\frac{1}{2\pi\hbar}(\overline{b})^2{\sum_{jj^\prime}}\int\langle {j^\prime},{j}\rangle\text{exp}(-\text{i}\omega t)\text{d}t \nonumber \
\\&+\frac{k_\text{f}}{k_\text{i}}\frac{1}{2\pi\hbar}\left\{\overline{b^2}-(\overline{b})^2\right\}{\sum_{j}}\int\langle {j},{j}\rangle\text{exp}(-\text{i}\omega t)\text{d}t,
\label{eq:CrossSectionCohAndIncoh}
\vspace{-0pt}\end{align}
where the first term represents the coherent and the second term the incoherent scattering cross-sections \cite{Squires96}. Accordingly, the total scattering cross section from eq.\,\ref{eq:DifferentialCrossSection}, $\sigma\!=\!\int_{0}^{4\pi}\frac{\text{d}\sigma}{\text{d}\Omega}\text{d}\Omega$, can be expressed as a sum of two contributions, $\sigma=\sigma_{\text{coh}}+\sigma_{\text{inc}}$, where
\vspace{-0pt}\begin{equation}
\sigma_{\text{coh}}=4\pi(\overline{b})^2,
\hspace{10pt}
\sigma_{\text{inc}}=4\pi\{\overline{b^2}-(\overline{b})^2\}.
\label{eq:CrossSectionCohInc}
\vspace{-10pt}\end{equation}
It becomes clear that the coherent contribution arises from the correlation between the positions of all atoms in the sample. Since it gives interference effects, it is responsible for the occurrence of the Bragg reflections. At the same time the incoherent contribution represents an effect due to the random distribution of scattering
lengths between different nuclei \cite{LechnerRichter83,Kostorz79,Squires96, LucatortoParr13}. This means that for a material that consists of single isotopes, the incoherent scattering cross-section can only originate from  nuclear-spin interaction, and therefore vanishes for the elements with zero nuclear spin, like $^{4}_{2}\text{He}$ or $^{28}_{14}\text{Si}$.

\begin{wraptable}{r}{8cm}\vspace{-37pt}
\begin{tabular}{c@{~~~~~~}l@{~~~~~~}l@{~~~~~~}l@{~~}}\\
\toprule
Element & $\sigma_{\text{coh}}$ & $\sigma_{\text{inc}}$ & $\sigma_{\text{abs}}$\\
\toprule
$^{1}_{1}\text{H}$      & 1.7583        & 80.27         & 0.3326\\
$^{2}_{1}\text{H}$      & 5.592        & 2.05         & 0.000519\\
\midrule
$^{3}_{2}\text{He}$      & 4.42        & 1.6         & 5333\\
$^{4}_{2}\text{He}$      & 1.34        & 0         & 0\\
\midrule
$^{6}_{3}\text{Li}$      & 0.51        & 0.46         & 940\\
$^{7}_{3}\text{Li}$      & 0.619        & 0.78         & 0.0454\\
\midrule
$\subsup{\text{B}}{5}{10}$      & 0.144        & 3.0         & 3835\\
$\subsup{\text{B}}{5}{11}$      & 5.56        & 0.21         & 0.0055\\
\midrule
$^{27}_{13}\text{Al}$      & 1.495        & 0.0082         & 0.231\\
\midrule
$^{28}_{14}\text{Si}$      & 2.120        & 0         & 0.177\\
\midrule
$^{51}_{23}\text{V}$      & 0.0203        & 5.07         & 4.9\\
\midrule
$^{64}_{29}\text{Cu}$      & 5.2        & 0.006        & 4.50\\
\midrule
$\subsup{\text{Cd}}{48}{113}$      & 0.3        & 12.4        & 20600\\
\midrule
$\subsup{\text{Gd}}{64}{157}$      & 650        & 394        & 259000\\
\midrule
\bottomrule
\end{tabular}
\caption{Neutron scattering cross-sections for various elements. $\sigma_{\text{coh}}$, $\sigma_{\text{inc}}$ and $\sigma_{\text{abs}}$ represent coherent, incoherent, and absorption cross-sections respectively given in barn (1 barn = 100 fm$^2$). Values are taken from ref.\,\citenum{Sears92}. }
\label{Tab:SummaryCohIncAbs}\vspace{-22pt}
\end{wraptable}

Thus when studying a certain material it is necessary to take into account all possible results of the interaction. In table \ref{Tab:SummaryCohIncAbs}  cross-section values of important elements for neutron scattering are summarized. In particular $^{2}_{1}\text{H}$ in comparison with $^{1}_{1}\text{H}$ has a smaller absorption cross-section, and therefore $^{2}_{1}\text{H}_{2}\text{O}$ is used in  nuclear reactors that utilize low-enriched uranium as its fuel. At the same time due to the large incoherent cross-section of $^{1}_{1}\text{H}$, materials like water, polyethylene, and concrete are used for shielding against neutrons. Besides, elements with huge  absorption cross-sections like  $\subsup{\text{B}}{5}{10}$, $\subsup{\text{Cd}}{48}{113}$,  $\subsup{\text{Gd}}{64}{157}$, can be used as well. Polycrystalline  alloys of Cu and Al, due to their neutron ``transparency'' and low incoherent cross-section, are used for construction of instruments and sample environments. Apart from this vanadium should be mentioned: because of a relatively large incoherent cross-section and an almost absent coherent scattering, polycrystalline vanadium is used for instrument calibration and as a material for powder containers in diffraction experiments.

\begin{wrapfigure}[12]{r}{0.35\textwidth}\vspace{-15pt}
\includegraphics[width=0.35\textwidth]{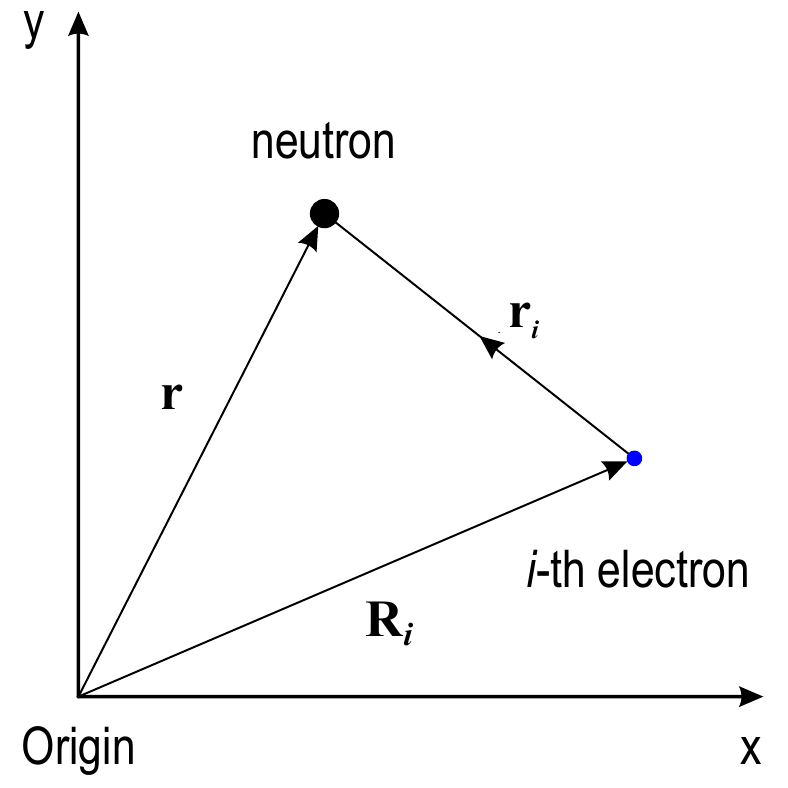}\vspace{-5pt}
\caption{Neutron and electron position vectors.}
\label{Fig:PositionVectors}\vspace{-0pt}
\end{wrapfigure}

We have discussed possible outcomes of interaction processes between the neutron and a nucleus, however due to the presence of the neutron magnetic moment it can also interact with unpaired electrons. Interaction between the neutron in a spin state ${\boldsymbol{\upsigma}}$ and a moving electron of momentum ${\mathbf{p}}$ in a spin state ${\mathbf{s}}$ can be written as:
\vspace{-5pt}\begin{equation}
V=-{\boldsymbol{\upmu}_{\text{n}}}\cdot{\mathbf{B}}=-{2\gamma\mu_{\text N}\mu_{\text B}} {\boldsymbol{\upsigma}} \cdot { \left[\nabla\times\left(\frac{\mathbf{s}\times\mathbf{\hat{r}}}{r^2}\right)+\left(\frac{{\mathbf{p}}\times\mathbf{\hat{r}}}{\hbar r^2}\right)\right]},
\label{eq:ElectronNuetronScatteringPotential}
\vspace{-5pt}\end{equation}
where $\mu_{\text N}$ $\mu_{\text B}$ are the nuclear and Bohr magnetons, and $\mathbf{\hat{r}}$ is a unit vector in the direction $\mathbf{{r}}$ of the electron as shown in fig.\,\ref{Fig:PositionVectors}. In the scattering process, the target can change its state $\tau_{\text{i}}$ to a state $\tau_{\text{f}}$, and the neutron changes its spin from $\sigma_{\text{i}}$ to $\sigma_{\text{f}}$. The transition probability per unit of time, based on the first Born approximation, is given by:
\vspace{-5pt}\begin{equation}
W_{\mathbf{k}_\text{i}\sigma_{\text{i}}\tau_{\text{i}}\rightarrow\mathbf{k}_\text{f}\sigma_{\text{f}}\tau_{\text{f}}}=
\frac{2\pi}{\hbar}|\langle\mathbf{k}_\text{f}\sigma_{\text{f}}\tau_{\text{f}}|V|\mathbf{k}_\text{i}\sigma_{\text{i}}\tau_{\text{i}}\rangle|^2\rho_{\mathbf{k}_\text{f}\sigma_{\text{f}}}(E_{\text{f}}),
\label{eq:TransitionPprobability}
\vspace{-5pt}\end{equation}
where $\rho_{\mathbf{k}_\text{f}\sigma_{\text{f}}}(E_{\text{f}})$ is the density of final states.

\enlargethispage{5pt}

\begin{wrapfigure}[12]{r}{0.45\textwidth}\vspace{-10pt}
\includegraphics[width=0.45\textwidth]{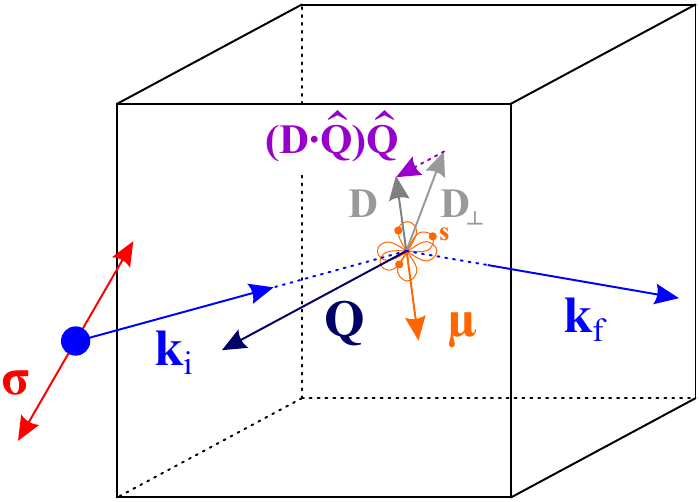}\vspace{-5pt}
\caption{Relation between the scattering vector $\mathbf{Q}$ and magnetic interaction vector $\mathbf{D}_{\bot}$, that emphasizes the fact that neutron scattering is only sensitive
to the magnetization component of the crystal (schematically shown with orange color), that are perpendicular to the scattering vector $\mathbf{Q}$.}
\label{Fig:PositionVectorsDandQ}\vspace{-0pt}
\end{wrapfigure}

Now let us consider a basic neutron scattering experiment. A neutron beam with a flux $\Phi$, characterized by a wave vector $\mathbf{k}_\text{i}$ and a corresponding spin state  $\sigma_{\text{i}}$, scatters on the sample which consists of $N$ atoms. Scattered neutrons are characterized by $\mathbf{k}_\text{f}$ and $\sigma_{\text{f}}$ and measured by the detector with efficiency $\eta$ that subtends a solid angle $\Updelta\Omega$. The count rate $C$ on the detector can be written as:
\vspace{-0pt}\begin{equation}
C=\eta\Phi N (\frac{\text{d}\sigma}{\text{d}\Omega}) \Updelta\Omega.
\label{eq:CountRate}
\vspace{-0pt}\end{equation}
Since $C/\eta$ is equal to transition probability $W_{\mathbf{k}_\text{i}\sigma_{\text{i}}\tau_{\text{i}}\rightarrow\mathbf{k}_\text{f}\sigma_{\text{f}}\tau_{\text{f}}}$, the differential cross-section is:
\vspace{-0pt}\begin{equation}
\left(\frac{\text{d}\sigma}{\text{d}\Omega}\right)_{\mathbf{k}_\text{i}\sigma_{\text{i}}\tau_{\text{i}}\rightarrow\mathbf{k}_\text{f}\sigma_{\text{f}}\tau_{\text{f}}}=
\frac{1}{N\Phi\Updelta\Omega}W_{\mathbf{k}_\text{i}\sigma_{\text{i}}\tau_{\text{i}}\rightarrow\mathbf{k}_\text{f}\sigma_{\text{f}}\tau_{\text{f}}}.
\label{eq:DifferentialCrossSectionForMagneticScattering}
\vspace{-0pt}\end{equation}
Using eq.\,\ref{eq:TransitionPprobability} for  transition probability as well as introducing the density of final states as $\rho_{\mathbf{k}_\text{f}\sigma_{\text{f}}}(E_{\text{f}})=({V_{0}m_{\text{n}}k_\text{f}}/{8\pi^3\hbar^2})\Updelta\Omega$
and incident beam flux
$\Phi={\hbar k_\text{i}}/{V_{0}m_{\text{n}}}$,
we can rewrite eq.\,\ref{eq:DifferentialCrossSectionForMagneticScattering} as
\vspace{-0pt}\begin{equation}
\left(\frac{\text{d}\sigma}{\text{d}\Omega}\right)_{\mathbf{k}_\text{i}\sigma_{\text{i}}\tau_{\text{i}}\rightarrow\mathbf{k}_\text{f}\sigma_{\text{f}}\tau_{\text{f}}}=
\frac{1}{N}\frac{k_\text{f}}{k_\text{i}} {\left(\frac{m_{\text{n}}V_{0}}{2\pi\hbar^2}\right)^2} |\langle\mathbf{k}_\text{f}\sigma_{\text{f}}\tau_{\text{f}}|V|\mathbf{k}_\text{i}\sigma_{\text{i}}\tau_{\text{i}}\rangle|^2.
\label{eq:DifferentialCrossSectionForMagneticScattering2}
\vspace{-0pt}\end{equation}
Taking into account the energy conservation law $E=E_\text{i}-E_\text{f}=E_{\tau_{\text{f}}}-E_{\tau_{\text{i}}}$, summing over all final states, and averaging over all initial states that have probabilities $p_{\tau_{\text{i}}}$ and neutron spin states $p_{\sigma_{\text{i}}}$, we can derive the double-differential cross-section \cite{LucatortoParr13}:
\vspace{-0pt}\begin{equation}
\left(\frac{\text{d}^2\sigma}{\text{d}\Omega\text{d}E_{\text{f}}}\right)_{\mathbf{k}_\text{i}\rightarrow\mathbf{k}_\text{f}}=
\frac{1}{N}\frac{k_\text{f}}{k_\text{i}} {\left(\frac{m_{\text{n}}V_{0}}{2\pi\hbar^2}\right)^2}
\sum_{\sigma_{\text{i}}\tau_{\text{i}}}p_{\tau_{\text{i}}} p_{\sigma_{\text{i}}}
\sum_{\sigma_{\text{f}}\tau_{\text{f}}}|\langle\mathbf{k}_\text{f}\sigma_{\text{f}}\tau_{\text{f}}|V|\mathbf{k}_\text{i}\sigma_{\text{i}}\tau_{\text{i}}\rangle|^2
\delta(E+E_{\tau_{\text{i}}}-E_{\tau_{\text{f}}}).
\label{eq:MasterFormula}
\vspace{-0pt}\end{equation}
Substituting the neutron-electron interaction potential $V$, defined by eq.\,\ref{eq:ElectronNuetronScatteringPotential}, into eq.\,\ref{eq:MasterFormula}, double-differential cross-section becomes:
\vspace{-0pt}\begin{equation}
\frac{\text{d}^2\sigma}{\text{d}\Omega\text{d}E_{\text{f}}}=
\frac{1}{N_{\text{m}}}\frac{k_\text{f}}{k_\text{i}} (\gamma r_{0})^2
\sum_{\alpha}\sum_{\tau_{\text{i}}}p_{\tau_{\text{i}}}
\sum_{\tau_{\text{f}}}
\langle \tau_{\text{i}} |\mathbf{D}_{\bot\alpha}^{\dag}|\tau_{\text{f}}\rangle
\langle \tau_{\text{f}} |\mathbf{D}_{\bot\alpha}|\tau_{\text{i}}\rangle
\delta(E+E_{\tau_{\text{i}}}-E_{\tau_{\text{f}}}),
\label{eq:MasterFormulaWithPotential}
\vspace{-0pt}\end{equation}
where $r_{0}=e^2/4\pi\varepsilon_{0}m_\text{e}c^2=\text{2.8179}\,\mathrm{fm}$ is the classical electron radius and $\mathbf{D}_{\bot}$, the magnetic interaction vector, is the projection of a generalized vector $\mathbf{D}$ onto the plane perpendicular to the scattering vector $\mathbf{Q}$ (see fig.\,\ref{Fig:PositionVectorsDandQ}), defined as:
\vspace{-2pt}\begin{equation}
\mathbf{D}_{\bot}(\mathbf{Q})=\sum_{j}
\left[\hat{\mathbf{Q}}\times(\mathbf{s}_j\times\hat{\mathbf{Q}})+\frac{\text{i}}{\hbar Q}\mathbf{p}\times\hat{\mathbf{Q}}\right]
\text{e}^{\text{i}\mathbf{Q}\cdot\mathbf{r}_j}.
\label{eq:MagneticInteractionOperator}
\vspace{-2pt}\end{equation}
The vector $\mathbf{D}$ represents the Fourier transform of the magnetization density $\boldsymbol{\upmu}$ and usually can be written in the following form:
\vspace{-2pt}\begin{equation}
\mathbf{D}(\mathbf{Q},t)=-\frac{1}{2\mu_{\text{B}}}\boldsymbol{\upmu}(\mathbf{Q},t)=
-\frac{1}{2\mu_{\text{B}}}\int\boldsymbol{\upmu}(\mathbf{r},t)\text{e}^{\text{i}\mathbf{Q}\cdot\mathbf{r}}\text{d}\mathbf{r}.
\label{eq:MagneticInteraction}
\vspace{-2pt}\end{equation}
In a simple case when the atom has no orbital angular momentum, magnetism  arises only from the electron's spin. Therefore, the magnetic moment of the $j^{\text{th}}$ atom at the position $\mathbf{R}_j$ can be rewritten using electron's spin $\mathbf{S}$ as $\boldsymbol{\upmu}_{j}=2\mu_{\text{B}}\mathbf{S}_{j}$. Using an assumption that the magnetization density can be divided into parts that correspond to the atomic site $j$ at position $\mathbf{R}_j$, the magnetization density becomes $\boldsymbol{\upmu}(\mathbf{r})={\sum}{\boldsymbol{\upmu}(\mathbf{R}_j+\mathbf{r}_j)}$. Finally eq.\,\ref{eq:MagneticInteraction} can be written in the following form:
\vspace{-2pt}\begin{equation}
\mathbf{D}(\mathbf{Q},t)=\sum_{j}f_{j}(\mathbf{Q})\mathbf{S}_{j}(t)\text{e}^{\text{i}\mathbf{Q}\cdot\mathbf{R}_j}.
\label{eq:MagneticInteractionWithFF}
\vspace{-2pt}\end{equation}
The component $f_{j}(\mathbf{Q})$ is known as the magnetic form factor and is defined as the Fourier transform of the magnetization density \cite{Collins89}:
\vspace{-2pt}\begin{equation}
f_{j}(\mathbf{Q})=\frac{\int{\boldsymbol{\upmu}(\mathbf{R}_j+\mathbf{r}_j)\text{e}^{\text{i}\mathbf{Q}\cdot\mathbf{r}_j}}\text{d}\mathbf{r}_j}
{\int{\boldsymbol{\upmu}(\mathbf{R}_j+\mathbf{r}_j)}\text{d}\mathbf{r}_j}.
\label{eq:MagneticFF}
\vspace{-2pt}\end{equation}

According to  fig.\,\ref{Fig:PositionVectorsDandQ}, we can write that $\mathbf{D}_{\bot}=\hat{\mathbf{Q}}\times\mathbf{D}\times\hat{\mathbf{Q}}=\mathbf{D}-(\mathbf{D}\cdot\hat{\mathbf{Q}})\hat{\mathbf{Q}}$, and therefore rewrite $\mathbf{D}_{\bot}^{\dag}\cdot\mathbf{D}_{\bot}$ from eq.\,\ref{eq:MasterFormulaWithPotential} as \cite{Squires96}:
\vspace{-0pt}\begin{align}
\mathbf{D}_{\bot}^{\dag}\cdot\mathbf{D}_{\bot}^{} &=\{\mathbf{D}^{\dag}-(\mathbf{D}^{\dag}\cdot\hat{\mathbf{Q}})\hat{\mathbf{Q}}\}\cdot\{\mathbf{D}-(\mathbf{D}\cdot\hat{\mathbf{Q}})\hat{\mathbf{Q}}\} \nonumber\
\\&=\mathbf{D}^{\dag}\cdot\mathbf{D}-(\mathbf{D}^{\dag}\cdot\hat{\mathbf{Q}})(\mathbf{D}\cdot\hat{\mathbf{Q}}) \nonumber\
\\&=\sum_{\alpha,\beta}(\delta_{\alpha,\beta}-\hat{Q}_{\alpha}\hat{Q}_{\beta}){D}^{\dag}_{\alpha}{D}^{}_{\beta},
\label{eq:DirectionalTensor}
\vspace{-0pt}\end{align}
where  $\alpha$ and $\beta$ stand for $x, y, z,$ and $\delta_{\alpha,\beta}$ is the Kronecker delta. Substituting the results obtained in eq.\,\ref{eq:MagneticInteractionOperator}, eq.\,\ref{eq:MagneticInteractionOperator}  and eq.\,\ref{eq:DirectionalTensor} into eq.\,\ref{eq:MasterFormulaWithPotential}, and since electron's  magnetic moments have a negligible effect on interatomic forces, nuclear and electronic coordinates can be summed up separately, so the double-differential scattering cross-section becomes \cite{LucatortoParr13}:
\vspace{-0pt}\begin{align}
\frac{\text{d}^2\sigma}{\text{d}\Omega\text{d}E_{\text{f}}}
&=\frac{1}{N_{\text{m}}}\frac{k_\text{f}}{k_\text{i}} (\gamma r_{0})^2
\sum_{\alpha,\beta}(\delta_{\alpha,\beta}-\hat{Q}_{\alpha}\hat{Q}_{\beta})
\sum_{j,j^\prime}f_{j}^{\ast}(\mathbf{Q})f_{j^\prime}^{}(\mathbf{Q})\times \nonumber\
\\&\times {\frac{1}{2\pi\hbar}}\int\langle\mathbf{S}_{j\alpha}(0)\mathbf{S}_{j^{\prime}\beta}(t)\rangle
\langle\text{e}^{-\text{i}\mathbf{Q}\cdot\mathbf{R}_j(0)}\text{e}^{\text{i}\mathbf{Q}\cdot\mathbf{R}_{j^\prime}(t)}\rangle
\text{e}^\frac{-\text{i}Et}{\hbar}\text{d}t.
\label{eq:MasterFormulaMagneticScattering}
\vspace{-0pt}\end{align}

Now let us consider each component of the differential scattering cross-section in details:
\begin{itemize}
  \item $(\delta_{\alpha,\beta}-\hat{Q}_{\alpha}\hat{Q}_{\beta})$ is the tensor that is responsible for the directional dependence of the neutron scattering. It shows that neutrons can only ``probe'' magnetization component orthogonal to the momentum transfer $\mathbf{Q}$. In particular this means that for the case of collinear magnets (e.g. ferro- and antiferromagnets), elastic scattering gains weighting factor ${1-(\mathbf{Q}\cdot\boldsymbol{\upeta})^2}$, where $\boldsymbol{\upeta}$ is the spin direction, and therefore magnetic scattering vanishes when magnetization and momentum transfer are parallel. But for the inelastic magnetic scattering, the result is slightly different. In case of scattering by the spin waves in local-moment ferromagnets, spin  deviations take place, and the displacement that exists in the orthogonal direction with respect to the quantisation axis  is important. This significantly changes the weighting factor, and now it becomes ${1+(\mathbf{Q}\cdot\boldsymbol{\upeta})^2}$. Therefore, although magnetic moments of the system are ``parallel'' to the momentum transfer $\mathbf{Q}$, the spin-wave cross section is nonzero \cite{Squires96}.

  \item $\langle\mathbf{S}_{j\alpha}(0)\mathbf{S}_{j^{\prime}\beta}(t)\rangle \langle\text{e}^{-\text{i}\mathbf{Q}\cdot\mathbf{R}_j(0)}\text{e}^{\text{i}\mathbf{Q}\cdot\mathbf{R}_{j^\prime}(t)}\rangle$ is the correlation function that accounts for  the atomic positions and electron's spin at different moments of time, and therefore ``probes'' structure and dynamics of the magnetic properties in the crystal.

  \item $f_{j}(\mathbf{Q})$ is the magnetic form factor previously mentioned. According to its definition with eq.\,\ref{eq:MagneticFF}, this function is normalized, so that $f_{j}(0)=1$ and falls off sharply as the scattering vector increases. This result has an important impact on the experiment planning, as it significantly limits possible $\mathbf{Q}$ space to lower values.
\end{itemize}

\subsection{Form-factor analytical approximation}\label{SubSec:FormFactorAnalyticalApproximation}

As it was shown previously, the magnetic form factor $f_{j}(\mathbf{Q})$ is an important part of the expression for the double-differential scattering cross-section. Its $\mathbf{Q}$ dependence can be derived from first principles using the dipole approximation \cite{Lovesey86}, however, because of the extremely complex derivation procedure, the final result is represented in a fairly complex form which is inconvinient for practical use. For the form factor defined by  \cite{LucatortoParr13}:
\vspace{-0pt}\begin{equation}
f(\mathbf{|Q|})=\frac{g_{\text{S}}}{g}\langle\,\text{j}_{\text{0}}(\mathbf{|Q|})\rangle+\frac{g_{\text{L}}}{g}\bigl(\langle\,\text{j}_{\text{0}}(\mathbf{|Q|})\rangle+\langle\,\text{j}_{\text{2}}(\mathbf{|Q|})\rangle\bigr),
\label{eq:MagneticFFdef}
\vspace{-0pt}\end{equation}
where $\langle\,\text{j}_{\text{L}}(\mathbf{|Q|})\rangle=\int\text{j}_{\text{L}}(|\mathbf{Q}|r)|\psi(r)|^2\text{d}\mathbf{r}$ and $\text{j}_{\text{L}}$ are the spherical Bessel functions, integrals $\langle\,\text{j}_{\text{L}}(\mathbf{|Q|})\rangle$ can be approximated by an analytical expression for simplification. This greatly streamlines work with the form factor and in particular allows one to fit it to the neutron-scattering data. The approximation has the form:
\vspace{-0pt}\begin{align}
\langle\text{j}_{0}(s)\rangle & =A\,\text{exp}(-as^2)+B\,\text{exp}(-bs^2)+C, \label{eq:MagneticFFSimpleFormula0}\\
\langle\text{j}_{2}(s)\rangle & =[A\,\text{exp}(-as^2)+B\,\text{exp}(-bs^2)+C]s^2,\label{eq:MagneticFFSimpleFormula2}
\vspace{-0pt}\end{align}
where $s=\text{sin}(\theta/\lambda)$ \cite{WatsonFreeman61,LisherForsyth71}.

\begin{wrapfigure}[12]{r}{0.45\textwidth}\vspace{-12pt}
\includegraphics[width=0.45\textwidth]{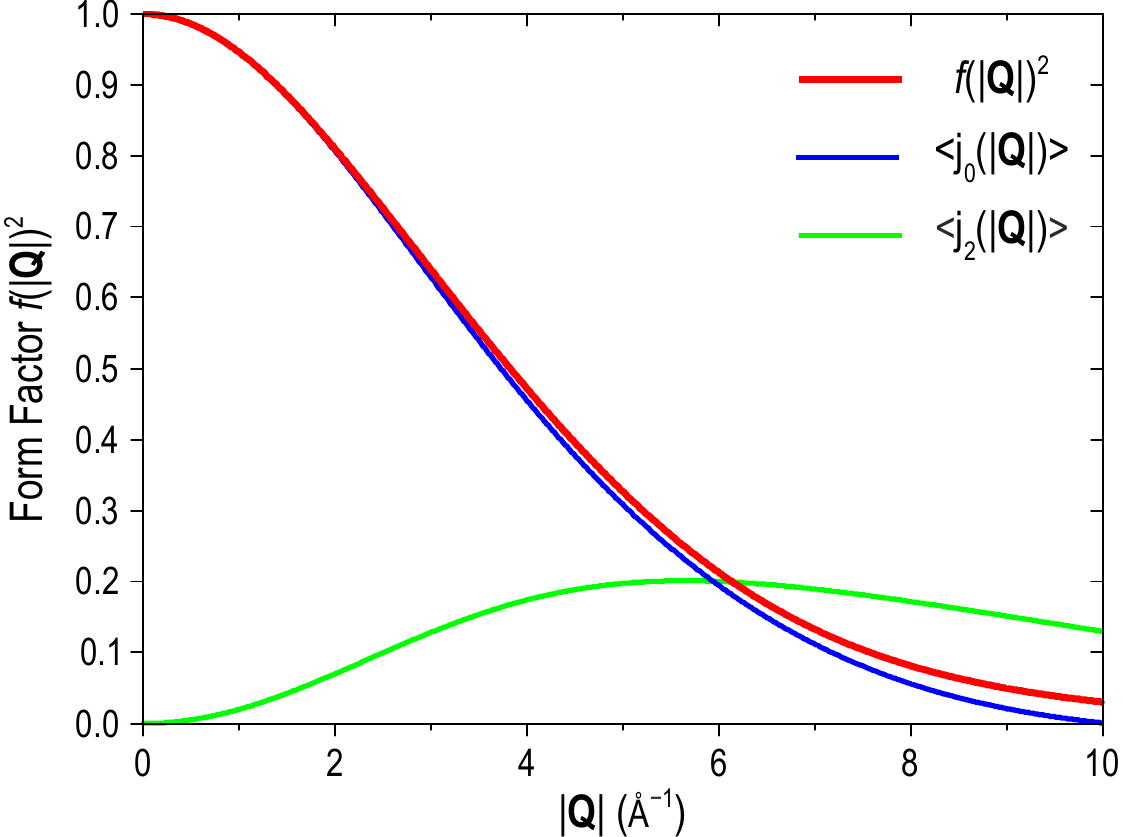}\vspace{-5pt}
\caption{The magnetic form factor $f(\mathbf{|Q|})^2$ for $\text{Ce}^{3+}$ and its components $\langle\,\text{j}_{0}(\mathbf{|Q|})\rangle$ and $\langle\,\text{j}_{2}(\mathbf{|Q|})\rangle$ according to eq.\,\ref{eq:MagneticFFforCe}.}
\label{Fig:FormFactor}\vspace{-0pt}
\end{wrapfigure}

For the purpose of my thesis, I will  compare experimental results with the predicted form factor for $\text{Ce}^{3+}$. The corresponding values of $A,a,B~\text{etc.}$ were taken from ref.\,\citenum{LisherForsyth71}. Because $\text{Ce}^{3+}$ is a rare-earth element, it has strong spin-orbit coupling and the total angular momentum $\mathbf{J}=\mathbf{L}+\mathbf{S}$. It has a single $4f^1$ electron, thus $\mathbf{L}=3$, $\mathbf{S}=1/2$, and $\mathbf{J}=5/2$. A simplified expression for the form factor is \cite{BaruchelHodeau94}
\vspace{-0pt}\begin{equation}
f(\mathbf{|Q|})=\langle\,\text{j}_{0}(\mathbf{|Q|})\rangle+{C}_2\langle\,\text{j}_{2}(\mathbf{|Q|})\rangle
\label{eq:MagneticFFforCe}
\vspace{-0pt}\end{equation}
with
\vspace{-0pt}\begin{equation}
{C}_2=\frac{ {J}({J}+1)+{L}({L}+1)-{S}({S}+1)}{ 3{J}({J}+1)-{L}({L}+1)+{S}({S}+1)}.
\label{eq:C2coeff}
\vspace{-0pt}\end{equation}
is shown in fig.\,\ref{Fig:FormFactor}.

\section{Neutron Scattering Instruments}\label{Sec:NeutronScatteringInstruments}

\begin{wrapfigure}[12]{r}{0.45\textwidth}\vspace{-25pt}
\includegraphics[width=0.45\textwidth]{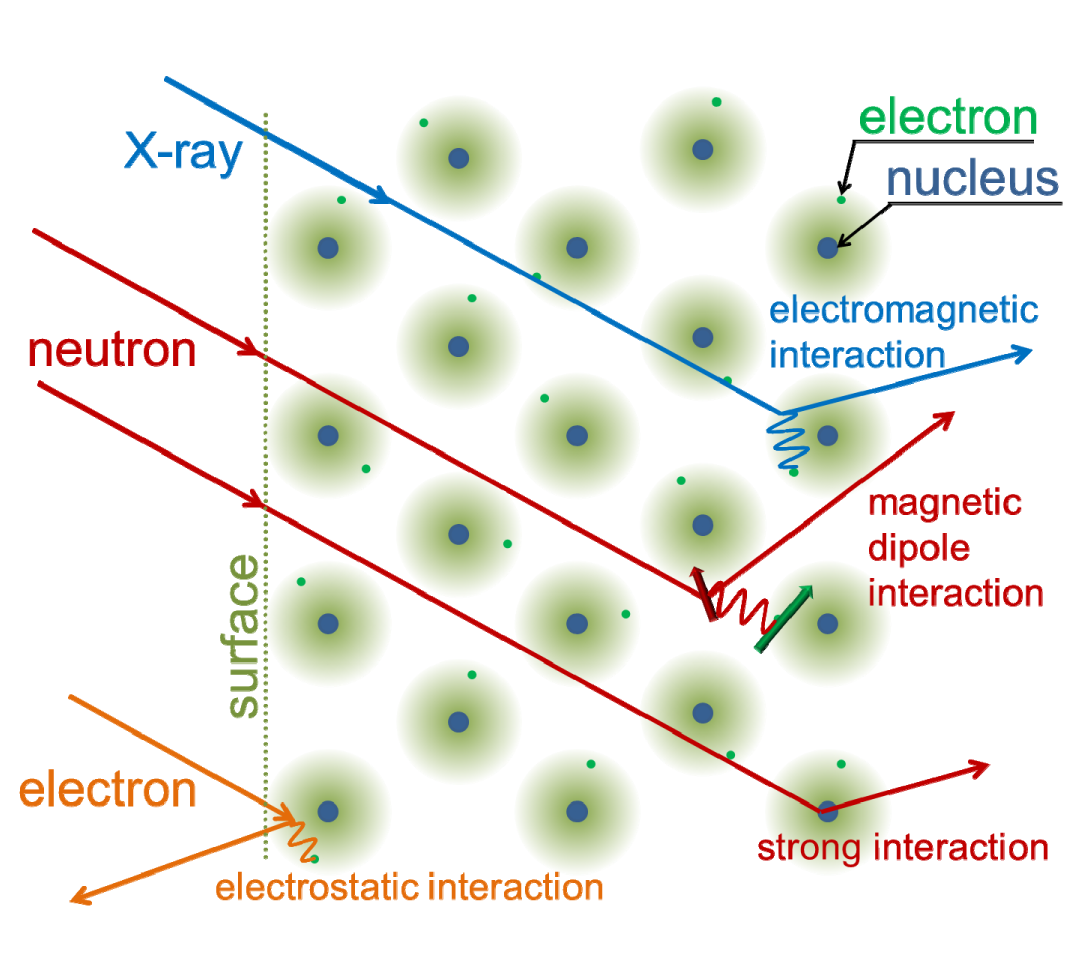}\vspace{-14pt}
\caption{Comparison between possible interactions for various particles, showing important properties in the case of neutrons. Reproduced from ref.\,\citenum{AngstBrueckel12}.}
\label{Fig:NeutronInteractionWithMatter}\vspace{0pt}
\end{wrapfigure}

As it was previously mentioned, the neutron is a unique probe that was granted its wide application due to important properties. Fig.\,\ref{Fig:NeutronInteractionWithMatter} illustrates different types of interaction between neutrons and matter as well as compares them with other particles. Neutrons penetrate deep into the material, because they carry no electric charge, and as mentioned earlier in sec.\,\ref{SubSec:NuclearAndMagneticScattering}, they probe any elements irrespectively of atomic mass, unlike electrons or X-rays that are more sensitive to heavy elements. The magnetic moment of the neutron is a key feature in neutron scattering experiments, giving us the possibility to determine magnetic structures directly. However, despite such strong benefits, neutron experiments cannot be carried out in every scientific lab. The widespread use of neutron scattering techniques is severely limited by several important shortcomings. Mentioned before, the neutron cross-section does not vary so much for various elements, unlike the X-ray cross-section which is much larger in the case of heavy atoms. In particular, the cross-section of light elements like $^{4}_{2}\text{He}$ is 1.34\,barn and 2.65\,barn for neutrons and X-rays, respectively, and does not differ significantly. At the same time for heavier elements like $^{64}_{29}\text{Cu}$ these values already differ by orders of magnitude: 5.2\,barn and 559\,barn \cite{HubbellOverbo79}. Taking into account that the typical photon flux can reach $\sim10^{15}\,\text{ph/s/0.1BW}$, with a spot size on a sample $100\times100\,\upmu \text{m}^2$, it becomes clear that to achieve reasonable counting time for a neutron scattering experiment, a  high-intensity neutron source is required.

A high-flux neutron source can be only obtained at large-scale facilities that use reactors or spallation sources to produce neutrons. And even in this case, a typical neutron flux on the sample is of the order of $\sim10^{7}-10^{8}\,\text{n/cm}^{2}\text{s}$. Another disadvantage, which was previously  referred to as a positive argument, because neutron is highly penetrating, it is extremely difficult to manipulate the neutron beam as well as to create a neutron detector. Therefore, the neutron beam spot has relatively large dimensions (several cm), so in order to use as many neutrons as possible, neutron scattering requires large samples.

Thanks to the hard work of many scientists, the described difficulties have been overcome, and the basic instrument for neutron scattering, known as a \textit{triple-axis spectrometer} (TAS), was invented by B.\,Brockhouse. The name of this instrument arises from the fact that there are three different axes that can be rotated in order to probe specific energy-momentum coordinates. In the next sections I will give an explanation of most important elements of TAS (sec.\,\ref{SubSec:TAS}) and time-of-flight (TOF) (sec.\,\ref{SubSec:TOF}) spectrometers as well as give a short overview of possible spurious signals in app.\,\ref{Appendix:Spurions} that may arise during the experiment, thus significantly complicating interpretation of the results.

\begin{figure}[t]\vspace{-5pt}
\includegraphics[width=\textwidth]{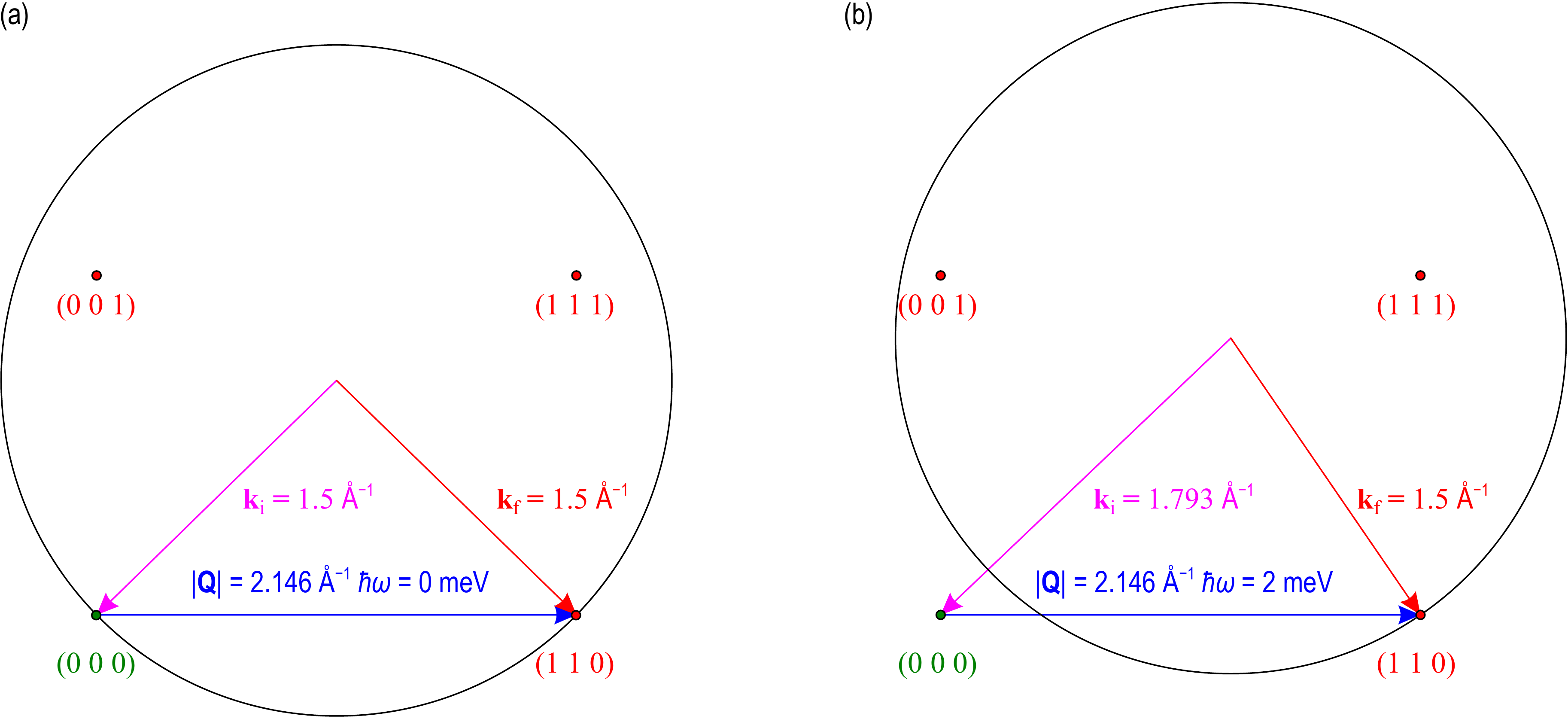}\vspace{0pt}
\caption{ Scattering geometry in case of (a) elastic $\hbar\omega=0\,\text{meV}$ and  (b) inelastic $\hbar\omega=2\,\text{meV}$ neutron scattering for CeB$_6$, showing the relative position of the $\mathbf{k}_\text{f}$ and $\mathbf{k}_\text{i}$ vectors. The black circle represents the cut through the Ewald sphere with the radius $|\mathbf{k}_\text{f}|$ in the scattering plane, spanned by the (001) and (110) basis vectors and therefore named $(HHL)$. The image was created using TAKIN \cite{WeberGeorgii16}.}
\label{Fig:ScatterinGeometryElasticInelastic}\vspace{-5pt}
\end{figure}

\subsection{Triple-axis spectrometer}\label{SubSec:TAS}

Every scattering experiment is governed by the two conservation laws. In case of neutron scattering, energy and momentum conservation laws can be written as
\vspace{-0pt}\begin{equation}
\hbar\omega=E_{\text{i}}-E_{\text{f}},
\label{eq:EnergyConservation}
\vspace{-8pt}\end{equation}
\vspace{-8pt}\begin{equation}
\mathbf{Q}=\mathbf{k}_\text{f}-\mathbf{k}_\text{i},
\label{eq:ScatteringVector}
\vspace{-0pt}\end{equation}
respectively. Usually indices `i' and `f' refer to the incident and final neutron beams, forming the scattering plane, which in the case of a TAS is usually parallel to the floor. The dispersion relation for the neutron can be written as $E=(\hbar^2 k^2)/(2m_{\text{n}})$, and therefore the energy conservation law becomes:
\vspace{-0pt}\begin{equation}
\hbar\omega=\frac{\hbar^2}{2m_{\text{n}}}(|\mathbf{k}_\text{i}|^2-|\mathbf{k}_\text{f}|^2).
\label{eq:EnergyConservationWithKiKf}
\vspace{-0pt}\end{equation}
Depending on the length of incident and final wave vectors, different energy transfer values can be probed, while positioning defines momentum transfer. For example, fig.\,\ref{Fig:ScatterinGeometryElasticInelastic} schematically shows a typical scattering experiment, where the (110) Bragg position is measured at elastic (a), and inelastic (b) value of energy transfer. According to eq.\,\ref{eq:EnergyConservationWithKiKf}, the condition for inelastic scattering is satisfied when $|\mathbf{k}_\text{i}|\neq |\mathbf{k}_\text{f}|$. Also as compared to the case of elastic scattering, the angle between $\mathbf{k}_\text{f}$ and $\mathbf{k}_\text{i}$ has changed. Thus to carry out neutron scattering experiment length of wave vectors and their relative positions have to be changed.

\begin{figure}[t]\vspace{-5pt}
\includegraphics[width=\textwidth]{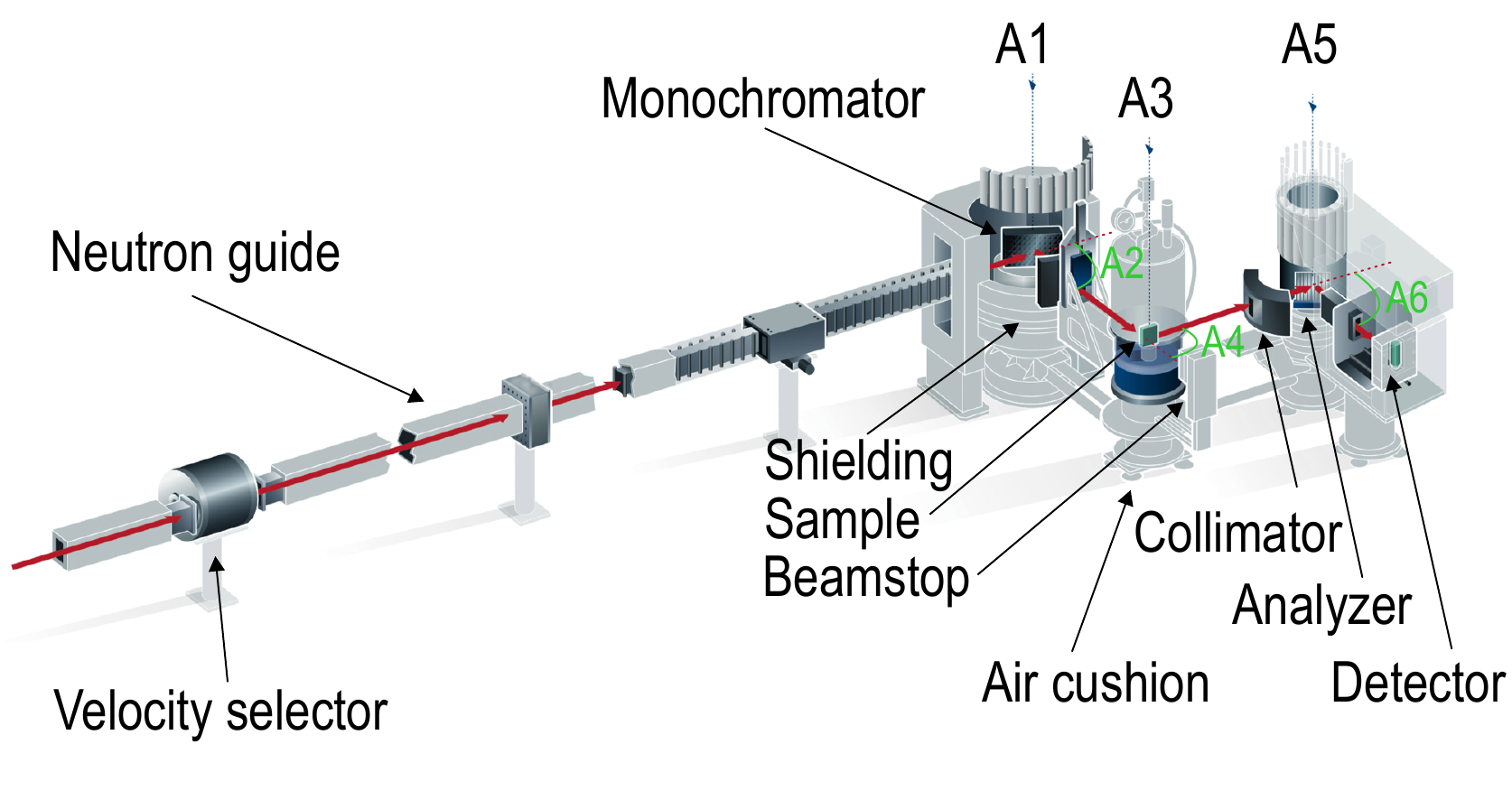}\vspace{0pt}
\caption{ A schematic drawing of the key components of the cold neutron three-axis spectrometer ``FLEXX''. The white beam, which consists of neutrons thermalized in a moderator  with a  Maxwellian wavelength distribution, requires  monochromatization. This is done at the velocity selector and at the monochromator. To achieve high flux on the sample, neutrons are directed through special neutron guides, whose surfaces reflect neutrons. After scattering on the sample, neutrons with the wave vectors $\mathbf{k}_\text{f}$, selected by the analyzer crystal, are measured by the detector. The figure was taken from the HZB web page.}
\label{Fig:InstrumentTAS}\vspace{-5pt}
\end{figure}

\begin{wrapfigure}[17]{l}{0.65\textwidth}\vspace{-15pt}
\includegraphics[width=0.65\textwidth]{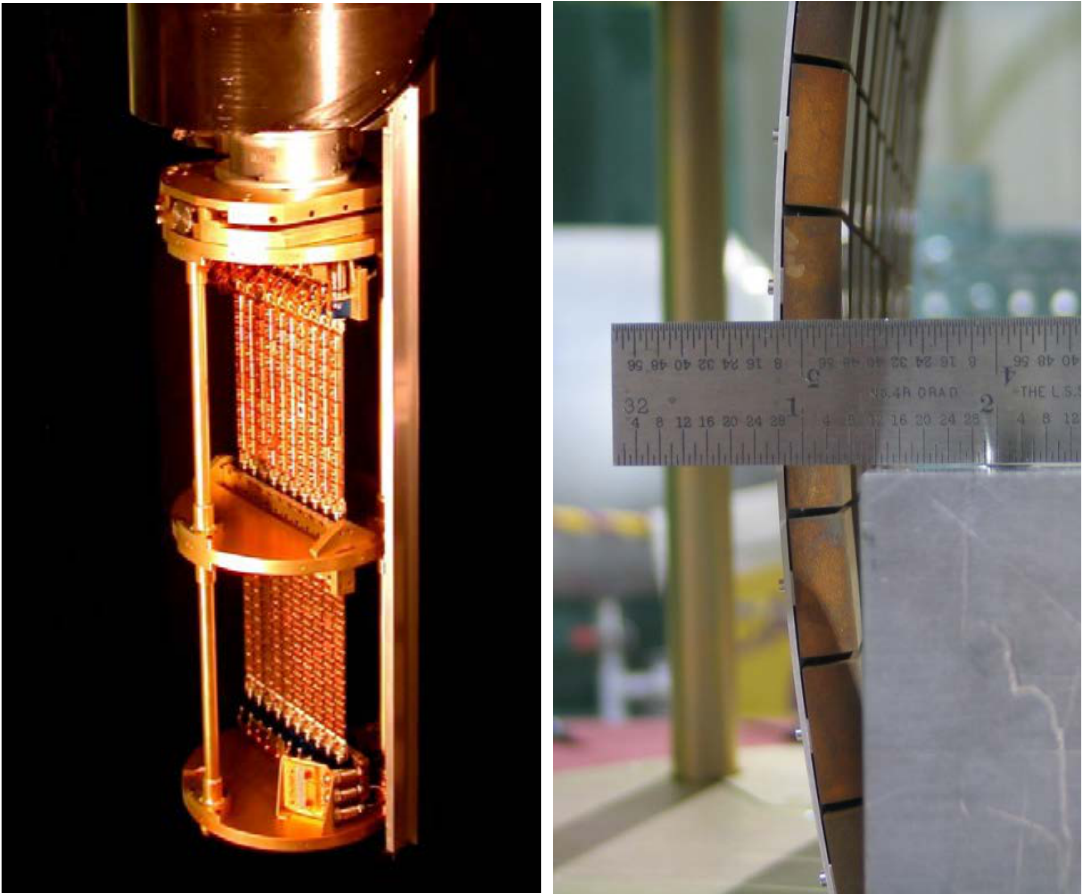}\vspace{-0pt}
\caption{Left: double stack, made of PG(002) and cold-pressed Cu(220) crystals, mounted on the Al plates for a monochromator system used at the BT-7 instrument (NIST). Right:  bent Al plate showing how vertical focusing is achieved. Figure from ref.\,\citenum{LynnChen12}}.
\label{Fig:Monochromator}\vspace{-5pt}
\end{wrapfigure}

A typical TAS spectrometer is schematically shown in fig.\,\ref{Fig:InstrumentTAS}. It is necessary to assign three main axes, associated with the rotation of the monochromator, the sample table, and the analyzer. The monochromator is one of the most important component of the TAS. It consists of a large single crystal or array of properly aligned crystals and diffracts neutrons with a wavelength $\lambda_\text{i}$ that fulfills the corresponding Bragg condition. The white beam, incident on the monochromator, has a natural divergence. Usually it is much larger than the angular width of the Bragg-diffracted beam from a good single crystal. Therefore to increase the number of neutrons diffracted by the monochromator  and increase the neutron flux on the sample, the monochromator is made of a single-crystal mosaic. In order to gain even more intensity, the single-crystal mosaic is made movable, thus allowing  the neutron beam to be focussed. As shown in fig.\,\ref{Fig:Monochromator}, Cu and pyrolytic graphite (PG) monochromator crystals are mounted on an aluminum plate, which can be  rotated to any angle to achieve the desired crystal alignment and horizontal focusing or bent for vertical focusing. The desire to maximize flux and to keep the background level reasonably low imposes strong constrains on to the monochromator material. It should have large Bragg peak reflectivity and as low as possible incoherent cross section. Low absorption and a large Debye temperature in order to avoid  phonon excitations are also desirable.

However, when using a monochromator, it is impossible to avoid a serious drawback. The Bragg condition satisfied with a neutron wavelength $\lambda_\text{i}$ is also fulfilled by the higher-order harmonics. This higher-order contamination might significantly spoil the result of the experiment, and therefore requires suppression. It can be achieved using various filters, or through the initial white beam monochromatization with a velocity selector. This technique is used for example in the FLEXX spectrometer (fig.\,\ref{Fig:InstrumentTAS}). The velocity selector is a cylinder with helical channels covered with a neutron absorber. When rotating at a constant speed, it transmits only neutrons with a specific velocity and therefore cuts off higher-order contamination in a natural way.

In order to  be able to measure the desired point in energy-momentum space, it is necessary to rotate the sample. Therefore, the sample is mounted in a way that allows rotation and covers a wide angular range within the scattering plane. Depending on the experimental setup, rotation can be done by the cryostat insert or by rotating the cryostat together with the sample inside, mounted on a sample table. Each of the options has its own advantages, in particular rotation of the insert is typically used when the magnet is mounted on the sample table. Magnet coils have to be split, in order to let the neutron beam scatter from the sample. This significantly complicates their cooling and results in some limitation of the accessible horizontal angular range. Typically $\sim\!30^\circ$ are assumed to be ``dark'' angles of the vertical-field magnet since they are engaged by technical channels for cooling the lower coil. In order to avoid blocking of the incident or scattered beam by the ``dark'' angles, orientation of the sample table together with the magnet is usually fixed with respect to the incident beam, and sample rotation is done by the insert. In case of the normal cryostat, which is usually fully symmetric, rotation can be done by the sample table. This configuration allows us to use a two-circle goniometer in order to compensate single-crystal misorientation.

After interaction  with the sample, neutrons scatter in all directions with various energies. In order to probe one specific point of energy-momentum coordinates, one has to select neutrons with a certain energy and momentum direction in a similar way as it was done with a monochromator for the white beam. This is done by the analyzer. It selects neutrons with a  wavelength $\lambda_\text{f}$ in the same way as it is done by the monochromator, and directs the  scattered beam to a detector. Because the analyzer and monochromator work on the same principle, they both meet very similar requirements to the material selection. Thus, everything described above for the monochromator
is also true for the analyzer.

Finally after discarding unnecessary scattered neutrons, appropriate neutrons with the momentum $\mathbf{k}_\text{f}$, selected by the analyzer, can be counted with the detector. As already mentioned, neutrons are relatively difficult to detect, therefore the detector is a  fairly complex device and deserves a more detailed explanation given in sec.\,\ref{SubSec:NeutronDetector}. It is obvious that the detector has to be sensitive to neutrons over a wide range of energies. Because we are interested in measuring only the signal originating from the sample, the detector has to be shielded against stray neutrons. Taking into account that the typical neutron flux on the sample is within the order of $\sim10^{7}-10^{8}\,\text{n/cm}^{2}\text{s}$, while the signal during the inelastic scattering experiment is as low as $\sim1-100\,\text{counts/min}$, a heavy shielding is placed around the detector. Background that is measured by the detector when the neutron beam is incident on a sample, but the pass to the analyzer is blocked, is usually $\sim1\,\text{count/min}$ for instruments located in the reactor hall and $\sim0.1\,\text{count/min}$ in the guide hall. However, the described background should not be confused with the actual background measured during the experiment, because the sample, holder, and sample environment produce additional background due to the incoherent or multiple scattering.

Every axis of a triple axis spectrometer is usually characterized by two angles. In case of monochromator, A1 angle defines the rotation of the monochromator with respect to the white beam, while A2 is the corresponding $2\Theta$ angle that fulfills Bragg's condition for the selected wavelength $\lambda_\text{i}$ whenever A1=A2/2. The sample installed on the sample table can rotate freely around the vertical axis determined by the A3 angle. Corresponding $2\Theta$ scattering angle on the sample, spanned by the incident and scattered wave-vectors, is given by the A4 angle. The analyzer is oriented with A5 and A6 angles, which are also responsible for the rotation and $2\Theta$ angle as in the case of a monochromator.

At this point it becomes clear that in order to operate a triple-axis spectrometer, all described parts have to be moved independently. All angles must rotate smoothly and have to be set with a great accuracy of up to $\pm0.01^{\circ}$. They also need to provide a relatively large accessible range. Given that the monochromator and analyzer are heavily shielded, and massive equipment can be mounted on the sample table, precise positioning of the spectrometer turns into a nontrivial problem. These requirements are implemented by setting the spectrometer on  air cushions. After supplying the air pads with  compressed air, positioning of heavy equipment can be literally done with one hand, and is therefore easily achieved  with electric motors that control all of the axes. As soon as the spectrometer reaches the correct position, the air supply is turned off and the spectrometer becomes~immovable.

\subsection{Time-of-flight spectrometer}\label{SubSec:TOF}\vspace{5pt}

\begin{wrapfigure}[14]{r}{0.5\textwidth}\vspace{-15pt}
\includegraphics[width=0.5\textwidth]{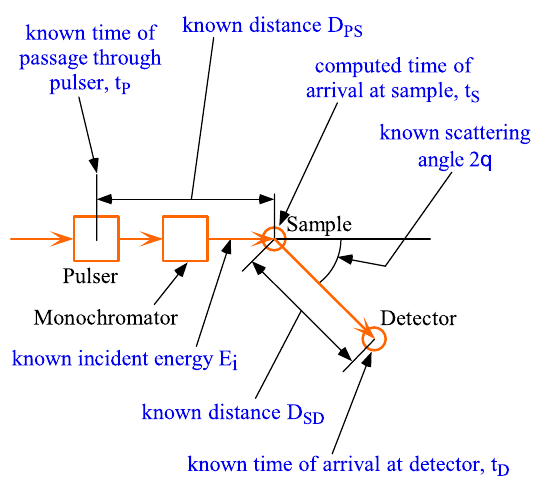}\vspace{-0pt}
\caption{ Illustration of the main components and basic principle of a time-of-flight measurement.  Figure was taken from the NIST web page.}
\label{Fig:IdeaTOF}\vspace{0pt}
\end{wrapfigure}

The TAS setup described above is not always practical, especially when mapping various magnetic excitations. This arises from the fact that TAS is capable of measuring only one specific point in the energy-momentum space. The simplest way to overcome this drawback would be to install additional analyzers and detectors. Based on this idea, TAS can be upgraded with the  multianalyzer option. This allows us to measure several $2\Theta$ angles simultaneously, as it combines a set of analyzers and detectors in a very compact design. In particular \textit{The Multi Axis Crystal Spectrometer} (MACS) available at NIST or the FlatCone at ILL  can be used for reciprocal-space mapping. The described instruments can be improved even further if several analyzers for various $\mathbf{k}_\text{f}$ are used. This is done for example with the new multianalyzer option: MultiFLEXX at HZB. It offers 5 different final energies and 31 different $2\Theta$ channels, therefore significantly facilitating overview measurements. Accordingly there is a tendency, instead of discarding unnecessary scattered neutrons, to detect scattered neutrons with all possible energies and momentum transfers. This simple idea was developed into a TOF spectrometer. \vspace{5pt}

\begin{wrapfigure}[8]{r}{0.5\textwidth}\vspace{-12pt}
\includegraphics[width=0.5\textwidth]{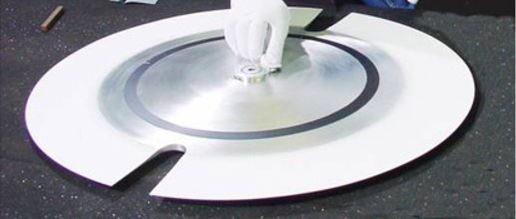}\vspace{-5pt}
\caption{ IN5 chopper, the white part is coated with gadolinium oxide Gd$_2$O$_3$. Figure was taken from the ILL web page.}
\label{Fig:ChopperILL}\vspace{-5pt}
\end{wrapfigure}

The TOF method is widely used for numerous techniques with the main idea to measure the time that it takes for an object to travel a certain distance. For the first time this method was used just 3 years after the discovery of the neutron. In order to prove the energy distribution of the neutrons from a moderated Rn-Be source, four rotating disks with Cd masks were used to select the energy of the neutrons, thus acting as a velocity selector \cite{DunningPegram35}. Further development of this technique was back to back related to the development of the neutron reactors and pulsed sources, but the basic principles remained the same.

\clearpage
\begin{figure}[!t]\vspace{-5pt}
\includegraphics[width=\textwidth]{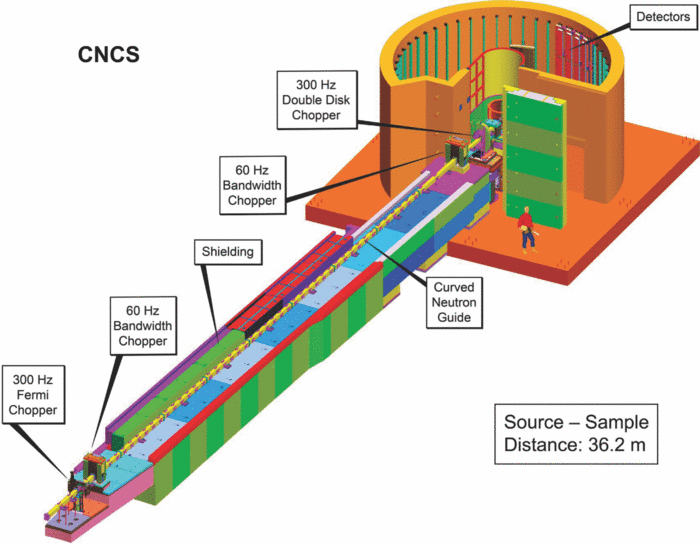}\vspace{-5pt}
\caption{ General layout of the CNCS (SNS) spectrometer that shows main components of the instrument. The roof of the detector enclosure has been removed to show the detectors \cite{EhlersPodlesnyak11}. }
\label{Fig:CNCS}\vspace{-5pt}
\end{figure}

The main idea of the TOF spectrometer  is shown in fig.\,\ref{Fig:IdeaTOF}. In order to be able to determine the time that the scattered neutron requires to cover the distance between the sample and an array of detectors, the neutron beam has to be first of all monochromatized  and divided into pulses. It is usually done by a set of choppers. A neutron chopper is a rotating aluminum disk covered with an absorber that has physical slits for the beam, shown in fig.\,\ref{Fig:ChopperILL}. When the cut is in line with the neutron beam, neutrons can freely pass through. The other part of the disk (the absorbing part) is covered with gadolinium oxide Gd$_2$O$_3$, a highly neutron-absorbing material (see table\,\ref{Tab:SummaryCohIncAbs}). It is important to mention that the chopper is mounted on the special contactless magnetic bearings without lubrication, that allows it to rotate at speeds of up to $\sim20000$\,rpm for a long time without maintenance.

The first set of choppers creates a neutron pulse at known time $t_\text{P}$. The pulse still consists of neutrons with various energies, therefore while traveling on their way to the sample through the guide with a distance $D_\text{PS}$, they are passing additional choppers in order to monochromatize the neutron beam, remove unnecessary higher order contamination as well as frame overlap. Finally a monochromatic neutron pulse, with well defined energy $E_{\text{i}}$, reaches the sample position at the known time $t_\text{P}$. Neutrons scatter from the sample and fly toward the detector with various speeds, and therefore they are detected at different times $t_\text{D}$. Dividing the distance between the detector and the sample $D_\text{SD}$  by the time of flight $t=t_\text{D}-t_\text{S}$ of each neutron, the scattered energy $E_{\text{f}}$ is determined. As a detector, position-sensitive tubes filled with $^{3}\text{He}$ gas are  usually used. Depending on the instrument, the detector covers different  scattering angles in plane as well as some region perpendicular to it (for details see \cite{OllivierMutk11,EhlersPodlesnyak11}). Each scattering event can be defined by the known scattering angle $2\Theta$,  the out-of-plane component $\varphi$, and time $t$, which are easily converted to $\mathbf{Q}$ and~$\hbar\omega$.

\vspace{7pt}
Fig.\,\ref{Fig:CNCS} shows a schematic diagram of the CNCS spectrometer \cite{EhlersPodlesnyak11} at the  Spallation Neutron Source, Oak Ridge National Laboratory in the USA, as an example of the latest generation of direct geometry spectrometers. In particular this instrument was used for the measurements described in Chapter\,\ref{Chap:ResultsCeLaB6}. Figs.\,\ref{Fig:CeB6Map2p5Tand5T}b and \ref{Fig:He_sub} were prepared from the experiments that were done with this instrument. The spectrometer can provide low incident neutron energies (1--50\,meV), and its detector bank is made of  400 two-meter-long tubes, covering  scattering angles between $-50^\circ$ and $+135^\circ$ in the horizontal plane and $\pm16^\circ$ perpendicular to the scattering plane, which corresponds to a  solid angle of 1.7\,sr in total. The instrument resolution for the fixed value of the incident energy $E_{\text{i}}$ can be adjusted by the last pair of 300\,Hz choppers with two counter-rotating disks. These choppers have three slits with various opening widths and can be synchronized in different ways, providing better resolution or higher neutron flux. After the sample is mounted in the center of the detector bank, it is rotated around the vertical axis (analog of A3 rotation in case of TAS) in order to cover the desired range of reciprocal space. For every sample orientation, a 3D manifold in the four dimensional (4D) energy-momentum space is measured. Then, eventually, the sequence of collected data is combined and a 4D data-set is constructed. In order to operate and visualize data, special software such as \emph{Horace} \cite{EwingsButs16} has to be used.

\vspace{7pt}
It may seem that TOF is a much better and more versatile spectrometer than TAS, as large detectors simultaneously collect all the scattered neutrons. The TAS technique being equipped in the classical case with only one detector is therefore preferably suited to study specific points of the energy-momentum space, which is especially suitable for parametric studies as a function of magnetic field or temperature. However TOF pays a fairly high price for its capacity: the incident beam has to be pulsed, therefore maximum achievable intensity on the sample is significantly reduced in comparison with TAS. So both of these methods perfectly complement each other.

\clearpage
\subsection{Neutron sources}\label{SubSec:NeutronSources}

As it was mentioned before in sec.\,\ref{Sec:Neutronproperties}, because of the relatively short lifetime, free neutrons do not occur in nature. Therefore in order to do a neutron scattering experiment it is necessary to have a neutron source with high intensity. The general idea of any neutron source can be formulated as follows: to separate neutrons from the nuclei where they are  particularly loosely bound. In some elements the binding energy is relatively small, and neutrons can be released after an interaction of the nuclei with $\alpha$-,particles,  protons, or $\gamma$-rays. In particular, small neutron sources, used for the first experiments with neutrons, were based on the ($\alpha$, $n$) reaction. Irradiation of beryllium with $\alpha$-particles leads to the formation of neutrons as a result of the following reaction:
\vspace{-8pt}\begin{equation}
^{9}\text{Be}+ ^{4}\!\text{He}\,\longrightarrow \,^{12}\text{C}+n+5.704\,\text{MeV}.
\label{eq:BeReactionAlfa}
\vspace{-8pt}\end{equation}
Similar reactions are valid for $^{11}\text{B}$ or $^{7}\text{Li}$, and as a source of $\alpha$-particles radium, plutonium or americium are used. This reaction  is used as a weak neutron source with an energy spectrum reminiscent of that from the fission source. By parity of reasoning to the photoelectric effect, the nuclear photoeffect by ($\gamma$, $n$) reaction is used to produce nearly monoenergetic neutrons. Usually as a source of $\gamma$-rays, antimony-124 is used, as it is unstable and undergoes $\beta^{-}$ decay to an exited state of $^{124}\text{Te}$: $\subsup{\text{Sb}}{51}{124}\,\longrightarrow \,\subsup{\text{Te}}{52}{124}^\star+\!e^{-}\!+\!\overline{\nu}_{e}$. This excited state decays by emitting $\gamma$-rays with energies 1.692\,MeV and 2.091\,MeV, which are further absorbed:
\vspace{-8pt}\begin{equation}
^{9}\text{Be}+\gamma\,\longrightarrow \,^{8}\text{Be}+n-1.666\,\text{MeV}.
\label{eq:BeReactionGamma}
\vspace{-8pt}\end{equation}
As a result, nearly monochromatic neutrons with $E_{\text{n}}=24\pm2.2\,\text{keV}$  are produced. However, the described above methods for obtaining neutrons are extremely inefficient and also suffer from a serious drawback, as the initial radioactive sources have a short half-life \cite{BeckurtsWirtz64}.

\begin{wrapfigure}[14]{r}{0.5\textwidth}\vspace{-12pt}
\includegraphics[width=0.5\textwidth]{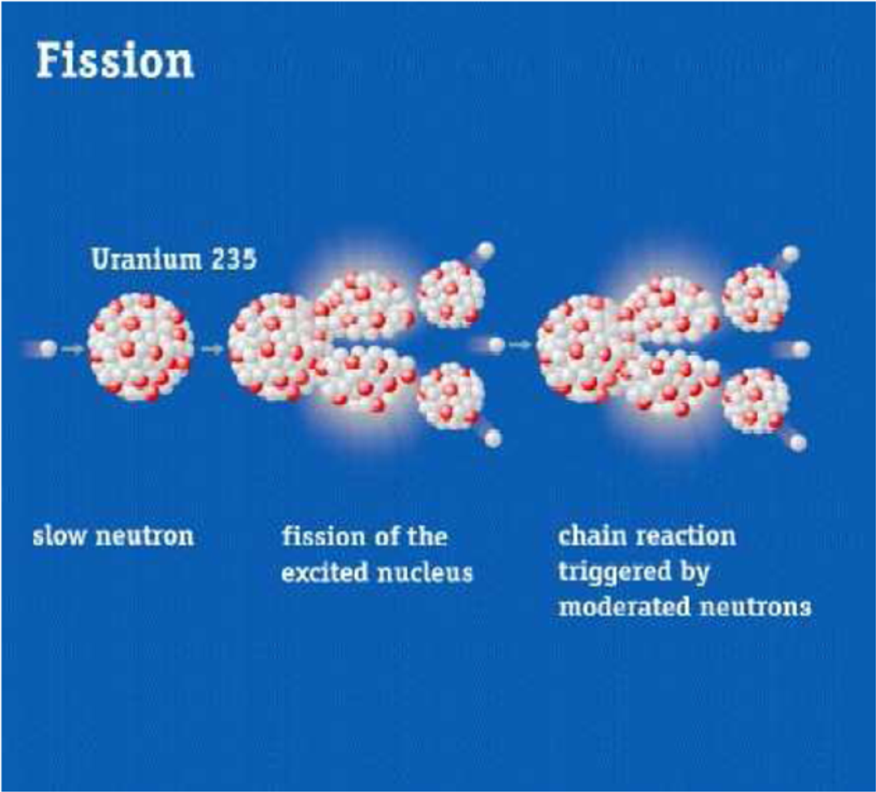}\vspace{-5pt}
\caption{ Nuclear fission of the $^{235}\text{U}$ nucleus. Figure was taken from \href{http://www.formulation.org.uk/196-meetings/clever-characterisation-for-smarter-formulation/524-ccsf-neutrons.html}{\textit{http://www.formulation.org.uk}}.}
\label{Fig:U235Fission}\vspace{-5pt}
\end{wrapfigure}

Another approach is to use heavy atomic nuclei. Some of the transuranium elements decay by spontaneous fission, and in particular $^{233}\text{U}$, $^{235}\text{U}$, or $^{239}\text{Pu}$ are widely used \cite{ReillyEnsslin91}. During the reaction these unstable nuclei capture a slow neutron and then split into daughter nuclei and several neutrons, in particular:
\vspace{-0pt}\begin{equation}
\subsup{\text{U}}{92}{235}+\subsup{\text{n}}{0}{1}\longrightarrow \,\subsup{\text{Ba}}{56}{141}+\subsup{\text{Kr}}{36}{92}+3\subsup{\text{n}}{0}{1}.
\label{eq:U235Reaction}
\vspace{-0pt}\end{equation}
These additional neutrons move too fast to create a  controlled self-sustaining chain reaction, and therefore require moderation. After slowing down, the neutron becomes capable of interaction with other unstable nuclei. This creates a self-sustaining chain reaction that is controlled. Because during fission, on average more than one neutron is released, and only one neutron is required to sustain the reaction further, excess neutrons are able to leave the reactor core and can represent a neutron source. The described process is shown in fig.\,\ref{Fig:U235Fission}.

\begin{wrapfigure}[26]{r}{0.5\textwidth}\vspace{-15pt}
\includegraphics[width=0.5\textwidth]{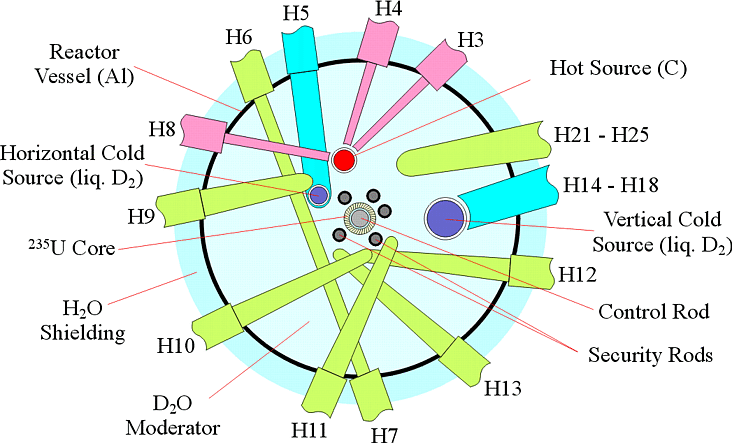}\vspace{-5pt}
\caption{Core and reflector tank layout of the high flux reactor at the ILL, Grenoble, France. Figure was taken from \href{http://pd.chem.ucl.ac.uk/pdnn/inst3/reactors.htm}{\textit{http://pd.chem.ucl.ac.uk}}.}
\label{Fig:ILLCore}\vspace{-5pt}
\includegraphics[width=0.5\textwidth]{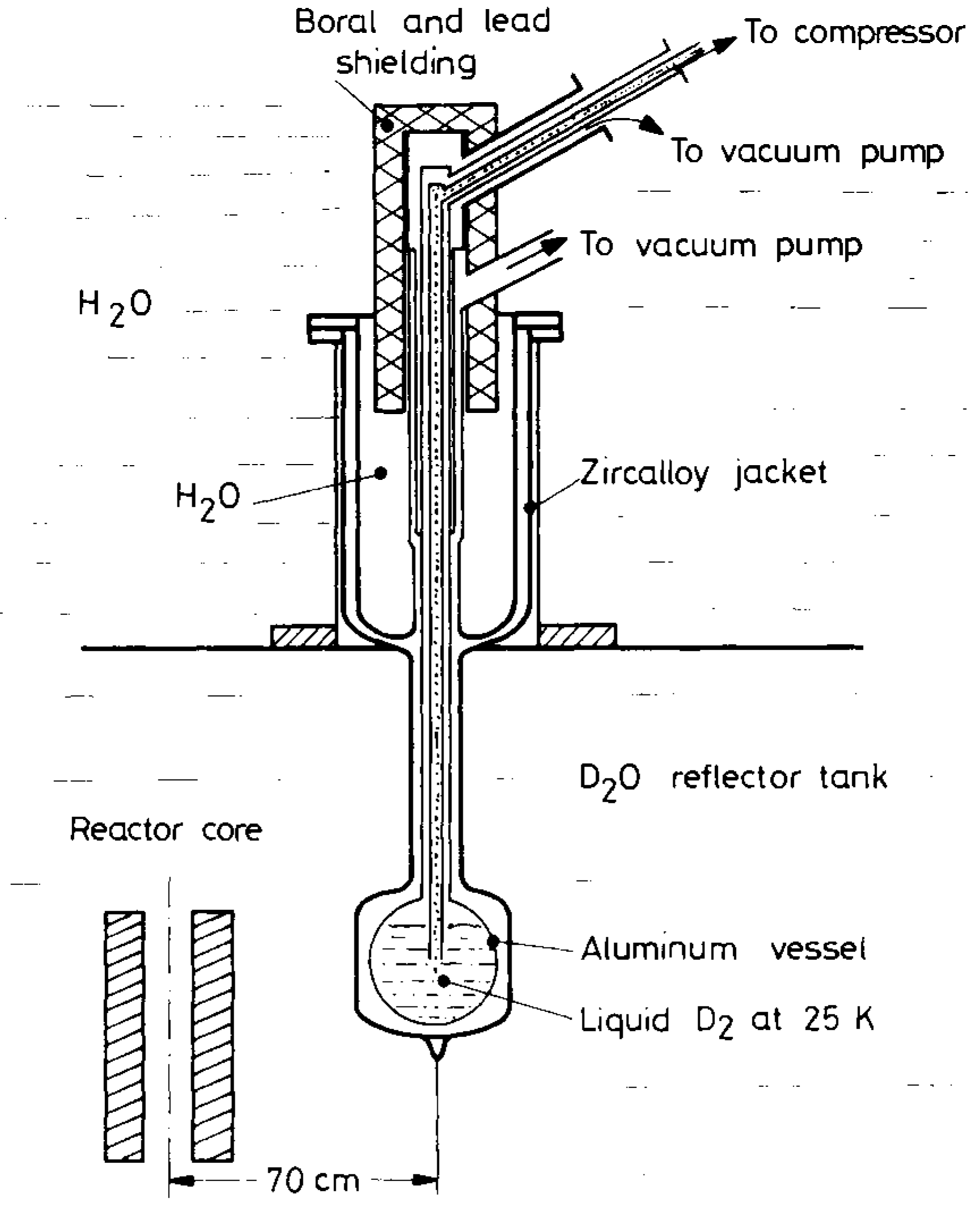}\vspace{-5pt}
\caption{Schematic view of the D$_2$ cold source installed at ILL \cite{Kostorz79}.}
\label{Fig:ColdSource}\vspace{-5pt}
\end{wrapfigure}

The layout of the high flux reactor of the  Institut Laue-Langevin is shown in fig.\,\ref{Fig:ILLCore}. The active zone of the reactor contains 8.6\,kg of highly enriched (97\%) $\subsup{\text{U}}{92}{235}$ and is surrounded by a 2.5\,m tank filled with heavy water that cools down the reactor by  heat exchange with light water, which in turn is cooled by the local river, and also acts as a moderator. As the temperature of the cooling heavy water is about $35^\circ\text{C}$, thermally moderated neutrons have a Maxwellian wavelength distribution with a maximum that corresponds to the temperature of the coolant. These neutrons leave the reactor core through the beam tubes, labeled H1 to H25. In order to avoid enhanced fast-neutron background in the tube, they are mounted in positions, directed away from the fuel element at the center of the vessel.

\begin{wrapfigure}[29]{r}{0.5\textwidth}\vspace{-15pt}
\includegraphics[width=0.5\textwidth]{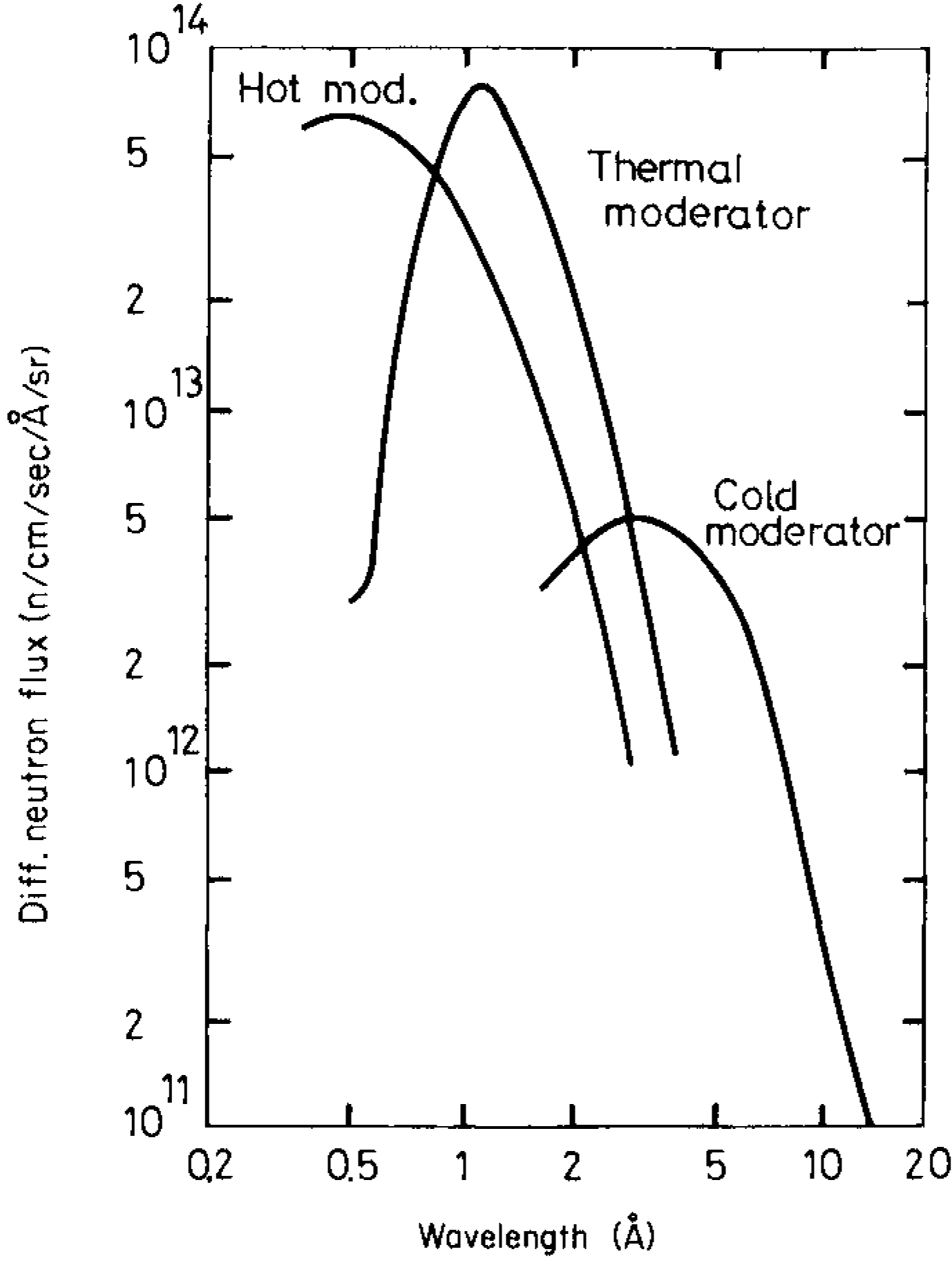}\vspace{-5pt}
\caption{Differential neutron flux as a function of wavelength for different moderators \cite{Kostorz79}.}
\label{Fig:HotColdThermalFlux}\vspace{-0pt}
\includegraphics[width=0.5\textwidth]{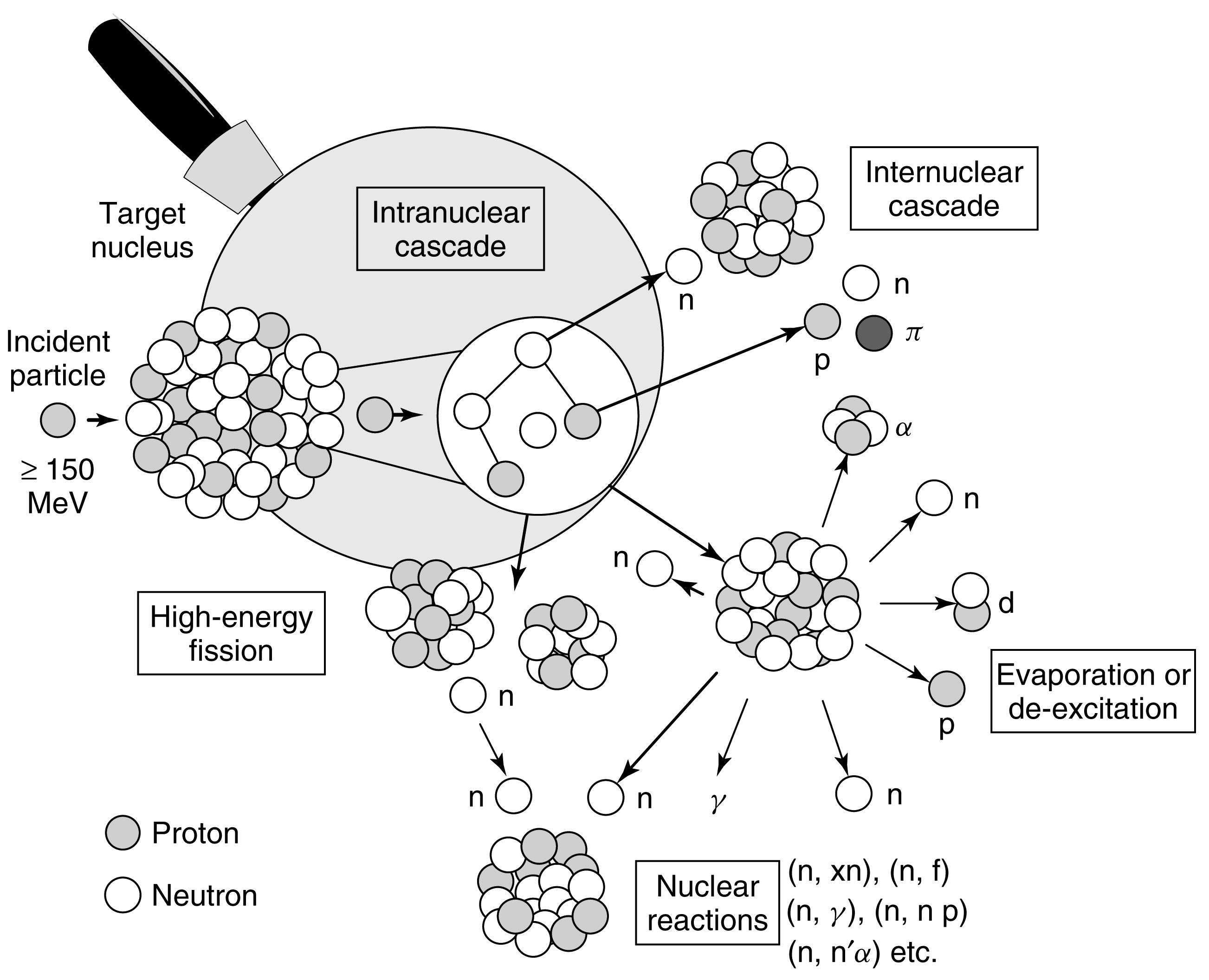}\vspace{-5pt}
\caption{ The principal scheme of spallation. Figure was taken from ref.\,\citenum{FilgesGoldenbaum10}.}
\label{Fig:Spallation}\vspace{-0pt}
\end{wrapfigure}

Tubes shown with a green color in fig.\,\ref{Fig:ILLCore} are so-called thermal beam tubes. They provide instruments with neutrons that have wavelengths of approximately 1--2\,\AA. However, some experiments require neutrons with either shorter or longer wavelengths. This is done by additional moderators that have a different temperature. As a source of cold neutrons, a vessel containing 25\,l of liquid D$_2$ is used (shown in fig.\,\ref{Fig:ColdSource}). A low temperature (25\,K) is achieved by a 10\,kW helium refrigerator. A hot source utilizes a graphite cylinder 20\,cm in diameter. Unlike the cold source, which requires the removal of heat, graphite requires heating in order to moderate neutrons with higher energies. But thanks to tremendous number of collisions of high-energy neutrons within the graphite, its  temperature reaches 2000\,K without additional heating. Thus, it is possible to distinguish three sources of neutrons with various wavelength, originating from different moderators.  The differential flux for each source is shown in fig.\,\ref{Fig:HotColdThermalFlux}.

The maximal thermal flux of the ILL reactor is $1.2\times10^{15}\,\text{n/cm}^2\text{s}$, measured at about 15\,cm from the core surface. This value is already at the limit  in thermal neutron flux for the steady-state reactor neutron sources. Pulsed reactors, like IBR-30 or IBR-2 at Dubna (Russia) can provide higher peak flux, that becomes especially effective for the TOF experiments, which naturally require pulses. However, the realization of such  source is extremely complex: the nuclear chain reaction is controlled by the mechanical modulation of the reactor's reactivity, done by the pair of movable reflectors. At a certain moment, when both reflectors are aligned, a power pulse is generated.

Another concept to produce pulsed neutron sources is based on the use of particle accelerators. When an electron beam hits a target, intense $\gamma$-rays are produced as \textit{bremsstrahlung}. They can make neutrons by the ($\gamma$,\,$n$) reaction described earlier \cite{SkoeldPrice86,Kostorz79}. In particular, the ORELA electron accelerator-based neutron source located in the USA produces neutrons by \textit{bremsstrahlung} from a tantalum radiator with a production rate of about $10^{14}\,\text{n}/\text{s}$. Further increase in intensity has reached its limit because of the heat transfer at the target.

A different result occurs when high-energy proton, deuteron, neutron or other subatomic particles interact with a target nucleus. When the particle has high enough energy, that the de Broglie wavelength $\lambda\approx10^{-15}\,\text{m}$ becomes smaller than the typical diameter of the heavy nucleus ($\sim10^{-14}\,\text{m}$), it can interact with nucleons in the nucleus as if they were free. This is the case for proton energies $E_\text{p}\geq150\,\text{MeV}$, and the corresponding process is schematically shown in fig.\,\ref{Fig:Spallation}. Interaction between the proton and the nucleus can be divided into two stages:

\vspace{-6pt}\begin{itemize}
  \item after the initial collision between the incident particle and the target, an intranuclear cascade is created. The particle interacts with various nucleons and as a result produces high-energy secondary particles. Depending on the energy and impact parameter of the incident particle, various particles leave the target nucleus as secondary particles and the rest of the energy is distributed among nucleons, thus leaving nucleus in an excited  state. This process is extremely fast, about $\sim10^{-22}\,\text{s}$. \vspace{-5pt}
  \item at the second stage, excited nucleus relaxes by emitting low-energy particles, in particular neutrons, protons and $\alpha$-particles, and it takes  $\sim10^{-16}\,\text{s}$  for the equilibrium stage to come up. This process is called de-excitation or \textit{evaporation}.
\end{itemize}\vspace{-6pt}

\vspace{-10pt}\begin{figure}[!hb]
\minipage{0.42\textwidth}
    \vspace{0pt}\includegraphics[width=\linewidth]{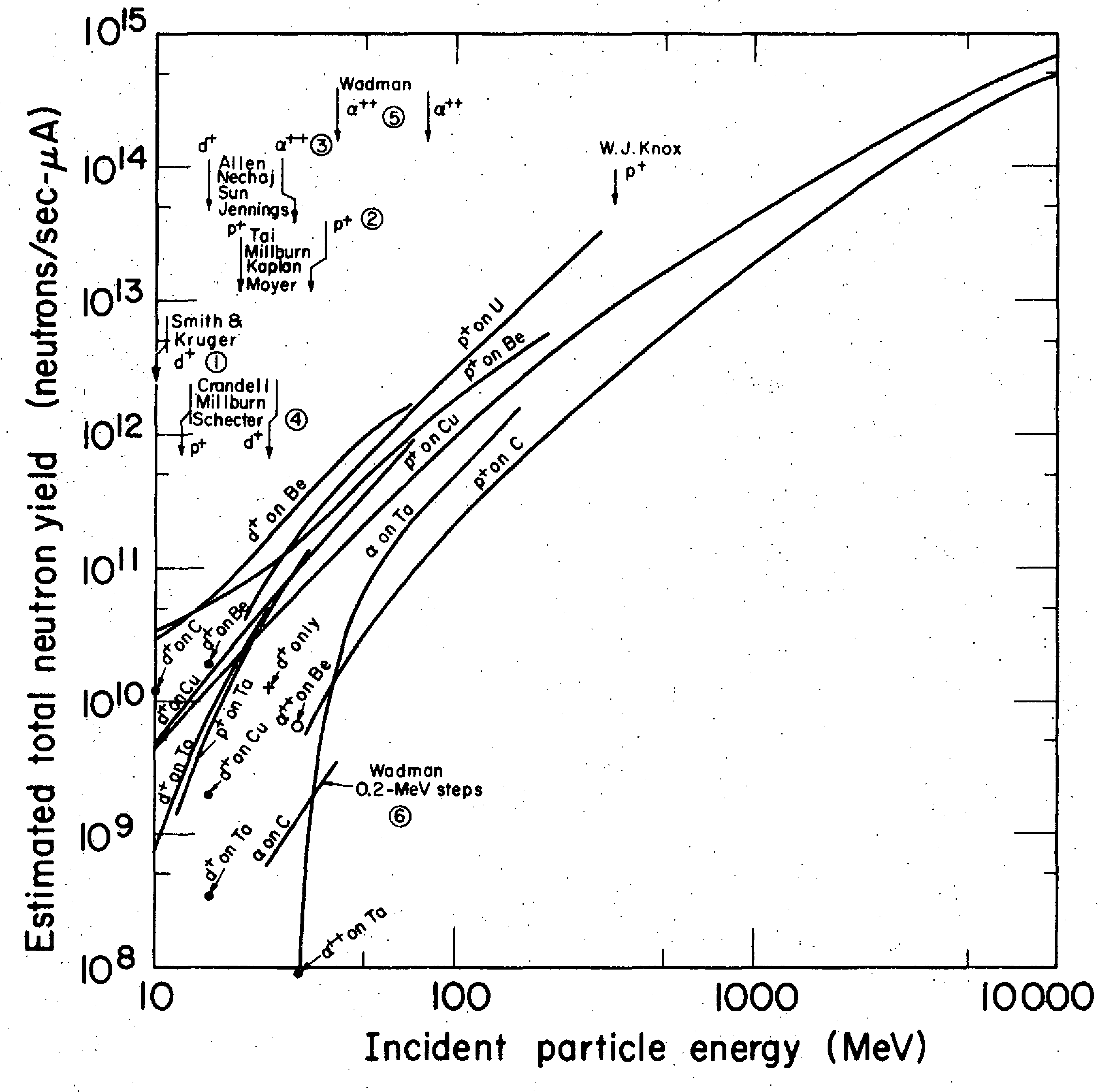}\vspace{0pt}
    \vspace{-30pt}\caption{Plot of neutron yields vs incident particle energy for several combinations of targets and ions. Figure was taken from ref.\,\citenum{StephensMiller69}.}\label{Fig:NeutronsVsEnergyParticles}\vspace{-0pt}
\endminipage\hfill
\hspace{0.03\textwidth}\minipage{0.55\textwidth}
    \vspace{-16pt}\includegraphics[width=\linewidth]{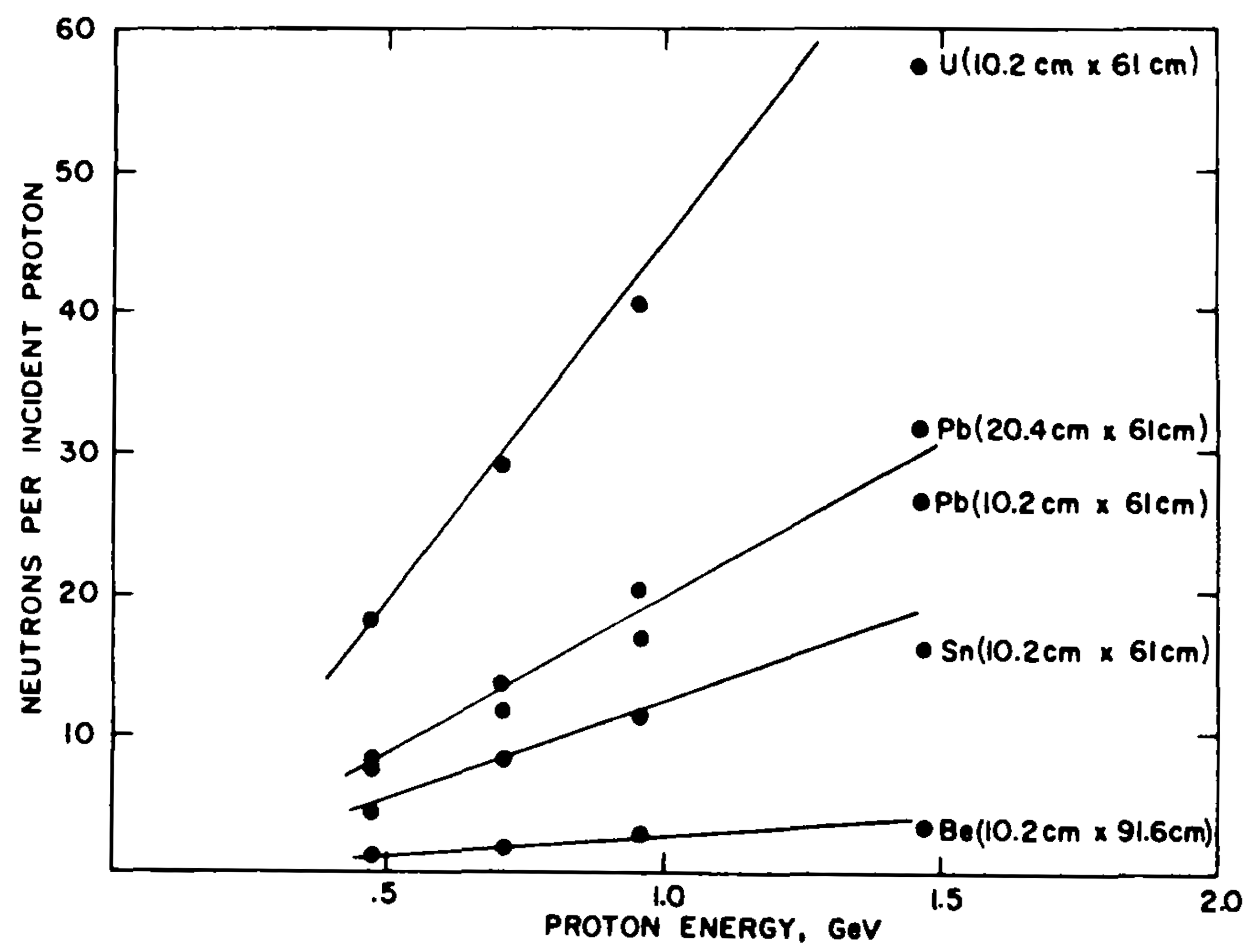}\vspace{0pt}
    \vspace{-30pt}\caption{Measured neutron yield versus proton energy for various targets. Figure was taken from ref.\,\citenum{SkoeldPrice86}.}    \label{Fig:NeutronsVsTargetMaterial}\vspace{-0pt}
\endminipage\hfill
\end{figure}\vspace{-20pt}

\enlargethispage{15pt}
This results in a broad energy-angular dependence of the  emitted particles. The high-energy cascade is strongly focused in the direction of the incident beam while the low-energy part, that originates from \textit{evaporation}, is isotropic \cite{FilgesGoldenbaum10,David15}. This process is known as a spallation reaction, and was for the first time observed to occur naturally in Earth's atmosphere in  experiments that were done in 1912, and now is known to be the origin of cosmic rays. Later in 1947 at the Berkeley cyclotron, a high-energy deuteron and $\alpha$-particle beam was used to demonstrate the emission of neutrons and charged particles \cite{BrobeckLawrence47}. Immediately after that, the spallation process was described by two stages \cite{Serber47}.

Spallation should not be interpreted as  fission, as they differ in several ways. First of all, the remaining nuclei are usually just about 15 atomic mass units lighter than the original target nucleus, while fission usually divides nuclei into two fragments. But the main difference overall is the  number of released neutrons. The overall amount of neutrons related with each fission event is about 2 to 3, which is considerably less than in the case of spallation. Fig.\,\ref{Fig:NeutronsVsEnergyParticles} shows that the neutron yield is strongly dependent on the incident particle, its energy, and certainly on the target material. In particular, it was found that for the case of a $^{238}\text{U}$ target, neutron yield for the the same kinetic energy per incident particle is about 25\% greater with deuterons than with protons \cite{BauerConrad82}. Dependence on the target material is shown in fig.\,\ref{Fig:NeutronsVsTargetMaterial}, in particular it shows a clear advantage of $^{238}\text{U}$ as a target material. However, for real neutron sources, other materials like Pb, W, Hg are usually  chosen as a target. Not least this is due to the fact that the target material must have good thermal conductivity in order to provide adequate cooling. Therefore, despite the fact that $^{238}\text{U}$ target has the highest neutron yield, its low thermal conductivity is a huge disadvantage~\cite{Carpenter77}.

Finally, we discuss the background of a time-of-flight instrument. For the time-of-flight experiments, which are naturally best suited for pulsed neutron sources, the technique  of time-independent background (TIB) subtraction is successfully used. The idea is that in the intervals between the pulses it is possible to distinguish the time frame when no ``useful'' neutrons are counted. These neutrons are responsible for a certain part of the background and therefore can be subtracted. In fissionable materials, delayed neutrons can arise, in particular the amount of such neutrons produced by the spallation source with a $^{238}\text{U}$ target  can reach up to 1\% \cite{Carpenter77}, and is relatively large if we recall the typical values for incident beam flux and amount of counts in a typical inelastic experiment (see section\,\ref{SubSec:TAS}).

\begin{figure}[t]
\minipage{0.45\textwidth}
    \vspace{0pt}\includegraphics[width=\linewidth]{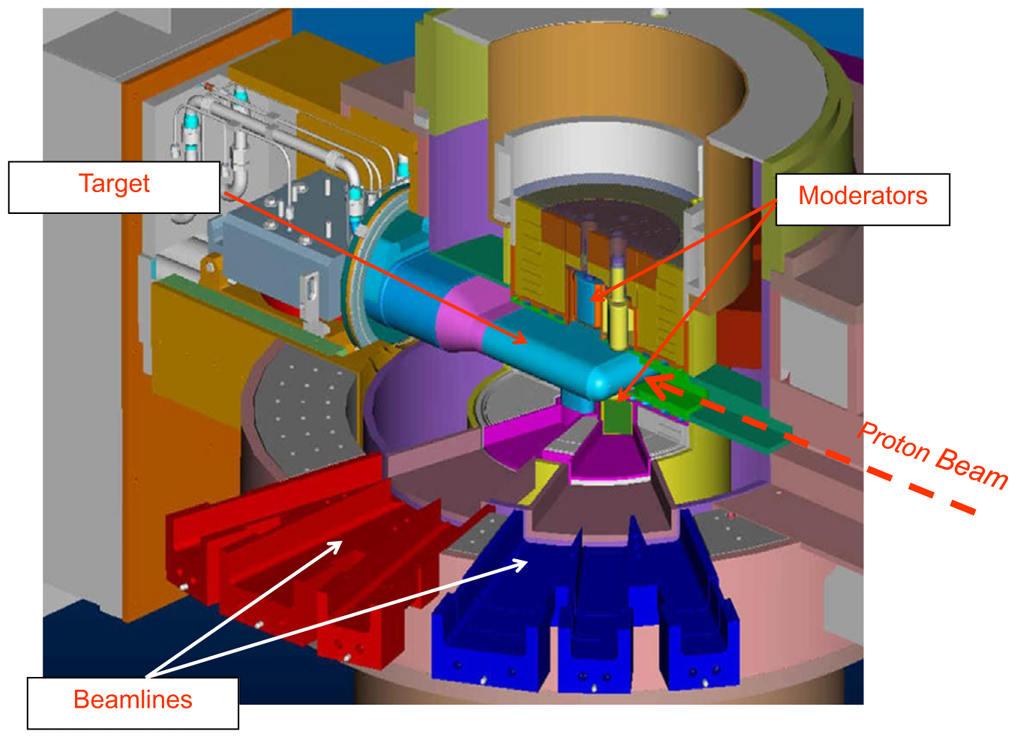}\vspace{-5pt}
    \caption{A cutaway view of the mercury target in the SNS, surrounded by moderators inside of a beryllium reflector. Figure was taken from ref.\,\citenum{FominGreene15}.}\label{Fig:SNS}\vspace{-0pt}
\endminipage\hfill
\hspace{0.03\textwidth}\minipage{0.52\textwidth}
    \vspace{2pt}\includegraphics[width=\linewidth]{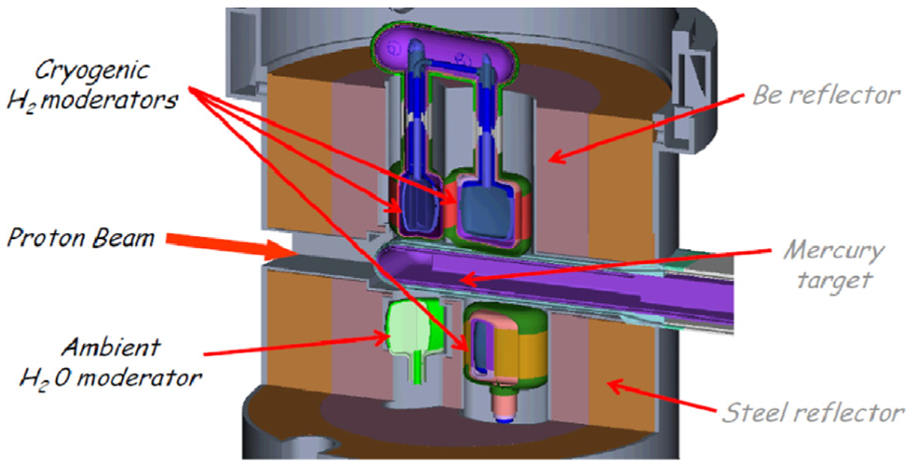}\vspace{0pt}
    \vspace{-0pt}\caption{The mercury target vessel surrounded by three cold and one thermal source moderators inside of a beryllium reflector. Figure was taken from ref.\,\citenum{FominGreene15}.}\label{Fig:SNSTarget}\vspace{-0pt}
\endminipage\hfill\vspace{-15pt}
\end{figure}

\enlargethispage{10pt}
Two  vivid examples of spallation sources are SNS (\textit{Spallation Neutron Source}) located in USA and ISIS near Oxford in the United Kingdom. I want to briefly describe some of the important technical solutions that were applied during development of the SNS. Fig.\,\ref{Fig:SNS} shows a cutaway view through the ``heart'' of the spallation source: the target. As it was previously mentioned, the target material has to meet various requirements, and what is very important, must have high thermal conductivity. In case of SNS, liquid mercury is used as a target. This material is very suitable as a target as it has high atomic number. As a liquid metal, mercury does not suffer from radiation damage, and therefore can be used for the whole lifetime of a facility. The mercury target is surrounded by two concentric stainless steel containers, in order to prevent a possible mercury leak from the inner vessel. The outer container is cooled by water, while the inner vessel is cooled down in a natural way by the 20\,tons of mercury that flows through it with a flow rate of $\sim24\,\,\text{l/s}$.

An incident proton beam has an energy up to 1.0\,GeV and pulses on the target at a repetition rate of 60\,Hz, with sub-microsecond length of each pulse. As a result, emitted neutrons have a broad distribution of energies and therefore require moderation, in a similar way as it was described before for  the high-flux reactor at the ILL. In order to decrease the number of neutrons that leave the target without being moderated, thus being lost as a neutron leakage, a water-cooled beryllium reflector is used. It acts as a neutron mirror and scatters back neutrons, that would otherwise escape.  Three moderators that contain hydrogen at $\sim20\,\text{K}$ and act as a cold neutron source around the target are shown in fig.\,\ref{Fig:SNSTarget}. As a source of thermal neutrons, a liquid water moderator at $\sim320\,\text{K}$ is used \cite{FominGreene15}.

\vspace{0pt}\vspace{-0pt}

\subsection{Neutron detector}\label{SubSec:NeutronDetector}

As it was mentioned in sec.\,\ref{Sec:Neutronproperties}, due to the absent electric charge, neutrons cannot directly ionize atoms, and therefore they are much more difficult to count. Some of the absorbtion reactions, mentioned in sec.\,\ref{SubSec:NeutronAbsorption}, can lead to the production of charged particles, which  are then recorded with an appropriate detector. Here I will focus on the gas-filled $^{10}\text{B}$ and $^{3}\text{He}$  detectors and describe basic principles of their work.

\begin{figure}[t]\vspace{-0pt}
\includegraphics[width=\textwidth]{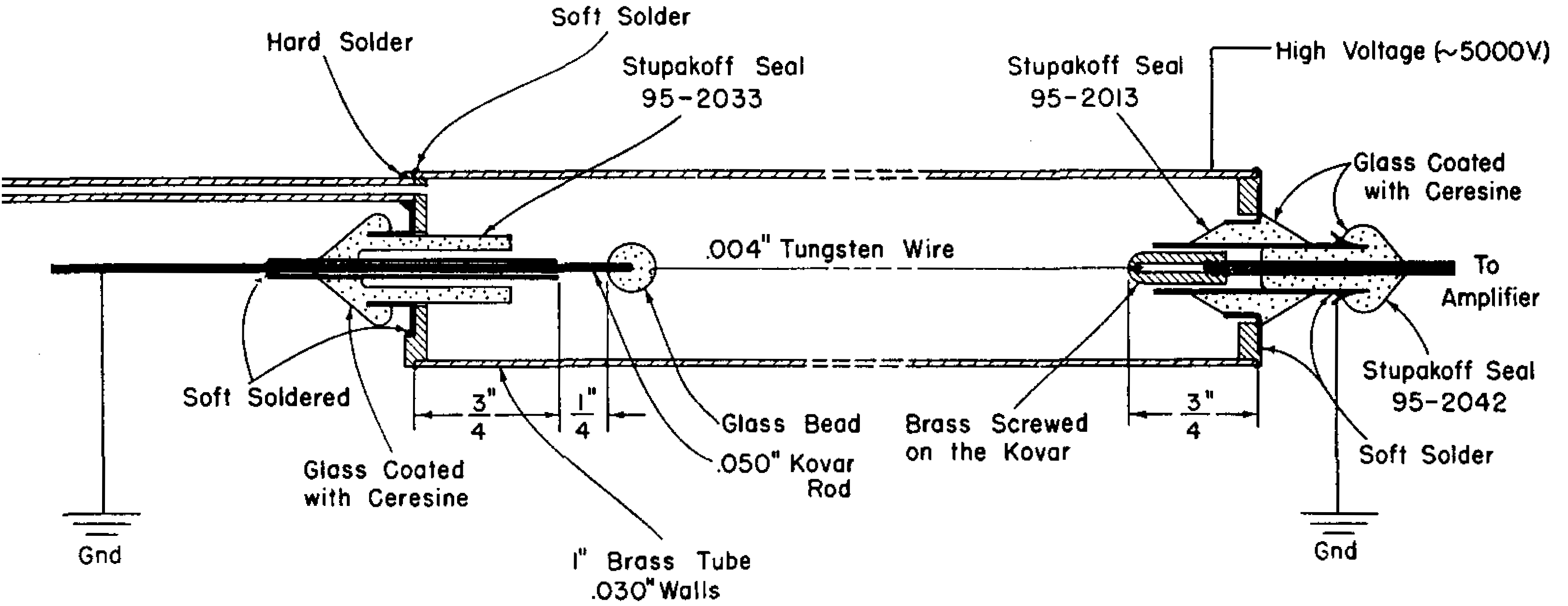}\vspace{-5pt}
\caption{ Details of a neutron counter construction. Figure was taken from  ref.\,\citenum{TongiorgiHayakawa51}.}
\label{Fig:GasFilledDetector}\vspace{-18pt}
\end{figure}

The boron counter is one of the simplest detectors for counting neutrons, as it has a relatively large reaction cross-section. A detector is usually filled with boron triflouride  BF$_3$ or boron trimethyl B(CH$_3$)$_3$ gas, contained inside an aluminium cylinder that acts as the cathode, and a thin wire going through the tube as the anode. In fig.\,\ref{Fig:GasFilledDetector} a typical gas-filled detector schematic is shown. When a neutron is absorbed by the boron, a lithium and an $\alpha$-particle are created. There are two possible outcomes from this interaction, as the lithium nucleus can be left in an excited (eq.\,\ref{eq:BeReactionGamma1}) or ground state (eq.\,\ref{eq:BeReactionGamma2}):
\vspace{-8pt}\begin{equation}
\subsup{\text{B}}{5}{10}\,+^{1}_{0}\!\text{n}\,\longrightarrow \,^{4}_{2}\text{He}\!+\!^{7}_{3}\text{Li}^\star\!+\!2.31\,\text{MeV},
\label{eq:BeReactionGamma1}
\vspace{-14pt}\end{equation}
\vspace{-14pt}\begin{equation}
\subsup{\text{B}}{5}{10}\,+^{1}_{0}\!\text{n}\,\longrightarrow \,^{4}_{2}\text{He}\!+\!^{7}_{3}\text{Li}\!+\!2.79\,\text{MeV},
\label{eq:BeReactionGamma2}
\vspace{-5pt}\end{equation}
and the excited state goes to the ground state through the relaxation process  by emitting a gamma ray with an energy $0.48\,\text{MeV}$. In case of the lithium ground state, its kinetic energy becomes 1.01\,MeV, while the $\alpha$-particle gets 1.78\,MeV.  The reaction where the lithium nucleus is left in the ground state occurs about 6\% of the time, while another reaction occurs 94\% of the time, and the kinetic energies of the particles become 0.84\,MeV and 1.47\,MeV \cite{SchieberKhakhan10}. Emitted charged particles share relatively large kinetic energy, which is easy to detect, regardless of the energy of the incident neutron \cite{Tsoulfanidis95}.

\begin{wrapfigure}[14]{r}{0.5\textwidth}\vspace{-12pt}
\includegraphics[width=0.5\textwidth]{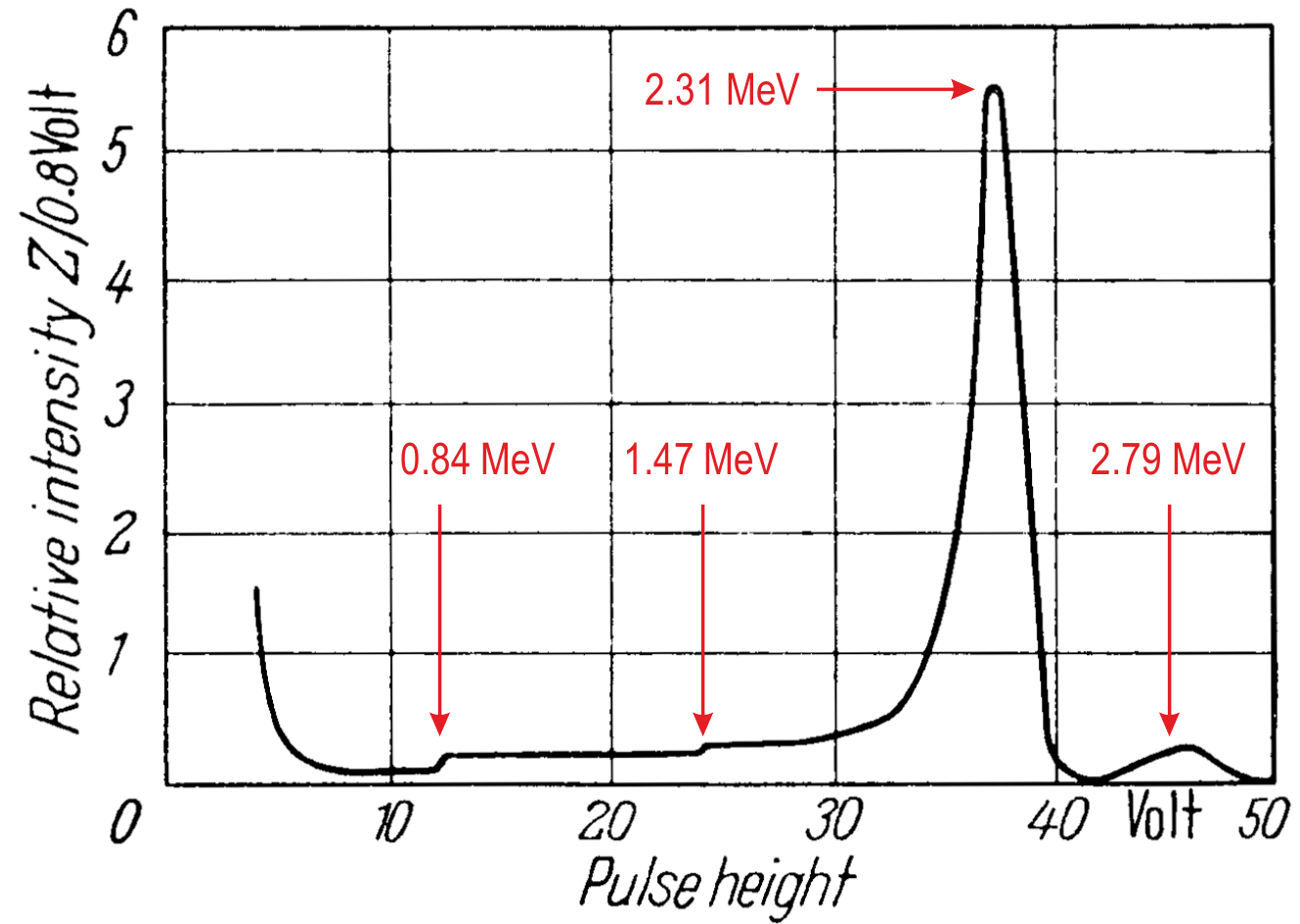}\vspace{-5pt}
\caption{ Neutrons spectrum detected by the BF$_3$ tube detector, that shows two peaks that corresponds to the reaction with the lithium nucleus left in the excited state (2.31\,MeV), and in the ground state (2.79\,MeV). Two steps on the left side are due to the ``wall effect'', discussed in the text. Figure source obtained from  ref.\,\citenum{BeckurtsWirtz64}.}
\label{Fig:CounterPulseB}\vspace{-0pt}
\end{wrapfigure}
A typical pulse height spectrum is shown in fig.\,\ref{Fig:CounterPulseB}, with two maxima that correspond to the possible outcomes of the reaction with the lithium nucleus left either in the ground or excited state. It is also necessary to pay attention to the fact that the lower steps on the left side of the spectrum arise from an interesting effect. Because either the Li nucleus or $\alpha$-particle, in the situation when they are created close to the detector wall, may deposit its energy not in the sensitive volume of the gas but rather in the wall. Thus, for example, upon absorption of the $\alpha$-particle on the detector wall, a pulse that corresponds to 0.84\,MeV is observed, while when the Li nucleus is absorbed, a 1.47\,MeV step is seen. Occurrence  of the spectrum increase for very small pulses is explained by $\gamma$-rays. A detailed description of the boron counter can be found in refs.\, \citenum{Tsoulfanidis95, BeckurtsWirtz64, ReillyEnsslin91, FowlerTunnicliffe50,SchieberKhakhan10,TongiorgiHayakawa51,FergusonJablonski57}.

\begin{wrapfigure}[11]{r}{0.65\textwidth}\vspace{-5pt}
\includegraphics[width=0.65\textwidth]{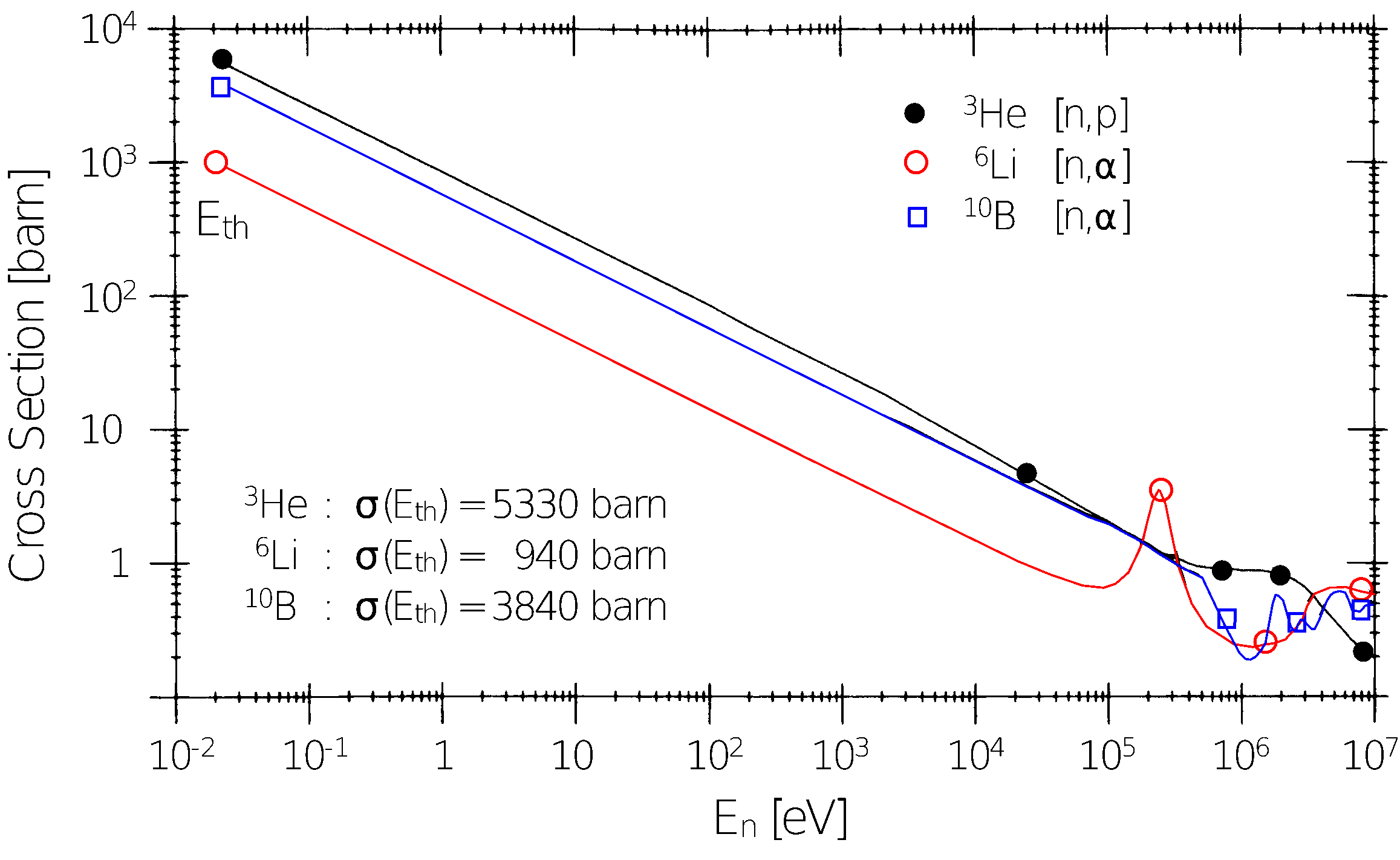}\vspace{-11pt}
\caption{ Cross-section as a function of incident neutron energy for some reactions. Figure was taken from ref.\,\citenum{JohannaStachel15,Knoll10}.}
\label{Fig:He3vsBvsLi}\vspace{-0pt}
\end{wrapfigure}
Another type of detector that is widely used is the $^{3}\text{He}$  detector. The cross-section of  $^{3}\text{He}$ is larger then the cross-section for  $^{10}\text{B}$, and as it was explained earlier in sec.\,\ref{SubSec:NeutronAbsorption}, also follows $1/\surd{E_{\text{n}}}$ dependence on the incident neutron energy $E_{\text{n}}$, shown in fig.\,\ref{Fig:He3vsBvsLi}. Neutron detection is based on the reaction $^{3}_{2}\text{He}\,+^{1}_{0}\!\text{n}\,\longrightarrow \,^{1}_{1}\text{H}\!+\!^{3}_{1}\text{H}\!+\!765\,\text{keV},$ and works on the same principle as the boron detector. Tritium produced by the reaction is unstable, with a half-life of 12.33 years, and undergoes $\beta^{-}$ decay into $^{3}\text{He}$:
$^{3}_{1}\text{H}\,\longrightarrow \,^{3}_{2}\text{He}\!+\!e^{-}\!+\!\overline{\nu}_{e}+18.6\,\text{keV}$, thus $^{3}\text{He}$ gas is self-regenerated.

\begin{wraptable}{!hr}{8cm}\vspace{-37pt}
\begin{tabular}{c@{~~~~}c@{~~~~}c@{~~~}}\\
\toprule
Detector & Thermal neutron & 1\,MeV $\gamma$-ray\\
\toprule
$^{3}\text{He}$      & 0.77        & 0.0001     \\
$^{10}$BF$_3$      & 0.29        & 0.0006     \\
\bottomrule
\end{tabular}
\vspace{-8pt}\caption{Interaction probability given for the 2.5\,cm diameter tube filled with 4\,atm $^{3}\text{He}$. Corresponding values for $^{10}$BF$_3$ detector are 5 cm and 0.66\,atm. Values are taken from ref.\,\citenum{ReillyEnsslin91}. }
\label{Tab:InteractionProbabilityHeB}\vspace{-32pt}
\end{wraptable}

A typical pulse-height spectrum from the detector filled with the $^{3}\text{He}$ gas represents a strong signal that originates from the $765\,\text{keV}$ kinetic energy released in the reaction, as well as a low-energy tail that reflects the ``wall effect'' \cite{BatchelorAves55}. In order to increase the probability of the neutron detection, the ``wall effect'' has to be reduced, as smaller size pulses sometimes do not reach the nominal threshold setting for the neutron detector. This can be achieved by using tubes of bigger diameter, so that  the number of reactions that occur near the wall is smaller, or by  increasing in stopping power of the gas, usually done by higher gas pressure or by introduction of a small fraction of krypton \cite{Tsoulfanidis95}. In table\,\ref{Tab:InteractionProbabilityHeB},  interaction probabilities for both type of the detectors for thermal neutrons and $\gamma$-rays are compared. Obviously, despite the fact that the B-filled detector is cheaper than the $^{3}\text{He}$, it is toxic and less efficient in neutron detection while simultaneously more sensitive to  $\gamma$-rays. These are significant drawbacks that lead to wider prevalence of the $^{3}\text{He}$ detectors.
\enlargethispage{15pt}
Further development of the detector tube was related with the position-sensitive technique, which is used in TOF spectrometers (see sec.\,\ref{SubSec:TOF}). The position of the incident neutron is encoded through  measurement of the charges, collected at each end of the detector tube, as the total resistance of the resistive wire may be represented as two parts through which the charge is measured separately. This makes it possible to determine the position where the neutron reaches the tube accurately, as well as the arrival time \cite{FrostEccleston98}.
\chapter{Heavy Fermions}\label{Chap:HeavyFermions}\vspace{-24pt}

\begin{wrapfigure}[14]{r}{0.4\textwidth}\vspace{-12pt}
\includegraphics[width=0.4\textwidth]{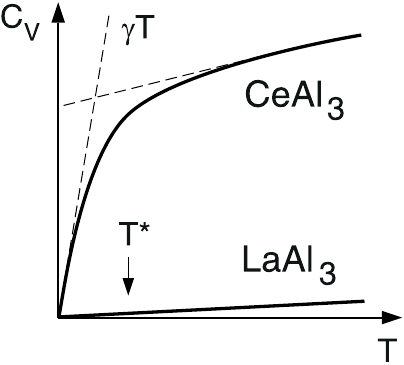}\vspace{-5pt}
\caption{Comparison of the specific heat for the heavy fermion system CeAl$_3$ and a normal metal LaAl$_3$, that shows enhanced value of the linear specific-heat term. The characteristic temperature (eq.\,\ref{eq:FLevelEnergy}) indicates hybridization \cite{Roessler09}.}
\label{Fig:SpecificHeatCeLaAl3}\vspace{-0pt}
\end{wrapfigure}

Since the late 70s, heavy-fermion compounds have attracted more and more attention. This term was  first  introduced in 1975 by Andres, Graebner and Ott to describe the unusually large value of the linear specific-heat term for the low-temperature specific heat $C=\gamma T$ in CeAl$_3$ \cite{AndresGraebner75}. The reported value of  $\gamma=1620$\,mJ/(mol$\cdot$K$^2$) below 150\,mK was thousand times larger than for conventional metals, as expected from the free-electron theory. This discrepancy can be overcome by introducing heavy quasiparticles with an  effective mass several hundred times higher than the free electron mass in the Drude--Sommerfeld model. The discovery of superconductivity below $T_\text c = 0.5$\,K in CeCu$_2$Si$_2$ \cite{SteglichAarts79}, where heavy fermion quasiparticles presumably formed Cooper-pair states,  was keeping the issue on the boil.  Despite the fact that heavy fermion compounds have been intensively studied for more than four decades, we are still far from a complete understanding of the various ordering phenomena that arise from the localized-itinerant dichotomy of $f$\!-electrons.

Heavy fermion systems are intermetallic compounds which usually consist of 4$f$\!- or 5$f$\!-electron ions, like rare earth Ce and Yb, or actinides like U, Np, Pu. Partially filled  orbitals on each atom, considered as localized magnetic moments, are hybridized with the conduction electrons forming heavy fermion bands below the Kondo temperature. These bands are strongly localized in space, so that direct interaction between them is negligible, but they can interact via the conduction electrons. Thereby properties of these systems are mainly governed by the competition between the Kondo screening and the Ruderman-Kittel-Kasuya-Yosida (RKKY) coupling mechanism via the conduction electrons.

\section{Anderson Model and Formation of the ``Heavy'' Bands}\label{Sec:AndersonModel}

The description of the interaction between the localized $f$\!-electrons with a conduction band is usually done with the Anderson lattice Hamiltonian. In this case, the Hamiltonian is represented by four terms \cite{Anderson61}:
\vspace{-8pt}\begin{equation}
H=H_{0c}+H_{cf}+H_{0f}+H_{\text{corr}}.
\label{eq:AndersonHamiltonianFourTerms}
\vspace{-8pt}\end{equation}
Here $H_{c0}$ is the unperturbed energy of the free electron that describes the wide conduction band:
\vspace{-8pt}\begin{equation}
H_{0c}=\sum_{\mathbf{k}n\sigma}\varepsilon(\mathbf{k})a^{\dag}_{\mathbf{k}n\sigma}a^{}_{\mathbf{k}n\sigma}
\label{eq:FreeElectronTerm}
\vspace{-0pt}\end{equation}
where $n$ in a band index, while $\varepsilon(\mathbf{k})$ is the energy of the delocalized electron with the wave vector $\mathbf{k}$ and spin $\sigma$. The conduction electron creation and annihilation operators are denoted here by $a^{\dag}_{\mathbf{k}n\sigma}$ and $a^{}_{\mathbf{k}n\sigma}$.

Hybridization of the $f$-electrons with the conduction band is described by $H_{cf}$:
\vspace{-8pt}\begin{equation}
H_{cf}=\frac{1}{\surd N_0}\sum_{i m \mathbf{k} n \sigma}V^{}_{m \sigma}(\mathbf{k},n)
\left( a^{\dag}_{\mathbf{k}n\sigma}f^{}_{m}(i)\text{e}^{-i \mathbf{k}\cdot\mathbf{R}_{i}}+
a^{}_{\mathbf{k}n\sigma}f^{\dag}_{m}(i)\text{e}^{i \mathbf{k}\cdot\mathbf{R}_{i}}\right),
\label{eq:FElectronBandHybridization}
\vspace{-8pt}\end{equation}
where $V^{}_{m \sigma}(\mathbf{k},n)$ is a hybridization matrix element that accounts for the overlap between a localized $f$\!-state and a Bloch wave.

\begin{figure}[!hb]
\includegraphics[width=\textwidth]{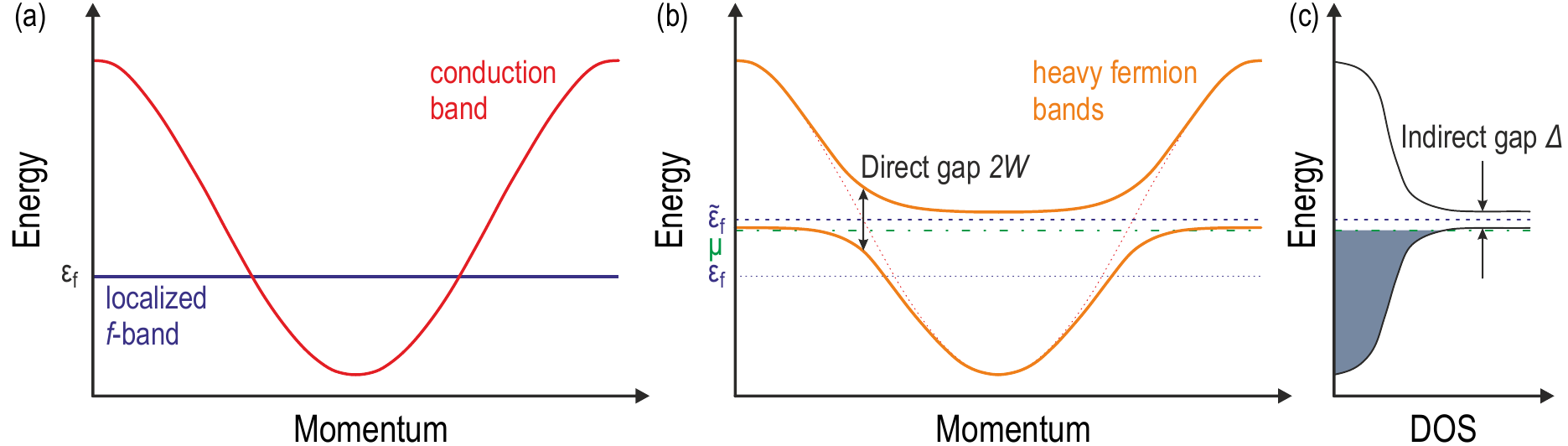}\vspace{-5pt}
\caption{a) Schematic representation of the hybridization between the conduction band and the dispersionless narrow $f$\!-electron band. b) The hybridized bands derived from the Anderson model. c) Renormalized density of states, showing ``hybridization gap''.}
\label{Fig:HeavyFermionBand}\vspace{-10pt}
\end{figure}

A dispersionless narrow $f$\!-electron band with the energy ${\varepsilon{_{\!f}}}$ is defined by the term $H_{0f}$:
\vspace{-8pt}\begin{equation}
H_{0f}=\sum_{i,m}{\varepsilon{_{\!f}}}{\!^{}_{m}}f^{\dag}_{m}(i)f^{}_{m}(i),
\label{eq:FElectronBand}
\vspace{-8pt}\end{equation}
and it is shown together with the conduction band in fig.\,\ref{Fig:HeavyFermionBand}(a).

The term $H_{\text{corr}}$ is the on-site interaction between $f$\!-electrons due to the Coulomb repulsion, written as:
\vspace{-8pt}\begin{equation}
H_{\text{corr}}=\frac{U}{2}\sum_{i,m\neq m^\prime}n^{f}_{m}(i)n^{f}_{m^\prime}(i),
\label{eq:FElectronRepulsion}
\vspace{0pt}\end{equation}
with corresponding creation, $f^{\dag}_{m}$, and annihilation, $f^{}_{m}$, operators that define  the number operator  $n^{f}_{m}=f^{\dag}_{m}f^{}_{m}$. The index $i$ denotes the $f$\!-sites at the position $\mathbf{R}_{i}$, and $m$ is the $z$ component of the total angular momentum $\mathbf{j}$.

The solution of the Hamiltonian is further considered for the case when the $f$\!-shell has only one or zero electrons. Double occupancies of the $f$\!-orbitals are forbidden because of the large Coulomb repulsion \cite{Fulde88}. Because the conduction electron can hop into an empty $f$\!-orbital only when it is free, the proper hybridization matrix element therefore becomes:
\vspace{-8pt}\begin{equation}
V^{}_{m \sigma}(\mathbf{k},n)\rightarrow rV^{}_{m \sigma}(\mathbf{k},n)=\widetilde{V}^{}_{m \sigma}(\mathbf{k},n),
\label{eq:Renormalization}
\vspace{-8pt}\end{equation}
where $r^2=1-n_f$ is the probability to find the $f$-site being empty. The hopping rate of the conduction electrons with the $N(0)$ density of states at the Fermi level onto the $f$\!-shell is defined as:
\vspace{-8pt}\begin{equation}
\widetilde{\Gamma}=(1-n_f)\pi V^{2}N(0)=\pi\widetilde{V}^{2}N(0).
\label{eq:HoppingRate}
\vspace{0pt}\end{equation}

Using the mean-field approach, the Hamiltonian in eq.\,\ref{eq:AndersonHamiltonianFourTerms} can be rewritten as:
\vspace{-0pt}\begin{align}
H_{\text{MF}} &=
\sum_{\mathbf{k}n\sigma}\varepsilon(\mathbf{k})a^{\dag}_{\mathbf{k}n\sigma}a^{}_{\mathbf{k}n\sigma} +\sum_{m \mathbf{k}}{{\widetilde{\varepsilon}{_{\!f}}}{\!^{}_{m}}}f^{\dag}_{\mathbf{k}m}f^{}_{\mathbf{k}m}\nonumber\
\\&+\sum_{nm\mathbf{k}\sigma}rV^{}_{m\sigma}(\mathbf{k},n)(a^{\dag}_{\mathbf{k}n\sigma}f^{}_{\mathbf{k}m}\!+a^{}_{\mathbf{k}n\sigma}f^{\dag}_{\mathbf{k}m})
+\Lambda N_{0}(r^2-1),
\label{eq:AndersonHamiltonianMF}
\vspace{-0pt}\end{align}
where ${{\widetilde{\varepsilon}{_{\!f}}}{\!^{}_{m}}}={{\varepsilon{_{\!f}}}{\!^{}_{m}}}+\Lambda$, $f^{}_{\mathbf{k}m}$ is the Fourier transformation of the operator $f^{}_{m}(i)$, and $\Lambda$ is the Lagrangian parameter. After appropriate mathematical transformations, described in detail in refs.\,[\citenum{Roessler09,Fulde03}], two quasiparticle bands with energies defined as:
\vspace{-0pt}\begin{align}
E^{}_{l}(\mathbf{k}) &=\frac{1}{2}\bigl((\varepsilon(\mathbf{k})+\widetilde{\varepsilon}{_{\!f}})\mp v(\varepsilon(\mathbf{k})) \bigr), \nonumber \\
v(\varepsilon(\mathbf{k}))&=\sqrt{(\varepsilon(\mathbf{k})-\widetilde{\varepsilon}{_{\!f}})^2+4\widetilde{V}^2},
\label{eq:TwoHFBands}
\vspace{-0pt}\end{align}
shown in fig.\,\ref{Fig:HeavyFermionBand}(b) with an orange solid line, can be derived. The renormalized energy of the $f$\!-level can be obtained from the condition  that the lower quasiparticle branch crosses the chemical potential for $k=k_{\text{F}}$. Therefore $E_{1}(k_{\text{F}})=\mu$, and the corresponding energy becomes:
\vspace{-0pt}\begin{equation}
\widetilde{\varepsilon}{_{\!f}}=\mu+\frac{\widetilde{V}^2}{\varepsilon(k_{\text{F}})-\mu}=\mu+k_{\text{B}}T^{*},
\label{eq:FLevelEnergy}
\vspace{0pt}\end{equation}
where $T^{*}$ is the characteristic temperature, below which hybridization between the $f$\!-band and conduction electrons occurs (see fig.\,\ref{Fig:SpecificHeatCeLaAl3}).

The occupation number is
\vspace{-0pt}\begin{equation}
n_{f}=1-\frac{k_{\text{B}}T^{*}}{v_{f}N(0)V^2},
\label{eq:OccupationNumber}
\vspace{0pt}\end{equation}
where $v_{f}$ is the degeneracy of the $f$\!-orbital, shows that because of  hybridization and corresponding $f$\!-level shift above $\mu$, its occupation is reduced. The conduction band is flattened close to the chemical potential, which leads to an increase in the density of states at this energy.

\vspace{-0pt}\vspace{0pt}

Another effect related with the increase of the density of states  at the Fermi level is the enhancement of the effective mass. As it has been shown in ref.~\citenum{Fazekas99} the effective ``heavy fermion'' mass becomes:
\vspace{-8pt}\begin{equation}
\frac{m^{*}}{m}\propto\frac{n_{f}}{2(1-n_{f})}.
\label{eq:HFMass}
\vspace{0pt}\end{equation}

Due to the hybridization, the system gains energy $E^{}_{0}-E^{(0)}_{0}=-k_{\text{B}}T^{*}$, with respect to the total energy without hybridization $E^{(0)}_{0}$. That leads to the formation of an upper and lower Fermi band separated by a ``hybridization gap'' $\Delta\sim T^{*}$, observed as an indirect gap in the renormalized density of states \cite{HelmutKronmueller07}, shown in fig.\,\ref{Fig:HeavyFermionBand}(c), while direct gap between the upper and lower heavy fermion bands is $2W$.

\section{Kondo Screening}\label{Sec:KondoScreening}

\begin{wrapfigure}[11]{r}{0.5\textwidth}\vspace{-28pt}
\includegraphics[width=0.5\textwidth]{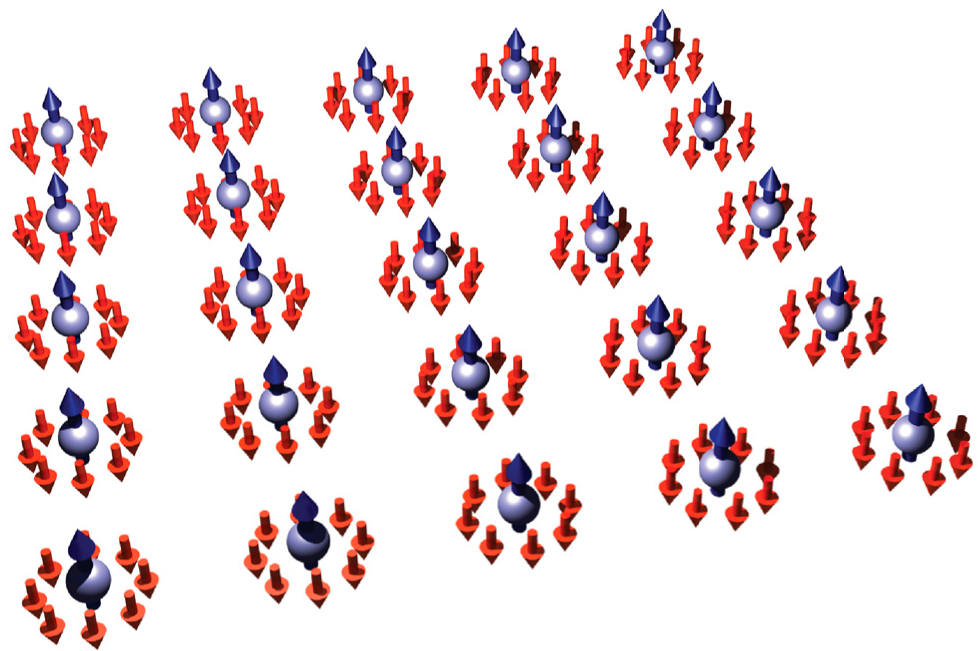}\vspace{-5pt}
\caption{Schematic of a Kondo lattice. The spins shown in blue represent the magnetic atoms. The conduction electrons screen these moments at low temperature are shown in red. Figure from ref.\,\citenum{HamidianSchmidt11}.}
\label{Fig:KondoLattice}\vspace{-0pt}
\end{wrapfigure}

The formation of the hybridized bands was derived using the  Anderson Hamiltonian, and the question how a localized spin that originates from the magnetic impurity interacts with a spin of a conduction electron is covered by the Kondo lattice model. Introduction of the internal degree of freedom, the spin of the magnetic impurity, makes the problem very different from one without.

A typical Kondo lattice is shown in fig.\,\ref{Fig:KondoLattice}, which consists of the magnetic impurity with spin $\mathbf{S}$ and an electron gas with spin $\boldsymbol{\upsigma}$. The spin-spin interaction is driven by virtual charge fluctuations. When conduction electrons are scattered by localized spins of impurity atoms, various pairs of the initial--final states that lead to the spin-flip or non spin-flip scattering have to be considered. You can imagine the process when the conduction electron with wave number $\mathbf{k}_1$ and spin $\downarrow$ scatters on the impurity spin $\uparrow$ into a state with wave number $\mathbf{k}_2$ and spin $\uparrow$, while the impurity changes in a state with spin $\downarrow$. This process can go through various perturbational processes with creation of $f^{0}$ or $f^{2}$ intermediate states. All these contributions are collected into $c$--$f$ coupling, given by:
\vspace{-8pt}\begin{align}
H_{c-f} &=
\frac{1}{L}\sum_{\mathbf{k}_1}\sum_{\mathbf{k}_2}\sum_{\mathbf{R}_{i}}
v_{\mathbf{k}_1}^{}\text{e}^{-i\mathbf{k}_1\cdot\mathbf{R}_{i}}
v_{\mathbf{k}_2}^{*}\text{e}^{-i\mathbf{k}_2\cdot\mathbf{R}_{i}}
\left(\frac{1}{\varepsilon_{\mathbf{k}_1}-\varepsilon_{\!f}}-
\frac{1}{\varepsilon_{\mathbf{k}_2}-\varepsilon_{\!f}-U_{\!f}}\right)\cdot\nonumber
\\&\cdot \left(S_{\mathbf{R}_{i}}^{-}c^{\dag}_{\mathbf{k}_1\uparrow}c^{}_{\mathbf{k}_2\downarrow}
+S_{\mathbf{R}_{i}}^{+}c^{\dag}_{\mathbf{k}_1\downarrow}c^{}_{\mathbf{k}_2\uparrow} +S_{\mathbf{R}_{i}}^{z}(c^{\dag}_{\mathbf{k}_1\uparrow}c^{}_{\mathbf{k}_2\uparrow}-c^{\dag}_{\mathbf{k}_1\downarrow}c^{}_{\mathbf{k}_2\downarrow})\right),
\label{eq:HamiltonianCF}
\vspace{-0pt}\end{align}
where $S^{z}$ and $S^{\pm}=S^{x}\pm i S^{y}$ are the spin operators for a state of spin $S$, while $L$ is the number of localized magnetic impurities. Assuming that the hybridization band $v_{\mathbf{k}}$ is actually $\mathbf{k}$-independent and $v_{\mathbf{k}}\equiv v$, the effective Kondo Hamiltonian that combines the conduction electron band kinetic energy and spin-spin interaction can be written as \cite{Fazekas99}:
\vspace{-0pt}\begin{align}
H_{\text{KL}} &=
\sum_{\mathbf{k}\sigma}\varepsilon(\mathbf{k})c^{\dag}_{\mathbf{k}\sigma}c^{}_{\mathbf{k}\sigma}+
\frac{J}{2}\sum_{\mathbf{R}_{i}}\sum_{\sigma\sigma^\prime}\mathbf{S}_{\mathbf{R}_{i}}
\left(c^{\dag}_{\mathbf{R}_{i}\sigma}\vec{\sigma}_{\sigma\sigma^\prime} c^{}_{\mathbf{R}_{i}\sigma^\prime}\right),\nonumber
\\J&=v^2\sum_{\mathbf{k}}\left(\frac{\Uptheta(\varepsilon_{\mathbf{k}}-\mu)}{\varepsilon_{\mathbf{k}}-\varepsilon_{f}}
+\frac{\Uptheta(\mu-\varepsilon_{\mathbf{k}})}{\varepsilon_{f}+U_{f}-\varepsilon_{\mathbf{k}}} \right),
\label{eq:KondoHamiltonian}
\vspace{-0pt}\end{align}
where $\vec{\sigma}$ is the vector of Pauli matrices, $\Uptheta$ is the Heaviside step function, and $J$ is the antiferromagnetic Kondo coupling. Another way to derive the Kondo Hamiltonian is to apply the Schrieffer-Wolff transformation to the Anderson Hamiltonian (eq.\,\ref{eq:AndersonHamiltonianFourTerms}) \cite{Fazekas99,Hewson97}.

\begin{wrapfigure}[18]{r}{0.5\textwidth}\vspace{-12pt}
\includegraphics[width=0.5\textwidth]{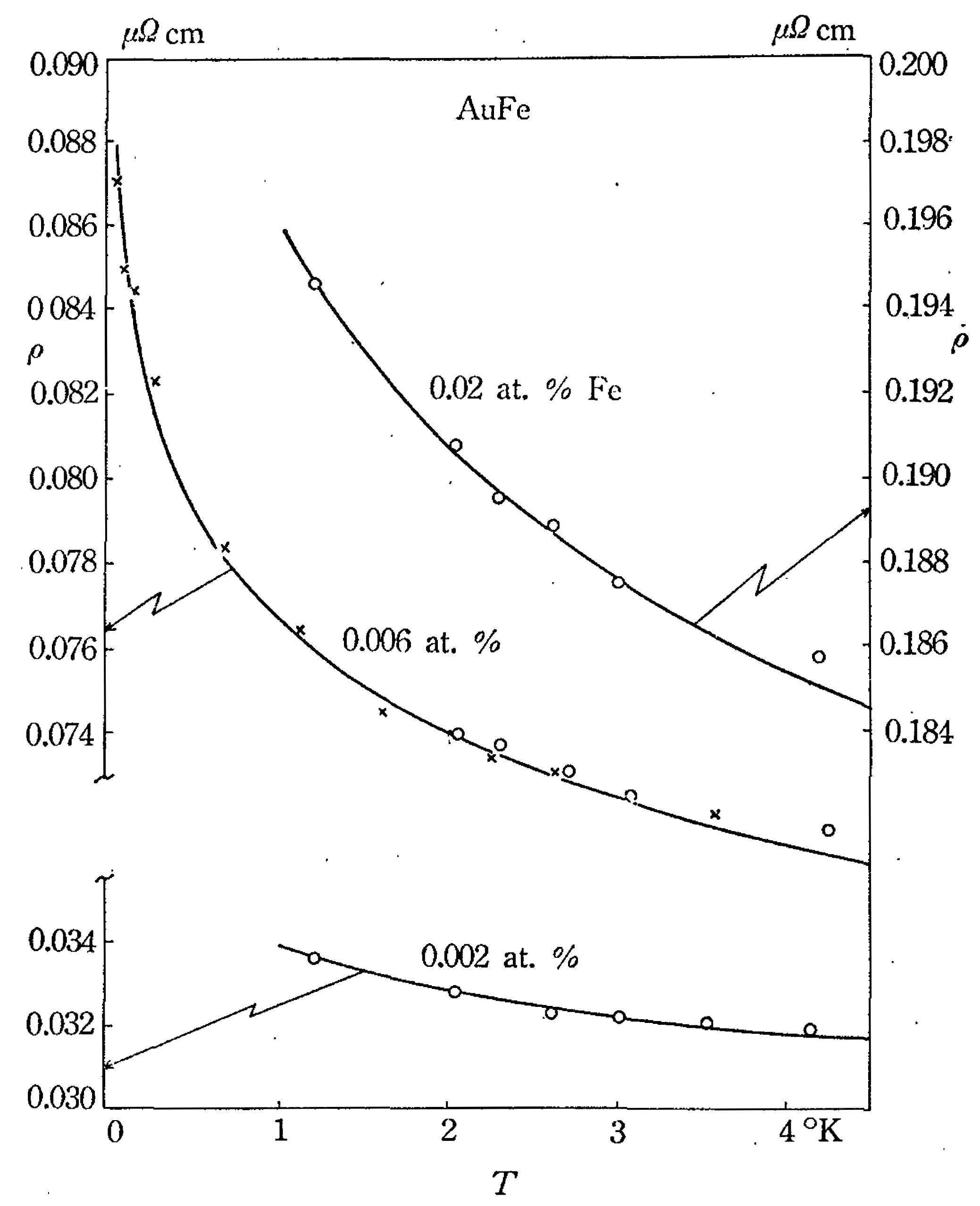}\vspace{-5pt}
\caption{Comparison of experimental and theoretical $\rho-T$ curves for dilute AuFe alloys. Figure from ref.\,\citenum{Kondo64}.}
\label{Fig:KondoResistivityTemperature}\vspace{-0pt}
\end{wrapfigure}

The presence of the exchange coupling $J$ has strong consequences. In a number of various magnetic alloys, because of the presence of magnetic impurities, conduction electrons exhibit an unusual scattering mechanism. It was found that as the temperature is lowered, electrical resistivity, shown in fig.\,\ref{Fig:KondoResistivityTemperature},  increases logarithmically, while one would expect a monotonic decrease. This effect was observed in 1934 \cite{HaasBoer34}, and it took 30 years to find an appropriate explanation to this longstanding puzzle. The work, carried out in 1964 by Jun Kondo, showed that the coupling between the conduction electrons and the $f$\!-electrons can lead to a term proportional to ln($T$) in the low-temperature resistivity \cite{Kondo64,Hewson97}:
\vspace{-0pt}\begin{equation}
R(T)=aT^5+c_{i}R_0-c_{i}R_1\text{ln}\left(\frac{k_{\text{B}}T}{D}\right),
\label{eq:Resistivity}
\vspace{0pt}\end{equation}
where $c_i$ is the concentration of impurity atoms and $D$ is the bandwidth of the conduction electron band. The first term in eq.\,\ref{eq:Resistivity} is the contribution due to the lattice vibrations, the second term is the temperature independent residual resistivity from the impurity potential, while the last term originates from the spin scattering. The presence of the last term is responsible for the observed minimum in the electrical resistivity, known as the Kondo effect. The resistivity minimum is observed at a temperature $T_{\text{min}}=\left(R_{1}c_{i}/5a\right)^{1/5}$ \cite{Hewson97}.

It is necessary to introduce the ``Kondo temperature'', below which the impurity and the conduction electron spins bind very strongly to form an overall non-magnetic singlet state:
\vspace{-0pt}\begin{equation}
T_\text{K}=\frac{D}{k_\text{B}}\text{exp}\left({-\frac{1}{2N(0)J}}\right).
\label{eq:KondoTemperature}
\vspace{0pt}\end{equation}

\begin{wrapfigure}[9]{r}{0.5\textwidth}\vspace{-12pt}
\includegraphics[width=0.5\textwidth]{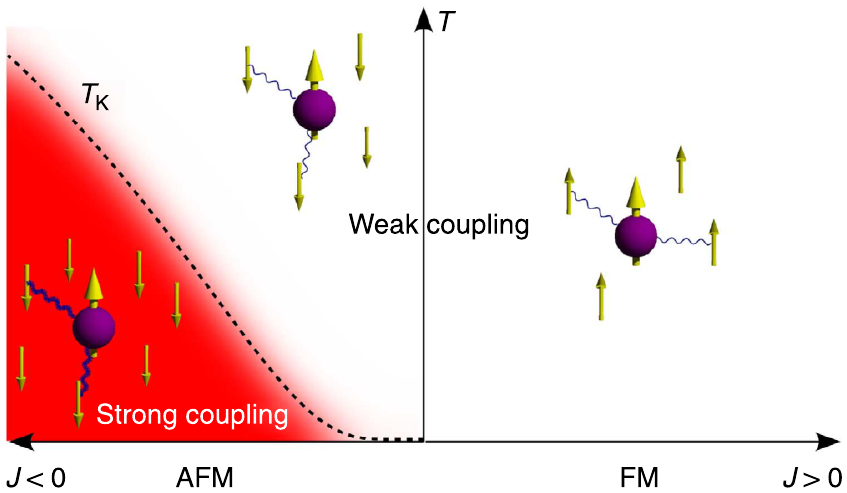}\vspace{-5pt}
\caption{The different regimes of the Kondo effect. Figure from ref.\,\citenum{ZhangKahle13}.}
\label{Fig:KondoModelJ}\vspace{-0pt}
\end{wrapfigure}

If we consider the case when $J=0$, the ground state of the Kondo Hamiltonian (eq.\,\ref{eq:KondoHamiltonian}) becomes $2^L$-fold degenerate, and localized $f$\!-electron spins, surrounded by a sea of conduction electrons, are free to point up or down. When $J>0$, which indicates ferromagnetic coupling, impurity spins are always weakly coupled and becomes asymptotically free in the limit of low temperature \cite{ZhangKahle13}. However when the coupling between the conduction electrons and the $f$\!-electrons is  antiferromagnetic and $J<0$, formation of a local $c$--$f$ singlet is preferable. This leads to the formation of the ``polarized'' conduction electron cloud that surrounds the $f$\!-electron spin and effectively screens the magnetic moment of the impurity and creates the non-magnetic heavy fermion state, schematically shown in fig.\,\ref{Fig:KondoModelJ}. The singlet state is destroyed  at high temperatures, when $T\gg T_\text{K}$.

It may seem that the Kondo lattice model does not create any coupling between the $f$\!-electrons, since below the characteristic Kondo temperature $T_\text{K}$, a singlet with the total spin $S=0$ is created. However this is actually not the case, as the ``polarized'' conduction electron cloud in dense Kondo systems interacts with the other $f$\!-site, as the $f$\!-spin can ``feel'' this disturbance of the surrounding electron sea. This indirect exchange between $f$\!-electrons is called the RKKY interaction and is described in the next section.

\section{RKKY Interaction}\label{Sec:RKKY}

\begin{wrapfigure}[6]{r}{0.5\textwidth}\vspace{-12pt}
\includegraphics[width=0.5\textwidth]{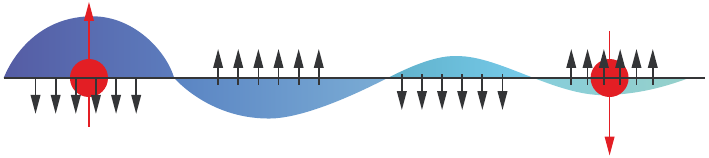}\vspace{-5pt}
\caption{Friedel oscillations of the spin polarization around magnetic impurity, which induce an RKKY interaction. Figure from ref.\,\citenum{Coleman16}.}
\label{Fig:RKKYFriedelOscillations}\vspace{-0pt}
\end{wrapfigure}

To understand how different $f$\!-sites interact with each other, one has to understand the polarization effect of the conduction electrons. The localized spin $\mathbf{S}$ acts as an external magnetic field that polarizes electrons, the Fourier component $H_{\mathbf{q}}$ into which the spatially varying magnetic field is decomposed, $\mathbf{H}(\mathbf{r})=\sum \mathbf{H}_{\mathbf{q}}\text{e}^{-i\mathbf{q}\cdot\mathbf{r}}$, is given by:
\vspace{-0pt}\begin{equation}
H_{\mathbf{q}}=\frac{2J}{Ng\mu_{\text{B}}}S_{z}.
\label{eq:FourierComponentHq}
\vspace{0pt}\end{equation}
\clearpage
\noindent The corresponding spin-magnetization density of the conduction electrons is given by:
\vspace{-0pt}\begin{equation}
\sigma(\mathbf{r})=\frac{1}{V}\sum_{\mathbf{q}}\chi_{\mathbf{q}}H_{\mathbf{q}}\text{e}^{-i\mathbf{q}\cdot\mathbf{r}}=
\frac{12\pi}{V}\frac{N_e}{N}\frac{J}{g\mu_{\text{B}}}S_{z}\chi_{\text{Pauli}}F(2k_\text{F}r),
\label{eq:SpinMagnetizationDensity}
\vspace{0pt}\end{equation}
where $F(x)=\left(-x\text{cos}(x)+\text{sin}(x)\right)/x^4$. The induced Friedel oscillations of the electron spin density \cite{HelmutKronmueller07}, shown in fig.\,\ref{Fig:RKKYFriedelOscillations}, decay as $1/r^3$ and influence the neighbor $f$\!-electrons. The interaction between two sites with spins $\mathbf{S}_n$ and $\mathbf{S}_m$, localized at points $\mathbf{R}_n$ and $\mathbf{R}_m$ respectively, can be written as an interaction between the spin-magnetization density that originates from the polarization by the spin $\mathbf{S}_n$, and the spin $\mathbf{S}_m$. Such interaction gives a contribution: \vspace{-0pt}\begin{equation}
-2J\frac{V}{N}\int\delta(\mathbf{r}-\mathbf{R}_m)\frac{1}{g\mu_{\text{B}}}\sigma(\mathbf{r}-\mathbf{R}_n)\mathbf{S}_m\text{d}\mathbf{r}=
-9\pi\frac{J^2}{\varepsilon_{f}}\left(\frac{N_e}{N}\right)^2F(2k_\text{F}|\mathbf{R}_n-\mathbf{R}_m|)\mathbf{S}_n\mathbf{S}_m,
\label{eq:RKKYContribution}
\vspace{0pt}\end{equation}
which in the case of an incomplete 4$f$\!-shell of rare-earth metals with strong $LS$ coupling should be written as:
\vspace{-0pt}\begin{equation}
H_{\text{RKKY}}=-\frac{9\pi}{2}\left(\frac{N_e}{N}\right)^2\frac{J^2}{\varepsilon_{f}}(g_J-1)^2 \sum_{n\neq m}
F(2k_\text{F}|\mathbf{R}_n-\mathbf{R}_m|)(\mathbf{J}_n\mathbf{J}_m),
\label{eq:RKKYHamiltonian}
\vspace{0pt}\end{equation}
where $\mathbf{J}=\mathbf{L}+\mathbf{S}$, $k_\text{F}$ is the Fermi  wave vector, and $N_e$ is the density of conduction electrons \cite{Yosida96}. The corresponding temperature scale, below which interaction between different $f$\!-electrons arises, can be written in form of \cite{Fazekas99,HelmutKronmueller07}:
\vspace{-0pt}\begin{equation}
T_\text{RKKY}=J^2\frac{N(0)}{Dk_\text{B}}.
\label{eq:RKKYTemperature}
\vspace{0pt}\end{equation}

The RKKY interaction is therefore responsible for the occurrence of the long range magnetic order. The magnetic ground state can be antiferromagnetic or  ferromagnetic, depending on $J$ and the distance between the magnetic impurities.

\section{Doniach Phase Diagram}\label{Sec:DoniachDiagram}

Magnetism in heavy-fermion metals, which can be represented in the form of the Kondo lattice, is governed by the competition between the two effects. On the one hand, Kondo screening tends to quench localized magnetic moments and to form a non-magnetic ground state. On the other hand, the RKKY interaction leads to an ordered ground state. To understand which phenomenon prevails, one needs to compare their energy scales. Let us estimate the relative strength of each component by comparing their characteristic temperature scales.

\begin{wrapfigure}[11]{r}{0.5\textwidth}\vspace{0pt}
\includegraphics[width=0.5\textwidth]{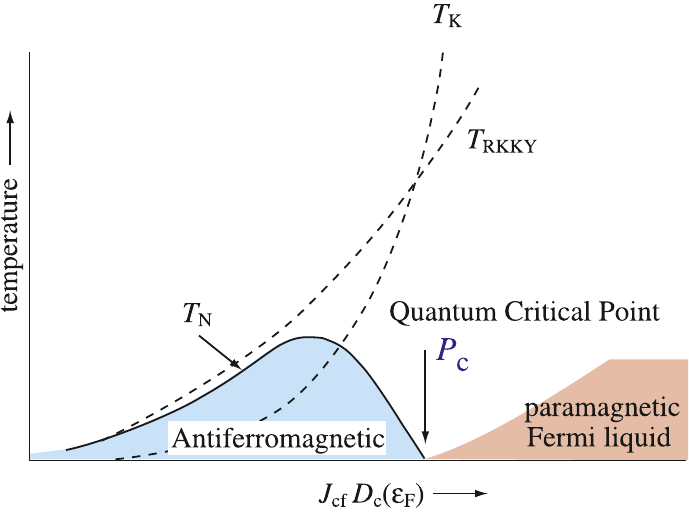}\vspace{-5pt}
\caption{Doniach phase diagram for the Kondo lattice. Figure from ref.\,\citenum{OnukiSettai09}.}
\label{Fig:DoniachPhaseDiagram}\vspace{-0pt}
\end{wrapfigure}
This question was discussed for the first time  by Doniach, where he suggested that a transition from an antiferromagnetic to a Kondo spin compensated
ground state will occur as the exchange coupling constant $J$ is increased to a critical value \cite{Doniach77}. More detailed investigation of the ground-state phase diagram for the periodic Anderson model can be found in ref.\,\citenum{DoradzinskiSpalek98}. The exchange constant $J$ enters both the Kondo temperature $T_{\text{K}}$ (eq.\,\ref{eq:KondoTemperature}) and $T_{\text{RKKY}}$ (eq.\,\ref{eq:RKKYTemperature}) with different dependencies. For the large values of $J$ the exponential dependence of $T_{\text{K}}$ dominates the RKKY energy scale, and therefore a non-magnetic singlet is formed. A small value of $J$ on the contrary makes $T_{\text{RKKY}}$ the largest scale, and a magnetically ordered state is formed. As a result, the Doniach phase diagram, shown in fig.\,\ref{Fig:DoniachPhaseDiagram}, that illustrates the described interplay, can be composed. The strength of the exchange interaction can be tuned by the change in composition or pressure, and  the transition between these two regimes results in a quantum critical point $P_{\text{c}}$ \cite{Fazekas99,HelmutKronmueller07}.

\vspace{-15pt}
\section{Crystal-Field Effect}\label{Sec:CrystalFieldEffect}

\begin{wrapfigure}[10]{l}{0.4\textwidth}\vspace{-25pt}
\centerline{
    \begin{tikzpicture}[
      scale=0.7,
      level/.style={thick},
      virtual/.style={thick,densely dashed,green},
      connect/.style = { dashed },
      trans/.style={thick,<->,shorten >=2pt,shorten <=2pt,>=stealth},
      classical/.style={thin,double,<->,shorten >=4pt,shorten <=4pt,>=stealth}
    ]
    \draw[level] (0,0) -- node[above] {} (0.9,0);
    \draw[level] (1,0) -- node[above] {} (1.9,0);
    \draw[level] (2,0) -- node[above] {$d$} (2.9,0);
    \draw[level] (3,0) -- node[above] {} (3.9,0);
    \draw[level] (4,0) -- node[above] {} (4.9,0);
    \draw[level] (6,-2) -- node[below] {$d_{xy}$} (6.9,-2);
    \draw[level] (7,-2) -- node[below] {$d_{xz}$} (7.9,-2);
    \draw[level] (8,-2) -- node[below] {$d_{yz}$} (8.9,-2);
    \draw[level] (6.4,+3) -- node[above] {$d_{x^2-y^2}$} (7.3,+3);
    \draw[level] (7.6,+3) -- node[above] {$d_{z^2}$} (8.5,+3);
    \draw[virtual] (5,0) -- node[above] {} (9,0);
    \draw[trans] (6.85,-2) -- (6.85,3) node[midway,left] {$\Delta_{0}$};
    \draw[trans] (8.05,-2) -- (8.05,0) node[midway,left] {};
    \draw[trans] (8.05,0) -- (8.05,3) node[midway,left] {};
    \draw[connect] (4.9,0)  -- (6,-2) (4.9,0) -- (6.4,3);
    \node[label] at (6,3)  {$e_g$};
    \node[label] at (5.5,-2) {$t_{2g}$};
    \end{tikzpicture}
  }
\caption{A level diagram of the $d$\!-orbitals that shows $t_{2g}$ and $e_g$ groups with different energies due to the CEF splitting.}
\label{Fig:CEFdOrbitalSplitting}\vspace{-0pt}
\end{wrapfigure}
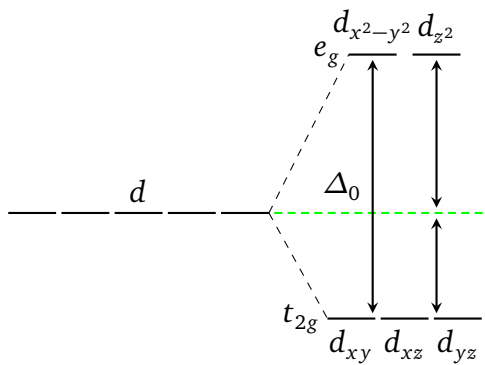

Atoms in crystals experience an interaction with the electrostatic field created by surrounding ions. If one considers an isolated metal ion, localized  $d$ or $f$ orbitals have the same energy and are therefore degenerate. But when the ion is located in the crystal, its orbitals interact with the potential that originates from  surrounding atoms (ligands) in the crystal lattice. For example, $d_{z^2}$, $d_{xz}$, $d_{yz}$, $d_{xy}$ and $d_{x^2-y^2}$ orbitals experience different interaction with the ligands arranged along $x$, $y$ and $z$. It is obvious that $d$ orbitals, which lie along ligand axes, experience greater repulsion, and therefore it is less energetically favourable to put an electron in one of the $d_{z^2}$ or $d_{x^2-y^2}$ orbitals. Therefore the energies of these orbitals increase with respect to their normal energy, while $d_{xy}$ $d_{xz}$ and $d_{yz}$ decrease their energy and become more stable. This process is shown in fig.\,\ref{Fig:CEFdOrbitalSplitting} and is known as crystal field splitting.

For the purpose of my thesis we need to consider the CEF effects for the $4f^1$ electron of Ce$^{3+}$. The interaction between the $f$\!-electron and surrounding ligands can be described by the effective potential $V_{\text{CEF}}$ that should be added to the atomic Hamiltonian:
\vspace{-0pt}\begin{equation}
H=\underbrace{\sum_{i=1}^{Z}\frac{p_{i}^2}{2m}}_{H_0}+
\underbrace{\sum_{i<j}^{Z}\frac{e^2}{|\mathbf{r}_i-\mathbf{r}_j|}-\sum_{i=1}^{Z}\frac{Ze^2}{r_i}}_{V_1}+
\underbrace{\sum_{i=1}^{Z}g(r_i)\mathbf{l}_i\cdot\mathbf{s}_i}_{V_2}+
e\sum_iV_{\text{CEF}}(\mathbf{r}_i)
\label{eq:AtomicHamiltonian}
\vspace{0pt}\end{equation}
where $V_1$ contains corrections to the self-consistent potential, and $V_2$ is the spin-orbit coupling. $V_{\text{CEF}}$ describes the potential at the point $\mathbf{r}$ of the charge $q_j$ located at the point $\mathbf{R}_j$. It is convenient to write it using spherical harmonics of the unit point charge potential \cite{Majlis01}:
\vspace{-0pt}\begin{equation}
V_{\text{CEF}}(\mathbf{r})=\sum_{j=1}^{N}\frac{q_j}{|\mathbf{R}_j-\mathbf{r}|}=
\sum_{l=0}^{\infty}\frac{q_jr^l}{R_j^{l+1}}
\sum_{m=-l}^{l}\frac{4\pi}{2l+1}Y_{lm}^{\star}(\theta_r\phi_r)Y_{lm}^{}(\theta_{R_j}\phi_{R_j}).
\label{eq:VCEF}
\vspace{0pt}\end{equation}
It is also possible to rewrite eq.\,\ref{eq:VCEF} with tesseral harmonics, defined in terms of spherical harmonics: $Z_{l0}^{}=Y_{l}^{0}$ , $Z_{lm}^{c}=(Y_{l}^{-m}+Y_{l}^{m})/\sqrt{2}$, $Z_{lm}^{s}=i(Y_{l}^{-m}+Y_{l}^{m})/\sqrt{2}$, and also by introducing a coefficient $\gamma_{l\alpha}$ related with the charge distribution $\rho(\mathbf{r})=\sum q_j\delta(\mathbf{R}_j-\mathbf{r})$ \cite{Majlis01,FuldeLoewenhaupt85}:
\vspace{-0pt}\begin{equation}
V_{\text{CEF}}(r,\theta,\phi)=
\sum_{l=0}^{\infty}r^l_{}\sum_{\alpha}^{}{\underbrace{\left(\frac{4\pi}{2l+1}\int\frac{\rho(\mathbf{r})}{r^{l+1}}Z_{l\alpha}^{}(\theta,\phi)\text{d}^3r\right)}_{\gamma_{l\alpha}}}Z_{l\alpha}^{}(\theta,\phi).
\label{eq:VCEF2}
\vspace{0pt}\end{equation}

In the case of rare-earth ions, the crystal-field splitting is small, as the  perturbation given by eq.\,\ref{eq:VCEF2} is small compared to the spin-orbit coupling, and therefore intermixing of different $J$ values is insignificant. This allows us to neglect off-diagonal elements in the Hamiltonian and to determine the matrix elements of certain potential functions in representations, in which the states are eigenstates of the total angular momentum. This method was fully described in \cite{Stevens52}, and using the  Stevens operator equivalents $\pazocal{O}_l^m$, created from the operators $J$, the CEF Hamiltonian can be written as:
\vspace{-0pt}\begin{equation}
H_{\text{CEF}}=\sum_{l,m}B_{l}^{m}\pazocal{O}_l^m.
\label{eq:HCEF}
\vspace{0pt}\end{equation}
Operator equivalents are obtained by replacing $x$, $y$ and $z$ in the Cartesian function by different combinations of $J_x,J_y,J_z$. For example, some important operators are \cite{Hutchings64,LeaLeask62}:
\vspace{-0pt}\begin{align}
\beta_{J}^{}\langle r^4\rangle\pazocal{O}_{4}^{4}&=\beta_{J}^{}\langle r^4\rangle(J_{+}^{4}+J_{-}^{4})/2 \equiv \sum_{i}\left( (x_{i}^{}+iy_{i}^{})^4+(x_{i}^{}-iy_{i}^{})^4\right)/2 = \sum_{i}(x_{i}^{4}-6x_{i}^{2}y_{i}^{2}+y_{i}^{4})
\nonumber \\
\pazocal{O}_{4}^{0}&=35J_{z}^{4}-(30J(J+1)-25)J_{z}^{2}-6J(J+1)+3J_{}^{2}(J+1)^2
\nonumber \\
\pazocal{O}_{6}^{4}&=[(11J_{z}^{2}-J(J+1)-38)(J_{+}^{4}+J_{-}^{4})+(J_{+}^{4}+J_{-}^{4})(11J_{z}^{2}-J(J+1)-38)]/4
\nonumber \\
\pazocal{O}_{6}^{0}&=231J_{z}^{6}-315J(J+1)J_{z}^{4}+735J_{z}^{4}+105J_{}^{2}(J+1)^2J_{z}^{2}-525J(J+1)J_{z}^{2}+294J_{z}^{2} \nonumber \\
&-5J_{}^{3}(J+1)^3+40J_{}^{2}(J+1)^2-60J(J+1).
\label{eq:StevensOperator}
\vspace{-0pt}\end{align}

\begin{wrapfigure}[27]{r}{0.4\textwidth}\vspace{-12pt}
\centerline{
    \begin{tikzpicture}[
      scale=0.9,
      level/.style={thick},
      virtual/.style={thick,densely dashed,green},
      connect/.style = { dashed },
      trans/.style={thick,<->,shorten >=2pt,shorten <=2pt,>=stealth},
      classical/.style={thin,double,<->,shorten >=4pt,shorten <=4pt,>=stealth}
    ]
    \draw[level] (0,-0.25) -- node[above] {} (0.9,-0.25);
    \draw[level] (0,-0.15) -- node[above] {} (0.9,-0.15);
    \draw[level] (0,-0.05) -- node[above] {} (0.9,-0.05);
    \draw[level] (0,0.05) -- node[above] {} (0.9,0.05);
    \draw[level] (0,0.15) -- node[above] {} (0.9,0.15);
    \draw[level] (0,0.25) -- node[above] {$^{2}F_{5/2}$} (0.9,0.25);
    \draw[level] (4,3) -- node[above] {} (4.9,3);
    \draw[level] (4,3.1) -- node[above] {} (4.9,3.1);
    \draw[level] (4,-1) -- node[below] {} (4.9,-1);
    \draw[level] (4,-1.1) -- node[below] {} (4.9,-1.1);
    \draw[level] (4,-1.2) -- node[below] {} (4.9,-1.2);
    \draw[level] (4,-1.3) -- node[below] {} (4.9,-1.3);
    \draw[trans] (4.45,3) -- (4.45,-1) node[midway,left] {$\Delta$};
    \draw[connect] (1,-0.05)  -- (3.9,-1.2) (1,0.05) -- (3.9,3.05);
    \node[label] at (5.5,3.05)  {$|\Upgamma_7\rangle$};
    \node[label] at (5.5,-1.15)  {$|\Upgamma_8\rangle$};
    \node[label] at (0.45,-1.2)  {Spin-orbit};
    \node[label] at (0.45,-1.7)  {interaction};
    \node[label] at (4.45,-1.7)  {CEF};
    \end{tikzpicture}
  }
\vspace{-5pt}\caption{A level diagram that shows CEF splitting  in CeB$_6$. Crystal field transition between the $|\Upgamma_8\rangle$ ground state and $|\Upgamma_7\rangle$  is $\Delta=46$\,meV \cite{LoewenhauptCarpenter85}.}
\label{Fig:CEFfOrbitalSplitting}\vspace{-0pt}

    \vspace{-5pt}\includegraphics[width=\linewidth]{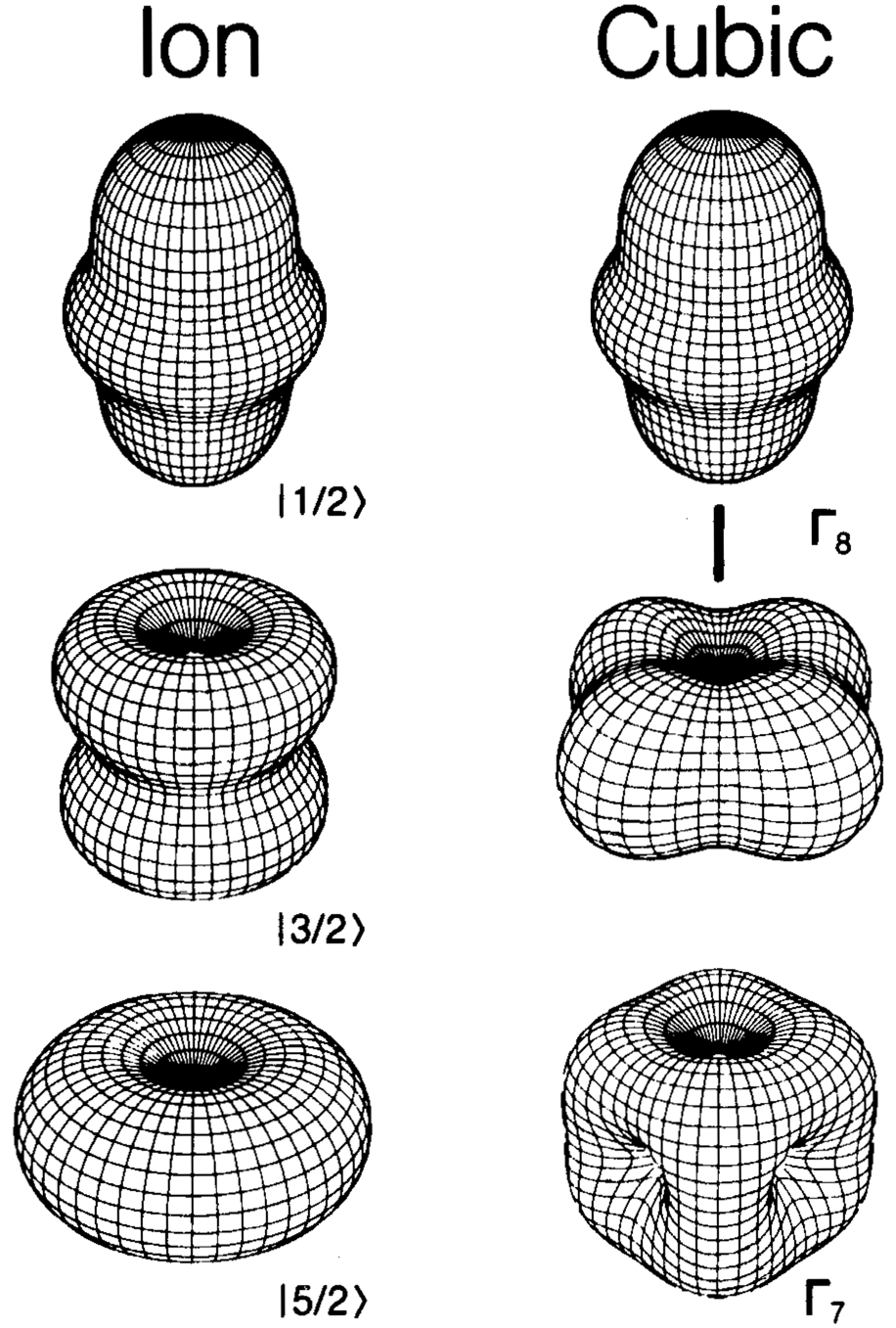}\vspace{0pt}
    \vspace{-20pt}\caption{Contours of equal 4$f$ charge density for a Ce$^{3+}$ ion and for Ce$^{3+}$ in cubic symmetry. Figure was taken from ref.\,\citenum{FuldeLoewenhaupt85}.}\label{Fig:G7G8}\vspace{-0pt}
\end{wrapfigure}

In the case of CeB$_6$, which will be discussed in Chap.\,\ref{Chap:ResultsCeLaB6}, the CEF can be evaluated considering the Ce$^{3+}$ ion that sits at the center of the simple cubic unit cell surrounded by eight B$_6$ anion groups along the $[111]$ directions. According to Hund's rule, the total angular momentum for the $f$\!-orbital with $L=3$ and one electron $S=1/2$ is $J=L-S=5/2$. Therefore the ground state of an isolated ion is sixfold degenerate. It can be shown \cite{Hutchings64} that, for the crystal field in a cubic symmetry, the Hamiltonian given by eq.\,\ref{eq:HCEF} becomes:
\vspace{-0pt}\begin{equation}
H_{\text{CEF}}=B_{4}^{0}[\pazocal{O}_{4}^{0}+5\pazocal{O}_{4}^{4}]+B_{6}^{0}[\pazocal{O}_{6}^{0}-21\pazocal{O}_{6}^{4}],
\label{eq:HCEFCubic}
\vspace{0pt}\end{equation}
and the corresponding coefficients $B_{4}^{0}$ and $B_{6}^{0}$ are
\vspace{-0pt}\begin{equation}
B_{4}^{0}=\frac{7}{18}\frac{|e|q}{d^5}\beta_J\langle r^4\rangle;
\hspace{10pt}
B_{6}^{0}=-\frac{1}{9}\frac{|e|q}{d^9}\gamma_J\langle r^6\rangle.
\label{eq:B40B60}
\vspace{0pt}\end{equation}
Here $d$ is the distance to the surrounding point charges $|e|q$, and the radial integral defined as $\langle r^n\rangle=\int r^{n+2}[f(r)]^2\text{d}r$ contains the electron radial wave function $f(r)$. The eigenfunctions of the Hamiltonian, given by eq.\,\ref{eq:HCEFCubic}, are written in terms of the $J_z$ to be \cite{LeaLeask62}:
\vspace{-0pt}\begin{align}
|\Upgamma_7\rangle&= \sqrt{\frac{1}{6}}\left|\pm\frac{5}{2}\right\rangle-\sqrt{\frac{5}{6}}\left|\mp\frac{3}{2}\right\rangle
\nonumber \\
|\Upgamma_8\rangle&=\begin{cases}
                        |\pm\frac{1}{2}\rangle \\
                        \sqrt{\frac{5}{6}}|\pm\frac{5}{2}\rangle+\sqrt{\frac{1}{6}}|\mp\frac{3}{2}\rangle.
                    \end{cases}
\label{eq:EigenfunctionsG7G8}
\vspace{0pt}\end{align}\vspace{0pt}

This means that, in the presence of a cubic crystal field environment, the sixfold degenerate ${^2}F_{\frac{5}{2}}$ multiplet  of $J=5/2$  splits into two sets: a $|\Upgamma_8\rangle$ quartet and $|\Upgamma_7\rangle$ doublet, shown in fig.\,\ref{Fig:CEFfOrbitalSplitting}. Therefore, there are no longer pure states with $J_z$ being $|\pm\frac{1}{2}\rangle$, $|\pm\frac{3}{2}\rangle$ or $|\pm\frac{5}{2}\rangle$, but linear combinations of these states. Charge clouds are thereby markedly different with respect to those for a free ion, and are shown in fig.\,\ref{Fig:G7G8}. It is easy to notice that the lobes of the $|\Upgamma_7\rangle$ configuration are pointing towards B$_6$ anion groups, and therefore should experience greater repulsion, being unfavourable in energy. At the same time, the $|\Upgamma_8\rangle$ configuration, with lobes that avoid these directions, should be lower in energy. This has been confirmed by direct measurements \cite{SundermannChen17}.  Strictly speaking, beyond the simple geometric explanation given here, the ground state of Ce$^{3+}$ in this case is determined only by $B_{4}^{0}$, as the coefficient $\gamma_J$ is actually zero and $B_{6}^{0}=0$. Therefore the ground state is either $|\Upgamma_7\rangle$ when $B_{4}^{0}>0$ or $|\Upgamma_8\rangle$  for $B_{4}^{0}<0$. The splitting between the excited and the ground states is determined as $\Delta=360|B_{4}^{0}|$ \cite{FuldeLoewenhaupt85}.

\section{Hidden Order and the Multipole Interaction}\label{Sec:HOandMI}

\subsection{Multipole expansion }\label{SubSec:MultipoleExpansion}
In the previous section, we found that in the presence of the CEF there are no longer states with the pure $J$, but new eigenstates emerge, being their linear combinations. In $f$\!-electron systems, spin and orbital degrees of freedom are tightly coupled. In order to describe such a complicated system, $f$\!-electron states can be represented in terms of a ``multipole'' degree of freedom. As a basis of the expansion,  spherical harmonics mentioned above (see sec.\,\ref{Sec:CrystalFieldEffect}) are used. Therefore one can define the electric multipole moments and also, due to the presence of the current density that originates from the orbital and the spin currents of the electron, magnetic multipole moments. They can be expressed as \cite{KuramotoKusunose09}:
\vspace{-0pt}\begin{equation}
Q_{lm}^{}=\int r_{}^{l}Z_{lm}^{\star}(\hat{\mathbf{r}})\rho_{\text{ele}}^{}(\mathbf{r})\text{d}\mathbf{r};
\hspace{30pt}
M_{lm}^{}=\int r_{}^{l}Z_{lm}^{\star}(\hat{\mathbf{r}})\rho_{\text{mag}}^{}(\mathbf{r})\text{d}\mathbf{r},
\label{eq:EandMMultipole}
\vspace{0pt}\end{equation}
where $\rho_{\text{ele}}^{}(\mathbf{r})$ is the charge density and $\rho_{\text{mag}}^{}(\mathbf{r})$ is the magnetization density of the $f$\!-electron, that satisfy the Poisson equation for the scalar and vector potentials correspondingly. The spherical harmonics are given by $Z_{lm}^{}(\hat{\mathbf{r}})\equiv\sqrt{4\pi/(2l+1)}Y_{lm}^{}(\hat{\mathbf{r}})$. An electric multipole is defined in a way that it is  even under time reversal and has parity $(-1)^{l}$. A magnetic multipole is odd and has parity $(-1)^{l+1}$. Assuming the presence of inversion symmetry, an odd-rank multipole is always magnetic while even-rank is electric.

These classical multipole moments can be expressed as the average of corresponding quantum mechanical operators over localized $f$\!-electron states:
\vspace{-0pt}\begin{equation}
Q_{lm}^{}=\langle\hat{Q}_{lm}^{}\rangle_{f}^{};
\hspace{30pt}
M_{lm}^{}=\langle\hat{M}_{lm}^{}\rangle_{f}^{}.
\label{eq:EandMMultipoleOperator}
\vspace{0pt}\end{equation}
According to the Wigner-Eckart theorem \cite{Sakurai93,KuramotoKusunose09}, operators $\hat{Q}_{lm}^{}$ and $\hat{M}_{lm}^{}$ can be expressed by the spherical tensor operator $\hat{J}_{lm}^{}$, which represents the polynomial of the total angular momentum $J$, and obtained by replacing $x$, $y$, and $z$ with  $J_x,J_y,J_z$  in  the spherical harmonic expression. Introduced earlier, Stevens operators are actually proportional to the tensor operator $\hat{J}_{lm}^{}$ \cite{Kusunose08}.

\subsection{Interaction between multipoles}\label{SubSec:InteractionBetweenMultipoles}
In an $f$\!-electron crystal, multipole degrees of freedom are restricted, since the point-group symmetry of the CEF in a solid is lower than the spherical symmetry. Therefore, if we consider a system with the CEF multiplet as a ground state that has degeneracy $d$, tensor operators up to the rank $d-1$ can be assigned in order to express multipole operators $\hat{X}_{i}^{\alpha}$ at one site. This can also result in an equivalency between multipolar operators of different rank. The lattice Hamiltonian, which accounts for the interaction  between different sites, and therefore describes a generalized exchange model with a uniform external field, can be written as:
\vspace{-0pt}\begin{equation}
H=-\frac{1}{2}\sum_{ij}\sum_{\alpha\beta}D_{ij}^{\alpha\beta}\hat{X}_{i}^{\alpha}\hat{X}_{j}^{\beta}-\sum_{i\alpha}\hat{X}_{i}^{\alpha}h_{}^{\alpha},
\label{eq:MPHamiltonian}
\vspace{-0pt}\end{equation}
where $D_{ij}^{\alpha\beta}$ is the exchange coupling, $\hat{X}_{i}^{\alpha}$ is an operator constructed from the multipole operators $\hat{Q}_{lm}^{}$ and $\hat{M}_{lm}^{}$, and the last term determines coupling to the external field \cite{Kusunose08,KuramotoKusunose09}.

\renewcommand{\arraystretch}{0.9}
\begin{wraptable}{r}{7.5cm}\vspace{-30pt}
\centering
\resizebox{7.5cm}{!}{
\begin{tabular}{l@{~}c@{~}c@{}c@{}}\\
\toprule
$O_{h}$ & Rank & Notation &  Angle dependence\\
\toprule
$\Upgamma_{1g}^{}$ & 4 & $\hat{\pazocal{H}}_{}^{0}$ & $\frac{5\sqrt{21}}{12}(x^4+y^4+z^4-\frac{3}{5}r^4)$\\
\midrule
$\Upgamma_{2u}^{}$ & 3 & $\hat{\pazocal{T}}_{xyz}^{}$ & $\sqrt{21}xyz$\\
\midrule
$\Upgamma_{3g}^{(1)}$ & 2 & $\hat{\pazocal{O}}_{2}^{0}$ & $(3z^2-r^2)/2$\\
                      &   & $\hat{\pazocal{O}}_{2}^{2}$ & $\sqrt{3}(x^2-y^2)/2$\\
\midrule
$\Upgamma_{3g}^{(2)}$ & 4 & $\hat{\pazocal{H}}_{1}^{e}$ & $\frac{\sqrt{15}}{12}(7(2z^4-x^4-y^4)-6(3z^2-r^2)r^2)$\\
                    &   & $\hat{\pazocal{H}}_{2}^{e}$ & $\sqrt{5}(x^2-y^2)(r^2-7z^2)/4$\\
\midrule
$\Upgamma_{4u}^{(1)}$ & 1 & $\hat{\pazocal{J}}_{x}^{}$ & $x$\\
                      &   & $\hat{\pazocal{J}}_{y}^{}$ & $y$\\
                      &   & $\hat{\pazocal{J}}_{z}^{}$ & $z$\\
\midrule
$\Upgamma_{4u}^{(2)}$ & 3 & $\hat{\pazocal{T}}_{x}^{\alpha}$ & $x(5x^2-3r^2)/2$\\
                      &   & $\hat{\pazocal{T}}_{y}^{\alpha}$ & $y(5y^2-3r^2)/2$\\
                      &   & $\hat{\pazocal{T}}_{z}^{\alpha}$ & $z(5z^2-3r^2)/2$\\
\midrule
$\Upgamma_{4g}^{   }$ & 4 & $\hat{\pazocal{H}}_{x}^{\alpha}$ & $\sqrt{35}yz(y^2-z^2)/2$\\
                      &   & $\hat{\pazocal{H}}_{y}^{\alpha}$ & $\sqrt{35}zx(z^2-x^2)/2$\\
                      &   & $\hat{\pazocal{H}}_{z}^{\alpha}$ & $\sqrt{35}xy(x^2-y^2)/2$\\
\midrule
$\Upgamma_{5u}^{   }$ & 3 & $\hat{\pazocal{T}}_{x}^{\beta}$ & $\sqrt{15}x(y^2-z^2)/2$\\
                      &   & $\hat{\pazocal{T}}_{y}^{\beta}$ & $\sqrt{15}y(z^2-x^2)/2$\\
                      &   & $\hat{\pazocal{T}}_{z}^{\beta}$ & $\sqrt{15}z(x^2-y^2)/2$\\
\midrule
$\Upgamma_{5g}^{(1)}$ & 2 & $\hat{\pazocal{O}}_{yz}^{}$ & $\sqrt{3}yz$\\
                      &   & $\hat{\pazocal{O}}_{zx}^{}$ & $\sqrt{3}zx$\\
                      &   & $\hat{\pazocal{O}}_{xy}^{}$ & $\sqrt{3}xy$\\
\midrule
$\Upgamma_{5g}^{(2)}$ & 4 & $\hat{\pazocal{H}}_{x}^{\beta}$ & $\sqrt{5}yz(7x^2-r^2)/2$\\
                      &   & $\hat{\pazocal{H}}_{y}^{\beta}$ & $\sqrt{5}zx(7y^2-r^2)/2$\\
                      &   & $\hat{\pazocal{H}}_{z}^{\beta}$ & $\sqrt{5}xy(7z^2-r^2)/2$\\
\bottomrule
\end{tabular}
}
\vspace{-5pt}\caption{Multipoles under cubic symmetry $O_{h}$  up to fourth rank. Values are taken from ref.\,\citenum{KuramotoKusunose09}. }\vspace{-0pt}
\label{Tab:Multipoles}\vspace{-90pt}
\end{wraptable}\vspace{-0pt}

Now let us consider the case of Ce$^{3+}$ ion in CeB$_6$. As we have seen above in sec.\,\ref{Sec:CrystalFieldEffect}, under a CEF with the cubic symmetry $O_{h}$, the sextet of $J=5/2$ splits into a $|\Upgamma_8\rangle$ quartet and a $|\Upgamma_7\rangle$ doublet. In order to obtain irreducible multipole representations under $O_{h}$, we have to do decomposition of the two-electron state basis, obtained by the direct product of the representations in the single-electron state. It can be shown that \cite{KuramotoKusunose09}:
\begin{align}
\Upgamma_8 \otimes \Upgamma_8&=\Upgamma_{1g}\oplus\Upgamma_{2u}\oplus\Upgamma_{3g} \nonumber \\ &\oplus2\Upgamma_{4u}\oplus\Upgamma_{5u}\oplus\Upgamma_{5g} \nonumber \\
\Upgamma_7 \otimes \Upgamma_8&=\Upgamma_{3g}\oplus\Upgamma_{4u}\oplus\Upgamma_{5u} \nonumber \\
\Upgamma_7 \otimes \Upgamma_7&=\Upgamma_{1g}\oplus\Upgamma_{4u}.
\label{eq:IrreducibleRepresentations}
\end{align}
Subscripts  $g$ (gerade) and $u$ (ungerade) refer to even and odd parity under space inversion and therefore represent electric and magnetic multipole moments, respectively. The complete table up to the sixth rank can be found in \cite{Kusunose08}, while the angular dependences of the  multipoles up to rank 4 are listed in table\,\ref{Tab:Multipoles}. It should be mentioned that multipoles with different rank, for example quadrupoles $\hat{\pazocal{O}}$ and hexadecapoles $\hat{\pazocal{H}}$, or dipoles $\hat{\pazocal{J}}$ and octupoles $\hat{\pazocal{T}}$, belong to the same irreducible representation $\Upgamma_{5g}^{}$ and $\Upgamma_{4u}^{}$ respectively.

In order to understand the interaction between different multipoles, and what is more important to reveal which one will order upon crossing the phase transition from the high-temperature paramagnetic state into low-temperature ordered state, one has to  consider several important factors, like the influence of the field-induced moments, character of the interacting multipoles, and the Zeeman term. Numerous efforts were made in order to explain the rich temperature--magnetic-field phase diagram of CeB$_6$, based on the RKKY interaction between neighboring sites and mean-field analysis    \cite{Ohkawa83,Ohkawa85,UiminKuramoto96,ShiinaShiba97,SakaiShiina97,ShiinaSakai98,SakaiShiina99,SeraKobayashi99}. Ordering of the $\hat{\pazocal{O}}_{xy}^{}$, $\hat{\pazocal{O}}_{zx}^{}$, $\hat{\pazocal{O}}_{yz}^{}$ quadrupolar moments within the $\Upgamma_8$ quartet assumed to be realized at zero field \cite{ShiinaSakai98}, while upon application of external magnetic field induced dipolar and octupolar moments become responsible for the observed ``stabilization'' of the ordered phase \cite{ShiinaSakai98}.

\vspace{-0pt}\begin{figure}[!b]
\includegraphics[width=1\textwidth]{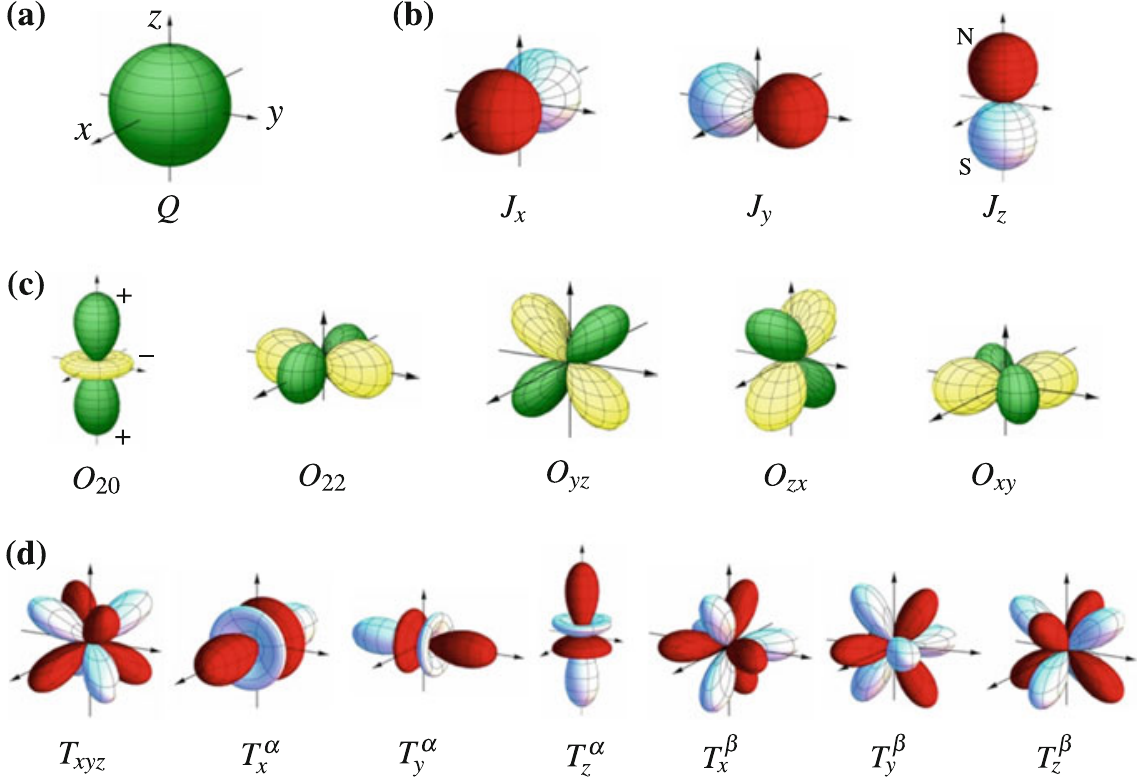}\vspace{-5pt}
\caption{ Angular dependences of the: a) spherical electric charge distribution; b) magnetic dipole; c) electric quadrupole; d) magnetic
octupole moments, listed in the table\,\ref{Tab:Multipoles}. Here $+$,$-$ and N, S represent the distribution of electric and magnetic charges respectively. Figure from ref.\,\citenum{MurakamiIshihara17}.}
\label{Fig:MultipoleAngularDependences}\vspace{-18pt}
\end{figure}

Let us consider which moments are induced in a magnetic field.  Angular dependences of the multipole moments are shown in fig.\,\ref{Fig:MultipoleAngularDependences}, where $\hat{\pazocal{Q}}$ represents a spherical part of the charge distribution. In case of the $\hat{\pazocal{O}}_{2}^{0}$ quadrupole, the charge distribution cloud is elongated along $z$, and therefore it becomes obvious that magnetic field applied along [001] induces $\hat{\pazocal{J}}_{z}^{}$ dipole and $\hat{\pazocal{T}}_{z}^{\alpha}$ octupole. For the $\hat{\pazocal{O}}_{xy}^{}$ charge distribution, similar arguments lead to the emergence of a $\hat{\pazocal{J}}_{z}^{}$ dipole and $\hat{\pazocal{T}}_{xyz}^{}$ octupole \cite{MurakamiIshihara17}. The reverse effect also takes place when magnetic field is applied  to the $\hat{\pazocal{T}}_{xyz}^{}$ octupole along [001], and a charge distribution of $\hat{\pazocal{O}}_{xy}^{}$ type accompanied by a $\hat{\pazocal{J}}_{z}^{0}$ dipole are induced. For the magnetic field applied along [110] or [111], the situation becomes more complicated, and the induced multipoles cannot be explained by the naive arguments described above for the field along [001]. In the table\,\ref{Tab:PossibleMultipoles} a summary of the possible quadrupolar phases and relevant induced multipoles is given. Therefore, in the $f$\!-electron system a strong coupling exists between electric and magnetic degrees of freedom.

\renewcommand{\arraystretch}{0.8}
\begin{table}[t]
\fontsize{9}{10}\selectfont
\vspace{-25pt}\begin{tabular*}{\textwidth}{@{\extracolsep{\fill}} cccc}\\
\toprule
\small{Field direction}   & \small{Quadrupolar moments} & \small{Induced dipolar moment} & \small{Induced octupolar moment} \\
\toprule
$[0 0 1]$   & $\hat{\pazocal{O}}_{2}^{0}$           & $\hat{\pazocal{J}}_{z}^{}$            & $\hat{\pazocal{T}}_{z}^{\alpha}$   \\
            & $\hat{\pazocal{O}}_{2}^{2}$           & ---\tablefootnote{\label{first} Absence of the induced $\hat{\pazocal{J}}_{z}^{}$ dipolar moment for $\hat{\pazocal{O}}_{2}^{2}$ and ${\pazocal{O}}_{xy}^{}$ quadrupoles in \cite{ShiinaShiba97} contradicts the described earlier arguments from \cite{MurakamiIshihara17}. }
                                                                                            & $\hat{\pazocal{T}}_{z}^{\beta}$                                                     \\
            & $\hat{\pazocal{O}}_{xy}^{}$           & ---\footref{first}                    & $\hat{\pazocal{T}}_{xyz}^{}$                                  \\
            & $\hat{\pazocal{O}}_{yz}^{}$           & $\hat{\pazocal{J}}_{y}^{}$         & $\hat{\pazocal{T}}_{y}^{\alpha}$, $\hat{\pazocal{T}}_{y}^{\beta}$      \\
            & $\hat{\pazocal{O}}_{zx}^{}$           & $\hat{\pazocal{J}}_{x}^{}$         & $\hat{\pazocal{T}}_{x}^{\alpha}$, $\hat{\pazocal{T}}_{x}^{\beta}$      \\
\midrule
$[1 1 0]$   & $\hat{\pazocal{O}}_{2}^{0}$, $\hat{\pazocal{O}}_{xy}^{}$      & $\hat{\pazocal{J}}_{x}^{}+\hat{\pazocal{J}}_{y}^{}$               & $\hat{\pazocal{T}}_{x}^{\alpha}+\hat{\pazocal{T}}_{y}^{\alpha}$, $\hat{\pazocal{T}}_{x}^{\beta}-\hat{\pazocal{T}}_{y}^{\beta}$
\\
            & $\hat{\pazocal{O}}_{2}^{2}$                               & $\hat{\pazocal{J}}_{x}^{}-\hat{\pazocal{J}}_{y}^{}$               & $\hat{\pazocal{T}}_{x}^{\alpha}-\hat{\pazocal{T}}_{y}^{\alpha}$, $\hat{\pazocal{T}}_{x}^{\beta}+\hat{\pazocal{T}}_{y}^{\beta}$
\\
            & $\hat{\pazocal{O}}_{yz}^{}+\hat{\pazocal{O}}_{zx}^{}$         & $\hat{\pazocal{J}}_{z}^{}$                                  &
$\hat{\pazocal{T}}_{xyz}^{}$, $\hat{\pazocal{T}}_{z}^{\alpha}$
\\
            & $\hat{\pazocal{O}}_{yz}^{}-\hat{\pazocal{O}}_{zx}^{}$         & ---\tablefootnote{\label{second} In \cite{MatsumuraYonemura12} (see fig.\,\ref{Fig:FieldDepMultipole}),  presence of the induced $\hat{\pazocal{J}}_{z}^{}$ dipolar moment for $\hat{\pazocal{O}}_{yz}^{}-\hat{\pazocal{O}}_{zx}^{}$  was shown.}               &
$\hat{\pazocal{T}}_{z}^{\beta}$
\\
\midrule
$[1 1 1]$   & $\hat{\pazocal{O}}_{2}^{0}$, $2\hat{\pazocal{O}}_{xy}^{}-\hat{\pazocal{O}}_{yz}^{}-\hat{\pazocal{O}}_{zx}^{}$                 & $2\hat{\pazocal{J}}_{z}^{}-\hat{\pazocal{J}}_{x}^{}-\hat{\pazocal{J}}_{y}^{}$                                                                 & $2\hat{\pazocal{T}}_{z}^{\alpha}-\hat{\pazocal{T}}_{x}^{\alpha}-\hat{\pazocal{T}}_{y}^{\alpha}$, $\hat{\pazocal{T}}_{x}^{\beta}-\hat{\pazocal{T}}_{y}^{\beta}$
\\
            & $\hat{\pazocal{O}}_{2}^{2}$, $\hat{\pazocal{O}}_{yz}^{}-\hat{\pazocal{O}}_{zx}^{}$                                        &
$\hat{\pazocal{J}}_{x}^{}-\hat{\pazocal{J}}_{y}^{}$                                                                                     &
$\hat{\pazocal{T}}_{x}^{\alpha}-\hat{\pazocal{T}}_{y}^{\alpha}$, $2\hat{\pazocal{T}}_{z}^{\beta}-\hat{\pazocal{T}}_{x}^{\beta}-\hat{\pazocal{T}}_{y}^{\beta}$
\\
            & $\hat{\pazocal{O}}_{xy}^{}+\hat{\pazocal{O}}_{yz}^{}+\hat{\pazocal{O}}_{zx}^{}$                                           &
$\hat{\pazocal{J}}_{x}^{}+\hat{\pazocal{J}}_{y}^{}+\hat{\pazocal{J}}_{z}^{}$                                                                  &
$\hat{\pazocal{T}}_{x}^{\alpha}+\hat{\pazocal{T}}_{y}^{\alpha}+\hat{\pazocal{T}}_{z}^{\alpha}$, $\hat{\pazocal{T}}_{xyz}^{}$
\\
\bottomrule
\end{tabular*}
\caption{ Possible order parameters for various magnetic field direction. The horizontal bar represents absence of induced moments. Values are taken from ref.\,\citenum{ShiinaShiba97}.}\vspace{-0pt}
\label{Tab:PossibleMultipoles}\vspace{-25pt}
\end{table}

As the next step it is necessary to introduce pseudo-spin operators  $\boldsymbol{\uptau}^\prime=(\tau_z,\tau_x)$ and $\boldsymbol{\upmu}=(\mu_x,\mu_y,\mu_z)$, consistent with  irreducible representations $\Upgamma_{3g}^{(1)}$ and $\Upgamma_{5g}^{(1)}$ and mixed moment operators between the $\Upgamma_{4u}^{}$ dipole and octupole,
$\boldsymbol{\upsigma}^\prime=(\sigma_x,\sigma_y,\sigma_z)$ and $\boldsymbol{\upeta}=(\eta_x,\eta_y,\eta_z)$. In addition, pure octupolar moments $\boldsymbol{\upzeta}=(\zeta_x,\zeta_y,\zeta_z)$ and $\xi=\tau_y$ with symmetries $\Upgamma_{5u}^{}$ and $\Upgamma_{2u}^{}$ respectively,  defined as \cite{KuramotoKusunose09,ShiinaShiba97,ShiinaSakai98}:
\vspace{-0pt}\begin{align}
\tau_z&\equiv\hat{X}_{2}^{3g}\rightarrow \frac{1}{8}\hat{\pazocal{O}}_{2}^{0},
&
\mu_z&\equiv\hat{X}_{z}^{5g}\rightarrow \frac{1}{8}\hat{\pazocal{O}}_{xy}^{}
\nonumber \\
\sigma_z&\equiv\hat{X}_{z}^{4u}\rightarrow \frac{7}{15}\hat{\pazocal{J}}_{z}^{}-\frac{2}{45}\hat{\pazocal{T}}_{z}^{\alpha},
&
\eta_z&\equiv\hat{X}_{z}^{4u}\rightarrow -\frac{2}{15}\hat{\pazocal{J}}_{z}^{}+\frac{7}{45}\hat{\pazocal{T}}_{z}^{\alpha}
\nonumber \\
\zeta_z&\equiv\hat{X}_{z}^{5u}\rightarrow \frac{\sqrt5}{30}\hat{\pazocal{T}}_{z}^{\beta},
&
\xi&\equiv\hat{X}_{}^{2u}\rightarrow \frac{\sqrt5}{45}\hat{\pazocal{T}}_{xyz}^{}.
\label{eq:SigmaZEtaZ}
\vspace{-0pt}\end{align}

\begin{figure}[!ht]
\includegraphics[width=1\textwidth]{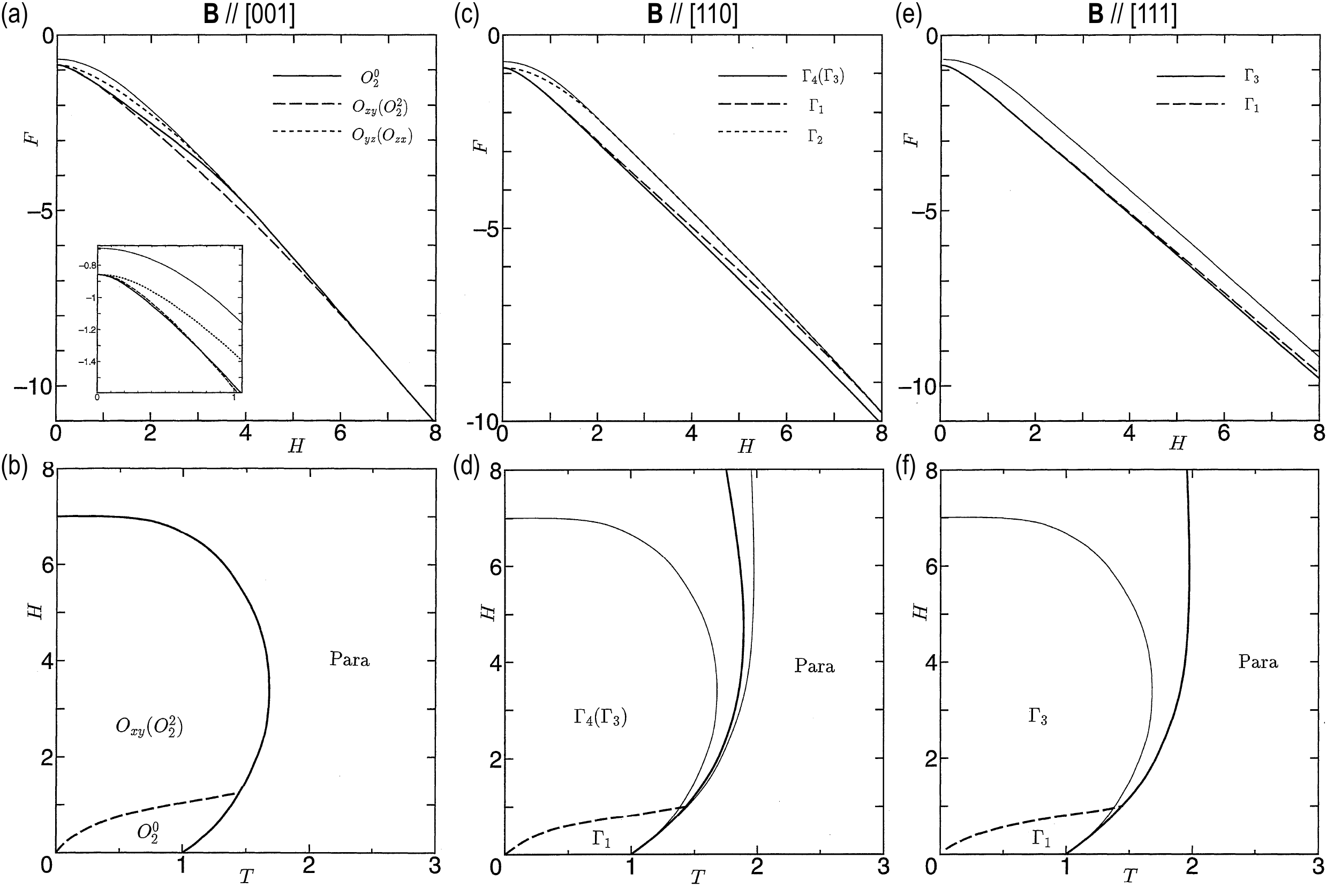}\vspace{-5pt}
\caption{ Free energies  of possible ordered phases at $T=0.5$ for field parallel to $[0 0 1]$, $[1 1 0]$, and $[1 1 1]$ are shown in (a), (c), and (e) respectively. In these panels thin lines represent the free energy of the paramagnetic phase. Phase diagrams for the same field directions are shown in (b), (d), and (f) respectively. In panel (d) the thin line shows for comparison  the transition between ordered and paramagnetic phases for field directions $[0 0 1]$ and $[1 1 1]$. Figure from ref.\,\citenum{ShiinaShiba97}.}
\label{Fig:GLFreeEnergyPhaseDiagram}\vspace{-20pt}
\end{figure}

Using the notations above, the generalized exchange model given by eq.\,\ref{eq:MPHamiltonian} can written as \cite{ShiinaShiba97}:
\vspace{-15pt}\begin{align}
H=D\sum_{i,j}\left(\boldsymbol{\uptau}_i^\prime\cdot\boldsymbol{\uptau}_j^\prime +\boldsymbol{\upmu}_i \cdot \boldsymbol{\upmu}_j +\boldsymbol{\upsigma}_i \cdot\boldsymbol{\upsigma}_j +\boldsymbol{\upeta}_i \cdot \boldsymbol{\upeta}_j +\boldsymbol{\upsigma}_i \cdot \boldsymbol{\upsigma}_j+\boldsymbol{\upsigma}_i \cdot\boldsymbol{\upsigma}_j\right)
\nonumber \\
-\frac{7}{3}g\mu_{\text{B}}\sum_{i}(\boldsymbol{\upsigma}_i+\frac{4}{7}\boldsymbol{\upeta}_i)\cdot\mathbf{H}.
\label{eq:PseudoSpinHamiltonian}
\vspace{-5pt}\end{align}
Analyzing the Hamiltonian in eq.\,\ref{eq:PseudoSpinHamiltonian} by mean-field approximation and comparing Ginzburg-Landau free energy of possible multipoles ordering against the field, the phase diagram for each magnetic field direction can be determined. According to the table\,\ref{Tab:PossibleMultipoles}, five different types of multipole ordering can exist for the field along $[0 0 1]$. However because of the high symmetry of the model, $\hat{\pazocal{O}}_{yz}^{}$ and $\hat{\pazocal{O}}_{zx}^{}$, as well as $\hat{\pazocal{O}}_{2}^{2}$ and $\hat{\pazocal{O}}_{xy}^{}$, are degenerate. When the field is applied along $[1 1 0]$, four types of phases, which are denoted as $\Upgamma_1$, $\Upgamma_4$, $\Upgamma_3$ and $\Upgamma_2$, are possible. Two $\Upgamma_3$ phases and one $\Upgamma_1$ are possible for the field $[1 1 1]$. For a weak field, applied parallel to $[0 0 1]$, the free energy of the $\hat{\pazocal{O}}_{2}^{0}$ phase is lower, as shown in fig.\,\ref{Fig:GLFreeEnergyPhaseDiagram}\,(a), therefore suggesting that $\hat{\pazocal{O}}_{2}^{0}$ ordering takes place. At higher fields the situation changes, and the free energy of the $\hat{\pazocal{O}}_{2}^{2}$ ($\hat{\pazocal{O}}_{xy}^{}$) becomes lower. Consequently, at intermediate field, a transition from the $\hat{\pazocal{O}}_{2}^{0}$ to $\hat{\pazocal{O}}_{2}^{2}$ ($\hat{\pazocal{O}}_{xy}^{}$) phases takes place, as shown in fig.\,\ref{Fig:GLFreeEnergyPhaseDiagram}\,(b). Free energies of possible ordered phases, together with corresponding phase diagram for other field directions $[1 1 0]$ and $[1 1 1]$, are shown in figs.\,\ref{Fig:GLFreeEnergyPhaseDiagram}\,(c,d) and (e,f), respectively. It is easy to notice the tendency of ordered phase stabilization with respect to the field direction, since $T_\text{Q}[001]<T_\text{Q}[110]<T_\text{Q}[111]$. This effect is explained by the presence of the field-induced dipolar and octupolar moments, which tend to stabilize quadrupolar ordering \cite{ShiinaShiba97}.

The described model is a good starting point for the analysis of the multipolar interaction, however it does not fully reproduce all experimental facts known about CeB$_6$. In particular, the large stable region of the $\hat{\pazocal{O}}_{2}^{2}$ or  $\hat{\pazocal{O}}_{xy}^{}$ phase for the field along $[0 0 1]$ or degeneracy of the zero-field ordered state. To achieve better consistency with the experiment an improved model was also proposed by authors in ref.\,\citenum{ShiinaShiba97}. By introducing additional parameters, which lift the degeneracy of the $\Upgamma_5$ type ordered phase in zero field and affect the stability of the $\hat{\pazocal{O}}_{2}^{2}$ or  $\hat{\pazocal{O}}_{xy}^{}$ phase, improved phase diagrams were obtained \cite{ShiinaShiba97,ShiinaSakai98}.

In addition to the mentioned above ordering of quadrupole moments in CeB$_6$ with a propagation vector $\mathbf Q=(\frac{1}{2}\,\frac{1}{2}\,\frac{1}{2})$, labeled as phase II (see sec.\,\ref{SubSec:OrderingPhenomenaCeB6}), another low-temperature phase III exists. It represents the AFM ordering with a double-$\mathbf{k}$ (or $2\mathbf{k}-\mathbf{k}^{\prime}$) commensurate structure with the propagation vectors $\mathbf k_1^{}=(\frac{1}{4}\,\frac{1}{4}\,0)$, $\mathbf k_1^{\prime}=(\frac{1}{4}\,\frac{1}{4}\,\frac{1}{2})$ and $\mathbf k_2^{}=(\frac{1}{4}\,\bar{\frac{1}{4}}\,0)$, $\mathbf k_2^{\prime}=(\frac{1}{4}\,\bar{\frac{1}{4}}\,\frac{1}{2})$ \cite{EffantinRossat-Mignod85}. However, the present understanding of ordering phenomena within the phase III is a bit obscure. Magnetic ordering with these propagation vectors should be related to the ordering that takes place within phase II, as $\mathbf k_1^{}+\mathbf k_1^{\prime}=\mathbf k_2^{}+\mathbf k_2^{\prime}=\mathbf Q$, and requires detailed calculations on the basis of mixed multipoles \cite{ShiinaSakai98}.
\clearpage
\vspace{-0pt}
It is known that a non-collinear spin orientation does not gain energy from the nearest-neighbor exchange interactions, therefore suggesting the investigation of conditions that stabilize such ordering under next-nearest neighbor interactions of the multipole moments given in table\,\ref{Tab:Multipoles} under the antiferromagnetic $\hat{\pazocal{O}}_{xy}^{}$ ordering \cite{KuramotoKusunose00,KusunoseKuramoto01,KusunoseKuramoto01a}. The Hamiltonian used in this case is written in the same way as it was introduced in eq.\,\ref{eq:MPHamiltonian} with the interaction given by:
\begin{wrapfigure}[11]{h}{0.40\textwidth}\vspace{-15pt}
\includegraphics[width=0.40\textwidth]{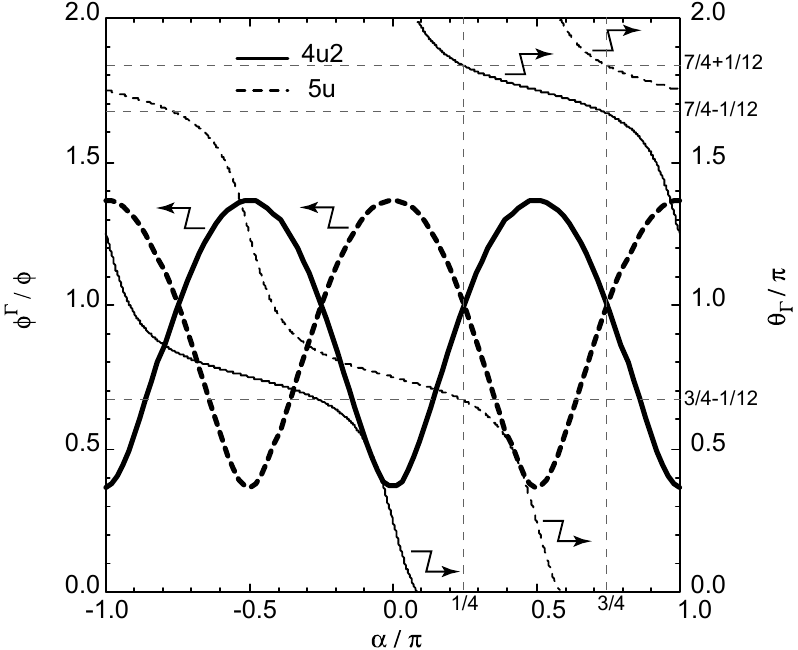}\vspace{-5pt}
\caption{ The molecular fields, where solid (dotted) line corresponds to the dipole (octupole) moment. Figure from ref.\,\citenum{KusunoseKuramoto01}.}
\label{Fig:MolecularFieldMultipole4u2and5u}\vspace{-0pt}
\end{wrapfigure}\vspace{-15pt}
\[
D_{\Upgamma}^{}(\mathbf{q})=
\left\{\vphantom{\begin{array}{@{}c@{}} \strut \\ \strut\end{array}}\right.\kern-\nulldelimiterspace
\begin{aligned}\vspace{-0pt}
    &J_{\Upgamma}^{}(\mathbf{q}),                              & &(\Upgamma=2u,\, 3g,\, 5g)\\
    &J_{\Upgamma}^{}(\mathbf{q})+K_{\Upgamma}^{}(\mathbf{q}),  & &(\Upgamma=4u1,\, 4u2,\, 5u), \vspace{0pt}
\label{eq:InteractionDij}\end{aligned}
\]
where $J_{\Upgamma}(\mathbf{q})_{\gamma \gamma^{\prime}}=-J_{\Upgamma}\delta_{\gamma \gamma^{\prime}}(\text{cos}\,q_x+\text{cos}\,q_y+\text{cos}\,q_z)/{3}$ defines the nearest neighbor interaction for the cubic lattice and $K_{ij}^{\Upgamma \gamma \gamma^{\prime}}=-K_{\Upgamma}^{}(\delta_{}^{\gamma \gamma^{\prime}}-n_{ij}^{\gamma}n_{ij}^{\gamma^{\prime}})/12$ account for the next-nearest neighbor $i$ and $j$ sites. The Ginzburg-Landau free energy functional consists of the entropy of the non-interacting system and mode coupling $f_\text{mc}$ between multipole fluctuation, given by:
\vspace{-0pt}\begin{equation}
f_\text{mc}=-\frac{1}{\beta N}\sum_{i}
\left(
\text{ln}\left[\text{Tr}_i\text{exp}(\beta\phi_{i}^{A}X_{i}^{A})/4\right]-\frac{\beta^2}{2}\phi_{i}^{A}\phi_{i}^{A}
\right)
\label{eq:MultipoleCoupling}
\vspace{-0pt}\end{equation}
where  $\phi$ is the molecular field and $N$ is the number of cites.

\begin{figure}[!hb]
\includegraphics[width=1\textwidth]{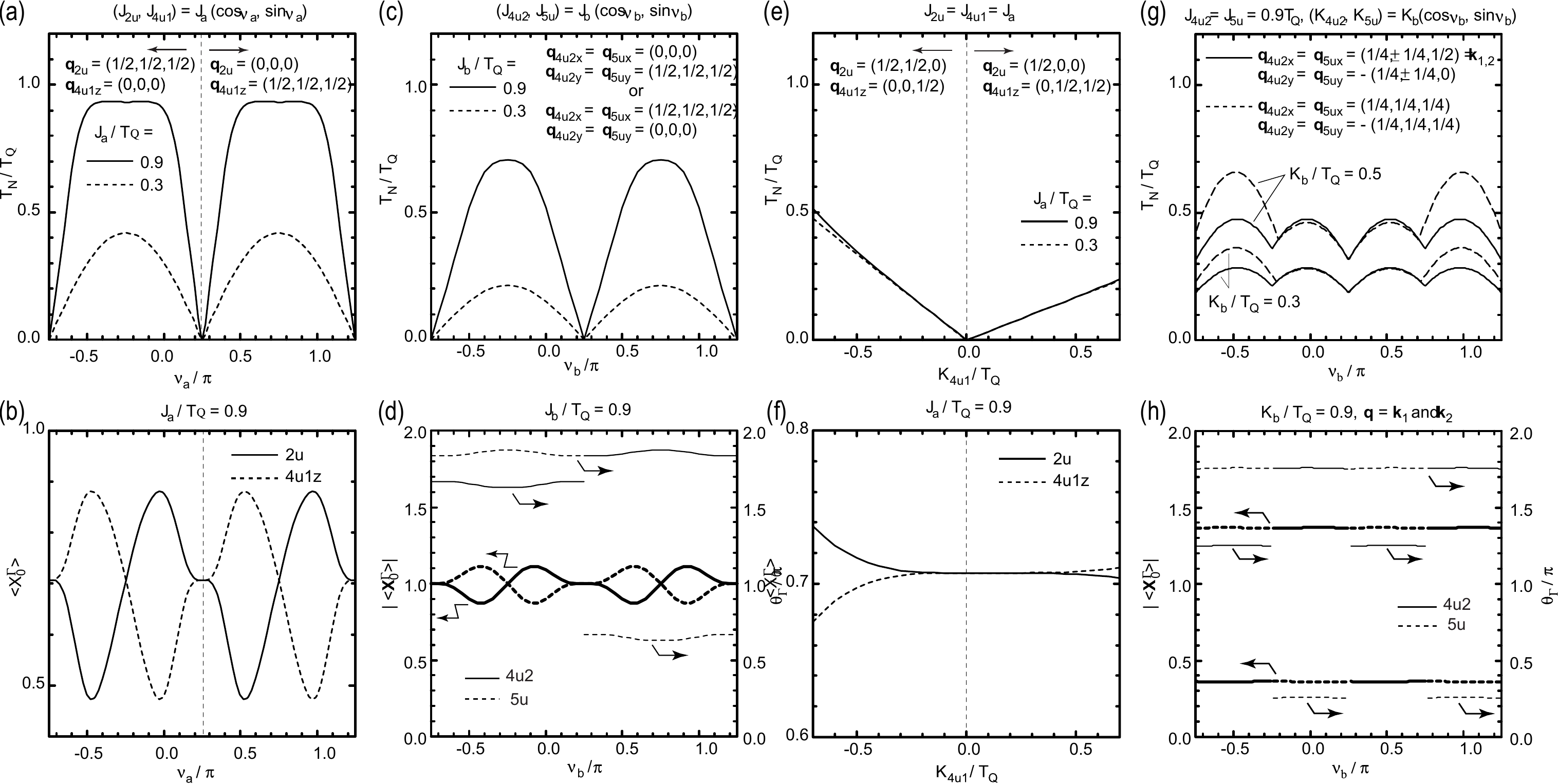}\vspace{-5pt}
\caption{ The lower transition temperature (a, c, e, g) and order parameter (b, d, f, h) for various interactions. Figure from ref.\,\citenum{KusunoseKuramoto01}.}
\label{Fig:TransitionTemperatureMO}\vspace{-15pt}
\end{figure}

As the first step, intersite interactions between the $\Upgamma_{4u2}$ dipole and $\Upgamma_{5u}$ octupole moments are considered. Molecular fields $\phi$ shown in fig.\,\ref{Fig:MolecularFieldMultipole4u2and5u} indicate that the easy axis of the dipole moment would correspond to the hard axis of the octupole moment. Specifying $K_{\Upgamma}=0$, the nearest neighbor interaction between the $\Upgamma_{2u}$ octupole and $\Upgamma_{4u1}$ dipole can be investigated. Parameterizing the free energy by $(J_{2u},J_{4u1})=J_{a}(\text{cos}\,\nu_a,\text{sin}\,\nu_a)$, a phase transition can be determined with two possible propagation vectors: $\mathbf{Q}$ or $\mathbf{0}=(0 0 0)$, as shown in fig.\,\ref{Fig:TransitionTemperatureMO}(a). Dependence of the order parameter $X_{i}^{\Upgamma}$,  shown in fig.\,\ref{Fig:TransitionTemperatureMO}(b),  suggests that when $\nu_a~<~\pi/4$, antiferro-octupolar  $X_{\mathbf{Q}}^{2u}$ and ferro-dipolar $X_{\mathbf{0}}^{2u1z}$ orders exist, while the ordering vector is exchanged when $\nu_a>\pi/4$. The nearest neighbor interaction between the $\Upgamma_{4u2}$ dipole and $\Upgamma_{5u}$ octupole, parameterized by $(J_{4u2},J_{5u})=J_{b}(\text{cos}\,\nu_b,\text{sin}\,\nu_b)$, determines a phase transition, shown in fig.\,\ref{Fig:TransitionTemperatureMO}(c,d).

\vspace{-0pt}

The phases described above would not lead to the observed non-collinear ordering with propagation vectors $\mathbf{k}_i$ and $\mathbf{k}_i^{\prime}$, therefore it is necessary to take into account next-nearest neighbor interactions. Considering the limiting case, when there is a perfect competition between the nearest neighbor interactions  ($J_{4u2}=J_{5u}$ and $J_{2u}=J_{4u1}$), another transition to an ordered phase with propagation vectors $\mathbf{Q}_z=(0\,0\,\frac{1}{2})$  or $\mathbf{Q}_{yz}=(0\,\frac{1}{2}\,\frac{1}{2})$ arises, as shown in fig.\,\ref{Fig:TransitionTemperatureMO}(e). Dipolar $\Upgamma_{4u1}$ order along $(0\,0\,\frac{1}{2})$ is implemented  for $K_{4u1}<0$, fig.\,\ref{Fig:TransitionTemperatureMO}(f). Although this type of magnetic order does not correspond to any static ordering in pure  CeB$_6$, a significant amount of spectral weight was found  around $\mathbf{Q}_z$ (see sec.\,\ref{SubSec:DiffusefluctuationsCeLaB6}). Indeed, AFM order with such a propagation vector was found  in Ce$_{1-x}$Nd$_{x}$B$_6$ for Nd doping levels $x\geq0.4$ \cite{MignotRobert09}. According to fig.\,\ref{Fig:TransitionTemperatureMO}(g), when the next-nearest neighbor interaction is parameterized as $(K_{4u2},K_{5u})=K_{b}(\text{cos}\,\nu_b,\text{sin}\,\nu_b)$, with $(J_{4u2},J_{5u})=0.9T_\text{Q}$, ordering with propagation vectors  $\mathbf{Q}^{\prime}=(\frac{1}{4}\,\frac{1}{4}\,\frac{1}{4})$, $\mathbf{k}_i$ or $\mathbf{k}_i^{\prime}$ is induced. Double-$\mathbf{k}$ ordering, found in CeB$_6$, is realized for $-\pi/4<\nu_b<\pi/4$, as shown in fig.\,\ref{Fig:TransitionTemperatureMO}(h).

\begin{wrapfigure}[13]{r}{0.45\textwidth}\vspace{-10pt}
    \vspace{-0pt}\includegraphics[width=\linewidth]{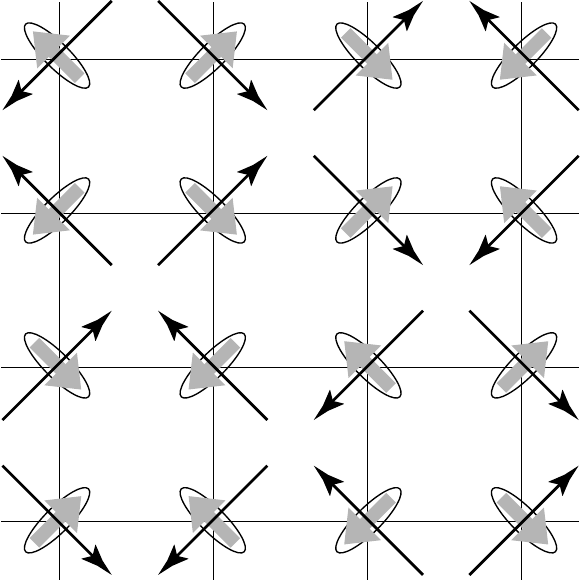}\vspace{0pt}
    \vspace{-20pt}\caption{The~ordering~pattern~for~dipole~and octupole moments. Fig. was taken from ref.\,\citenum{KusunoseKuramoto01}.}\label{Fig:PhaseIIIOrdering}\vspace{-0pt}
\end{wrapfigure}
It becomes clear that the presence of the pseudo-dipole-type interaction for the next-nearest neighbor dipole and octupole moments is the necessary condition for the existence of the ordering, observed within phase III in CeB$_6$. The real space ordering pattern, that corresponds to the double-$\mathbf{k}$ structure which is consistent with the AFQ ordering, is shown in fig.\,\ref{Fig:PhaseIIIOrdering} for the mean-field solution $-\pi/4<\nu_b<\pi/4$ in the plane $z=0$.  The ovals represent $\hat{\pazocal{O}}_{xy}^{}$ quadrupoles, and $\Upgamma_{4u2}$ dipole and $\Upgamma_{5u}$ octupole moments are shown with thin and thick arrows respectively. In the case of $\nu_b<-\pi/4$, the ordering pattern should remain the same, however the magnitude of the octupole moment is enlarged  due to the stronger octupole-octupole interaction.

\subsection{Hidden order}\label{SubSec:HiddenOrder}
In sec.\,\ref{SubSec:NuclearAndMagneticScattering} we proved that neutrons can interact with the  magnetic moment of the atom within the dipole approximation. Now we realize, that the interaction between multipoles plays a great role in some $f$\!-electron systems, and therefore it is necessary to understand the possible outcome from the interaction between the multipole and neutron and whether multipolar ordering is observable in experiments. First of all we have to recall that the double-differential cross-section, given by eq.\,\ref{eq:MasterFormulaWithPotential}, consists of the magnetic interaction vector $\mathbf{D}_{\bot}=\hat{\mathbf{Q}}\times\mathbf{D}\times\hat{\mathbf{Q}}$. Within the dipole approximation for small $\mathbf{Q}$, the unprojected operator $\mathbf{D}$ can be  expressed with the total angular momentum operator: $\mathbf{D}\sim\frac{g}{2}f(\mathbf{Q})\mathbf{J}$.  In case of higher order multipoles, the correction to the dipole approximation of the component $\mathbf{D}$ can be expressed as \cite{KuramotoKusunose09,ShiinaSakai07,SantiniCarretta09}:
\vspace{-0pt}\begin{equation}
\hat{D}_{i}^{}(\mathbf{Q})=\frac{g}{2}f(\mathbf{Q})\left(\hat{J}_{i}^{}+\sum_{\alpha }c_{i\alpha}^{}(\mathbf{Q})\hat{X}_{i}^{\alpha}\right),
\label{eq:MagneticInteractionMultipoles}
\vspace{0pt}\end{equation}
where $\hat{X}_{i}^{\alpha}$ denotes a component of magnetic multipoles, introduced above. However it does not mean that neutrons can probe all possible multipoles. Only time-odd multipoles, that represent magnetic multipoles, are visible in magnetic neutron scattering \cite{Lovesey15}. Therefore odd-rank multipoles can be probed, and in the first approximation dipolar (rank 1) and octupole (rank 3) moments have to be considered. Quadrupole ordering, represented with a rank 2 operator, is invisible to neutron scattering and is therefore usually referred to as a  ``magnetically hidden'' order{\footnotemark} \cite{JackeliKhaliullin09}.
\footnotetext{Such an order can be nevertheless clearly observed in thermodynamic probes. In particular in Sr$_{2}$VO$_{4}$, no magnetic order in the low-temperature phase could be confirmed with INS \cite{CyrotLambert-Andron90}, however, the observed anomalies in magnetic susceptibility \cite{ZhouConner07} suggested existence of this ``magnetically hidden'' order \cite{JackeliKhaliullin09}.}

\vspace{-20pt}
\subsection{Magnetic form factor in the case of multipolar ordering }\label{SubSec:MagneticFFMO}

\enlargethispage{15pt}
\vspace{-10pt}
\begin{wrapfigure}[23]{r}{0.4\textwidth}\vspace{-15pt}
    \vspace{-0pt}\includegraphics[width=\linewidth]{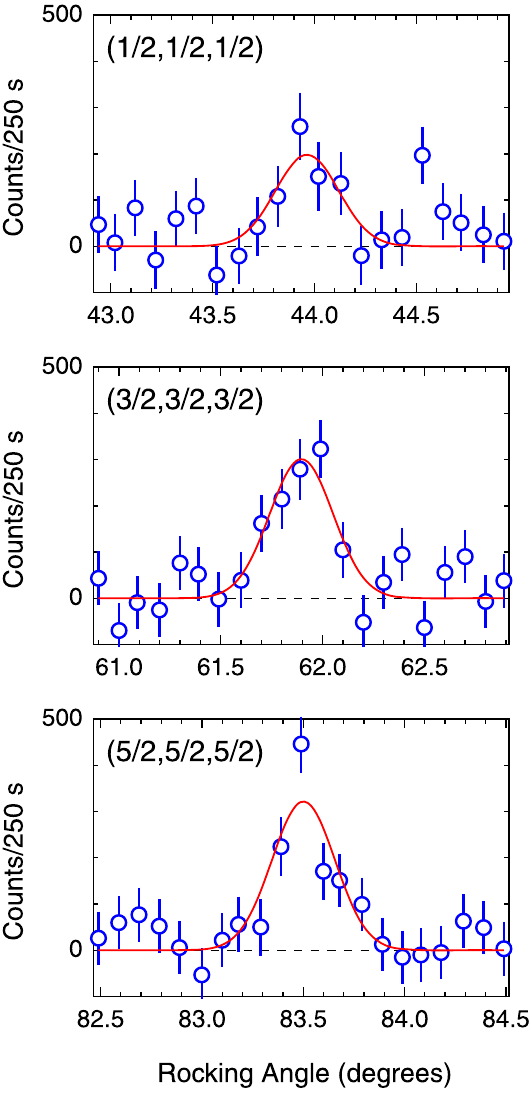}\vspace{0pt}
    \vspace{-25pt}\caption{ Rocking scans of the $(\frac{h}{2}\,\frac{h}{2}\,\frac{l}{2})$ reflections, which emphasize intensity increase at larger $\mathbf{Q}$. Solid lines are Gaussian fits. Figure was taken from ref.\,\citenum{KuwaharaIwasa07}.}\label{Fig:RXDE1E2}\vspace{-0pt}
\end{wrapfigure}
In  sec.\,\ref{SubSec:FormFactorAnalyticalApproximation} an analytical expression for the magnetic form factor was derived, as  it is an important part of neutron scattering and can be used as the easiest way to determine the magnetic nature of the signal. However there are experimental results where an unusual $\mathbf{Q}$ dependence of the $(\frac{h}{2}\,\frac{h}{2}\,\frac{l}{2})$ reflections in Ce$_{0.7}$La$_{0.3}$B$_6$ was observed, and was assumed to represent ordering of the $\hat{\pazocal{T}}_{}^{\beta}$ octupoles.  Unlike the expected monotonic suppression of the signal for the dipolar interaction towards larger scattering vectors,  enhancement of the scattering intensity was observed \cite{KuwaharaIwasa07}. Such a nonmonotonic $\mathbf{Q}$-dependence is expected for multipolar moments.

We have already shown that the scattering operator $\mathbf{D}$  can be approximated as a linear function of $\mathbf{J}$ only within the dipole approximation, while in order to study multipolar interactions it should be further expanded. This was done in eq.\,\ref{eq:MagneticInteractionMultipoles}, therefore in order to obtain magnetic form factor in case of multipoles it is important to derive $\langle \tau_i|\mathbf{D}_{\bot\alpha}^{\dag}| \tau_f \rangle$ from eq.\,\ref{eq:MasterFormulaWithPotential}. Detailed derivation of the corresponding expression can be found in textbooks or papers \cite{Lovesey86,ShiinaSakai07,Shiina12}, and the matrix elements can be written as:
\vspace{-0pt}\begin{align}
\langle jm|\mathbf{D}_{\bot}^{\dag}| jm^{\prime} \rangle&=\sum_{m_1 m_2}\sum_{\sigma_1 \sigma_2}\langle lm_1s\sigma_1|\mathbf{D}_{\bot}^{\dag}| lm_2s\sigma_2 \rangle
\times\langle jm|lm_1s\sigma_1\rangle
\langle lm_2s\sigma_2|jm^{\prime}\rangle,
\label{eq:MatrixElementMP}
\vspace{-0pt}\end{align}
where $\tau=\pm$ and $\sigma=\uparrow,\downarrow$ are orbital and spin indices of the $|\Upgamma_8\rangle$ wave functions of Ce$^{3+}$ ions given in eq.\,\ref{eq:EigenfunctionsG7G8}.

\begin{wrapfigure}[11]{r}{0.45\textwidth}\vspace{-15pt}
    \vspace{-0pt}\includegraphics[width=\linewidth]{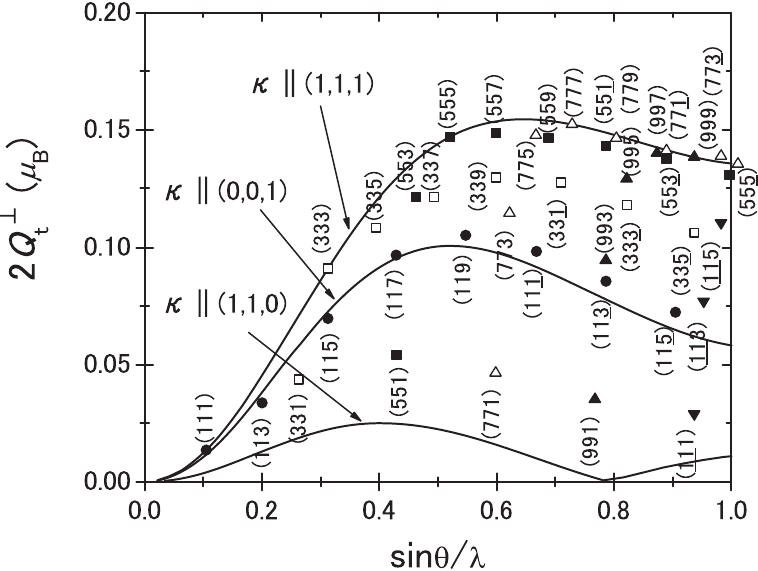}\vspace{0pt}
    \vspace{-20pt}\caption{ $\mathbf{Q}$ dependence of the magnetic structure factor for nondipolar order.  Figure was taken from ref.\,\citenum{ShiinaSakai07}.}\label{Fig:FFMultipole}\vspace{-0pt}
\end{wrapfigure}\vspace{-0pt}
In fig.\,\ref{Fig:FFMultipole} the magnetic structure factor of  magnetic reflections as a function of neutron momentum transfer is shown. As the magnetization density of octupoles and dipoles is significantly different,  the form factor for the octupolar ordering has a maximum value at finite $\mathbf{Q}$, while in case of dipolar ordering it should have a maximum at $\mathbf{Q}=0$ (see sec.\,\ref{SubSec:FormFactorAnalyticalApproximation}). This result can be used as an evidence for the presence of a high-rank interaction and can help to distinguish dipoles from octupoles with neutron scattering.

\vspace{-0pt}
\subsection{Resonant X-ray diffraction as a tool to directly probe multipolar orbital ordering}\label{SubSec:ResonantXrayDiffraction}\vspace{-10pt}
\begin{wrapfigure}[10]{r}{0.45\textwidth}\vspace{-15pt}
    \vspace{-0pt}\includegraphics[width=\linewidth]{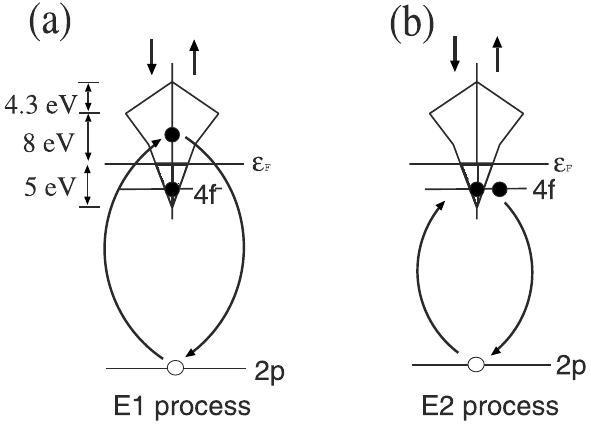}\vspace{0pt}
    \vspace{-20pt}\caption{E1 (a), and E2 (b) process at the Ce $L_2$ and $L_3$ edges. Figure was taken from ref.\,\citenum{NagaoIgarashi10}.}\label{Fig:RXDE1E2}\vspace{-0pt}
\end{wrapfigure}
As we can see, direct observation of the multipolar orbital ordering with neutron scattering is a difficult task. Another way to probe multipole ordering is to use resonant X-ray diffraction (RXD), as this technique is able to detect multipole moments up to the fourth rank \cite{LoveseyBalcar05,MurakamiIshihara17}. In case of RXS, the energy of incoming photons is tuned to the value of one of the absorption edges, therefore inducing a transition of the core-level electrons to some unoccupied levels. The decay process emits new photons with the same energy but with different polarization. This process is schematically shown in fig.\,\ref{Fig:RXDE1E2}. Detailed description of the interaction between photons and electrons goes beyond my dissertation and can be found in numerous text books, therefore here I will focus only on the results. The transition matrix which describes interaction of the photon with the electrons includes terms E1, E2, E3, M1, M2, T1,  which describe  contributions from the electric dipole, electric quadrupole, electric octupole, magnetic dipole, magnetic quadrupole and toroidal dipole respectively. It should be emphasized that the
\begin{wrapfigure}[30]{r}{0.45\textwidth}\vspace{-0pt}
    \vspace{-0pt}\includegraphics[width=\linewidth]{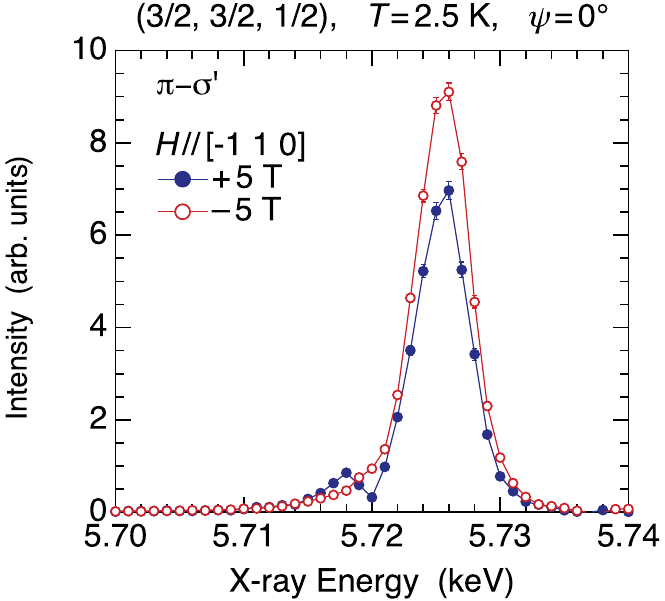}\vspace{0pt}
    \vspace{-25pt}\caption{ CeB$_6$ X-ray energy spectra at $H=\pm5\,\text{T}$. Figure was taken from \cite{MatsumuraYonemura12}.}\label{Fig:RXDSpectrum1}\vspace{-0pt}
    \vspace{-5pt}\includegraphics[width=\linewidth]{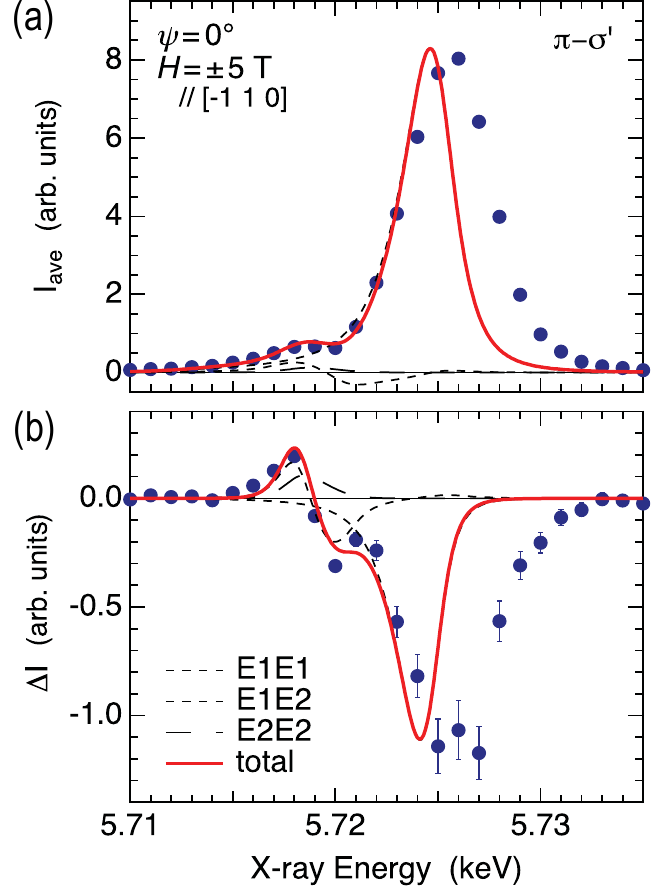}\vspace{0pt}
    \vspace{-25pt}\caption{ Average (a) and difference (b) spectra of the signal, shown in fig.\,\ref{Fig:RXDSpectrum1}. Fits are shown with the solid lines, while spectral components by the dotted, single-dotted, and double-dotted lines. Figure was taken from ref.\,\citenum{MatsumuraYonemura12}.}\label{Fig:RXDSpectrum2}\vspace{-0pt}
\end{wrapfigure}

\noindent  mentioned terms are multipoles of the vector potential of the x-ray beam interacting with a core electron \cite{Matteo12}, and should not be confused with the multipole expansion given by eq.\,\ref{eq:EandMMultipole}. The matrix elements E1 and E2 can be expanded into dipole and quadrupole components, and new dipole-dipole (E1-E1) rank\,2, dipole-quadrupole (E1-E2) rank\,3, and quadrupole-quadrupole (E2-E2) rank\,4 tensors are introduced \cite{BeaurepaireBulou10}.  Access to different contributions depends on the point symmetry of the material and is probed by polarization analysis near the absorption edges \cite{JolyMatteo12}.

Therefore RXD can be used to survey multipole ordering. In CeB$_6$ the ordering within phase II was studied around $L_3$ absorption edge, where E1 ($2p_{3/2}\leftrightarrow 5d$) and E2 ($2p_{3/2}\leftrightarrow 4f$) resonances were measured on the $(\frac{3}{2}\,\frac{3}{2}\,\frac{1}{2})$ reflection. Typical X-ray spectra, measured for $H=\pm5\,\text{T}$, are shown in fig.\,\ref{Fig:RXDSpectrum1}, where difference between signals for the reverse field direction can be found. This can be understood by considering interference between E1 and E2 terms, as the intensity of the signal is proportional to the square of the sum of the amplitudes, and not to the sum of squares of each component. In addition, the magnetic dipole and octupole tensors change their sign  with the field reversal, unlike electric quadrupole and hexadecapole tensors \cite{MatsumuraYonemura09}.

The observed signal is proportional to $|F_\text{E1}(\omega)+F_\text{E2}(\omega)|^2$, where $F_\text{E1}$ and $F_\text{E2}$ describe the resonant scattering amplitudes for the E1 and E2 processes, given by \cite{MatsumuraYonemura12}:
\vspace{-0pt}\begin{align}
F_\text{E1}(\omega)&=\sum_{\nu=0}^{2}\alpha_{\text{E1}}^{\nu}(\omega)\sum_{\mu=1}^{2\nu+1}P_{\text{E1},\mu}^{(\nu)}(\varepsilon,\varepsilon^{\prime}) \langle z_{\mu}^{(\nu)}\rangle, \nonumber
\\
F_\text{E2}(\omega)&=\frac{k^2}{9}\sum_{\nu=0}^{4}\alpha_{\text{E2}}^{\nu}(\omega)\sum_{\mu=1}^{2\nu+1}P_{\text{E2},\mu}^{(\nu)}(\varepsilon,\varepsilon^{\prime},\mathbf{k},\mathbf{k}^{\prime}) \langle z_{\mu}^{(\nu)}\rangle,
\label{eq:IrreducibleRepresentations}
\vspace{-0pt}\end{align}\vspace{0pt}
where $z_{\mu}^{(\nu)}$ is the operator equivalent of the rank-$\nu$ multipole tensor from table\,\ref{Tab:Multipoles}, and $P_{\text{E1},\mu}$, $P_{\text{E2},\mu}$ are the geometrical factors of the component $\mu$ and rank-$\nu$ E1 and E2 resonances, respectively. Rank-dependent energy profiles are given by $\alpha_{\text{E1}}^{\nu}(\omega)$, $\alpha_{\text{E2}}^{\nu}(\omega)$.

\begin{wrapfigure}[14]{r}{0.45\textwidth}\vspace{-5pt}
    \vspace{-5pt}\includegraphics[width=1.1\linewidth]{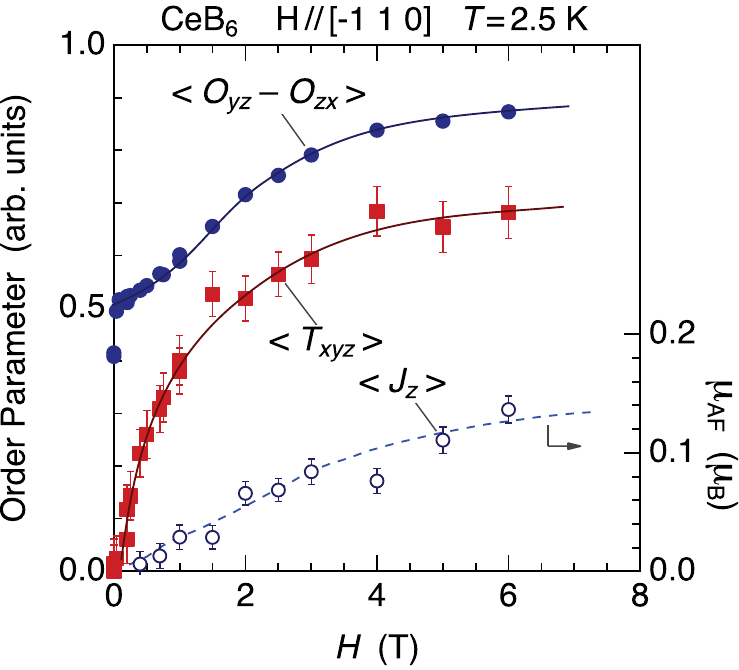}\vspace{0pt}
    \vspace{-25pt}\caption{Magnetic-field dependencies of the  multipole order parameters. Figure was taken from ref.\,\citenum{MatsumuraYonemura12}.}\label{Fig:FieldDepMultipole}\vspace{-0pt}
\end{wrapfigure}
Variation of $\langle z_{\mu}^{(\nu)}\rangle$ under the influence of external parameters, like magnetic field or its direction as well as  temperature, is at the heart of this technique, hence it allows to perform fitting of the measured spectra and extract the multipole order parameters. Accurate description of the spectral functions used for the fitting procedure can be found in ref.\,\citenum{MatsumuraYonemura12}, while fitting results for average and difference spectra are shown in figs.\,\ref{Fig:RXDSpectrum2}\,(a) and (b) respectively. Dependencies of the field-induced dipolar $J_z$, octupolar $T_{xyz}$ and quadrupole $O_{yz}-O_{zx}$ moments, determined from the RXS spectra fits, are shown in fig.\,\ref{Fig:FieldDepMultipole}. In agreement with the described above theories, magnetic field induces dipole and octupole moments, which stabilize the ordered phase and increase the transition temperature.

\section{Spin-Wave Excitations}\label{Sec:SpinExcitations}

In previous sections we have considered the importance of the interaction between electrons, which leads to the presence of static magnetic order with various propagation vectors. In addition, the expression for the neutron double-differential scattering cross-section, given by eq.\,\ref{eq:MasterFormulaMagneticScattering}, contains a component which accounts for dynamics of the magnetic properties in the crystal. In magnetic materials, time dependent spin fluctuations can form collective excitations, known as  spin waves or magnons, which are analogous to collective lattice  excitations known as phonons. The simplest way to understand the existence of spin-wave excitations is to consider the localized spin model, where interaction between spins can be written with the Heisenberg Hamiltonian:
\vspace{-0pt}\begin{equation}
H=-J\sum_{\mathbf{i},\mathbf{j}}\mathbf{S}_{\mathbf{i}}\cdot\mathbf{S}_{\mathbf{j}}.
\label{eq:HeisenbergModel}
\vspace{0pt}\end{equation}
Here $J$ is the coupling constant, and $\mathbf{S}_{\mathbf{i}}$ should not be restricted only to spins, but in case of rare-earth ions might also apply to the total angular momenta $\mathbf{J}$. Important contributions to the given Hamiltonian, like the single-ion anisotropy term $\sum D (S_{i}^{z})^2$, which can arise from spin-orbit coupling, or Zeeman energy of the spins in an external field $g\mu_{\text B}\sum \mathbf{H}\cdot\mathbf{S}_{\mathbf{i}}$ will be considered later, while now we will briefly discuss the dispersion expression of spin waves in  ferro- and antiferromagnets.

\subsection{Ferro- and antiferromagnets}\label{SubSec:FerroAntoferroMagnet}\vspace{-0pt}

Although the Heisenberg Hamiltonian given by eq.\,\ref{eq:HeisenbergModel} is spherically symmetric without external magnetic field, it is known to have a ferromagnetic or antiferromagnetic ground state, spontaneously lowering the symmetry. This situation is known as spontaneous symmetry breaking, and at first it is necessary to obtain the expression for elementary excitations out of this ground state. The total energy of the lattice can be obtained as a sum of bond terms $E_{\mathbf{ij}}$, where $\mathbf{i}$ and $\mathbf{j}$ are nearest neighbors. Taking into account that the energy can be minimized when two spins are parallel ($|\mathbf{S}_{\mathbf{i}}+\mathbf{S}_{\mathbf{j}}|=2S$), the ground state energy of the lattice can be written as \cite{Fazekas99}:
\vspace{-5pt}\begin{align}
\pazocal{E}_{\text{gr.st}}^{\text{FM}}\geq\sum_{\mathbf{i},\mathbf{j}}(E_{\mathbf{ij}})_{\text{min}}=
\sum_{\mathbf{i},\mathbf{j}}(J\mathbf{S}_\mathbf{i}\cdot\mathbf{S}_\mathbf{j})_{\text{min}}=
\sum_{\mathbf{i},\mathbf{j}}(|J|S(S+1)-\frac{|J|}{2}(\mathbf{S}_\mathbf{i}+\mathbf{S}_\mathbf{j})^2)_{\text{min}}
\nonumber \\
=\sum_{\mathbf{i},\mathbf{j}}\left(|J|[S(S+1)-S(2S+1)]\right)_{\text{min}}=
\sum_{\mathbf{i},\mathbf{j}}-|J|S^2=-\frac{1}{2}N|J|zS^2,
\label{eq:GroundState}
\vspace{-5pt}\end{align}
where $z$ is the number of nearest neighbors and $N$ is the number of lattice sites. In order to create the first excited state we have to select an arbitrary site $i$ and turn its spin one step down, from $S_\mathbf{i}^{z}=S$ to $S_\mathbf{i}^{z}=S-1$. As the Hamiltonian given by eq.\,\ref{eq:HeisenbergModel} can be rewritten with the operators $S_\mathbf{i}^{\pm}$, the spin state at the position $\mathbf{i}$ will increase to its ground state value while turn of a spin one step down occurs on site $\mathbf{j}$, when operators $S_{\mathbf{i}}^{+}S_{\mathbf{j}}^{-}$ act on the  described above spin state  wave function. This would lead to the sequential rotation of the spin at the next position and can be considered as a spin wave.

A collective excitation of the spin structure in a crystal lattice is called a magnon. Magnons are quantized magnetization density waves, just as phonons
are quantized lattice waves. Therefore it is possible to express spin operators using creation and destruction boson operators $a_{\mathbf{j}}^{\dag},\,a_{\mathbf{j}}^{}$, known as the \textit{Holstein-Primakoff} transformation \cite{HolsteinPrimakoff40}:
\vspace{-5pt}\begin{equation}
S_{\mathbf{j}}^{+}=\sqrt{2S}\left(1-\frac{a_{\mathbf{j}}^{\dag}a_{\mathbf{j}}^{}}{2S} \right)^{\frac{1}{2}}a_{\mathbf{j}}^{};
\hspace{20pt}
S_{\mathbf{j}}^{-}=\sqrt{2S}a_{\mathbf{j}}^{\dag}\left(1-\frac{a_{\mathbf{j}}^{\dag}a_{\mathbf{j}}^{}}{2S} \right)^{\frac{1}{2}}\mkern-9mu;
\hspace{20pt}
S_{\mathbf{j}}^{z}=S-a_{\mathbf{j}}^{\dag}a_{\mathbf{j}}^{}.
\label{eq:HolsteinPrimakoffTransformation}
\vspace{-0pt}\end{equation}
To describe propagating excitations, it is necessary to introduce the $\mathbf{k}$-space Bloch operators:
\vspace{-5pt}\begin{equation}
a_{\mathbf{j}}^{}=\frac{1}{\sqrt{N}}\sum_{\mathbf{k}}\text{e}^{-\mathbf{k}\cdot{\mathbf{j}}}b_{\mathbf{k}}^{};
\hspace{30pt}
a_{\mathbf{j}}^{\dag}=\frac{1}{\sqrt{N}}\sum_{\mathbf{k}}\text{e}^{\mathbf{k}\cdot{\mathbf{j}}}b_{\mathbf{k}}^{\dag}.
\label{eq:BlochRepresentation}
\vspace{-0pt}\end{equation}
Using the described transformations the Heisenberg Hamiltonian can be written as \cite{Roessler09,Fazekas99,PrabhakarStancil09}:
\vspace{-15pt}\begin{align}
\pazocal{H}=-J\sum_{\mathbf{i},\mathbf{j}}\left( \frac{1}{2}(S_{\mathbf{i}}^{+}S_{\mathbf{j}}^{-}+S_{\mathbf{i}}^{-}S_{\mathbf{j}}^{+})+S_{\mathbf{i}}^{z}S_{\mathbf{j}}^{z}\right)&\simeq
-2JS\sum_{\mathbf{i},\boldsymbol{\updelta}}\left(a_{\mathbf{j}}^{\dag}a_{\mathbf{j}+\boldsymbol{\updelta}}^{}+a_{\mathbf{j}}^{}a_{\mathbf{j}+\boldsymbol{\updelta}}^{\dag}+
a_{\mathbf{j}}^{\dag}a_{\mathbf{j}}^{}-a_{\mathbf{j}+\boldsymbol{\updelta}}^{\dag}a_{\mathbf{j}+\boldsymbol{\updelta}}^{}+S \right)
\nonumber \\
&=E_{0}-\frac{|J|zS}{2}\left( 2b_{\mathbf{k}}^{\dag}b_{\mathbf{k}}^{}-\gamma_{\mathbf{k}}^{}b_{\mathbf{k}}^{}b_{\mathbf{k}}^{\dag}-\gamma_{-\mathbf{k}}^{}b_{\mathbf{k}}^{\dag}b_{\mathbf{k}}^{} \right)
\label{eq:HeisenbergHamiltonianBosonOperators}
\vspace{-5pt}\end{align}
where $\boldsymbol{\updelta}$ is the vector pointing to a nearest-neighbor site, $\gamma_{\mathbf{k}}=\sum_{\boldsymbol{\updelta}}\text{e}^{i\mathbf{k}\cdot\boldsymbol{\updelta}}$ is the geometrical factor. Further simplifications of the expression can be obtained using $b_{\mathbf{k}}^{}b_{\mathbf{k}}^{\dag}=1+b_{\mathbf{k}}^{\dag}b_{\mathbf{k}}^{}$ as well as taking into account inversion symmetry $\gamma_{\mathbf{k}}^{}=\gamma_{-\mathbf{k}}^{}$ and noticing that $\sum_{\mathbf{k}}\gamma_{\mathbf{k}}^{}=0$ \cite{Fazekas99}. The spin wave Hamiltonian for ferromagnets takes the approximate form \cite{Roessler09,Fazekas99}:
\vspace{-5pt}\begin{equation}
\pazocal{H}\simeq E_{0}+2|J|zS\sum_{\mathbf{k}}(1-\gamma_{\mathbf{k}}^{})b_{\mathbf{k}}^{\dag}b_{\mathbf{k}}^{}=
E_{0}+\sum_{\mathbf{k}}\omega_{\mathbf{k}}^{}\hat{n}_{\mathbf{k}}^{},
\label{eq:SpinWaveHamiltonian}
\vspace{-5pt}\end{equation}
where $\hat{n}_{\mathbf{k}}^{}$ is the magnon occupation number. This result can be simply explained as a set of independent thermally excited harmonic oscillators, each labeled by a wavevector $\mathbf{k}$, called magnons, excited above the ground state with all spins pointing in the same direction.
\begin{wrapfigure}[5]{r}{0.50\textwidth}\vspace{-0pt}
    \vspace{-5pt}\includegraphics[width=1\linewidth]{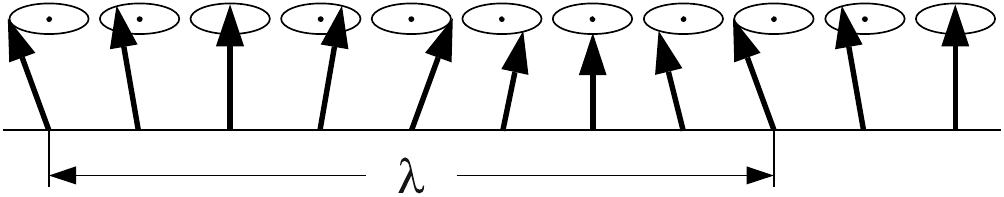}\vspace{0pt}
    \vspace{-25pt}\caption{Visualization of a ferromagnetic spin wave on a linear chain.  Figure was taken from ref.\,\citenum{Roessler09}.}\label{Fig:SpinWave}\vspace{-0pt}
\end{wrapfigure}
A schematic of a ferromagnetic spin wave is shown in fig.\,\ref{Fig:SpinWave} in which the angle of precession varies from spin to spin along the propagation vector.
We derived the spin-wave dispersion in a ferromagnet, which within the long-wave limit ($|\mathbf{k}\cdot\boldsymbol{\updelta}|\ll1$) results in an approximately quadratic dispersion:
\vspace{-5pt}\begin{equation}
\omega_{\mathbf{k}}^{}=|J|Sz(1-\gamma_{\mathbf{k}}^{})
\approx|J|Sz\left(1-\bigg[1-\frac{1}{2z}\sum_{\boldsymbol{\updelta}}(\mathbf{k}\cdot\boldsymbol{\updelta})^2\bigg]\right)
\approx S|J|(ak)^2,
\label{eq:MagnonDispersionFM}
\vspace{-5pt}\end{equation}
where $a$ is the lattice constant. A magnon with wavevector $\mathbf{k}$ costs an energy $\omega_{\mathbf{k}}^{}>0$, and the described quadratic dispersion shown in fig.\,\ref{Fig:FerromagnetAntiferromagnetDispersion}(a) has been first measured by inelastic neutron scattering in magnetite (Fe$_3$O$_4$) by B. Brockhouse \cite{Brockhouse57}. A review of early spin-wave measurements in various $3d$ metals can be found in ref.\,\citenum{ShiraneMinkiewicz68}.

In case of an  antiferromagnetic ground state, derivation of the spin-wave spectrum becomes more complicated as the unit cell is at least twice larger, because it consists of two sublattices $A$ and $B$. A simple strategy of the ground state energy calculation similar to eq.\,\ref{eq:GroundState} would not lead to energy  minimum when two spins point in opposite directions $|S\rangle_{i}$\,$|-S\rangle_{j}$, therefore this state is not an eigenstate. Unlike in a ferromagnet,  an antiferromagnetic ground state also gains energy from the spin-flip terms, therefore the antiferromagnetic Heisenberg model shows quantum fluctuations \cite{Fazekas99}. The Hamiltonian for this system can be written as:
\vspace{-5pt}\begin{equation}
H=J_{a}^{}\sum_{\mathbf{i},\mathbf{j}}\mathbf{S}_{A\mathbf{i}}\cdot\mathbf{S}_{B\mathbf{j}}=
J_{a}^{}\sum_{\mathbf{i},\mathbf{j}}\left( \frac{1}{2}(S_{A\mathbf{i}}^{+}S_{B\mathbf{j}}^{-}+S_{A\mathbf{i}}^{-}S_{B\mathbf{j}}^{+})+S_{A\mathbf{i}}^{z}S_{B\mathbf{j}}^{z}\right),
\label{eq:HeisenbergModelAFM}
\vspace{-5pt}\end{equation}
where $\mathbf{S}_{A\mathbf{i}}$ and $\mathbf{S}_{B\mathbf{j}}$ are the spin vector operators on the two sublattices. Taking into account the \textit{Holstein-Primakoff} transformation in a similar way as done before and the Bloch representation, the  Heisenberg Hamiltonian of the antiferromagnetic state becomes:
\vspace{-7pt}\begin{align}
\pazocal{H}&\simeq J_{a}^{}\sum_{\mathbf{i},\mathbf{j}}
\left( -S^2+S(a_{A\mathbf{i}}^{\dag}a_{A\mathbf{i}}^{}+a_{B\mathbf{i}}^{\dag}a_{B\mathbf{i}}^{}+a_{A\mathbf{i}}^{}a_{B\mathbf{j}}^{}+a_{A\mathbf{i}}^{\dag}a_{B\mathbf{j}}^{\dag})\right)
\nonumber \\
&\simeq E_a+J_{a}zS\sum_{\mathbf{k}}\left( (b_{A\mathbf{k}}^{\dag}b_{A\mathbf{k}}^{}+b_{B\mathbf{k}}^{\dag}b_{B\mathbf{k}}^{}+\gamma_{\mathbf{k}}^{}(b_{A\mathbf{k}}^{\dag}b_{B\mathbf{k}}^{\dag}+b_{A\mathbf{k}}^{}b_{B\mathbf{k}}^{})\right)
\label{eq:HeisenbergModelAFM2}
\vspace{-7pt}\end{align}
where the first summand $E_a=-J_{a}zNS^2/2$ gives the energy of the antiferromagnetic configuration, which contains only the contribution of the Ising $z$-$z$ term. However, unlike in the case of a ferromagnet, a $\mathbf{k}$-space transformation  is not sufficient to give a diagonal form of the Hamiltonian. Diagonalization can be done by a set of operators known as the \textit{Bogoliubov} transformation \cite{Bogoljubov58}:
\vspace{-7pt}\begin{align}
\alpha_{\mathbf{k}}^{}&=u_{\mathbf{k}}^{}b_{A\mathbf{k}}^{}-v_{\mathbf{k}}^{}b_{B\mathbf{k}}^{\dag};
&
\alpha_{\mathbf{k}}^{\dag}&=u_{\mathbf{k}}^{}b_{A\mathbf{k}}^{\dag}-v_{\mathbf{k}}^{}b_{B\mathbf{k}}^{};
\nonumber \\
\beta_{\mathbf{k}}^{}&=u_{\mathbf{k}}^{}b_{B\mathbf{k}}^{}-v_{\mathbf{k}}^{}b_{A\mathbf{k}}^{\dag};
&
\beta_{\mathbf{k}}^{\dag}&=u_{\mathbf{k}}^{}b_{B\mathbf{k}}^{\dag}-v_{\mathbf{k}}^{}b_{A\mathbf{k}}^{},
\label{eq:BogoliubovTransformation}
\vspace{-7pt}\end{align}
where the coefficients $u_{\mathbf{k}}^{}$ and $v_{\mathbf{k}}^{}$ are real. Using the transformations above, and taking into account conditions $u_{\mathbf{k}}^{2}- v_{\mathbf{k}}^{2}=1$ and $\gamma_{\mathbf{k}}^{}(u_{\mathbf{k}}^{2}+ v_{\mathbf{k}}^{2})+2u_{\mathbf{k}}^{}v_{\mathbf{k}}^{}=0$, which are necessary to get a diagonal form, the Hamiltonian can be rewritten as \cite{Roessler09,Fazekas99}:
\vspace{-7pt}\begin{multline}
\pazocal{H}\simeq E_a+J_{a}zS\sum_{\mathbf{k}}
([2\gamma_{\mathbf{k}}^{}u_{\mathbf{k}}^{}v_{\mathbf{k}}^{}+u_{\mathbf{k}}^{2}+v_{\mathbf{k}}^{2}](\alpha_{\mathbf{k}}^{\dag}\alpha_{\mathbf{k}}^{}+\beta_{\mathbf{k}}^{\dag}\beta_{\mathbf{k}}^{})+
2(\gamma_{\mathbf{k}}^{}u_{\mathbf{k}}^{}v_{\mathbf{k}}^{}+v_{\mathbf{k}}^{2}) \\
+(\gamma_{\mathbf{k}}^{}(u_{\mathbf{k}}^{2}+v_{\mathbf{k}}^{2})+2u_{\mathbf{k}}^{}v_{\mathbf{k}}^{})[\alpha_{\mathbf{k}}^{\dag}\beta_{\mathbf{k}}^{\dag}+\alpha_{\mathbf{k}}^{}\beta_{\mathbf{k}}^{}])
=E_0+J_{a}zS\sum_{\mathbf{k}}\omega_{\mathbf{k}}^{}(\alpha_{\mathbf{k}}^{\dag}\alpha_{\mathbf{k}}^{}+\beta_{\mathbf{k}}^{\dag}\beta_{\mathbf{k}}^{}),
\label{eq:HeisenbergModelAFM3}
\vspace{-7pt}\end{multline}
where $E_0=-J_{a}zNS(S+1)/2+\sum_{\mathbf{k}}\omega_{\mathbf{k}}^{}$ is the ground-state energy and $\omega_{\mathbf{k}}^{}=J_{a}Sz\sqrt{1-\gamma_{\mathbf{k}}^{2}}$ is the magnon energy. Unlike in the ferromagnetic case, the two boson operators $\alpha_{\mathbf{k}}^{\dag}\alpha_{\mathbf{k}}^{}$ and $\beta_{\mathbf{k}}^{\dag}\beta_{\mathbf{k}}^{}$ contain
two elementary excitations with the same energy. The dispersion is also different, as in case of the long-wavelength limit the antiferromagnetic spectrum, shown in fig.\,\ref{Fig:FerromagnetAntiferromagnetDispersion}(b), is linear like the phonon spectrum, and again gapless:
\vspace{-7pt}\begin{align}
\omega_{\mathbf{k}}^{}=J_{a}Sz\sqrt{1-\gamma_{\mathbf{k}}^{2}}&\approx
J_{a}Sz\sqrt{(a\mathbf{k})^{2}}
\nonumber \\
&\approx J_{a}Sza|\mathbf{k}|.
\label{eq:MagnonDispersionAFM}
\vspace{-15pt}\end{align}

\enlargethispage{10pt}
\vspace{-20pt}\subsection{Anisotropy effect}\label{SubSec:AnisotropyEffect}\vspace{-0pt}

In the previous section we have considered the case of spin-rotation symmetry and derived the  spin-wave spectrum. However, if we violate rotation symmetry, for example by introducing the single-ion anisotropy into the Heisenberg Hamiltonian as:
\vspace{-5pt}\begin{equation}
H=J\sum_{i,j}(S_{i}^{x}S_{j}^{x}+S_{i}^{y}S_{i}^{y}+DS_{i}^{z}S_{j}^{z}),
\label{eq:HeisenbergModelAnisotropy}
\vspace{-5pt}\end{equation}

\begin{wrapfigure}[21]{r}{0.40\textwidth}\vspace{0pt}
    \vspace{-0pt}\includegraphics[width=1\linewidth]{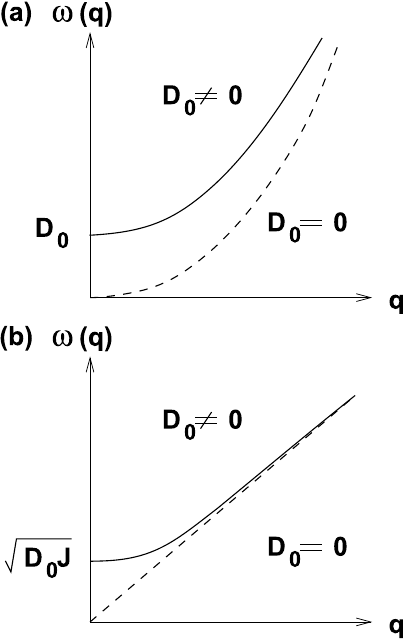}\vspace{0pt}
    \vspace{-25pt}\caption{ The energy transfer as a function of $\mathbf{k}$ for (a) ferromagnet and (b) antiferromagnet. Figure was taken from ref.\,\citenum{Henley08}.}\label{Fig:FerromagnetAntiferromagnetDispersion}\vspace{-0pt}
\end{wrapfigure}
\noindent where $D>0$, the spin wave spectrum given by eq.\,\ref{eq:MagnonDispersionFM} changes to
\vspace{-5pt}\begin{equation}
\omega_{\mathbf{k}}^{}=|J|Sz(D-\gamma_{\mathbf{k}}^{}),
\label{eq:MagnonDispersionFMwithAnisotropy}
\vspace{-5pt}\end{equation}
and the smallest excitation energy changes to  $\omega_{}^{}(\mathbf{k}=0)=D-1$, therefore  the spectrum has acquired a gap \cite{Roessler09}, as shown in fig.\,\ref{Fig:FerromagnetAntiferromagnetDispersion}(a). In addition, the opening of a gap in the case of a ferromagnet can be explained even for fully isotropic interactions. Presence of the classical dipole-dipole interaction that acts on a spin spins in the field of the other is given by \cite{Mattis81}:
\vspace{-5pt}\begin{equation}
\pazocal{H}_{\text{dip}}=\frac{1}{2}\sum_{i,j}\frac{g_{}^{2}\mu_{\text{B}}^{2}}{r_{ij}^{3}}\frac{\mathbf{S}_{i}^{}\cdot\mathbf{S}_{j}^{}\mathbf{r}_{ij}^{2}-3\mathbf{S}_{i}^{}\cdot\mathbf{r}_{ij}^{}\mathbf{S}_{j}^{}\cdot\mathbf{r}_{ij}^{}}{\mathbf{r}_{ij}^{2}}.
\label{eq:MagnonDispersionFMwithAnisotropy}
\vspace{-5pt}\end{equation}
Because of this interaction, the spin-wave spectrum starts to feel the demagnetization field and its energy becomes dependent on the shape of the sample. The long-range dipole-dipole interaction opens  a gap, which is for the ellipsoidal sample given by \cite{Fazekas99,Kittel96}:
\vspace{-5pt}\begin{equation}
\omega_{\mathbf{k}=0}^{}=\sqrt{(H+(N_{x}-N_{z})M_{0})(H+(N_{y}-N_{z})M_{0})},
\label{eq:MagnonDispersionFMwithAnisotropy2}
\vspace{-5pt}\end{equation}
where $N_{x}$, $N_{y}$, $N_{z}$ are the demagnetization factors, and $M_{0}$ is the magnetization density.

Following a similar line of reasoning when talking about anisotropy in an antiferromagnet, we find that the spin wave dispersion is also modified and can be written as \cite{Roessler09}:
\vspace{-5pt}\begin{equation}
\omega_{\mathbf{k}}^{}=J_{a}Sz\sqrt{(1+\rho)^2-\alpha^2\gamma_{\mathbf{k}}^{2}},
\label{eq:MagnonDispersionAFMwithAnisotropy}
\vspace{-5pt}\end{equation}
where $\rho=2D/J_{a}z$. The spin-wave gap, shown in fig.\,\ref{Fig:FerromagnetAntiferromagnetDispersion}(b), is   $\omega_{}^{}(0)=J_{a}Sz\sqrt{(1+\rho)^2-\alpha^2}$. These results suggest that the presence of the anisotropy has a strong influence on the magnon spectrum.

\subsection{Magnetic field dependence of spin-wave excitations}\label{SubSec:MagneticFieldDependenceSpinExcitation}

As we can see from the previous sections, both ferro- and antiferromagnets show correlated motion of the localized electrons. Below a certain temperature the exchange interaction becomes larger than the thermal energy, which results in a long-range order that we can probe with neutrons as magnetic Bragg peaks and spin excitations or with the  electron spin resonance. Application of an external magnetic field would introduce a Zeeman interaction, which should be also considered as an additional term to the Heisenberg Hamiltonian. The exchange interaction between spins and uniaxial anisotropy can be described using an effective molecular field $\mathbf{H}_{\text{E}}$ and anisotropy field $\mathbf{H}_{\text{A}}$. Thus, the external magnetic field $\mathbf{H}$, depending on the applied field direction, can help to uncover the underlying structure of the spin interaction in matter.

The energy density of a ferromagnet in an external magnetic field $\mathbf{H}$ is in terms of the local magnetization represented by $\mathbf{M}_{j}^{}(\mathbf{r})=\sum_{i}^{}\boldsymbol{\upmu}_{i}^{j}/v_{j}^{}$, where $\boldsymbol{\upmu}_{i}^{j}$ is the magnetic moment of the $i^{\text{th}}$ ion in the $j^{\text{th}}$ sublattice and the summation is done over all sites in a physically small volume $v_{j}^{}$ and $m_{j\alpha}=M_{j\alpha}/M_{j0}$, can be written as \cite{Turov65}:
\vspace{-0pt}\begin{align}
\mathcal{H}(\mathbf{r})=A\mathbf{m}_{}^{2}-K\mathbf{m}_{}^{2}+B_{\perp}^{}\left( \bigg(\frac{\partial m_{\alpha}}{\partial x}\bigg)^2+\bigg(\frac{\partial m_{\alpha}}{\partial y}\bigg)^2 \right)+B_{z}^{}\bigg(\frac{\partial m_{\alpha}}{\partial z}\bigg)^2
\nonumber \\
-\frac{1}{2}\mathbf{M}(\mathbf{r})\nabla
\left(
\int_{V}\frac{\text{div}\mathbf{M}(\mathbf{r}^{\prime})\text{d}\mathbf{r}^{\prime}}{|\mathbf{r}-\mathbf{r}^{\prime}|}
-\oint_{\sum}\frac{\mathbf{M}(\mathbf{r}^{\prime})\text{d}\sigma^{\prime}}{|\mathbf{r}-\mathbf{r}^{\prime}|}
\right)
-\mathbf{M}\cdot\mathbf{H}.
\label{eq:FerromagnetEnergyDensity}
\vspace{-0pt}\end{align}

The expression above is analogous to eq.\,\ref{eq:HeisenbergModel}. The first and second terms are the exchange and anisotropy energies respectively. The third, fourth and fifth terms are perturbations of the uniformity of magnetization, while the last term defines the Zeeman energy  in an external magnetic field. Detailed derivation of the spin excitation spectrum, similar to the one shown in sec.\,\ref{SubSec:FerroAntoferroMagnet}, can be found in ref.\,\citenum{Turov65}, while I will only give the results obtained for the spin-excitation spectra which strongly depend on the sign of the anisotropy constant $K$ and field direction with respect to the quantization axis.

For the case when the $\mathbf{H}\parallel\mathbf{Z}$, spin-wave spectrum of a ferromagnet consists of two solutions:
\vspace{-5pt}\begin{align}
1)\,\,\omega_{k}^{}&=\Bigg\{
\bigg[I_{k}+\mu\left(H+H_{\text{A}}-\frac{4\pi}{3}M_{0}\right) \bigg] \nonumber \\
&\times\bigg[I_{k}+\mu\left(H+H_{\text{A}}-\frac{4\pi}{3}M_{0}\right)+4\pi\mu M_{0}\frac{k_{x}^{2}+k_{y}^{2}}{k^2} \bigg]\Bigg\}^{\frac{1}{2}},
\label{eq:E1kParallel}
\\
2)\,\,\omega_{k}^{}&=\Bigg\{\bigg(I_{k}+\frac{4\pi}{3}\mu M_{0} \bigg)
\bigg[I_{k}-\mu H_{\text{A}}\left(1-\frac{H^2}{H_{\text{A}}^{2}}\right)-\frac{4\pi}{3}\mu \text{\footnotemark} M_{0}+4\pi\mu M_{0}\frac{k_{x}^{\prime2}+k_{y}^{2}}{k^2} \bigg]
\nonumber \\
&-\mu H_{\text{A}} \left(1-\frac{H^2}{H_{\text{A}}^{2}}\right)
4\pi \mu M_{0}\frac{k_{x}^{\prime2}}{k^2} \Bigg\}^{\frac{1}{2}}, \hspace{5pt}\text{with} \hspace{5pt}
k_{x}^{\prime}=-k_{x}\frac{H}{H_{\text{A}}}-k_{z}\sqrt{1-\frac{H^2}{H_{\text{A}}^{2}}}
\label{eq:E2kParallel}
\vspace{-5pt}\end{align}
\footnotetext{factor $\mu$ is absent in ref.\,\citenum{Turov65}, however it should be there.}
The first solution is valid for the whole field range when $K>0$. When $K<0$,  the first solution is valid  for $H\geq H_{\text{A}}$ , while the second solution is obtained for $H< H_{\text{A}}$.

When $\mathbf{H}\perp\mathbf{Z}$, the spin-wave spectrum is defined by the other two solutions:
\vspace{-15pt}\begin{align}
1)\,\,\omega_{k}^{}&=\Bigg\{
\bigg[I_{k}+\mu\left(H-\frac{4\pi}{3}M_{0}\right) \bigg] \nonumber \\
&\times\bigg[I_{k}+\mu\left(H-H_{\text{A}}-\frac{4\pi}{3}M_{0}\right)+4\pi\mu M_{0}\frac{k_{y}^{2}+k_{z}^{2}}{k^2} \bigg]-\mu H_{\text{A}}4\pi \mu M_{0}\frac{k_{z}^{2}}{k^2}\Bigg\}^{\frac{1}{2}},
\label{eq:E1kPerp}
\\
2)\,\,\omega_{k}^{}&=\Bigg\{
\bigg[I_{k}+\mu\left(H_{\text{A}}-\frac{4\pi}{3}M_{0}\right) \bigg]
\bigg[I_{k}+\mu H_{\text{A}}\left(1-\frac{H^2}{H_{\text{A}}^{2}}\right)
-\frac{4\pi}{3}\mu M_{0}
+4\pi\mu M_{0}\frac{k_{x}^{\prime2}+k_{y}^{2}}{k^2} \bigg]
\nonumber \\
&-\mu H_{\text{A}}\frac{H^2}{H_{\text{A}}^{}}
4\pi \mu M_{0}\frac{k_{x}^{\prime2}}{k^2} \Bigg\}^{\frac{1}{2}}, \hspace{5pt}\text{with} \hspace{5pt}
k_{x}^{\prime}=k_{x}\sqrt{1-\frac{H^2}{H_{\text{A}}^{2}}}-k_{z}\frac{H}{H_{\text{A}}}.
\label{eq:E2kPerp}
\vspace{-5pt}\end{align}
For this orientation the first solution is valid for any value of field when $K<0$.  For $K>0$ first solution applies for the field range $H\geq H_{\text{A}}$ and the second for $H<H_{\text{A}}$. As one can see, studying the field dependence of the energy gap can give a lot of valuable information. Comparing the observed dependence with the formulae given above , according to which results should be different for different states and field directions, provides a way to determine the physical constants of the material.

As one can see, \cref{eq:E1kParallel,eq:E2kParallel,eq:E1kPerp,eq:E2kPerp} contain information about the field, as well as the momentum dependence of the spin wave energy. The limiting case of the wave vector $|\mathbf{k}|=0$ is of particular importance, as the zone-center excitation can be directly probed with electron spin resonance (ESR). Main idea of the ESR experiment is to remove the degeneracy of the electron-magnetic moment of an atom with the help of an external magnetic field. Separation between the levels with different $m_{\text{s}}$, magnetic components  of the  spin quantum number,  can be probed with the absorbed or emitted photon, with the polarization $\mathbf{h}_{\omega}$, thus giving a fundamental equation $\Updelta E=\hbar\omega=g_{\text{e}}\mu_{\text{B}}B_{0}m_{\text{s}}$, where $g_{\text{e}}$ is the $g$-factor for the free electron, $\mu_{\text{B}}$ is the Bohr magneton, and $B_{0}$ is the strength of an external magnetic field. Usually, the sample is placed inside a resonant cavity, which in turn is placed inside a cryomagnet. When the magnetic field strength reaches such a value that the separation between the lower and the upper state $\Updelta E=\hbar\omega_{\text{res}}$, where $\omega_{\text{res}}$ is the resonance frequency of the cavity, a strong absorption line is observed.

In  the case of a ferromagnet, and discussed later an antiferromagnet, the observed resonance effect can also originate from the processional motion of the magnetization. As we have already learned in sec.\,\ref{SubSec:FerroAntoferroMagnet}, such motion is responsible for the appearance of spin waves. Therefore by measuring the dependence of $\hbar\omega_{\text{res}}(B)$, as a function of an  external magnetic field, one can  directly probe the spin waves. These measurements can be directly compared with the calculated field dependencies of the spin waves given by \cref{eq:E1kParallel,eq:E2kParallel,eq:E1kPerp,eq:E2kPerp}, expected for the ferromagnet, as well as antiferromagnet, in case of the wave vector $|\mathbf{k}|=0$.

\clearpage

\begin{wrapfigure}[21]{r}{0.6\textwidth}\vspace{-0pt}
    \vspace{-0pt}\includegraphics[width=1\linewidth]{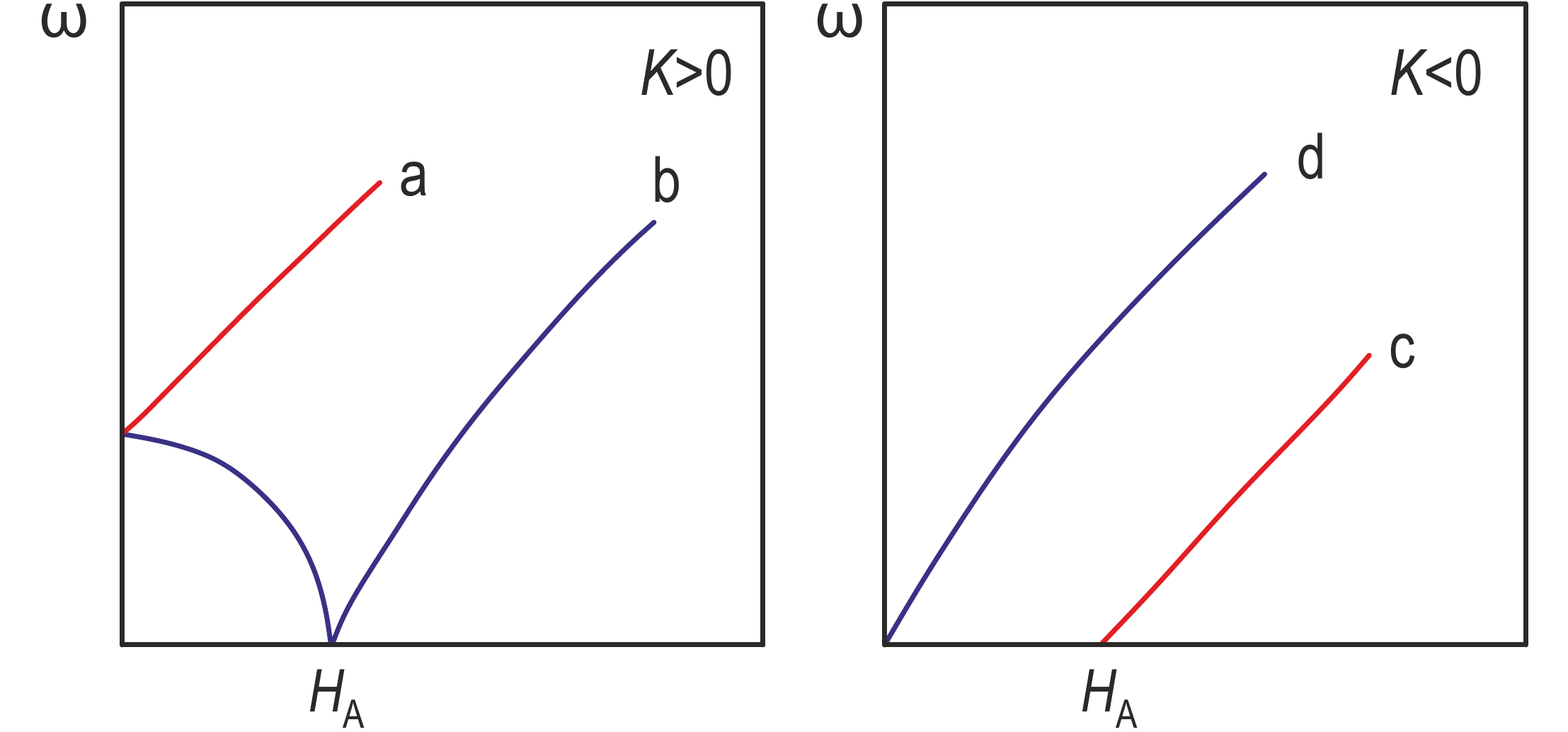}\vspace{0pt}
    \vspace{-28pt}\caption{ The dependence of the resonance frequency as a function of field. Figure was redrawn from ref.\,\citenum{Turov65}.}
    \label{Fig:SpinGapFieldDep1}\vspace{-0pt}
        \vspace{-8pt}\includegraphics[width=1\linewidth]{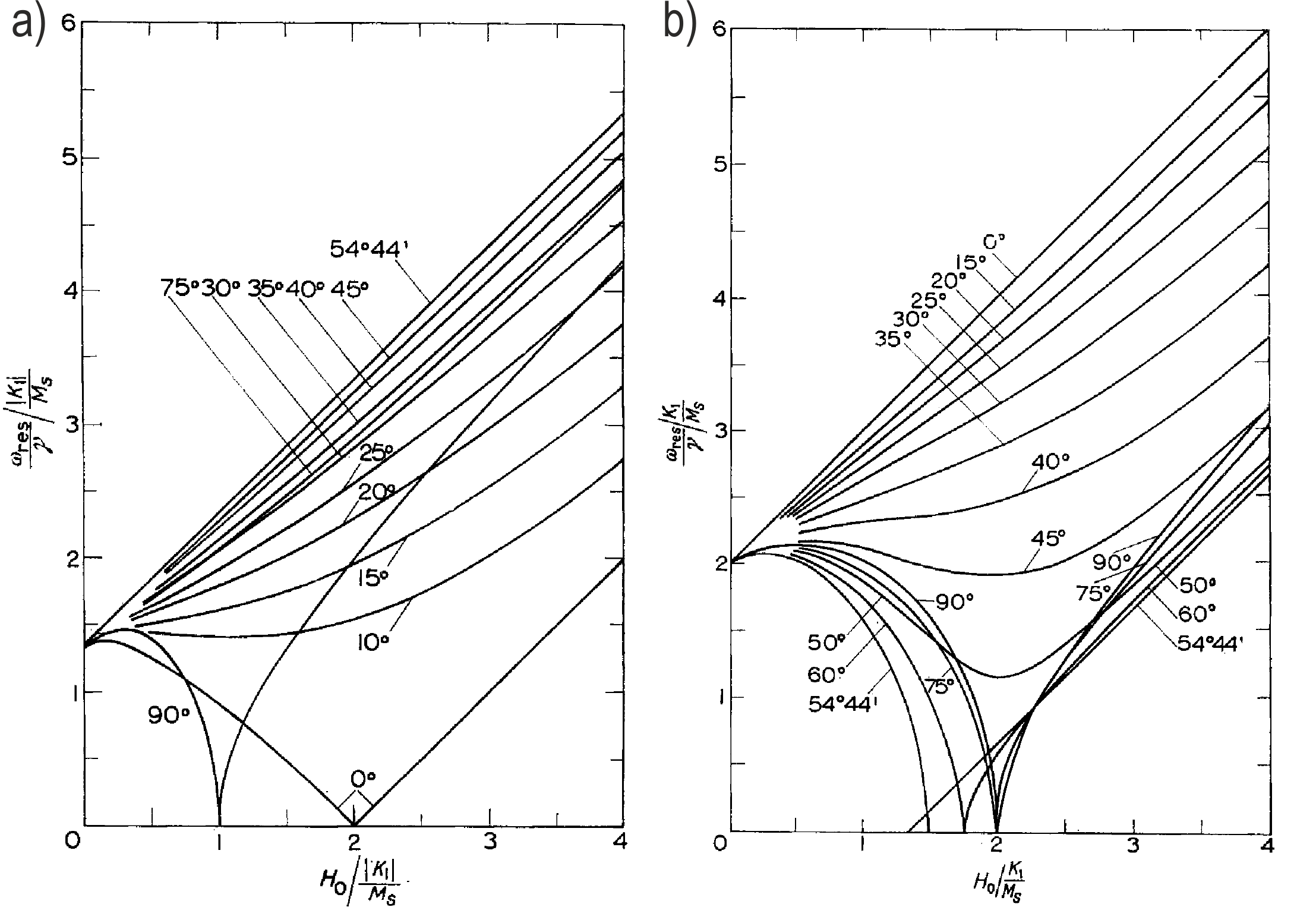}\vspace{0pt}
        \vspace{-28pt}\caption{The dependence of the resonance frequency as a function of field with different orientations for negative(a); and positive (b) anisotropy. Figure was taken from ref.\,\citenum{Vonsovskii66}.}
        \label{Fig:SpinGapFieldDep2}\vspace{-0pt}
\end{wrapfigure}
We can summarize the results as follows. When magnetic field is applied to a ferromagnet along the easy magnetization axis $\mathbf{Z}$ ($K>0$),  the resonance frequency is defined as linear function of the field and is shown in fig.\,\ref{Fig:SpinGapFieldDep1}(a). The case when magnetic field is perpendicular to the easy magnetization axis would mean that magnetization $\mathbf{M}$  will try to line up parallel to $\mathbf{H}$, that would lead to a decrease of the spin gap with the increasing field up to $H_{\text{A}}$, where spin-gap is zero, and then in the saturation region where $\mathbf{M}\parallel\mathbf{H}$  the gap will increase continuously, as shown in fig.\,\ref{Fig:SpinGapFieldDep1}(b). A slightly different result can be found when the field is applied along the hard  magnetization axis ($K<0$). Because the magnetization vector is degenerate with respect to the rotation along $\mathbf{Z}$,  the energy gap will be zero until the magnetization becomes parallel to $\mathbf{H}$, which happens when $H>|H_{\text{A}}|$. This case is shown in fig.\,\ref{Fig:SpinGapFieldDep1}(c). If the field is perpendicular to the hard  magnetization axis, rotational symmetry is instantly broken, and the energy gap shows a monotonic increase with field, as shown in fig.\,\ref{Fig:SpinGapFieldDep1}(d).  Dependence of the  spin gap for an arbitrary orientation of $\mathbf{H}$ parameterized by $\theta$, with $\theta=0^{\circ}$ is the field $\mathbf{H}\parallel\mathbf{Z}$ and $\theta=54^{\circ}44^{\prime}$ is the $\mathbf{X}\mathbf{Y}\mathbf{Z}$ axis and $\theta=90^{\circ}$ for $\mathbf{X}\mathbf{Y}$     was calculated in \cite{Vonsovskii66}  for crystals with negative and positive anisotropy and is shown in fig.\,\ref{Fig:SpinGapFieldDep2} (a) and (b) respectively. Thus, we can observe that the external magnetic field has a strong effect on the resonance energy, as it acts in addition to the internal effective field caused by magnetic crystallographic anisotropy.

\enlargethispage{-20pt}
As the next step let us consider the field dependence of the spin-wave resonance in an antiferromagnet. Again,  considering uniaxial crystals and assuming that the principal axis of symmetry is $\mathbf{Z}$ axis, the energy density of the antiferromagnet can be written as \cite{Turov65}:
\vspace{-5pt}\begin{align}
\mathcal{H}(\mathbf{r})=\frac{A}{2}\mathbf{m}_{}^{2}+\frac{A_{\gamma\delta}^{}}{2}\frac{\partial m_{\alpha}}{\partial r_{\gamma}}\frac{\partial m_{\alpha}}{\partial r_{\delta}}
+\frac{B_{\gamma\delta}^{}}{2}\frac{\partial l_{\alpha}}{\partial r_{\gamma}}\frac{\partial l_{\alpha}}{\partial r_{\delta}}
+\frac{a}{2}m_{z}^{2}+\frac{b}{2}l_{z}^{2}
\nonumber \\
-\frac{M_{0}^{2}}{2}(\mathbf{m}\nabla)
\int_{V}\frac{\text{div}\,\mathbf{m}(\mathbf{r}^{\prime})\text{d}\mathbf{r}^{\prime}}{|\mathbf{r}-\mathbf{r}^{\prime}|}
-\mathbf{m}\mathbf{h},
\label{eq:FerromagnetEnergyDensity}
\vspace{-15pt}\end{align}

\noindent where $\mathbf{m}$ and $\mathbf{l}$ are local magnetization and antiferromagnetic vectors, respectively. The first three terms define the exchange energy, while the fourth and fifth terms are the anisotropy energy. Last two terms give the magnetostatic energy and the Zeeman energy in an external magnetic field. Depending on whether $b$ is positive or negative, the spontaneous axis of antiferromagnetism would be perpendicular or parallel to $\mathbf{Z}$.

\begin{wrapfigure}[13]{r}{0.6\textwidth}\vspace{-15pt}
    \vspace{-0pt}\includegraphics[width=1\linewidth]{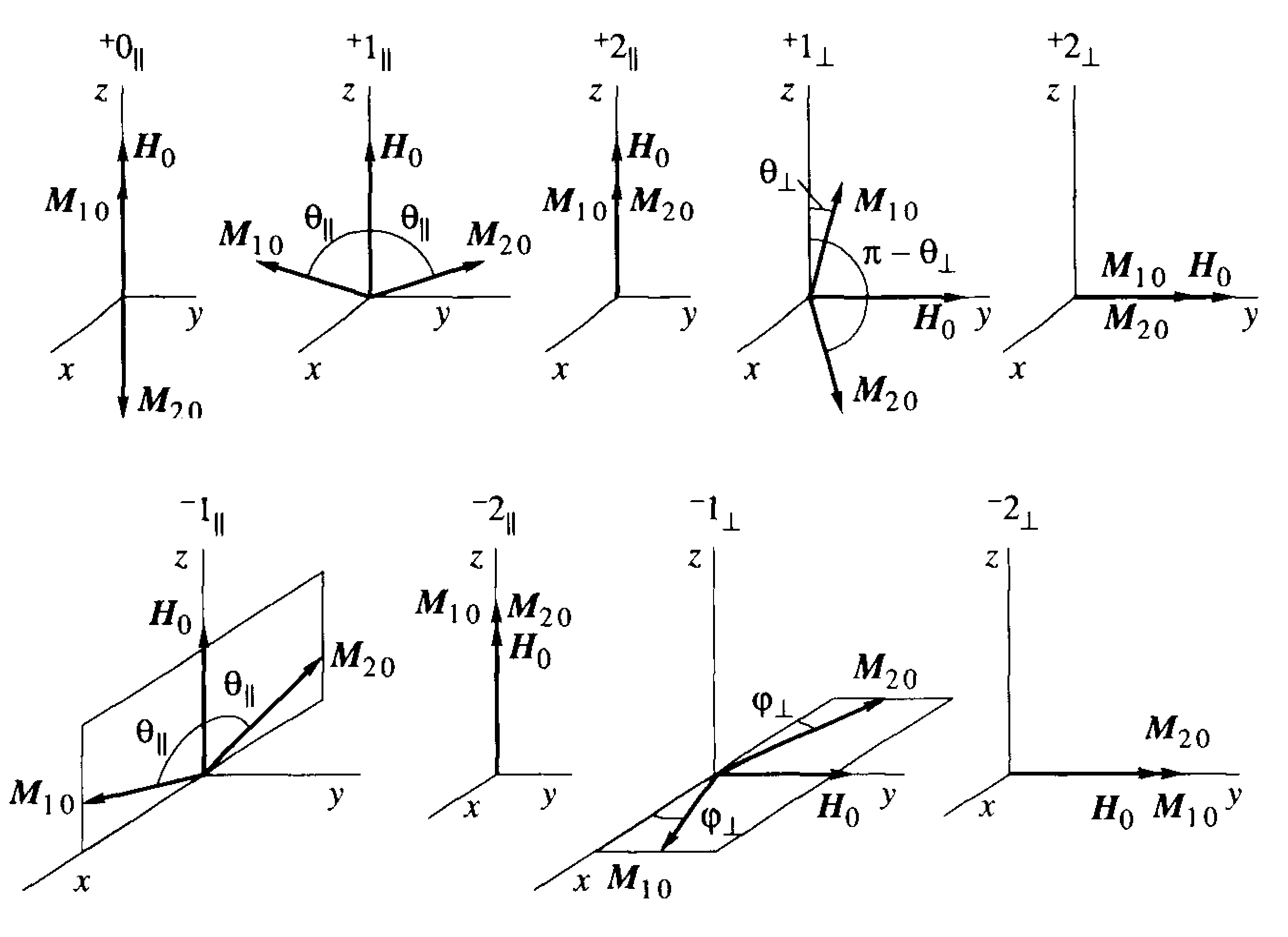}\vspace{0pt}
    \vspace{-28pt}\caption{Ground states of a two-sublattice antiferromagnet. Top: easy axis anisotropy with $b<0$. Bottom: easy plane anisotropy when $b>0$.  Figure was taken from ref.\,\citenum{GurevichMelkov96}.}
    \label{Fig:M1M2orientation}\vspace{-0pt}
\end{wrapfigure}
\begin{figure}[!t]
\includegraphics[width=1\textwidth]{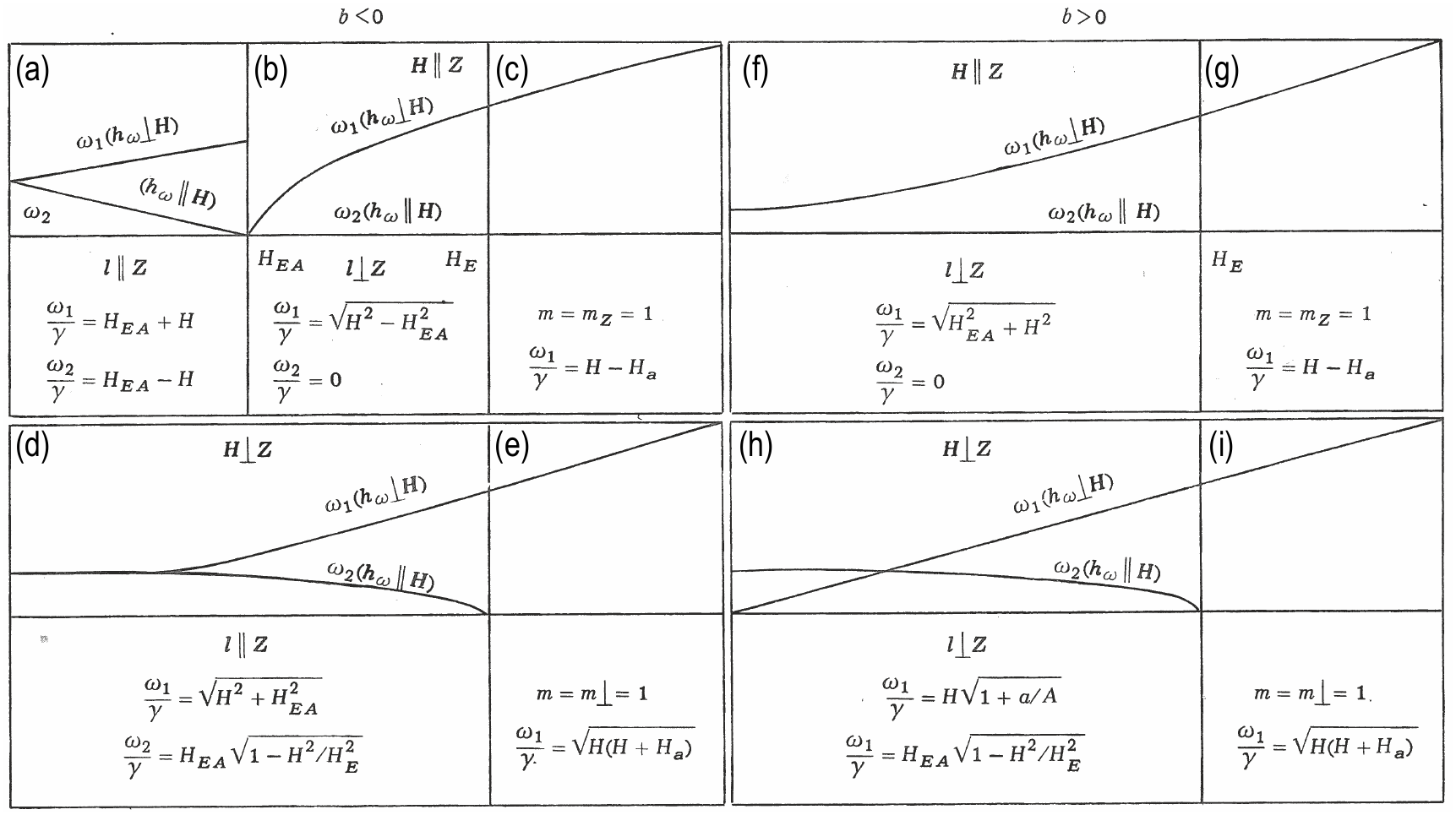}\vspace{-10pt}
\caption{Summary of possible magnetic field dependencies of the spin-resonance in the antiferromagnet. Observation of different modes with ESR depends on the orientation between the radio frequency used in ESR and the magnetic field, and it is indicated with $\mathbf{h}_{\omega}\parallel\mathbf{H}$ and $\mathbf{h}_{\omega}\perp\mathbf{H}$, where $\mathbf{h}_{\omega}$ is the direction of the radio-frequency field. Figure from ref.\,\citenum{Turov65}.}
\label{Fig:SpinGapFieldDepAFM}\vspace{-20pt}
\end{figure}

Detailed derivation of the spin-wave energy formulas depending on the direction of the magnetic field  can be found in ref.\,\citenum{Turov65}, while I will give a short summary of possible results. As the first step let us consider the case when $b<0$. This would force spins to align parallel to the $\mathbf{Z}$ axis. The corresponding antiparallel ground   state $^{+}0_{\parallel}$ is shown in fig.\,\ref{Fig:M1M2orientation}. As we already learned from the previous section, for this configuration two degenerate modes, which correspond to the two sublattice magnetizations, exist. Motion of each sublattice can be modeled by the classical equations of motion \cite{Kittel51,KefferKittel52}, where magnetization $\mathbf{M}_{1}$ leads the motion, while $\mathbf{M}_{2}$ just follows the molecular field $-\lambda\mathbf{M}_{1}$. As a result, right-hand rotation takes place, and this mode is a high-frequency $\omega_{1}$ mode, shown in fig.\,\ref{Fig:AFMModeAntiparallel}(a,c). Similar reasoning is also valid when  $\mathbf{M}_{2}$ leads the motion and the mode with frequency $\omega_{2}$ is a left-hand circular precession, as shown in fig.\,\ref{Fig:AFMModeAntiparallel}(b,d). In this case a weak external magnetic field $\mathbf{H}$, applied parallel to the $\mathbf{Z}$ axis, would result in a higher effective field for one magnetization, while for the other one, molecular and anisotropy fields would be antiparallel to $\mathbf{H}$, thus leading to a decrease in the effective field. As a result, the  resonance frequency would increase for one mode  and decrease for the other, as schematically shown in fig.\,\ref{Fig:SpinGapFieldDepAFM}(a).

\vspace{-0pt}\begin{figure}[!t]
\includegraphics[width=1\textwidth]{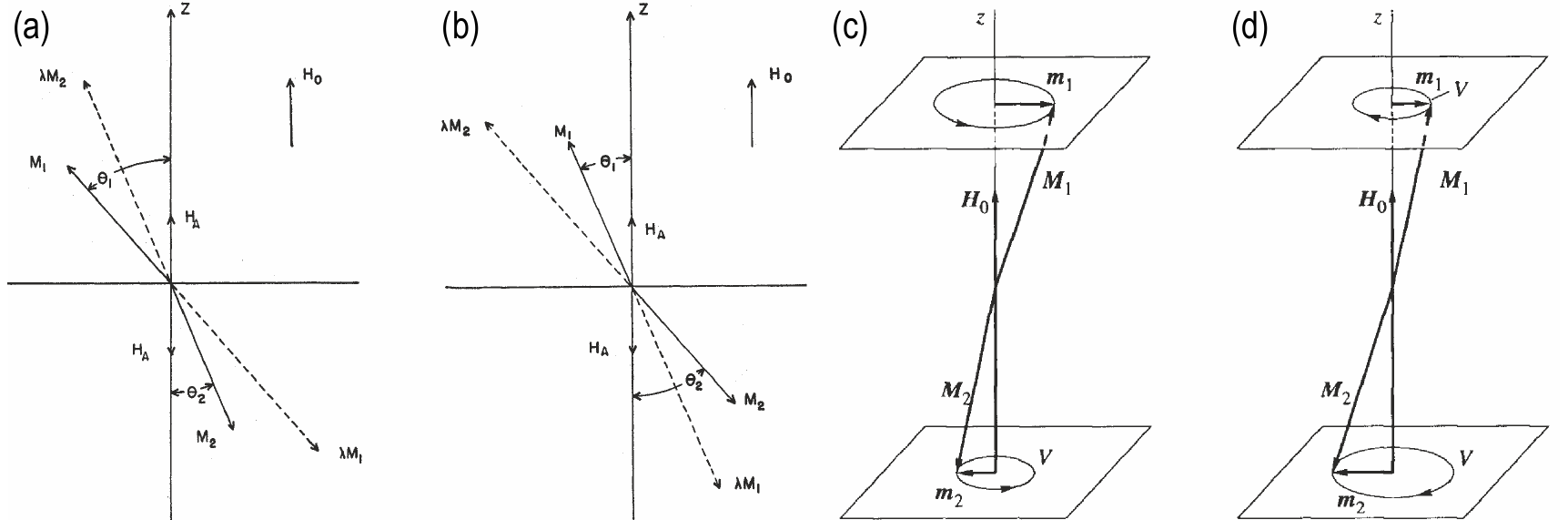}\vspace{-5pt}
\caption{ The normal modes of the antiferromagnet with the precession cone ratio (a) $\theta_{1}/\theta_{2}=\eta$ and (b) $\theta_{2}/\theta_{1}=\eta$.  Precession of the magnetization vectors that shows  mode with frequency (c) $\omega_{1}$ and  (d) $\omega_{2}$ in accordance with fig.\,\ref{Fig:SpinGapFieldDepAFM}(a). Figures (a), (b) and (c), (d) were taken from ref.\,\citenum{KefferKittel52} and ref.\,\citenum{GurevichMelkov96} respectively. }
\label{Fig:AFMModeAntiparallel}\vspace{-5pt}
\includegraphics[width=1\textwidth]{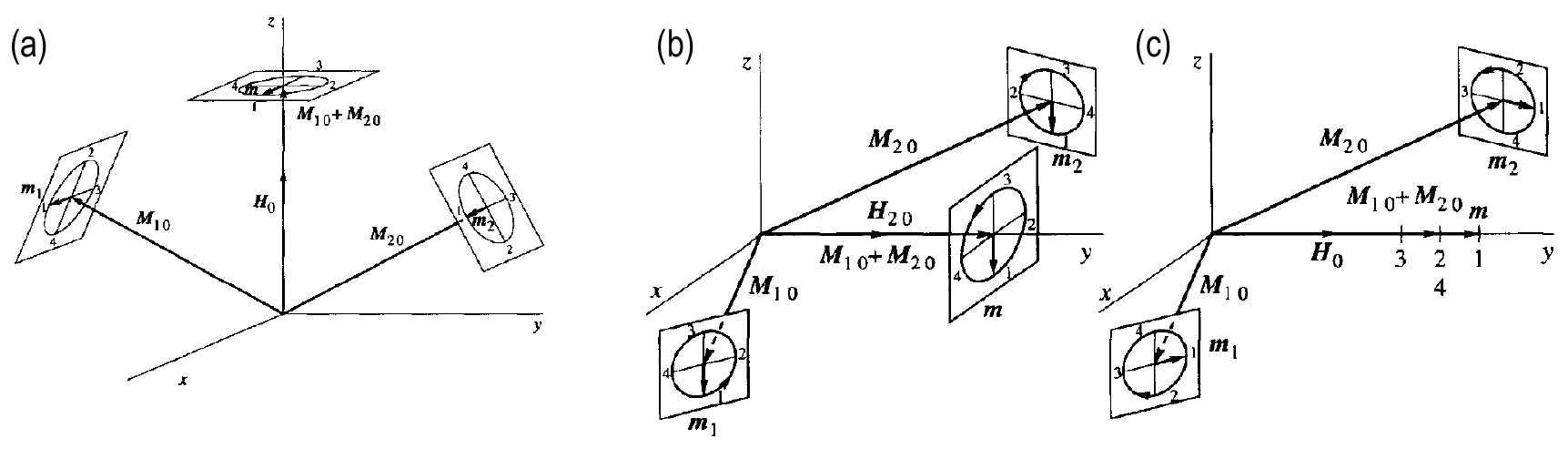}\vspace{-10pt}
\caption{Precession of the sublattice-magnetization and total-magnetization vectors with: (a) easy axis anisotropy in the noncollinear ground state $^{+}1_{\parallel}$; (b,c) easy plane anisotropy in the ground state  $^{-}1_{\perp}$ for modes  $\omega_{1}$  and $\omega_{2}$ respectively, according to  fig.\,\ref{Fig:SpinGapFieldDepAFM}(h). Figure from ref.\,\citenum{GurevichMelkov96}.}
\label{Fig:AFMModeAntiparallel}\vspace{-20pt}
\end{figure}

When the lower mode reaches zero energy, at a certain critical field $H_{EA}=\sqrt{A|b|}/M_{0}$, an interesting effect takes place. Instead of aligning magnetization parallel to the direction of the external magnetic field, it becomes energetically more favorable to flop the spin perpendicular to $\mathbf{H}$, as shown in fig.\,\ref{Fig:M1M2orientation} with a noncollinear ground state $^{+}1_{\parallel}$. This effect is called a spin flop transition. In this configuration, the loss of the anisotropy energy is compensated by the Zeeman energy, and now both magnetization vectors are degenerate and feel  the same effective field. Solving the system of equations which describe the motion of both  magnetization vectors, two modes can be found. The first solution $m_{1x}=m_{2x}$, $m_{1y}=m_{2y}$ and $m_{1z}=-m_{2z}$ corresponds to the precession of $\mathbf{M}_{1}$ and $\mathbf{M}_{2}$ as shown in fig.\,\ref{Fig:AFMModeAntiparallel}(a). Numbers 1--4 denote the position of the magnetization vectors at successive instants. From the geometrical arrangement of the magnetization vectors it becomes clear that the mode can be excited by the transverse component of the radio frequency $\omega_{1}$ with respect to the magnetic field $\mathbf{H}$, as shown in  fig.\,\ref{Fig:SpinGapFieldDepAFM}(b). The second mode has a trivial solution $m_{x}=m_{y}=m_{z}=0$, which corresponds to an infinitely slow rotation of vectors $\mathbf{M}_{1}$ and $\mathbf{M}_{2}$  around the $z$-axis, and since such rotation does not require energy, the spin gap $\omega_{2}$ for this mode is equal to zero. Of course, this is true only in the ideal case, while in real life anisotropy in the basal plane can lead to a nonzero value. The corresponding zero spin-gap dispersion is shown in fig.\,\ref{Fig:SpinGapFieldDepAFM}(b). If the magnetic field is increased further as shown in  fig.\,\ref{Fig:SpinGapFieldDepAFM}(c), the angle between the magnetization vectors decreases, and above $H_{A}=A/M_{0}$ a transition into a collinear field-polarized ground state $^{+}2_{\parallel}$ occurs (fig.\ref{Fig:M1M2orientation}).

\vspace{-5pt}

It becomes obvious that for the case when the external field is perpendicular to the  spontaneous axis, both sublattice magnetizations tilt in the direction of the magnetic field, as shown in fig.\,\ref{Fig:M1M2orientation} with a ground state $^{+}1_{\perp}$. The antiferromagnetic vector $\mathbf{l}$  still remains parallel to the $\mathbf{Z}$ axis until a field $H_{A}$ shown in fig.\,\ref{Fig:SpinGapFieldDepAFM}(d). However, the observation of both possible modes is restricted by the spatial arrangement of the  radio frequency and the magnetic field \cite{KefferKittel52,NagamiyaYosida55}. Above  $H_{A}$   magnetization $\mathbf{m}$ enters the saturation phase, shown in fig.\,\ref{Fig:SpinGapFieldDepAFM}(e). In this region, transition into the $^{+}2_{\perp}$ ground state, shown in fig.\,\ref{Fig:M1M2orientation}, happens.

When $b>0$, the antiferromagnetic vector $\mathbf{l}$ will be oriented in the basal plane. In this case application of a weak external magnetic field along the $\mathbf{Z}$ axis would end up with a noncollinear ground state $^{-}1_{\parallel}$, which with a further increase of the field transforms into collinear $^{-}2_{\parallel}$ ground state. They are both shown in fig.\,\ref{Fig:M1M2orientation}. Simply comparing the above-mentioned states with $^{+}1_{\parallel}$ and $^{+}2_{\parallel}$, it becomes obvious that the eigenfrequencies  are the same. However because the sigh of $H_{A}$ is different, the field dependence of the mode $\omega_{1}$, shown in fig.\,\ref{Fig:SpinGapFieldDepAFM}(f) is actually different from the one mentioned for the case $b<0$.

A qualitatively different picture is found for the field applied perpendicular to the $\mathbf{Z}$ axis. Ground states $^{-}1_{\perp}$ and $^{-}2_{\perp}$, which  correspond to weak and high magnetic fields, are shown in fig.\,\ref{Fig:M1M2orientation}. Solving the system of equations, two different modes $\omega_{1}$ and $\omega_{2}$ can be found, which represent oscillations of the  vectors $\mathbf{M}_{1}$ and $\mathbf{M}_{2}$ that are shown in fig.\,\ref{Fig:AFMModeAntiparallel}(b,c), and their dependence on the field is shown in fig.\,\ref{Fig:SpinGapFieldDepAFM}(h,i),  respectively. Looking at the precession of the magnetization vectors, it becomes obvious, which polarization is required in order to excite them.

\vspace{10pt}

In conclusion, let me give a brief summary of the peculiarities, expected in case of an antiferromagnetic resonance,  as compared with the ferromagnetic resonance. First of all in an antiferromagnet, the presence of two inequivalent sublattices leads to the existence of two, in the case of easy axis anisotropy, degenerate spin-wave modes. In contrast, in a ferromagnet only one mode is expected. Application of the external magnetic field lifts the degeneracy, and as a result, field splitting of the spin-wave modes can be found in an antiferromagnet. It is also important to mention that in an antiferromagnet  several different ground states exist  in different ranges of magnetic field.

\vspace{10pt}
Momentum dependence of the spin waves in the case of an antiferromagnet and ferromagnet is also significantly different. In the vicinity of $|\mathbf{k}|=0$, dispersion of the magnon excitation is either linear or quadratic, for the antiferromagnet and ferromagnet, respectively. Single-ion anisotropy  significantly influences the spin wave dispersion, and in both cases the magnon spectrum acquires a gap, which is determined by the magnitude of the anisotropy. At the same time, the gap size gradually saturates towards the zone boundary, and according to  \cref{eq:MagnonDispersionFM,eq:MagnonDispersionAFM}, mainly depends on the magnitude of the exchange interaction $|J|$. Thereby, measuring the dispersion curves along various wave vectors, one can obtain the information about the $\mathbf{Q}$-dependence of the exchange integral $J(\mathbf{q})$ (see sec.\,\ref{SubSec:InteractionBetweenMultipoles}).

\vspace{10pt}
As we have already discussed, external magnetic field  introduces the Zeeman interaction, which is considered as an additional term to the Hamiltonian, that should influence the  magnon spectrum equally, regardless of the $\mathbf{Q}$-vector. Therefore, if we consider the $g$-factor as a slope of the field dependence of the excitation energy, one would expect it to be the same at all points. However in real systems, its behavior may  significantly differ from the classical examples described above, and measurement of the dispersion relations for the magnetic excitations, and particularly their field dependence can give  valuable information on various exchange interactions in the Hamiltonian. For example, a set of the isotropic and anisotropic components of the exchange interactions between the ions on the hexagonal sites, crystal field, and magnetoelastic parameters were obtained in case of Pr metal by measuring field dependence of the dispersion relations \cite{HoumannRainford79}.

\clearpage
\section{Other Examples of Magnetic Excitations}\label{Sec:SpinExcitationExamples}

As it was shown in sec.\,\ref{Sec:SpinExcitations}, spin-wave excitation can demonstrate a large variety of responses upon application of the magnetic field. It may turn out that the signals, which look absolutely similar, can  demonstrate a significantly different response in field, as they might have various ground states, spin orientation or represent different phases. Before discussing the original result from my thesis, it would be helpful to recall some  well-known examples, in which various excitations were studied as a function of external magnetic field. Therefore, in this section I will give a brief overview of the magnetic field dependencies of the spin gap for various materials.

\vspace{-15pt}
\subsection{Classic example of $\mathbf{Q}$\,=\,0 resonances in anti- and ferromagnets}\label{SubSec:ESRforAFMandFM}

\begin{wrapfigure}[20]{r}{0.5\textwidth}\vspace{-10pt}
    \vspace{-0pt}\includegraphics[width=1\linewidth]{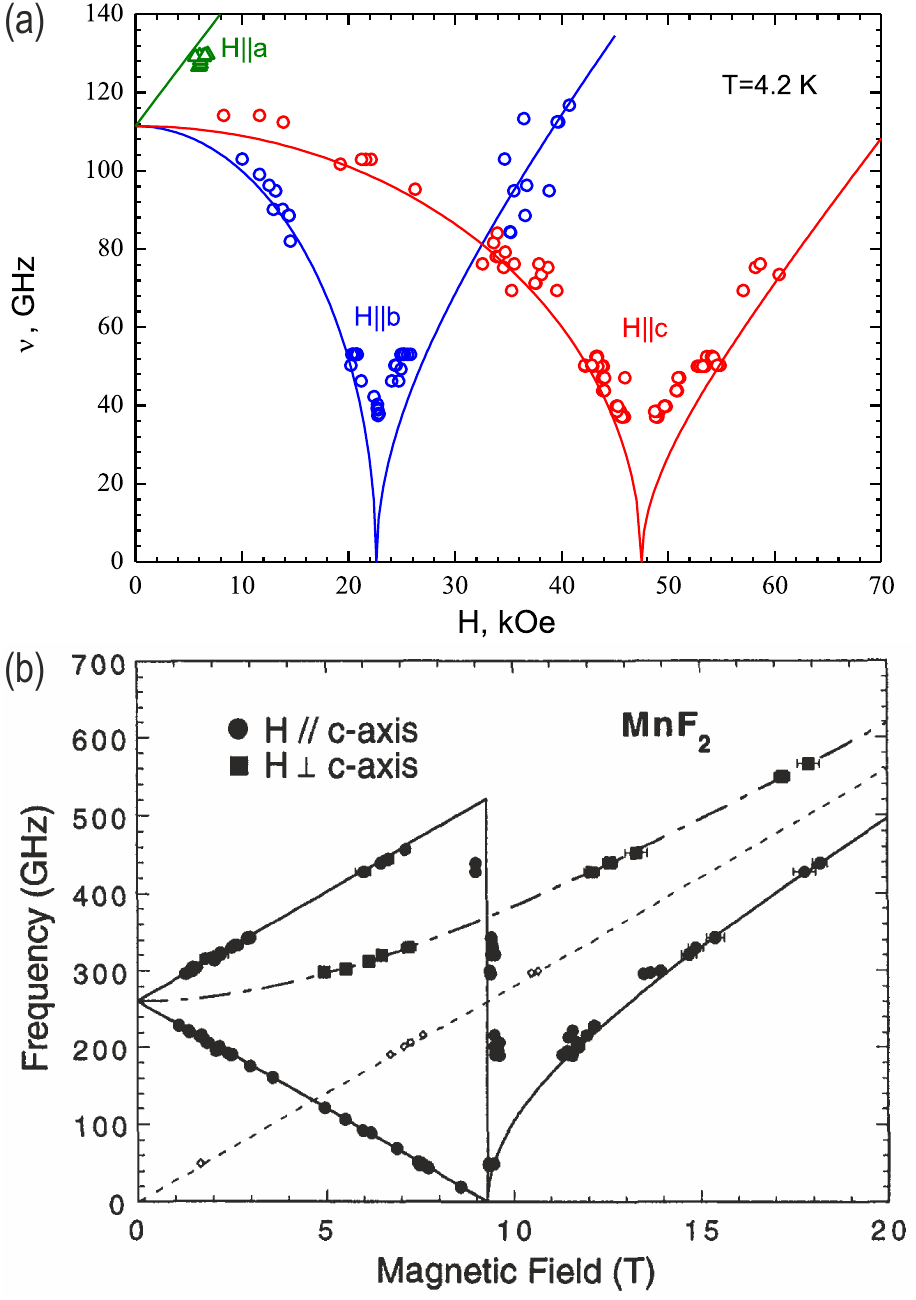}\vspace{0pt}
    \vspace{-28pt}\caption{Magnetic field dependence of the spin-gap measure with the ESR for: (a) ferromagnetic PbFeBO$_{4}$; (b) aniferromagnetic MnF$_{2}$. Figures were taken from refs.\,\citenum{PankratsSablina16,HagiwaraKatsumata99}, respectively.}
    \label{Fig:ExampleSpinWaveESR}\vspace{-0pt}
\end{wrapfigure}
First of all, let us compare field dependencies from the previous section with those actually measured. As shown in fig.\,\ref{Fig:ExampleSpinWaveESR}, measurements done on ferromagnetic PbFeBO$_{4}$ and antiferromagnetic MnF$_{2}$ single crystals at low temperatures, well below the ordering temperature, indicate the presence of a finite spin gap at zero field. As it is expected for the case of a ferromagnet in fig.\,\ref{Fig:SpinGapFieldDep1}(b), a magnetic field applied perpendicular to the easy magnetization axis $\mathbf{a}$, first softens the resonance energy until it becomes zero when the field reaches the value of the anisotropy field for this direction, and then it scales monotonically  with the field, as shown in fig.\,\ref{Fig:ExampleSpinWaveESR}(a) with red and blue colors. For the magnetic field applied parallel to the easy axis, one expects proportionality between the spin gap value and the field, which is confirmed by experiments and shown in fig.\,\ref{Fig:SpinGapFieldDep1}(a) in red and in fig.\,\ref{Fig:ExampleSpinWaveESR}(a) in green.

Applying magnetic field parallel to the easy $c$-axis of MnF$_{2}$, we can see a splitting of the zero field resonance into two branches, for one of which the energy increases with field while for the other branch it decreases. Above 9.3\,T a spin-flop transition takes place, and only one branch is observed. As we have already learned from the sec.\,\ref{SubSec:MagneticFieldDependenceSpinExcitation}, monotonic field dependence of one mode is expected for a configuration when the field is applied perpendicularly to the $c$-axis. All these experimental results, shown in fig.\,\ref{Fig:ExampleSpinWaveESR}(b), are in good agreement with theory described earlier and shown in fig.\,\ref{Fig:SpinGapFieldDepAFM}(a-d).

\vspace{-5pt}\enlargethispage{10pt}
\subsection{One-dimensional spin-$\frac{1}{2}$ antiferromagnet}\label{SubSec:s0p5antiferromagnet}
\begin{wrapfigure}[28]{r}{0.45\textwidth}\vspace{-15pt}
    \vspace{-0pt}\includegraphics[width=1\linewidth]{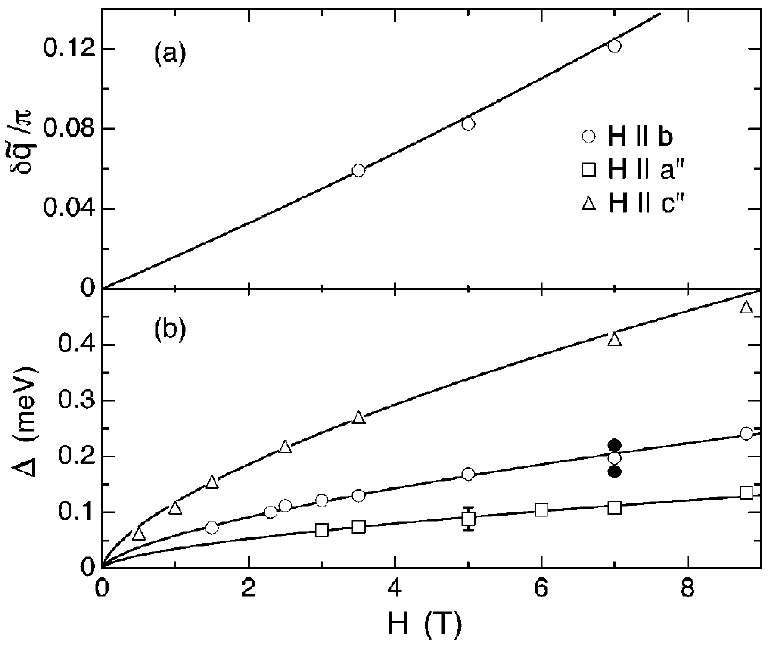}\vspace{0pt}
    \vspace{-20pt}\caption{ Field dependence of the: (a) displacement $\delta\tilde{q}$ of the peaks; (b) energy gap derived from fits to specific heat data in Cu(C$_6$D$_5$COO)$_{2}$$\cdot$D$_2$O. Filled symbols are the gaps measured by neutron scattering at $\tilde{q}=\pi$ and $\tilde{q}=1.12\pi$. Figure was taken from ref.\,\citenum{DenderHammar97}.}
    \label{Fig:ExampleSpinWaveNeutron1}\vspace{0pt}
    \vspace{-0pt}\includegraphics[width=1\linewidth]{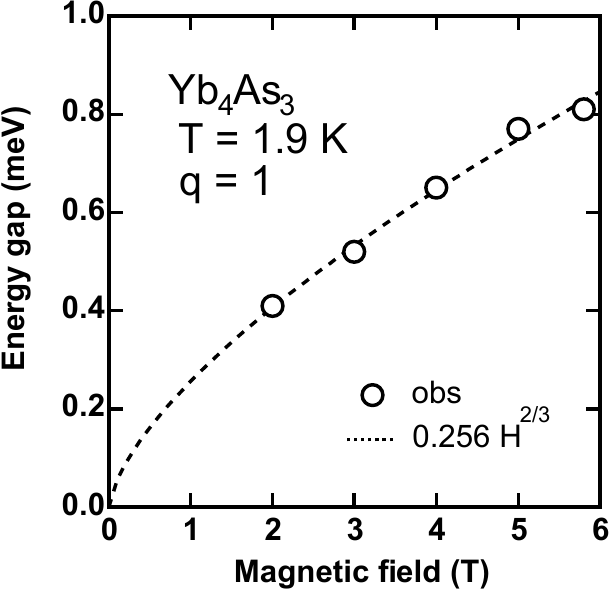}\vspace{0pt}
    \vspace{-20pt}\caption{Magnetic field dependence of the energy gap in Yb$_4$As$_3$. Figure were taken from ref.\,\citenum{KohgiIwasa01}.}
    \label{Fig:ExampleSpinWaveNeutron2p1}\vspace{-0pt}
\end{wrapfigure}
Fortunately or unfortunately, the magnetic field dependence of the spin gap is not limited to just such simple examples as mentioned above. Let me now show some of the examples where neutron scattering techniques were used to examine spin fluctuations. For example, in the case of a one-dimensional antiferromagnet with spin $S=1/2$,  new low-energy modes in the excitation spectrum at incommensurate wave vectors are predicted by theory, unlike in conventional antiferromagnets where a spin reorientation is expected by classical spin-wave theory \cite{MuellerThomas81}.  Direct experimental evidence for field-dependent incommensurate modes was found in copper benzoate Cu(C$_6$D$_5$COO)$_{2}$$\cdot$D$_2$O below $T_{\text{N}}$. Note that in order to be able to do neutron scattering experiments, deuterated single crystals were prepared (see sec.\ref{SubSec:NuclearAndMagneticScattering} to understand why). At zero field, a single peak at a low energy of $\hbar\omega=0.21\,\text{meV}$ can be found, while incommensurate values of the wave vector $\tilde{q}=\pi\pm\delta\tilde{q}(H)$ are induced by magnetic field, as shown in fig.\,\ref{Fig:ExampleSpinWaveNeutron1}(a). Dependence of the spin gap was deduced from specific heat measurements, which probe the momentum-space average of the density of states, and with neutron scattering for various field directions. A summary of the field dependence is shown in fig.\,\ref{Fig:ExampleSpinWaveNeutron1}(b), where solid lines are the power law $\Updelta(H)=AH^{\alpha}$ with $\alpha=0.65(3)$, and prefactors $A$ are in ratios $1:2.0:0.55$ for field applied along the $\mathbf{b}$, $\mathbf{c}^{\prime\prime}$, and $\mathbf{a}^{\prime\prime}$ directions respectively.

The inelastic neutron scattering experiments on a single-crystal sample of another one-dimensional spin $S=1/2$ antiferromagnet Yb$_4$As$_3$ shows a similar power law behavior of the spin gap value, as shown in fig.\,\ref{Fig:ExampleSpinWaveNeutron2p1}. However the excitation spectrum demonstrates a meaningful change upon application of the magnetic field only around $q=1$ \cite{KohgiIwasa01}, unlike those predicted for the pure Heisenberg system and mentioned above \cite{MuellerThomas81}.

\vspace{-5pt}
\subsection{Triplons in a spin-$\frac{1}{2}$ dimer system}\label{SubSec:s0p5dimer}
\begin{wrapfigure}[15]{l}{0.45\textwidth}\vspace{-15pt}
    \vspace{-0pt}\includegraphics[width=1\linewidth]{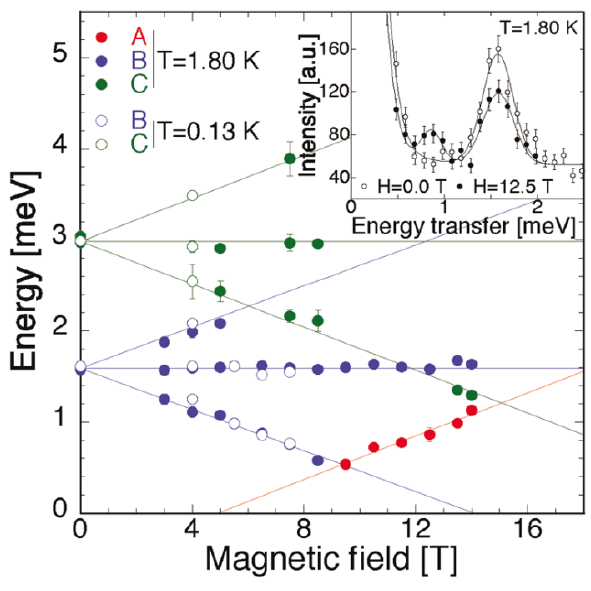}\vspace{0pt}
    \vspace{-25pt}\caption{Magnetic field dependence of the energy gap in NH$_4$CuCl$_3$.  Solid lines represent linear fits. Inset shows INS spectra measured at $H=12.5$\,T and in zero field. Figure were taken from ref.\,\citenum{RueeggOettli04}.}
    \label{Fig:ExampleSpinWaveNeutron2p2}\vspace{-0pt}
\end{wrapfigure}
Elastic and inelastic neutron scattering experiments were done on the dimer spin $S=1/2$ system NH$_4$CuCl$_3$. The low-temperature structure of NH$_4$CuCl$_3$ was assumed to consist of three inequivalent Cu$^{2+}$ dimer sites, however the observed long-range antiferromagnetic ordering below  $T_{\text{N}}=1.3$\,K presumably occurs only on one dimer site. In zero magnetic field two gaped excitations at $\hbar\omega\sim1.6$ and $3.0$\,meV, which arise from the remaining two Cu$^{2+}$ dimer sites left in a gapped singlet state, were reported. Application of the magnetic field splits the observed signal, and a linear increase of the Zeeman energy is shown in fig.\,\ref{Fig:ExampleSpinWaveNeutron2p2}. The observation of such a splitting directly proves that these two excitations have singlet-triplet nature \cite{RueeggOettli04}. It should be also noted that  the excitation energies above and  below the ordering temperature, $T_{\text{N}}=1.3$\,K, do not change within the experimental error. It is important to note that energy of the two reported  gapped excitations in zero field is in good agreement with the extrapolated zero-field frequencies measured with  ESR \cite{KurniawanTanaka99}.

\vspace{-5pt}
\subsection{Haldane gap in a one-dimensional spin-$1$ antiferromagnet}\label{SubSec:s1p0antiferromagnet}
\begin{wrapfigure}[11]{l}{0.45\textwidth}\vspace{10pt}
    \vspace{-10pt}\includegraphics[width=1\linewidth]{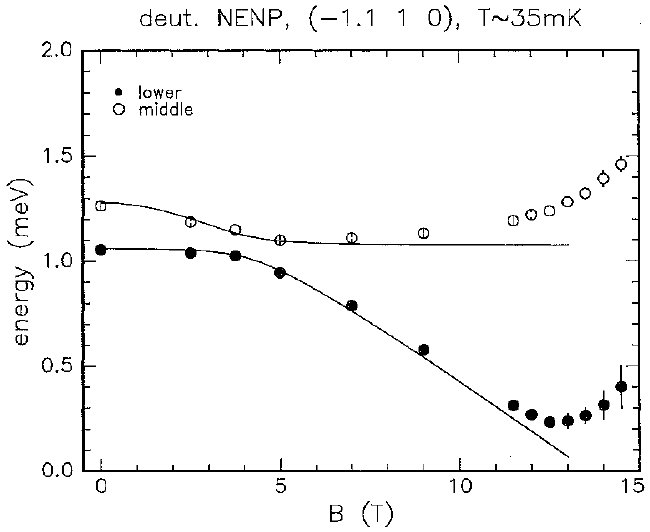}\vspace{0pt}
    \vspace{-30pt}\caption{Field dependence of the easy $a$\,--\,$c$ plane excitations. Figure was taken from  ref.\,\citenum{EnderleRegnault00}.}
    \label{Fig:ExampleSpinWaveNeutron3}\vspace{-5pt}
\end{wrapfigure}

It was predicted by F.~D.~M.~Haldane that one-dimensional Heisenberg antiferromagnets  with integer spin should have an energy gap between the singlet ground state and the first excited triplet states \cite{Haldane83,YamashitaIshii00}. One of the ideal system for studying physics of the Haldane gap is the one dimensional $S=1$ antiferromagnet Ni(C$_2$H$_8$N$_2$)$_2$NO$_2$ClO$_4$. In order to verify  some of the theoretical results it was also studied with neutron scattering. As it is shown in fig.\,\ref{Fig:ExampleSpinWaveNeutron3}, two (out of three, third gap with $\Updelta=2.5\,\text{meV}$ is not shown) distinct gaps can be
found in zero field. Splitting between them is determined by a sizable orthorhombic single-ion anisotropy \cite{RegnaultRenard97}. Presence of the  two different NO$_2$ and C$_2$H$_8$N$_2$ groups introduces a staggered field to the Ni$^{2+}$ sites upon application of an external field. Such configuration was studied theoretically \cite{ ChibaAjiro91,MitraHalperin94}, and should impose important consequences on the excitation spectrum.  In particular,  the spin-gap of the lowest excitation should not depend on the applied field. It was shown that the lower mode softens in increasing field, whereas the middle mode is found to be weakly field dependent. Above the characteristic field $H_{c}\approx13.6\,\text{T}$, the lower mode starts to increase monotonically with field, while the slight increase in energy of the  middle mode can be found at lower values of magnetic field. Absence of the gapless phase is explained by a crossover from a low-field phase with a Haldane gap to a high-field phase with a gap induced by the staggered field \cite{BerthierLevy02}. 

\subsection{CEF excitations}\label{SubSec:CEFExcitation}

Discussed in sec.\,\ref{Sec:CrystalFieldEffect} the crystal-field level scheme should exhibit some modification with application of a magnetic field. Considering a Zeeman term in the  Hamiltonian that  describes the system, it becomes possible to calculate the additional splitting of the CEF levels. In the case of CeB$_{6}$ it is difficult to observe such an effect, as the relatively large CEF splitting would require too large values of the magnetic field.

\begin{figure}[!t]
    \vspace{-0pt}\includegraphics[width=1\linewidth]{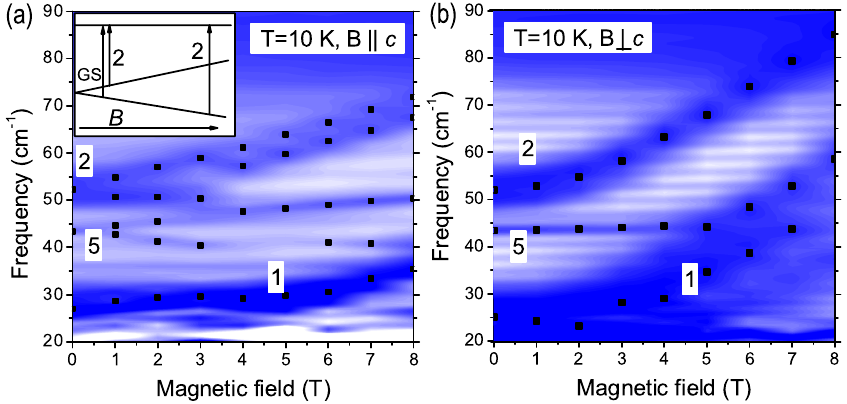}\vspace{0pt}
    \vspace{-25pt}\caption{Field dependence of the crystal-field excitations  Ho$^{3+}$ ions. Transmission maps were measured with far-infrared spectroscopy for (a) $B\parallel c$, and (b) $B\perp c$. Darker color indicates lower transmission. The inset shows a scheme of the transition 2, which happens from the ground state doublet. Figure was taken from ref.\,\citenum{TalbayevLaForge08}.}
    \label{Fig:ExampleSpinWaveNeutronCEF}\vspace{-25pt}
\end{figure}

Multiferroic hexagonal manganite HoMnO$_{3}$ was studied by means of far-infrared spectroscopy. In zero  magnetic field,  the crystal field of $C_{3v}$ and $C_{3v}$ symmetry splits the  degenerate $J=8$ manifold of the Ho$^{3+}$ ions into a set of singlet and doublet states. A magnetic-field applied along the $c$ axis, as shown in fig.\,\ref{Fig:ExampleSpinWaveNeutronCEF}(a), displays a typical  Zeeman splitting predicted for a doublet. At higher field  above 4\,T the upper branch of transition 2 exhibits an additional splitting due to the  hyperfine interactions. The absence of the lower branch in the second transition above 3\,T is explained by the thermal depopulation of the upper branch, shown in insert in fig.\,\ref{Fig:ExampleSpinWaveNeutronCEF}(a). As can be seen in fig.\,\ref{Fig:ExampleSpinWaveNeutronCEF}(b), the magnetic field applied  perpendicular to the $c$ axis produces no Zeeman splitting of the ground state doublet. An interesting observation is that in addition to the CEF splitting, a classical response from the antiferromagetic magnon (see fig.\,\ref{Fig:SpinGapFieldDepAFM}), can be found in figs.\,\ref{Fig:ExampleSpinWaveNeutronCEF}(a)\,and\,(b), marked as  transition 5 \cite{TalbayevLaForge08}.

\vspace{-0pt}

\vspace{-25pt}
\subsection{Spin resonance in heavy-fermion superconductors}\label{SubSec:SpinResonanceHF}
\begin{wrapfigure}[16]{l}{0.5\textwidth}\vspace{-15pt}
    \vspace{-0pt}\includegraphics[width=1\linewidth]{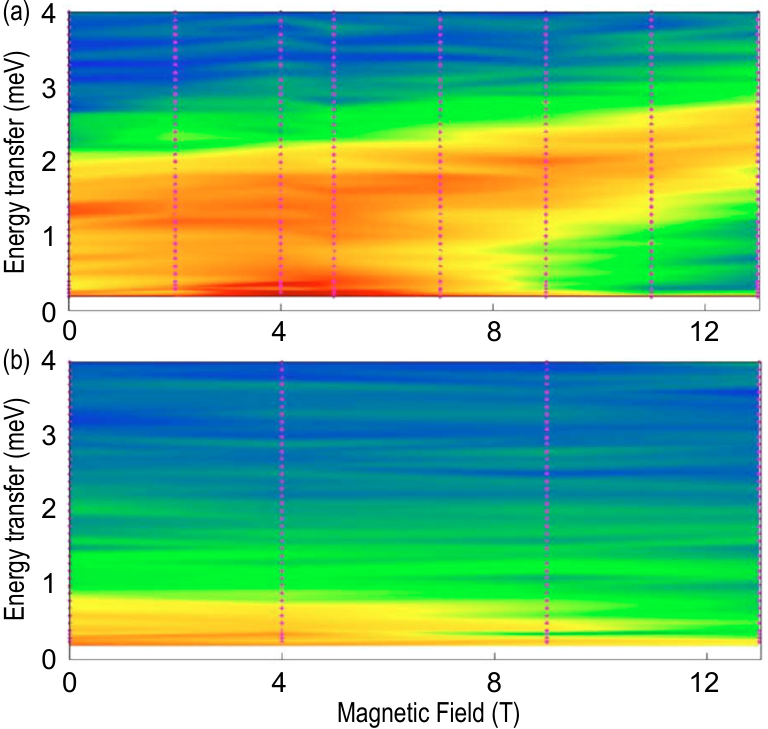}\vspace{0pt}
    \vspace{-20pt}\caption{Magnetic field dependence of the response at $\mathbf{Q}_{}=(1~0~\frac{1}{2})$ at: (a) 1.9\,K in the antiferromagnetic; (b) 20\,K in the paramagnetic state in UPd$_{2}$Al$_{3}$. Measured data points are marked by dots. The white space marks the resolution cutoff. Figure was redrawn from ref.\,\citenum{BlackburnHiess06}.}
    \label{Fig:ExampleSpinWaveNeutron7}\vspace{-0pt}
\end{wrapfigure}
Measurements at the antiferromagnetic zone center at $(1\,\,0\,\,\frac{1}{2})$ in UPd$_{2}$Al$_{3}$, which exhibits a superconducting  phase below $T_{c}\simeq1.8\,\text{K}$,  at various values of magnetic field below and above $T_{\text{N}}\simeq14.3\,\text{K}$ show the presence of two contributions to the excitation spectra. The inelastic response found within the AFM phase, centered at 1.4\,meV at zero field, gradually increases its energy up to 2.5\,meV at 15\,T, while an accompanying quasielastic response first increases its intensity up to 4\,T and then decreases, becoming very broad and flat, as shown in fig.\,\ref{Fig:ExampleSpinWaveNeutron7}(a). The observed decrease of the quasielastic intensity can also be found in the superconducting state at 0.3\,K, while the  paramagnetic regime at 20\,K  shows only purely quasielastic signal and no field induced excitations  \cite{BlackburnHiess06}. Intensity of the quasielastic response decreases  monotonically as a function of magnetic field as shown in fig.\,\ref{Fig:ExampleSpinWaveNeutron7}(b).

\begin{wrapfigure}[30]{r}{0.35\textwidth}\vspace{-0pt}
    \vspace{-0pt}\includegraphics[width=1\linewidth]{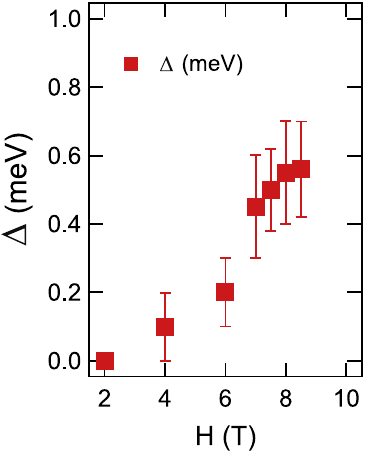}\vspace{0pt}
    \vspace{-30pt}\caption{Field dependence of the spin gap in CeCu$_{2}$Ge$_{2}$. Figure was taken from ref.\,\citenum{Singh14}.}
    \label{Fig:ExampleSpinWaveNeutron4}\vspace{-0pt}
    \vspace{-0pt}\includegraphics[width=1\linewidth]{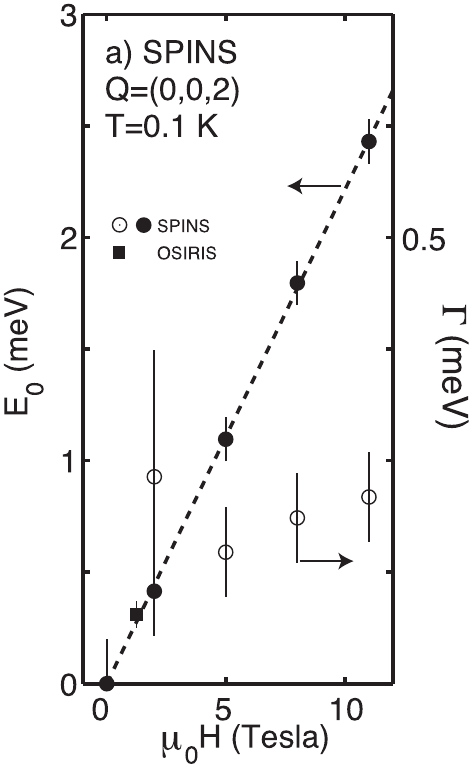}\vspace{0pt}
    \vspace{-25pt}\caption{Magnetic field dependence of the resonance energy and linewidth in YbRh$_{2}$Si$_{2}$. Reproduced from ref.\,\citenum{StockBroholm12}.}
    \label{Fig:ExampleSpinWaveNeutron8}\vspace{-0pt}
\end{wrapfigure}
Neutron spin resonance can be also observed in unconventional superconductors. In particular in CeCu$_{2}$Ge$_{2}$, which becomes superconducting below $T_{c}\simeq0.64\,\text{K}$ under application of $\simeq10\,\text{GPa}$ pressure. It was concluded that below  $T_{\text{N}}\simeq4\,\text{K}$,  Ce$^{3+}$ ions form a long-range spiral spin-density-wave ground  state, associated with the long-range incommensurate magnetic peaks at $\mathbf{Q}_{M}=\tau\pm\mathbf{k}$, where modulation vector $\mathbf{k}=(0.285,\,0.285,\,0.54)$ is defined  with respect to a simple commensurate position $\tau$. Upon application of magnetic field along $[\bar{1}\,1\,0]$, originally gapless at zero field the inelastic spectrum gradually develops a spin gap, as shown in fig.\,\ref{Fig:ExampleSpinWaveNeutron4}. As the system approaches QCP in a applied magnetic field of $\simeq8\,\text{T}$, the spin gap increases to its maximum value of $\Updelta=0.56\,\text{meV}$ \cite{Singh14}.

\vspace{-5pt}
Spin fluctuations in the heavy-fermion metal YbRh$_{2}$Si$_{2}$ show interesting behavior. At a high temperature of 15\,K, the ferromagnetic $\mathbf{Q}_{}=(002)$ wave vector demonstrates commensurate excitations, which form a ridge at the zone center. However, upon cooling, the ferromagnetic fluctuations evolve into two incommensurate correlations located at $\mathbf{Q}_{}=(\pm\delta\,\pm\delta\,2)$ with $\delta=0.14\pm0.04\,\text{r.l.u.}$ Application of a weak magnetic field merges the incommensurate signal into a resonance peak, well localized in $\mathbf{Q}$ space. The magnetic field dependence of the resonance peak is shown in fig.\,\ref{Fig:ExampleSpinWaveNeutron8}. Its energy scales linearly with magnetic field as $g\mu_{\text{B}}\mu_{0}H$ with $g=3.8\pm0.2$, while the linewidth almost does not change \cite{StockBroholm12}. The $g$-factor was also revealed in an ESR experiment \cite{DuqueBittar09} on the same compound and is in good agreement with the one obtained from neutron scattering.

\begin{wrapfigure}[20]{l}{0.5\textwidth}\vspace{-15pt}
    \vspace{-0pt}\includegraphics[width=1\linewidth]{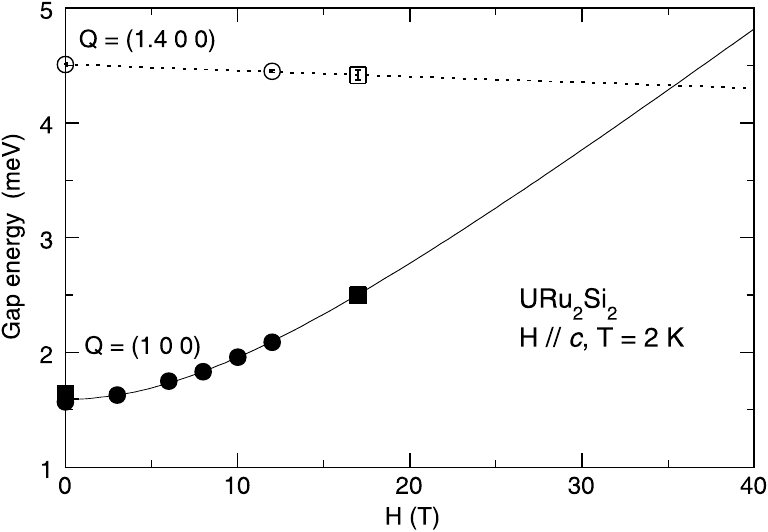}\vspace{0pt}
    \vspace{-25pt}\caption{The gap energies as a function of magnetic field in URu$_{2}$Si$_{2}$. Figure was taken from  ref.\,\citenum{BourdarotFak03}.}
    \label{Fig:ExampleSpinWaveNeutron6}\vspace{-0pt}
    \vspace{-0pt}\includegraphics[width=1\linewidth]{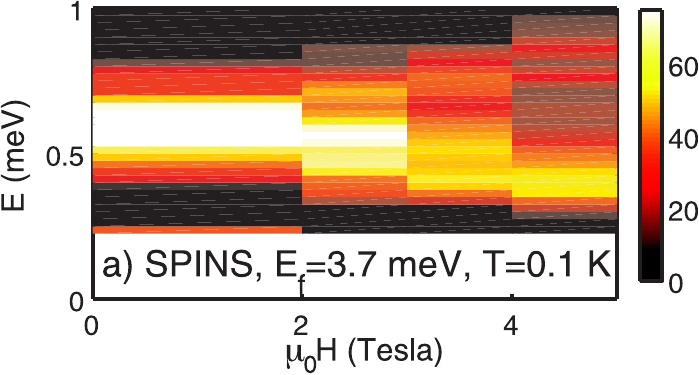}\vspace{0pt}
    \vspace{-25pt}\caption{Magnetic field dependence of the resonance energy after background correction in CeCoIn$_5$. Figure was taken from ref.\,\cite{StockBroholm12a}.}
    \label{Fig:ExampleSpinWaveNeutron9}\vspace{-0pt}
\end{wrapfigure}

\vspace{-5pt}
A heavy-fermion superconductor URu$_{2}$Si$_{2}$ ($T_{\text{c}}\simeq1.2\,\text{K}$) shows the static antiferromagnetic order below $T_{\text{N}}\simeq17.5\,\text{K}$. The magnetic excitation spectrum  at zero field exhibits two gapped modes with $\Updelta=1.59\,\text{meV}$ and $\Updelta=4.51\,\text{meV}$ at two distinct wave vectors, $\mathbf{Q}_{}=(1\,0\,0)$ and $\mathbf{Q}_{}=(1.4~0~0)$, respectively. Upon application of magnetic field, remarkably different behavior of  excitations, shown in fig.\,\ref{Fig:ExampleSpinWaveNeutron6}, was found. Similarly to the previous example, the energy of the lower mode increases strongly with applied field, without any sign of saturation at higher fields. Its behavior was also described with the same fit. On the other hand the second resonance remains nearly constant, with only a slight linear decrease in energy with increasing field \cite{BourdarotFak03}.

CeCoIn$_5$ is a heavy-fermion unconventional superconductor with $T_{\text{c}}\simeq2.3\,\text{K}$. Depending on the magnetic field orientation, superconducting phase can be suppressed either with $\sim12$ or $\sim5$\,T for the field along $a$-$b$ or $c$, respectively. An incommensurate magnetic order with the propagation vector $\mathbf{Q}_{}=(0.45\,0.45\,\frac{1}{2})$, which appears to be linked to the superconducting phase,  was found to exist in a narrow field range applied within $a$-$b$ plane in close proximity to the phase boundary \cite{KenzelmannStraessle08}. Upon entering the superconducting phase, the resonance peak, centered at $\mathbf{Q}_{}=(\frac{1}{2}\,\frac{1}{2}\,\frac{1}{2})$ at the energy transfer 0.6\,meV was found. The origin of the resonance mode remains an open question, as it can be considered as a magnon in a disordered N\'{e}el state or as an $S=1$ collective state \cite{RaymondKaneko12}. In the case of the latter, magnetic field should split observed resonance into three peaks, reflecting an excitation from a singlet ground state to a triplet excited state. Such kind of splitting was observed in cuprates, with the longitudinal  and transverse polarizations of the field dependent and independent peaks, respectively \cite{StockBroholm12a}. Therefore, in order to uncover the underlying structure, and particularly to investigate the multiplicity of the resonance peak, it was essential to perform  measurements under applied magnetic field.

The effect of magnetic fields is shown in \ref{Fig:ExampleSpinWaveNeutron9}, where a Zeeman splitting into two peaks, instead of theoretically predicted three \cite{AkbariThalmeier12},  with energies given by $\Updelta_{\pm}=\Updelta_{0}\pm g\mu_{\text{B}}\mu_{0}H$ with $g=0.96\pm0.05$ was found. Spectral weight is distributed unequally between two modes, as the $E_{+}$ peak shows a consistent trend to decreasing intensity while the $E_{-}$ peak intensity is constant within error \cite{StockBroholm12a}. Polarized inelastic neutron scattering under a magnetic field revealed that the low energy mode consists of two contributions. One of them is chiral and is the counterpart of the high energy line. This kind of splitting suggests that the observed spin resonance is a degenerate mode with three fluctuation channels, a Zeeman  contribution that exhibits splitting and an additional longitudinal mode \cite{RaymondKaneko12}.

As one can see, a considerable theoretical interest regarding the underlying physics of the observed resonances was in some cases satisfied. At first glance, similar peaks were observed in various materials. However, application of the magnetic field made it possible to determine essential difference between these signal. Cerium hexaboride is already been known for more than 50 years. Its physics is governed by the competing interactions, with both itinerant and local electrons playing important roles.  Theoretical models, that tried to describe  available data unfortunately do not allow accurate interpretation of the excitation spectra. Therefore, despite all attempts  to explain observed physical phenomena, we are still far away from the full understanding. In order to develop a reliable theoretical model and to validate existing calculations it is very important to check the response of the system to the application of a magnetic field along various directions, different phases and doping levels. This is the key idea of my dissertation, which is devoted to the study of the influence of the magnetic field, applied along different directions, as well as on the effects upon La or Nd-doping on the excitation spectrum in CeB$_{6}$ and Ce$_3$Pd$_{20}$Si$_6$.

\chapter{Ce$_{\text{1}-x}$La$_x$B$_\text{6}$ }\label{Chap:ResultsCeLaB6}

\vspace{-15pt}
\section{Introduction to Ordering Phenomena in CeB$_\text{6}$}\label{SubSec:OrderingPhenomenaCeB6}

\subsection{CeB$_\text{6}$ parent compound}

\begin{wrapfigure}[10]{r}{0.45\textwidth}\vspace{-10pt}
\vspace{-0pt}\includegraphics[width=0.45\textwidth]{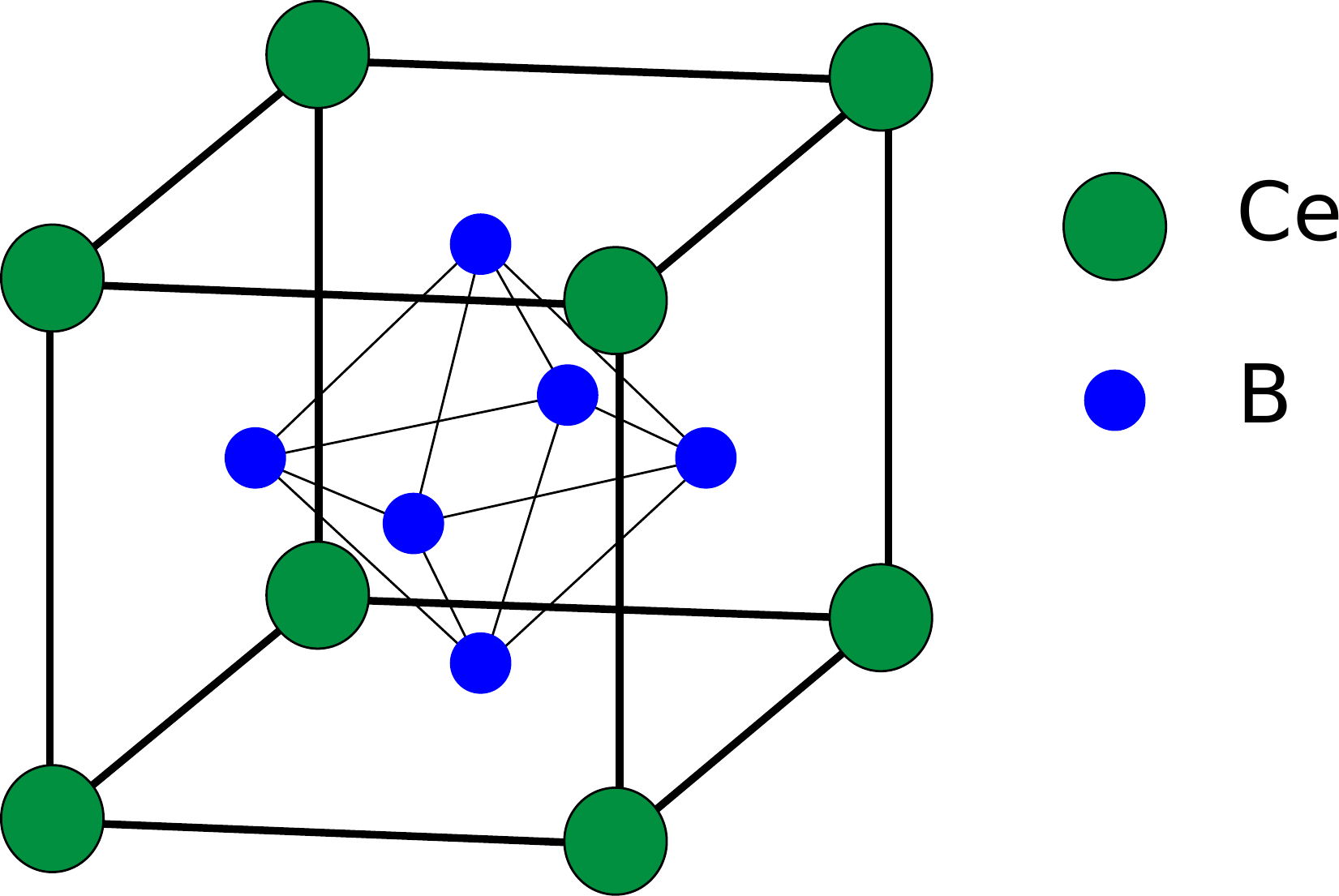}
\vspace{-20pt}\caption{ Sketch of the crystal structure of CeB$_\text{6}$, showing the cubic unit cell. Figure from ref.\,\citenum{CameronFriemel16}\label{Fig:CeB6CrystalStructure}.}
\end{wrapfigure}

The pure CeB$_6$ parent compound is a heavy-fermion metal with a simple cubic crystal structure characterized by the space group $Pm\overline{3}m$ and lattice constant 4.14\,{\AA} \cite{Alper76}. This compound has a unique crystal structure, where rare-earth \textit{Re} atoms sit in the corner and B$_6$ octahedron can be found in the body centre of a cubic unit cell, as shown in fig.\,\ref{Fig:CeB6CrystalStructure}. It is generally believed that 20 valence electrons are required to stabilize the B$_6$ octahedron, however six boron atoms can provide only 18 valence electrons. The remaining two electrons are donated by the metal atom. Accordingly in the case of cerium,  the electronic structure of which is [Xe]$4f^1 5d^1 6s^2$, the \textit{d}-electron forms the conduction band, two \textit{s}-electrons are donated to the boron octahedra and the single \textit{f}-electron remains localized    \cite{Longuet-HigginsRoberts54,GrushkoPaderno85,LiuYiu11}.

\begin{wrapfigure}[17]{r}{0.45\textwidth}\vspace{-15pt}
\vspace{-5pt}\includegraphics[width=0.45\textwidth]{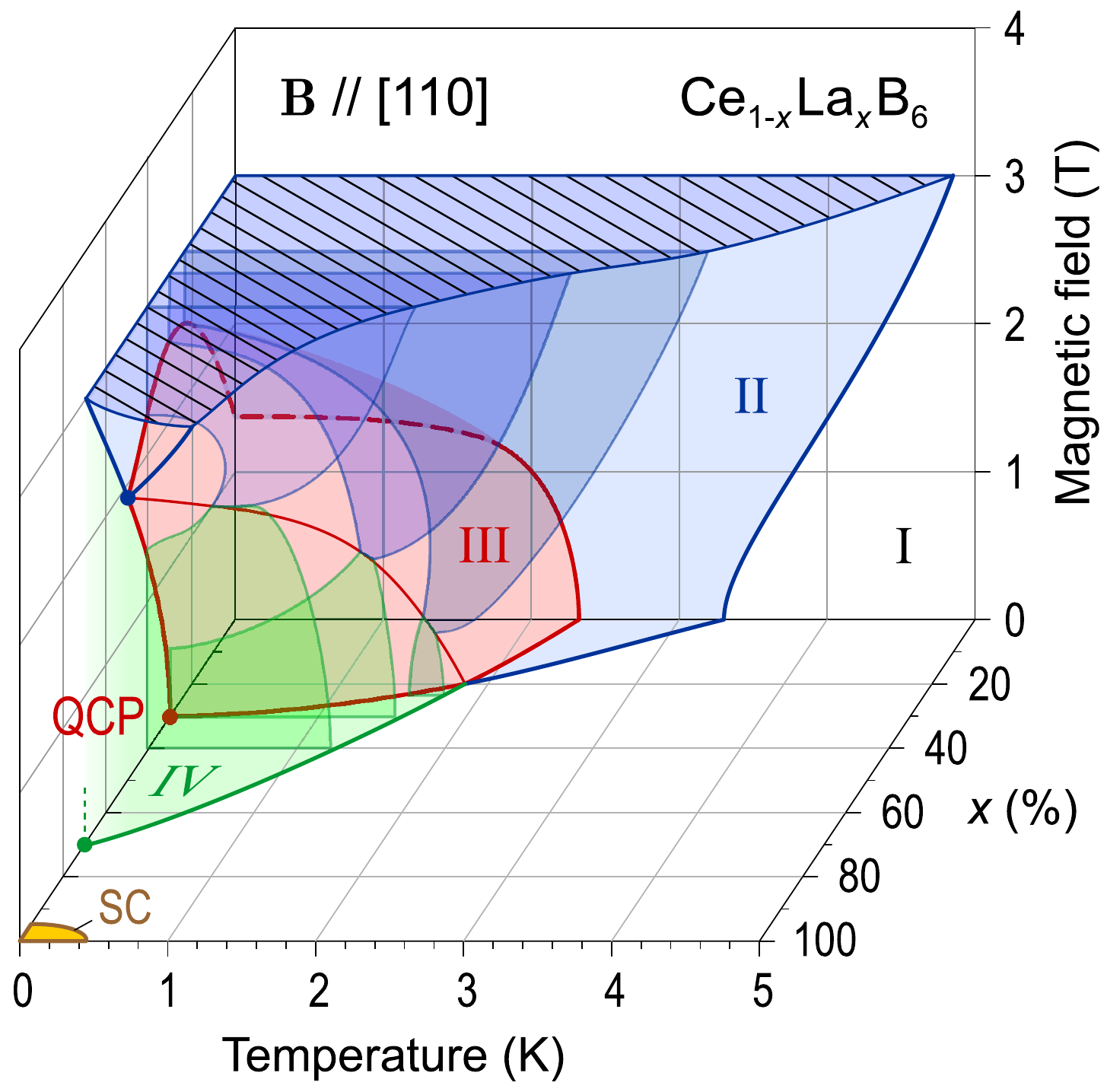}
\vspace{-28pt}\caption{ The phase diagram of Ce$_{\text{1}-x}$La$_x$B$_\text{6}$. Phase I stands for the paramagnetic normal state, phase II represents the AFQ phase, phase III is the AFM phase, and phase IV  is an elusive, less studied hidden-order phase presumably associated with the ordering of magnetic octupoles. The superconducting phase of the pure LaB$_\text{6}$ is marked as ``SC''. Figure from ref.\,\citenum{FriemelJang15}\label{Fig:PhaseDiagram3D}.}
\end{wrapfigure}

The magnetic phase diagram of Ce$_{\text{1}-x}$La$_x$B$_\text{6}$ solid solutions versus La concentration ($x$), temperature ($T$), and magnetic field ($\mathbf{B}$), which is shown in fig.\,\ref{Fig:PhaseDiagram3D}, demonstrates an amazingly rich interplay of magnetism, multipolar ordering phenomena, and superconductivity \cite{ErkelensRegnault87, TayamaSakakibara97, HiroiKobayashi98, KobayashiSera00, KondoTou06, KuwaharaIwasa07, ShiinaSakai07, KuwaharaIwasa09}. The most well-studied member of this family of compounds is the pure CeB$_\text{6}$ material, which is considered textbook example of a system with the so-called \textit{magnetically hidden order}, typically associated with an antiferromagnetic arrangement of magnetic quadrupole moments. This compound is a heavy-fermion metal characterized by the low-temperature specific-heat coefficient $\gamma\approx300$\,mJ/mol$\cdot$K$^2$ \cite{PeyssonAyache86}. In zero magnetic field, it exhibits an antiferromagnetic (AFM) ground state (phase~III) that persists in temperature up to $T_{\text N}=2.3$\,K. An antiferroquadrupolar (AFQ) ordered phase (phase~II) is found at somewhat higher temperatures, between $T_{\text N}$ and $T_{\text Q}=3.2$\,K. Above $T_{\text Q}$, the system enters a paramagnetic state. Application of weak magnetic fields completely suppresses phase~III and stabilizes phase~II over a broader temperature range. The exotic AFQ order remains hidden to neutron diffraction in zero magnetic field and can be therefore revealed only under an applied field \cite{SeraIchikawa01} or by using resonant X-ray diffraction, which is capable of directly probing the orbital degrees of freedom \cite{NakaoMagishi01, MatsumuraYonemura09, MatsumuraYonemura12}, as was mentioned in sec.\,\ref{SubSec:HiddenOrder} and sec.\,\ref{SubSec:ResonantXrayDiffraction} respectively.

Over recent years, an ever increasing amount of experimental evidence has accumulated, indicating that the commonly accepted localized description of magnetism in CeB$_\text{6}$ might be incomplete. Transport and thermodynamic properties were shown to deviate from the behavior expected within the Kondo-lattice model \cite{SluchankoBogach07}, whereas polarized neutron scattering revealed a weak magnetic Bragg peak in zero field at the AFQ wave vector, $\mathbf{Q}_{\text{AFQ}}=(\frac{1}{2} \frac{1}{2} \frac{1}{2})$, which was not expected within the multipolar ordering scenario and possibly represents a signature of a different order parameter of itinerant origin, similar to a spin density wave (SDW) \cite{PlakhtyRegnault05}. Finally, inelastic neutron scattering (INS) measurements \cite{FriemelLi12} also revealed an intense magnetic exciton mode at the same wave vector, which was fully unexpected in the framework of the prevailing theory. This led to a suggestion that the spin dynamics in CeB$_\text{6}$ is dominated by the itinerant magnetism of Ce\,5$d$ states \cite{AkbariThalmeier12}, which was previously neglected in theoretical calculations \cite{KusunoseKuramoto01, ThalmeierShiina03}.

The CeB$_\text{6}$ system bears many similarities to a number of actively studied ``hot-topic'' materials. Its electronic structure, composed of localized Ce\,4$f^1$ levels hybridized with itinerant Ce\,5$d$ and B\,2$p$ states \cite{GrushkoPaderno85} is closely analogous to that of heavy-fermion superconductors, such as CeCu$_2$Si$_2$ or CeCoIn$_5$. The spectrum of magnetic excitations exhibits a sharp exciton mode \cite{FriemelLi12} similar to the spin resonances found in most unconventional superconductors, including copper oxides \cite{ReznikBourges04}, iron pnictides \cite{InosovPark10, ParkFriemel11, FriemelLiu12}, and several heavy-fermion compounds \cite{StockBroholm08, StockertArndt11} with the notable exception of the noncentrosymmetric CePt$_\text{3}$Si \cite{InosovBourges11}. Multipolar ordering phenomena analogous to those observed in CeB$_6$ have been suggested to be responsible for the mysterious hidden-order phase in URu$_2$Si$_2$, which has so far avoided direct detection by conventional solid-state probes \cite{MydoshOppeneer11}. Intriguing parallels to other materials  with thus-far unidentified ``hidden'' order parameters can be found, such as CeRu$_2$Al$_{10}$ \cite{RobertMignot10}, or the small-gap ``Kondo insulators'' SmB$_6$ \cite{AlekseevMignot95} and YbB$_\text{12}$ \cite{MignotAlekseev05}, whose ground states remain poorly understood. Therefore the comprehensive description of the ordered phases and magnetic dynamics in CeB$_6$ on the background of its exceptionally simple crystal structure and chemical composition should have model character for a broad class of correlated-electron materials.

\subsection{Early inelastic neutron scattering measurements and their interpretation}

\begin{figure}[!t]
\includegraphics[width=\textwidth]{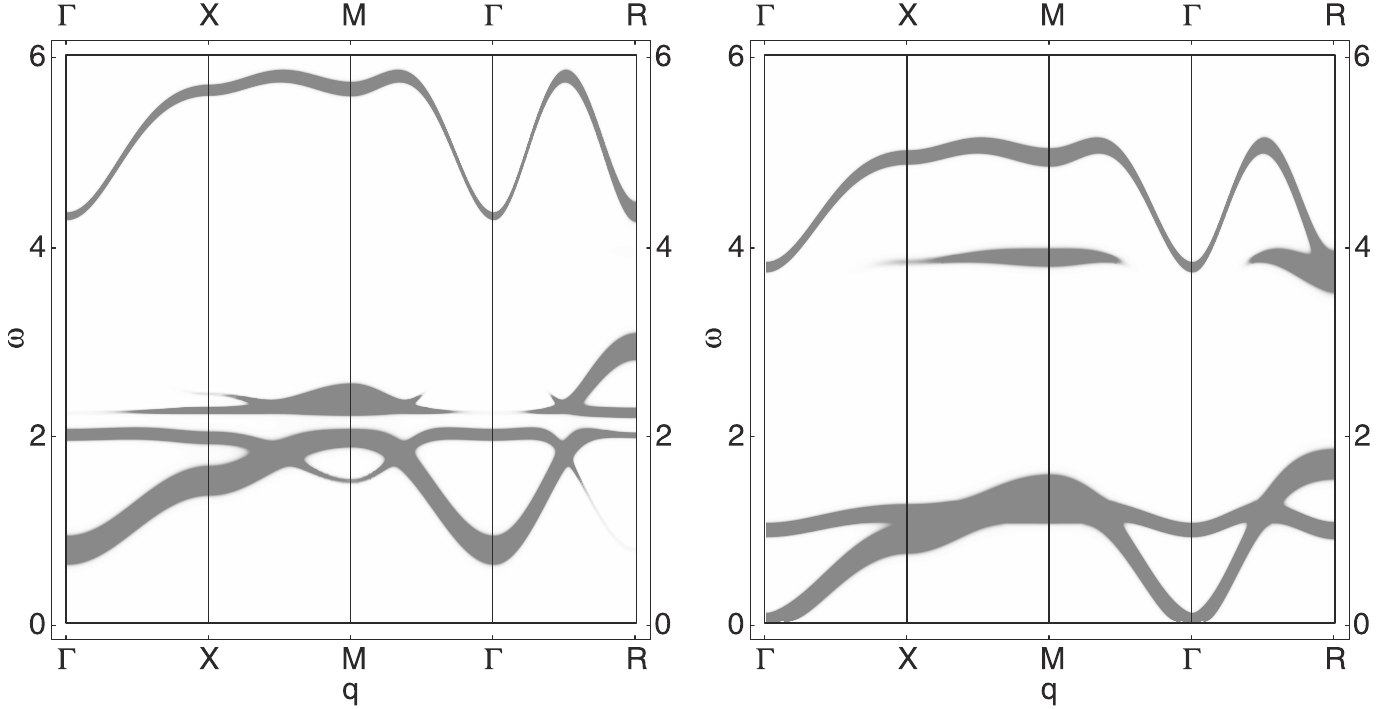}\vspace{-5pt}
\caption{\footnotesize Theoretically predicted dispersions of magnetic excitations in the AFQ phase in finite magnetic field (right) and in zero field (left). Figure from ref.\,\cite{ThalmeierShiina03}.}
\label{Fig:ThalmeierTheory}
\end{figure}

The first INS measurements on CeB$_6$, performed in the paramagnetic state, revealed low-energy quasielastic magnetic scattering, characterized by a temperature-dependent line width \cite{HornSteglich81}, and an intense high-energy crystal-field excitation, centered at 46\,meV \cite{ZirngieblHillebrands84}. The latter reportedly results from the splitting of the sixfold-degenerate 4$f^1$ ground state of the Ce$^{3+}$ ion into a $\Gamma_7$ doublet and $\Gamma_8$ quartet in its cubic crystal field. Momentum-resolved INS measurements on single crystals were first performed in phase~II \cite{RegnaultErkelens88}, demonstrating the appearance of an intense dispersive magnon branch induced by magnetic field. Unfortunately, these data were limited to only two high-symmetry directions of the reciprocal space and a single field value of 4\,T applied along $[001]$. No magnetic excitations were observed in zero field, and the nature of the field-induced magnon remained unclear. A more extensive dataset covering several magnetic fields and various positions of the reciprocal space (mostly in phase~II) has been presented by A.\,Bouvet \cite{Bouvet93}. In this work (most of it unpublished), it was suggested that dispersive magnetic modes could also be present in phase~II even in zero magnetic field. However, the energies of these modes were too low to be clearly resolved from the elastic line. As a result, the data analysis had to rely substantially on the line-shape analysis, which did not yield a reliable result. Nevertheless, these data motivated theoretical models in an attempt to describe the magnetic excitation spectrum in phase~II both with and without the application of magnetic field under the assumption of quadrupolar symmetry breaking \cite{KusunoseKuramoto01, ThalmeierShiina03}. Within these models, dispersive magnetic modes within phase~II can be understood as spin-orbital-wave excitations of the AFQ state, which are determined by multipolar interactions between localized entities involving both orbital and magnetic degrees of freedom. Fig.\,\ref{Fig:ThalmeierTheory} illustrates the theoretically calculated scattering function, $S(\mathbf{Q},\omega)$, which takes full account of intensities that would be expected in an INS experiment, as predicted by Thalmeier \textit{et al.} \cite{ThalmeierShiina03}. The degree of agreement between the outcomes of these models and the experimental data remained unsettled for a long time. Most recent experiments (see ref.\, \citenum{JangFriemel14}) indicate that the dispersive modes are in fact absent in zero field, which represents a major obstacle for the mentioned theoretical models.


\vspace{-0pt}

Until now, no systematic investigations of the spin dynamics within phase~III under the influence of magnetic field have been reported. There were also no high-quality data within phase~II that would cover the complete reciprocal space under various magnetic fields, thereby allowing for a detailed quantitative comparison with theoretical outputs. INS measurements on La-substituted samples have not been undertaken either. The described situation was a prerequisite for a more systematic investigation, readdressing the system both experimentally and theoretically. Such a study has benefited greatly from the substantial recent progress in neutron instrumentation (in particular, in cold-neutron time-of-flight spectrometers). In contrast to the earlier INS measurements, which yielded only a very fragmented dataset on the pure CeB$_6$ and lacked the energy resolution to resolve low-lying magnetic excitations in this system \cite{Bouvet93}, modern instrumentation allowed us for the first time to obtain systematic and conclusive high-resolution data over the multidimensional parameter space of high-quality Ce$_{\text{1}-x}$La$_x$B$_\text{6}$ single crystals, which will be presented in this thesis.

\section{Magnetic-Field Dependence of Collective Magnetic Excitations at the Zone Center in CeB$_{6}$}

Observation of a neutron spin resonance within a broad range of materials, in particular high-$T_{\text c}$ cuprates \cite{Eschrig06}, iron pnictides \cite{Dai15, Inosov16}, and heavy-fermion superconductors \cite{SatoAso01, StockBroholm08, StockertArndt11}, is recognized as an indicator of unconventional superconductivity. It was shown that sign-changing gap symmetry can lead to the existence of resonance behavior \cite{BulutScalapino96, EreminZwicknagl08, MazinSingh08, HirschfeldKorshunov11}. Of particular interest are inelastic neutron scattering (INS) results obtained on CeCoIn$_{5}$, where a sharp resonance peak was observed within the superconducting phase \cite{StockBroholm08, StockBroholm12a, RaymondKaneko12, RaymondLapertot15}. At first glance similar peaks were found in the antiferromagnetic (AFM) superconductor UPd$_{2}$Al$_{3}$ \cite{BlackburnHiess06, HiessBernhoeft06}, as well as in the normal state of the heavy fermion (HF) metal YbRh$_{2}$Si$_{2}$ \cite{StockBroholm12}, where superconductivity was recently discovered below $\sim2$\,mK \cite{SchuberthTippmann16}. Some of these results were summarized in sec.\,\ref{Sec:SpinExcitationExamples}. Another striking example of a resonant mode is given by the well known non-superconducting heavy-fermion antiferromagnet CeB$_{6}$ \cite{FriemelLi12,JangFriemel14}. The microscopic origins of such resonant magnetic excitations persisting in $f$\!-electron systems either with or without superconductivity may well differ among materials and are still hotly debated.

Application of an external magnetic field may help to unmask the differences between these various excitations. For instance, among $f\!$-electron compounds, a weak quasielastic signal gives rise to a field-induced ferromagnetic (FM) excitation in CeRu$_{2}$Si$_{2}$ \cite{SatoKoike04}. In YbRh$_{2}$Si$_{2}$, two incommensurate excitation branches merge into a commensurate FM resonance whose energy scales linearly with magnetic field \cite{StockBroholm12}, whereas in UPd$_{2}$Al$_{3}$ the energy gap initially remains almost constant inside the superconducting phase, but starts following a monotonic linear dependence at higher magnetic fields \cite{BlackburnHiess06}. The sharp resonance in CeCoIn$_{5}$ splits into a Zeeman doublet \cite{StockBroholm12a} rather than a theoretically predicted triplet~\cite{AkbariThalmeier12}, whereas in Ce$_{1-x}$La$_x$B$_6$ magnetic field reportedly leads to a crossover from itinerant to more localized behavior of spin fluctuations \cite{FriemelJang15}. Thus, application of external magnetic field is an important tool to distinguish different types of collective spin excitations and to develop microscopic theoretical models for the formation of the resonant modes.

A complementary way of probing spin dynamics is the electron spin resonance (ESR). For a long time it was believed that due to the effect of Kondo screening, no ESR signal could be observed in Kondo lattices, as the spin-orbit coupling significantly shortens electron spin relaxation times, leading to a broad and weak signal \cite{SichelschmidtIvanshin03}. This established opinion was impugned when for the first time a Yb$^{3+}$ resonance was observed in a dense Kondo lattice system \cite{SichelschmidtIvanshin03, SichelschmidtWykhoff07, SchaufusKataev09}. Various theoretical explanations proposed complementary models which explained the existence of the narrow ESR line \cite{AbrahamsWoelfle08, Schlottmann09, ZvyaginKataev09}, while further investigation of different Kondo lattice systems demonstrated that FM correlations are of principal importance for the observation of the ESR signal \cite{KrellnerFoerster08}. While this empirical result summarized observations from a limited number of $f$\!-electron compounds, no clear counterexamples are known to date.

\subsection{Magnetic-field dependence of the spin resonance studied with INS and ESR}

\begin{wrapfigure}[25]{r}{0.5\textwidth}\vspace{-25pt}
\noindent\begin{center}\includegraphics[width=0.5\textwidth]{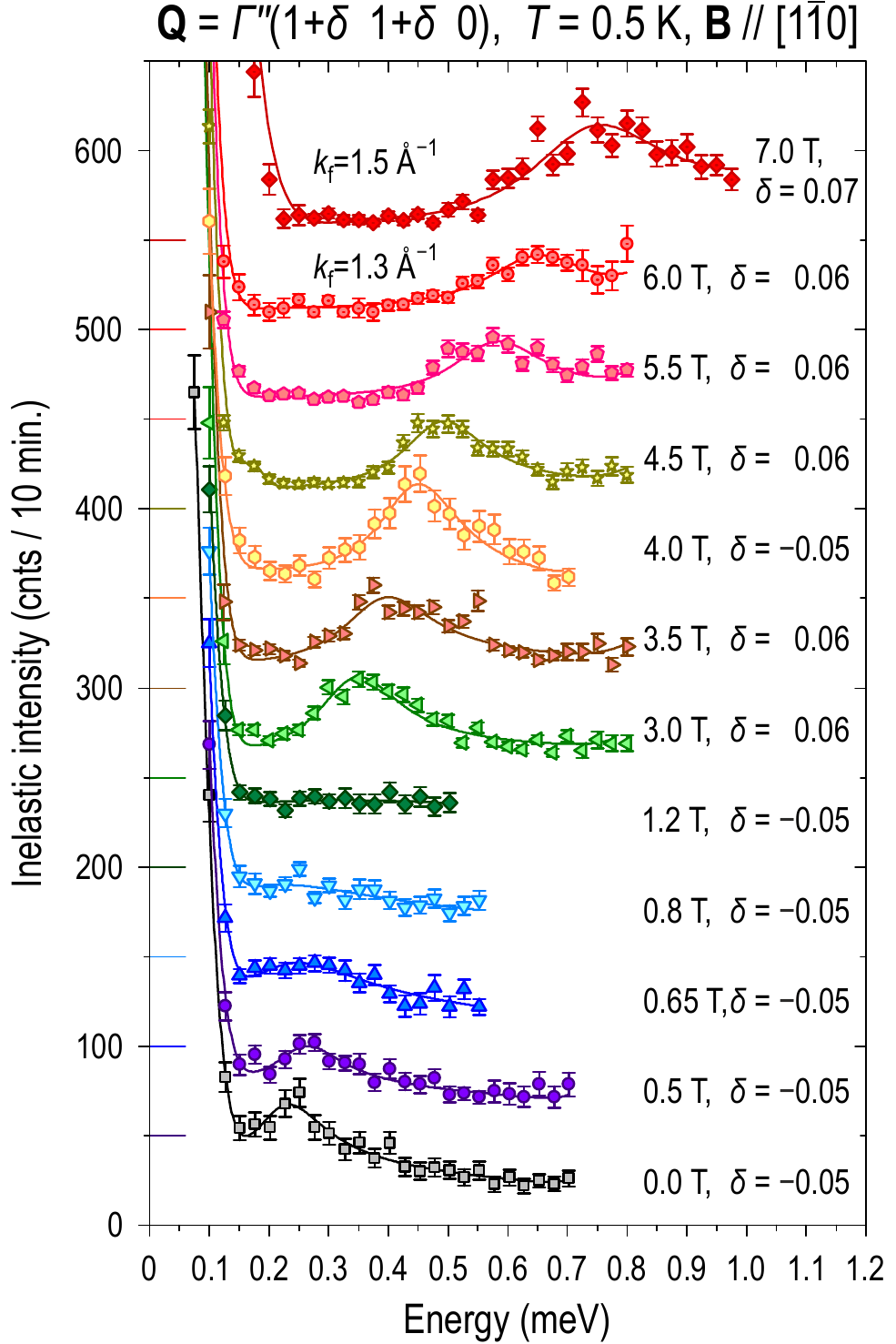}\end{center}
\vspace{-8pt}\caption{INS spectra measured near the zone center $\Gamma''(110)$ at a slightly incommensurate wave vector as indicated in the legend, to avoid the contamination from the Bragg tail. The spectra are shifted vertically for clarity with horizontal lines at the left indicating the background baseline for each spectrum. Solid lines represent Lorentzian fits on top of a nonmagnetic background. Figure from ref.\,\ref{PortnichenkoDemishev16}.}
\label{Fig:CeB6GammaMagneticFieldDependence}
\end{wrapfigure}
Cerium hexaboride is a non-superconducting HF metal with a simple-cubic crystal structure \cite{CameronFriemel16}. Competition between the Kondo screening and the RKKY coupling mechanism via the conduction electrons leads to a rich magnetic-field\,--\,temperature phase diagram. Its ground state below $T_{\text N}=2.3$\,K \cite{ZaharkoFischer03} is antiferromagnetic with a double-$\mathbf{q}$ structure, known as phase III, which undergoes a transition to single-$\mathbf{q}$ phase III$^\prime$ with the application of magnetic field \cite{EffantinRossat-Mignod85}. Another phase transition at $T_{\text Q}=3.2$\,K \cite{FujitaSuzuki80} corresponds to antiferroquadrupolar (AFQ) ordering in this compound (phase II), which was observed directly with resonant x-ray scattering \cite{NakaoMagishi01} as well as with neutron diffraction in magnetic field \cite{EffantinRossat-Mignod85}. As both order parameters are presumably driven by AFM interactions, the observation of a sharp ESR signal within the AFQ phase came as a surprise and was then explained with ferromagnetically interacting localized magnetic moments \cite{DemishevSemeno06, DemishevSemeno09, DemishevSemeno08, Schlottmann13}. Only recently, an INS study revealed a strong FM mode in the magnetic excitation spectrum of CeB$_6$ \cite{JangFriemel14}, yet these observations were done in the zero-field AFM state, whereas ESR measurements could be only performed at elevated magnetic fields within the AFQ phase, precluding a direct comparison. In addition, a sharp resonant mode similar to that of CeCoIn$_5$ was revealed below $T_{\text N}$ at the propagation wave vector of the AFQ phase \cite{FriemelLi12}, motivating a theoretical suggestion that the spin excitation spectrum of CeB$_6$ is dominated by the response of itinerant heavy quasiparticles rather than localized moments \cite{AkbariThalmeier12a, KoitzschHeming16}. However, a crossover to the localized-moment description was suggested for higher magnetic fields \cite{FriemelJang15}. Here we follow in detail the magnetic field dependence of spin excitations in CeB$_6$, including both FM and AFM spin resonances, across the quantum critical point (QCP) that separates the AFM and AFQ phases with the application of magnetic field. Thus, our present observations bridge the gap between previous zero-field INS and high-field ESR measurements and provide a consistent description of spin dynamics that is clearly distinct from that known for other $f$\!-electron systems.

\vspace{-5pt}
INS experiments were performed at the cold-neutron triple-axis spectrometer (TAS) PANDA at MLZ, Garching, the disk chopper time-of-flight (TOF) spectrometer IN5 \cite{OllivierMutk11} at ILL, Grenoble, and the cold-neutron chopper spectrometer (CNCS) \cite{EhlersPodlesnyak11} at the Spallation Neutron Source, ORNL. A rod-shaped single crystal of CeB$_{6}$ with a mass of 4\,g was grown by the floating-zone method from a 99.6\,\% isotope-enriched $^{11}$B powder (to minimize neutron absorption), as described elsewhere \cite{FriemelLi12}. We fixed the final wave vector of the neutrons to $k_{\text f}=1.3$ or $1.5$\,\AA$^{-1}$ and used a cold Be filter to avoid higher-order neutron contamination for TAS experiments. TOF measurements were done with the incident neutron wavelength fixed at 5\,\AA\:(3.27\,meV) for IN5 and at 5.1\,\AA\:(3.15\,meV) for CNCS experiments. The sample environment comprised a 7.5\,T vertical-field cryomagnet with a $^{3}$He insert, 2.5\,T ``orange'' cryostat based magnet, and a 5\,T cryomagnet for PANDA, IN5, and CNCS experiments, respectively. The ESR experiments were performed on a cavity spectrometer providing frequency range of 60--100\,GHz and magnetic field up to 7\,T (GPI, Moscow). Experiments at higher frequencies 100--360\,GHz using a 30\,T pulsed magnet were carried out at Kobe University with a quasioptical setup operating in reflection mode \cite{NakagawaYamada98}. The magnetic field for all experiments was aligned along the $[1\bar{1}0]$ \mbox{direction of the crystal}.

\enlargethispage{5pt}
\vspace{-5pt}
We first present the evolution of the FM resonance measured by TAS in magnetic fields up to 7\,T at $T=0.5$\,K. Fig.~\ref{Fig:CeB6GammaMagneticFieldDependence} shows unprocessed energy scans near the zone center $\Gamma''(1\!+\!\delta \;1\!+\!\delta\; 0)$. Slightly incommensurate wave vectors were chosen to avoid the contamination from phonons and the Bragg tail. The previously reported sharp resonance gets initially suppressed and broadens with the application of external magnetic field as long as the system remains in the AFM state. The observed signal can be described by a Lorentzian line shape \cite{GoremychkinOsborn00}

\begin{multline}\label{Eq:Quasielastic}
S(\mathbf{Q},\omega)\propto\,\frac{\omega}{1-\exp(-\hslash\omega/k_\text{B}T)}
\times\biggl(\frac{\Gamma}{\hslash^2(\omega-\omega_0)^2+\Gamma^2}+\frac{\Gamma}{\hslash^2(\omega+\omega_0)^2+\Gamma^2}\biggr),
\end{multline}

\vspace{5pt}
\noindent where $\Gamma$ is the half width at half maximum of the Lorentzians centered at $\pm\,\hslash\omega_0$, whereas $\hslash$ and $k_{\text B}$ are Plank and Boltzman constants. Upon entering phase III$^{\prime}$ at $\sim$\,1.2\,T~\cite{KunimoriKotani11}, the resonance is fully suppressed and becomes quasielastic with $\hslash\omega_0=0$. However, the excitation reappears at higher magnetic fields within the AFQ phase at an energy that continuously increases with the applied field.

\vspace{-0pt}\begin{figure}[!t]
\includegraphics[width=\textwidth]{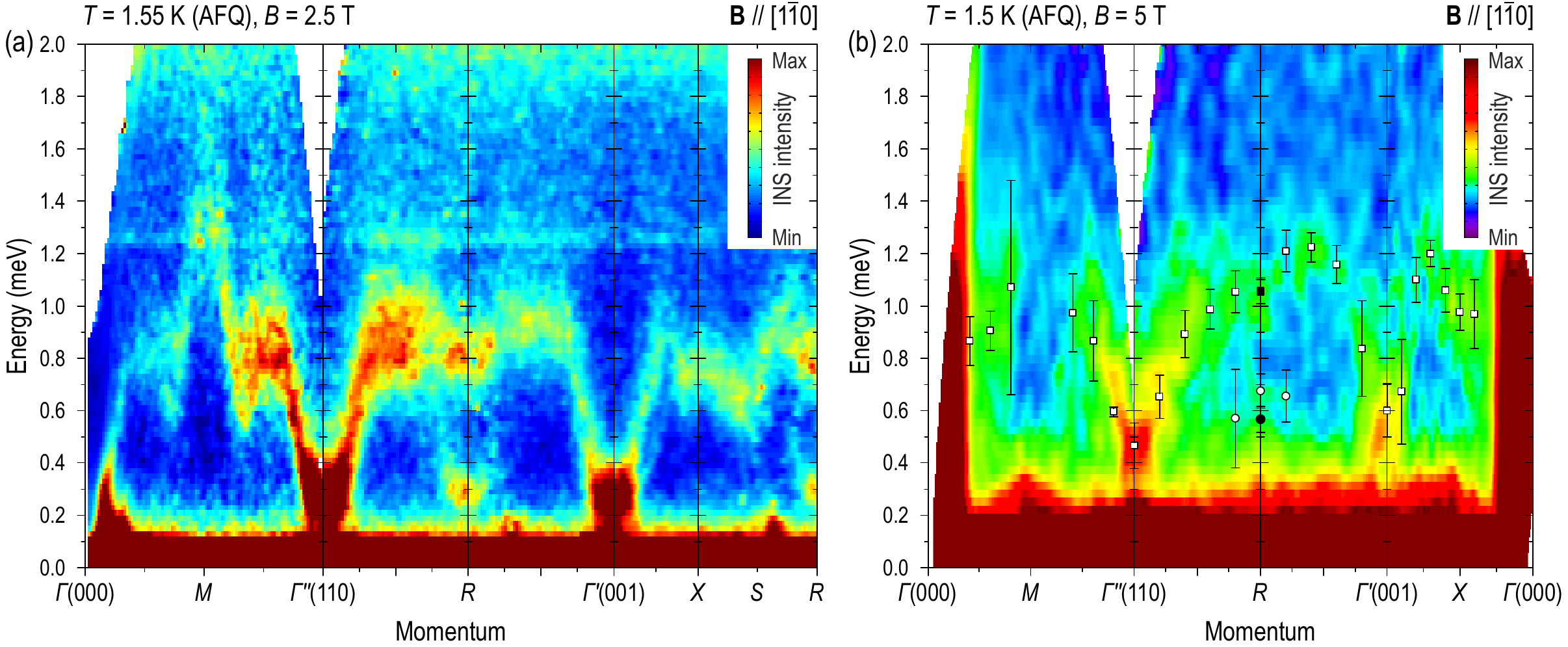}\vspace{-5pt}
\caption{Energy-momentum profiles along high-symmetry directions in the AFQ state: (a) $B = 2.5$\,T, (b) $B = 5$\,T. Empty markers are determined as peak maxima from the fits. Filled markers at the $R(\frac{1}{2}\frac{1}{2}\frac{1}{2})$ point were obtained from the interpolation of peak positions from ref.~\citenum{FriemelJang15}. Background contamination from the He exchange gas was subtracted from the data in panel (b) as explained in the app.\,\ref{Appendix:HeExchangeGasBackgroundSubstraction}. Because of the high-level background coming from the magnet, the field-induced low-energy magnetic excitation cannot be clearly resolved at this field. Figure from ref.\,\ref{PortnichenkoDemishev16}.}\vspace{-35pt}
\label{Fig:CeB6Map2p5Tand5T}
\end{figure}

To get a more complete picture about the field dependence of magnetic excitations, we also performed TOF measurements on the same sample. A continuous dispersive magnon band connecting the local intensity maxima at the zone center ($\Gamma$) and zone corner ($R$) was observed at 2.5 and 5\,T, and its intensity distribution along the main high-symmetry directions of reciprocal space is illustrated in fig.~\ref{Fig:CeB6Map2p5Tand5T}. It is remarkable that the magnon is more intense around the $\Gamma''(110)$ point than at the equivalent $\Gamma'(001)$ or $\Gamma(000)$ positions, suggesting an anomalous non-monotonic behavior of the dynamic form factor that is characteristic of multipolar moments (for conventional dipolar moments, it would decrease monotonically with $|\mathbf{Q}|$) \cite{KuramotoKusunose09, Shiina12, KuwaharaIwasa07},  as was mentioned earlier in sec.\,\ref{SubSec:MagneticFFMO} and \ref{SubSec:FormFactorAnalyticalApproximation} respectively. A magnetic field of 2.5\,T [fig.~\ref{Fig:CeB6Map2p5Tand5T}(a)] does not change the excitation energy at the zone center significantly, but increases the magnon bandwidth twofold, as the dispersion now reaches $\sim$\,1.4\,meV at the $M$ point in contrast to 0.7\,meV in zero field \cite{JangFriemel14}. In addition, a second field-induced low-energy magnetic excitation appears at the AFQ propagation vector, $R\,(\frac{1}{2}\frac{1}{2}\frac{1}{2})$. At first glance, the two modes at the $R$ point, separated by 0.5~meV, are reminiscent of the resonance-peak splitting in CeCoIn$_5$, yet our discussion hereinafter will demonstrate that the origin of this splitting is qualitatively distinct. An even higher magnetic field of 5\,T [fig.~\ref{Fig:CeB6Map2p5Tand5T}(b)] leads to a nearly twofold increase of the zone-center spin gap. We also note that the clear local maximum of intensity at the $R$ point \cite{FriemelLi12, JangFriemel14} is no longer seen at this field, indicating that the resonant exciton mode is suppressed and becomes part of the more conventional magnon spectrum emanating from the zone center.

\vspace{-0pt}\begin{figure}[t]
\includegraphics[width=\textwidth]{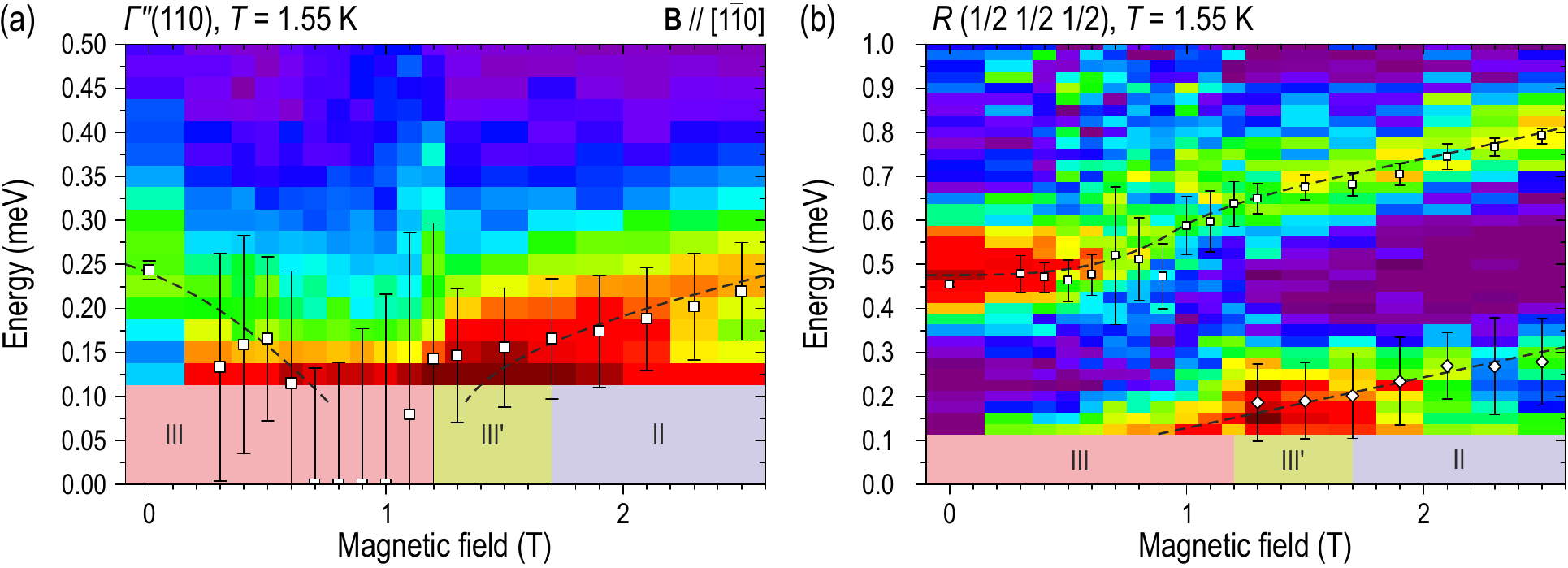}~\vspace{-5pt}
\vspace{-0pt}\caption{Magnetic field dependence of the resonance peaks at (a) $\Gamma$ and (b) $R$ points. Markers in both panels were determined as peak maxima from the fits. Dashed lines are guides to the eyes, and the shaded areas below each panel mark the resolution cutoff and indicate the field regions corresponding to the AFM (III, III$^\prime$) and AFQ (II) phases. Figure from ref.\,\ref{PortnichenkoDemishev16}.}
\label{Fig:CeB6BdepGammaRPoint}\vspace{-20pt}
\end{figure}

Evolution of the magnetic excitations at the $\Gamma$ and $R$ points as a function of field would complement our TAS data and reveal essential differences in the behavior of the resonances in comparison with other HF systems. Hence we focused our attention mainly on the $\Gamma$ and $R$ points and measured in detail the field dependence across the QCP using the TOF spectrometer IN5 equipped with a low-background 2.5 T cryomagnet. Energy-momentum profiles for each field along the $\Gamma R$ direction are shown as an animation in app.\,\ref{Appendix:DetailedMagneticFieldDependence}. We also present one-dimensional energy profiles obtained from the same data by integration within $\pm\,0.15$~r.l.u. around the $\Gamma$ and $R$ points as color maps in fig.~\ref{Fig:CeB6BdepGammaRPoint}. The data in fig.~\ref{Fig:CeB6BdepGammaRPoint}(a) illustrate the nonmonotonic behavior of the zone-center excitation as it initially softens to zero upon entering the phase III$^\prime$ and then reappears within phase II at an energy that continuously increases with the applied field. A qualitatively different picture is observed for the resonance peak at the $R$ point in fig.~\ref{Fig:CeB6BdepGammaRPoint}(b). Increasing the field within phase III keeps the resonance energy constant while it decreases in amplitude and broadens, transferring a significant part of its spectral weight to the second low-energy mode whose tail can be seen above the elastic line already above $\sim$\,0.5\,T. Upon crossing through the phase III--III$^\prime$ transition, the amplitude of the low-energy mode is maximized, whereas the higher-energy mode shifts up in energy. Both excitations then follow a linear trend with the same slope and approximately equal amplitudes in phase II, in agreement with previously report measurements \cite{FriemelJang15}. This behavior is completely different from the field-induced splitting of the neutron resonance in the SC state of CeCoIn$_{5}$, where the second mode emerges from the resonance energy and then shifts down monotonically with increasing field \cite{StockBroholm12a}.

\begin{wrapfigure}[18]{r}{0.57\textwidth}\vspace{-15pt}
\includegraphics[width=0.57\textwidth]{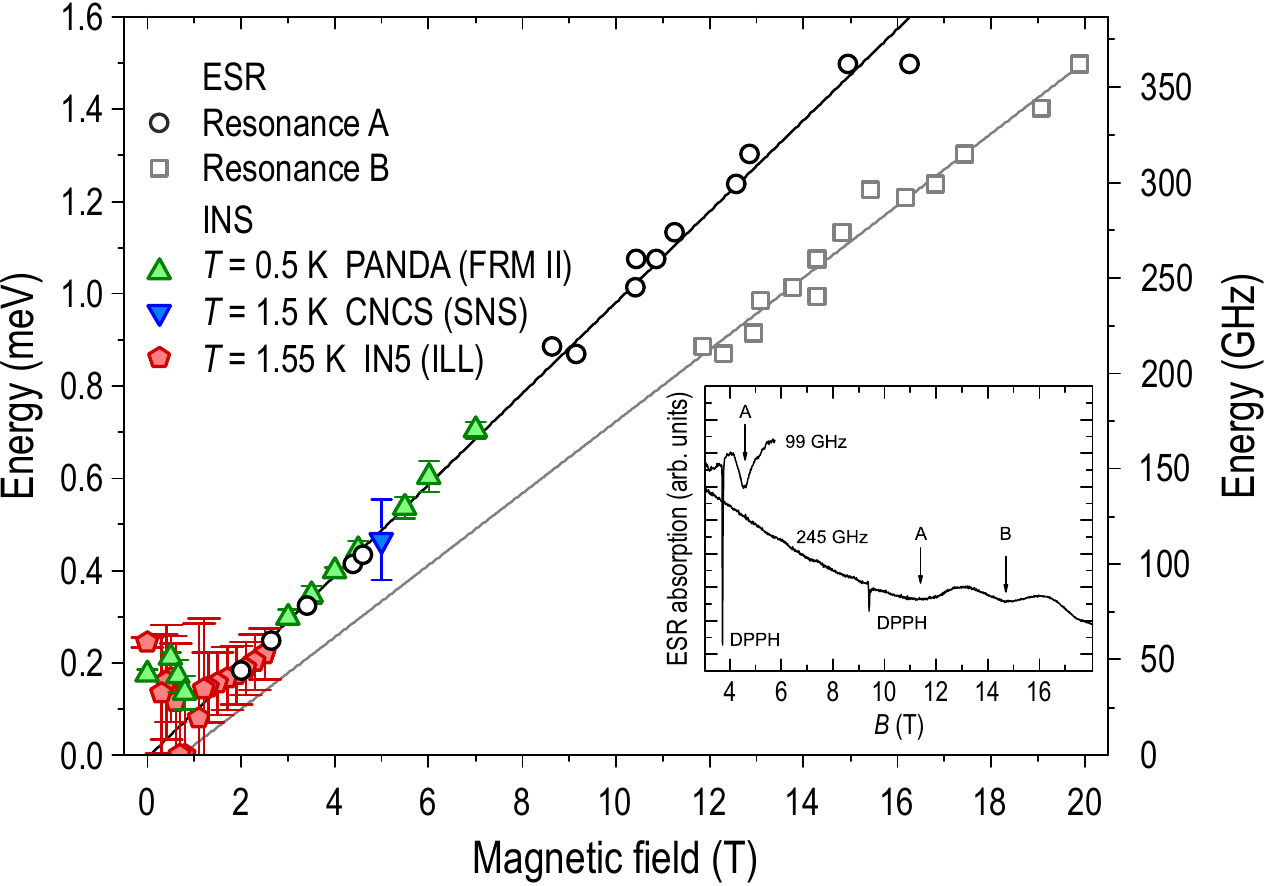}
\vspace{-15pt}\caption{Summary of the magnetic field dependence of zone-center excitations obtained from both INS and ESR spectra. Solid lines are linear fits of resonances A and B. The inset shows a field dependence of the cavity transmission at 99\,GHz and ESR spectrum obtained at 245\,GHz using a quasioptical technique as typical examples of unprocessed datasets from which the points in the main plot were obtained. Sharp lines marked as DPPH originate from a small 2,2-diphenyl-1-picrylhydrazyl reference sample. Figure from ref.\,\ref{PortnichenkoDemishev16}.\vspace{-0pt}}
\label{Fig:CeB6ESRandFieldDep}
\end{wrapfigure}
ESR measurements, which probe zone-center excitations, have shown that the frequencies of the two observed resonances A and B \cite{DemishevSemeno08} change linearly with field within phase II, as shown in fig.~\ref{Fig:CeB6ESRandFieldDep} with hollow symbols. The linear fits shown with solid lines, $\hslash\omega\kern-.2pt=\kern-.2pt\hslash\omega_{0}\kern-.3pt+{\kern-.3pt}g\mu_{\text B}B$, result in the slope of 0.098(2) and 0.077(4)\,meV/T for the resonances A and B \cite{DemishevSemeno08}, corresponding to $g$-factors of 1.70(4) and 1.35(7), respectively, as compared to that of 1.90(7) at the $R$ point \cite{FriemelJang15}. In fig.~\ref{Fig:CeB6ESRandFieldDep} we compare resonance energies obtained from ESR (hollow symbols) with the field-dependent energy of the zone-center INS excitation (filled symbols). We find perfect agreement between the INS data and the resonance A in the intermediate field range within phase II, where both datasets overlap, suggesting that the same FM excitation is probed in both experiments. This comparison nicely demonstrates the complementarity of the ESR and INS methods.\enlargethispage{6pt}

In summary, we have investigated magnetic field dependencies of collective magnetic excitations at the zone center ($\Gamma$) and zone corner ($R$), as well as the ESR signal in CeB$_{6}$. Unlike in CeCoIn$_{5}$, where the AFM resonance splits into a Zeeman doublet, in CeB$_{6}$ the second field-induced magnon at the $R$ point exhibits a monotonically increasing field dependence. The FM resonance at the $\Gamma$ point is initially suppressed with the magnetic field within the AFM phase, but reappears upon entering the AFQ phase. Its energy matches that of the resonance `A' seen in ESR, whereas the anomalous dynamic form factor of the zone-center excitation points towards its multipolar-wave character. This observation is consistent with the proposed orbital-ordering nature of the ESR response, resulting from the interplay of AFQ order with FM correlations \cite{DemishevSemeno08, Schlottmann13,Schlottmann18}, and with the multipolar character of phase II. The second ESR line observed in high fields (mode B) was interpreted as the result of a crossover of the excited state to the free-ion limit, as the field at which it appears is comparable with the condensation energy of the AFQ phase, $\sim\!1.75{\kern.5pt}k_{\text B}T_{\text Q}$ \cite{Schlottmann13}. In order to reach this regime and confirm appearance of the second resonance we conducted a separate experiment, with magnetic field up to 14.5\,T, discussed in sec.\,\ref{SubSec:SecondResonance}. Our current results are a rare example of a simultaneous observation of the FM resonance by INS and ESR, consistently over a broad range of magnetic fields, thus demonstrating the complementarity nature of these two spectroscopic probes.

\subsection{Discovery of the second resonance in CeB$_{6}$}\label{SubSec:SecondResonance}

In our previous measurements we studied the evolution of the FM resonance in magnetic fields up to 7\,T using neutron scattering, and up to 20\,T using electron spin resonance [\ref{PortnichenkoDemishev16}]. A weak magnetic field directed along $[1~\overline{1}~0]$ is sufficient to suppress the energy of an intense collective mode to zero together with the AFM order parameter. At higher fields inside the AFQ phase this excitation reappears and undergoes a linear increase with field, following the energy of the electron spin resonance.  The presence of only one resonance, rather than four,  within the fourfold degenerate $\Upgamma_8$ ground multiplet at the relatively low fields as well as appearance of the second resonance in the ESR spectrum at high fields above $\sim12$\,T \cite{DemishevSemeno08} was explained later \cite{Schlottmann13}. The antiferroquadrupolar ordering in CeB$_6$ introduces two interpenetrating sublattices, and simultaneously quenches three out of four{\footnotemark}
\footnotetext{According to P. Thalmeier \textit{et al}. \cite{ThalmeierShiina98}, for zero magnetic field antiferroquadrupolar order  splits the $\Upgamma_8$ multiplet into two Kramers doublets, thus allowing four transitions. Splitting of about  30\,K  into two doublets, $\Upgamma_{8,1}$ and $\Upgamma_{8,2}$, was confirmed by Raman spectroscopy by E. Zirngiebl \textit{et al}. \cite{ZirngieblHillebrands84}. Presence of two nonequivalent sites doubles the number of possible excitations. Breaking the symmetry with the application of external magnetic field removes all the degeneracies, increasing the number of possible transitions  up to six, which due to the existence of two sublattices allows us to expect twelve different multipolar excitation modes.}
resonances for each site. The coherence between sublattices prevents two distinct resonances for each site being occupied simultaneously, thus hybridizing them into a single resonance. This successfully explained the presence of only one resonance at 60\,GHz. Emergence of the second ESR line at frequencies above 200\,GHz was interpreted as the result of a crossover of the excited state to the free-ion limit, as the field at which it appears is comparable with the condensation energy of the AFQ phase, $\sim\!1.75{\kern.5pt}k_{\text{B}}T_{\text{Q}}$ \cite{Schlottmann12,Schlottmann13}. Since simultaneous observation of the resonance by neutron scattering and electron spin resonance was previously confirmed, we extended available neutron scattering data up to 14.5\,T in order to uncover the second resonance  that escaped direct observation  in our previous experiments due to the limited magnetic field range.

\begin{wrapfigure}[25]{l}{0.50\textwidth}\vspace{-15pt}
\includegraphics[width=0.50\textwidth]{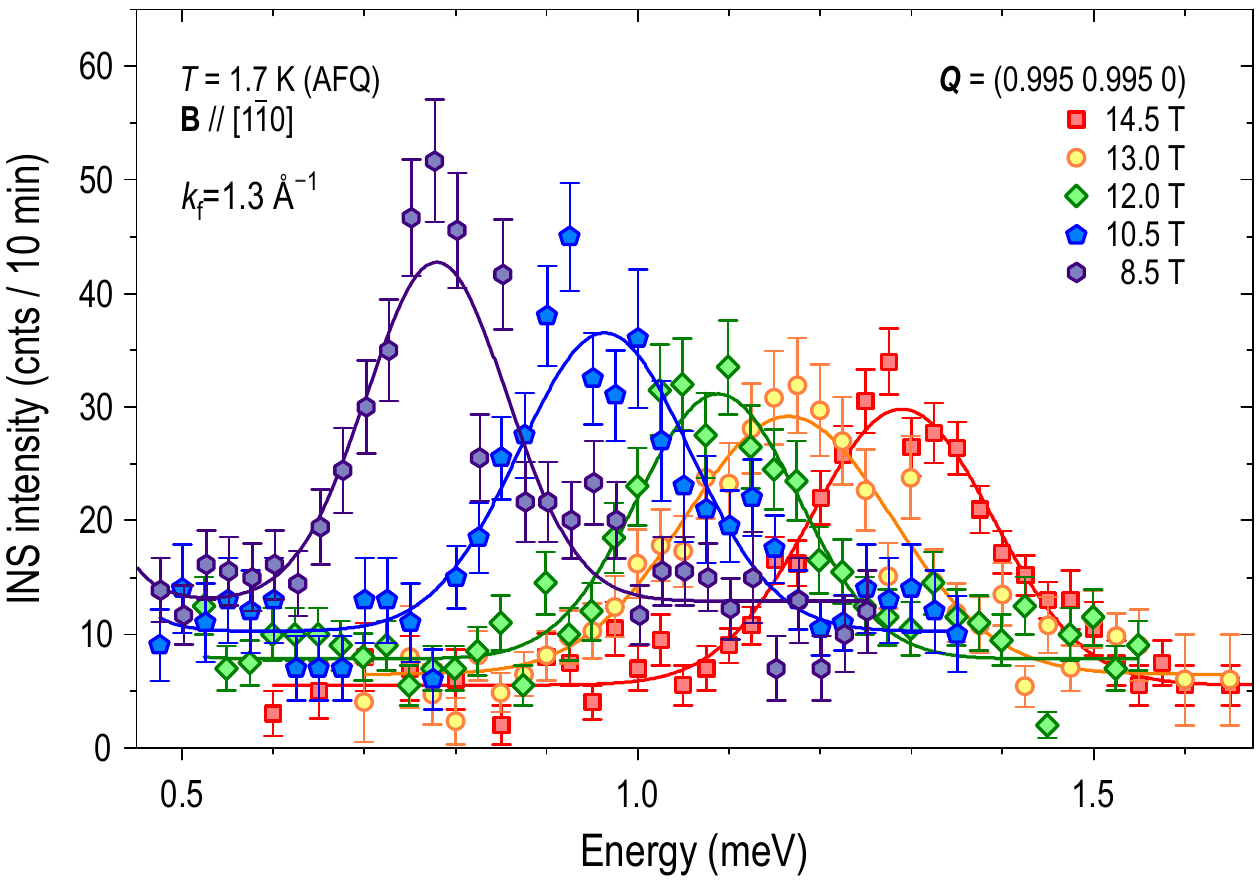}
\vspace{-25pt}\caption{INS spectra measured near the zone center $\Gamma''$ at a slightly incommensurate wave vector, to avoid contamination from the Bragg tail. Solid lines represent Gaussian fits on top of a nonmagnetic background.}
\label{Fig:CeB6HighFieldFLEXX}\vspace{-0pt}
\includegraphics[width=0.50\textwidth]{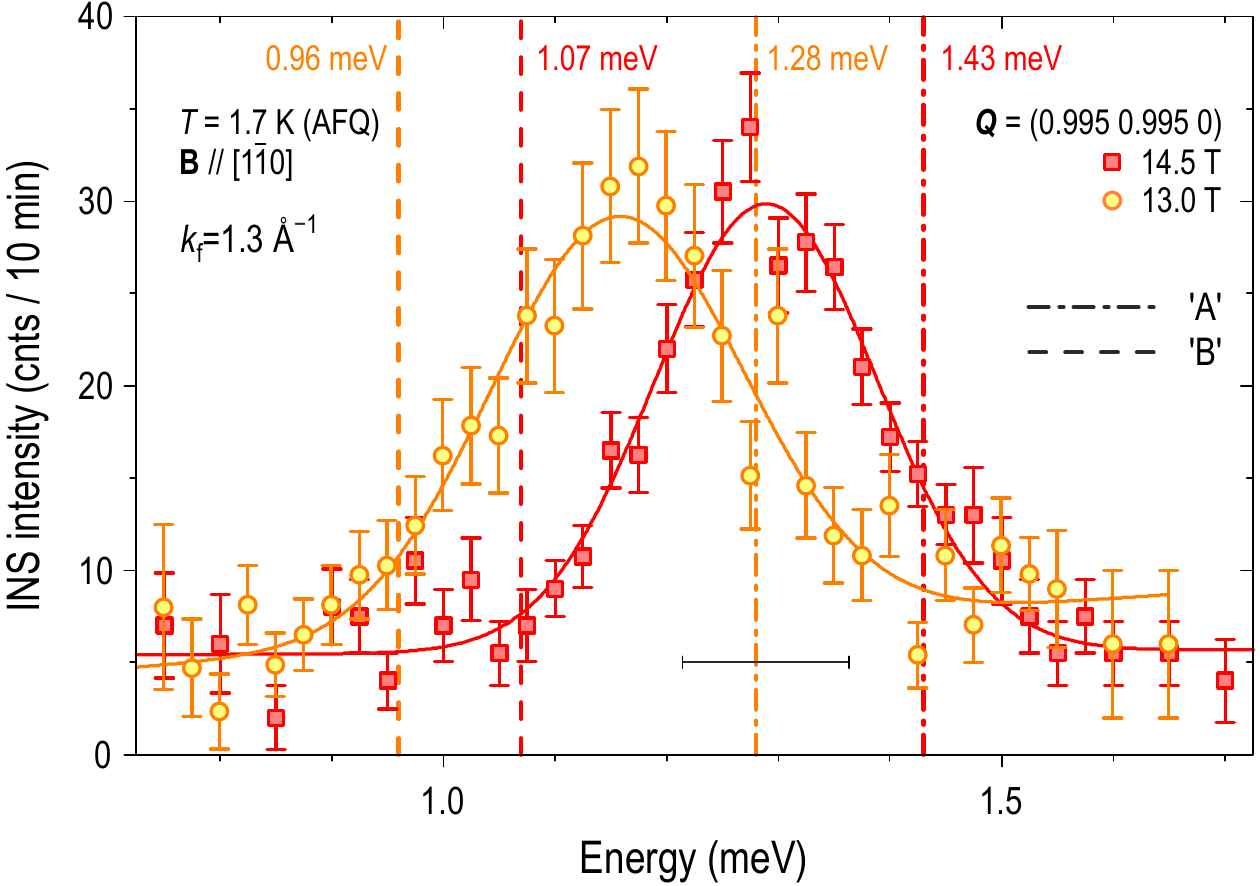}
\vspace{-25pt}\caption{Comparison of the INS spectra measured at 13\,T and 14.5\,T with the expected energies of the `A' and `B' resonances, shown with dashed and dash-dotted vertical lines.  The horizontal black bar inside the peak shows the calculated experimental energy resolution.}
\label{Fig:CeB6HighFieldFLEXXResolution}
\end{wrapfigure}
\begin{figure}[!t]
\includegraphics[width=\textwidth]{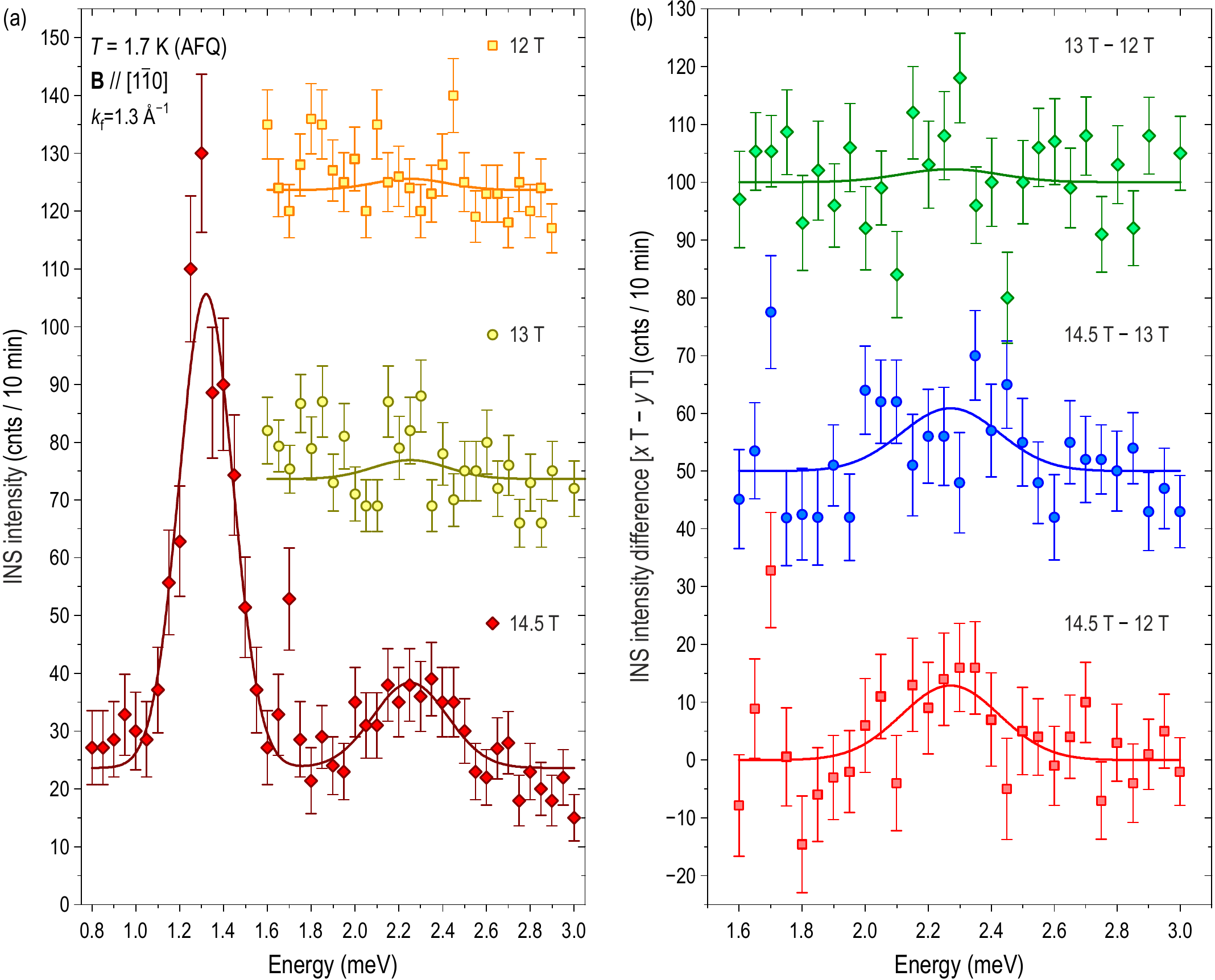}\vspace{+5pt}
\caption{ (a) Unprocessed INS spectra measured near the zone center $\Gamma\,(110)$, which shows the appearance of the second resonance. (b) Subtraction of various magnetic field combinations, as indicated in the legend, shown to emphasize the statistically significant field dependence of the second resonance and its magnetic nature. All spectra are offset by 50 counts vertically for clarity.}\vspace{-25pt}
\label{Fig:CeB6SecondModeIN14}
\end{figure}

INS experiments were performed at the three-axis spectrometer FLEXX \cite{HabichtQuintero-Castro15} at HZB, Berlin, the three-axis instrument for low energy spectrometry \textsc{ThALES} \cite{BoehmHiess08} and the disk chopper time-of-flight spectrometer IN5 \cite{OllivierMutk11} at ILL, Grenoble, the cold-neutron chopper spectrometer (CNCS) \cite{EhlersPodlesnyak11} at the Spallation Neutron Source, ORNL, and the cold three-axes spectrometer PANDA \cite{SchneidewindLink06} at MLZ, Garching.  For all TAS experiments the final wave vector of the neutrons was fixed  to  $k_{\text f}=1.3$\,\AA$^{-1}$ or $1.5$\,\AA$^{-1}$, as indicated in the legend in \cref{Fig:CeB6GammaMagneticFieldDependence,Fig:CeB6HighFieldFLEXX,Fig:CeB6HighFieldFLEXXResolution,Fig:CeB6SecondModeIN14}. A cold Be filter was used in all TAS experiments to avoid higher-order neutron contamination. In the case of the IN5 and CNCS experiments, the incident neutron wavelength was fixed at 5\,\AA\:(3.27\,meV) and at 5.1\,\AA\:(3.15\,meV) respectively. The sample environment for the experiments with the magnetic field aligned along the $[1~\bar{1}~0]$ direction of the crystal comprised similar 15\,T vertical-field cryomagnets VM-1B and 158OXHV26 manufactured by Oxford Instruments, a 2.5\,T ``orange'' cryostat based magnet, a 5\,T cryomagnet and a 7.5\,T vertical-field cryomagnet with a $^{3}$He insert for FLEXX, \textsc{ThALES}, IN5, CNCS, and  PANDA experiments respectively.  A rod-shaped single crystal of CeB$_{6}$ with a mass of 4\,g was grown by the floating-zone method from a 99.6\,\% isotope-enriched $^{11}$B powder (to minimize neutron absorption), as described elsewhere \cite{FriemelLi12}.\vspace{-3pt}

We start the presentation of our experimental results with the unprocessed energy scans measured near the zone center at the FLEXX spectrometer, shown in fig.\,\ref{Fig:CeB6HighFieldFLEXX}. In order to reveal the second resonance, observed in ESR as the resonance `B', we did a systematic study of the $\Gamma''(0.995\;0.995\;0)$ point in magnetic fields up to 14.5\,T at $T=1.7$\,K, with the crystallographic $[1\;\bar{1}\;0]$ axis aligned parallel to the magnetic field. The excitation energy continuously increases with the applied field, experiencing a slight broadening, however  no appearance of a second resonance can be observed above 12\,T. Observed resonance energies $\hbar\omega=1.09\pm0.01$\,meV, $\hbar\omega=1.17\pm0.01$\,meV, and $\hbar\omega=1.29\pm0.01$\,meV, which correspond to the 12\,T, 13\,T, and 14.5\,T data, result in
a rather smaller $g$-factor of $1.40\pm0.02$, as compared to that of $1.70\pm0.04$, previously reported in ref.\,\ref{PortnichenkoDemishev16} for the resonance `A'. The observed $g$-factor shows a much better agreement with the one determined from the resonance `B' slope: $1.35\pm0.07$.\vspace{-3pt}

\begin{wrapfigure}[24]{r}{0.55\textwidth}\vspace{-15pt}
\includegraphics[width=0.55\textwidth]{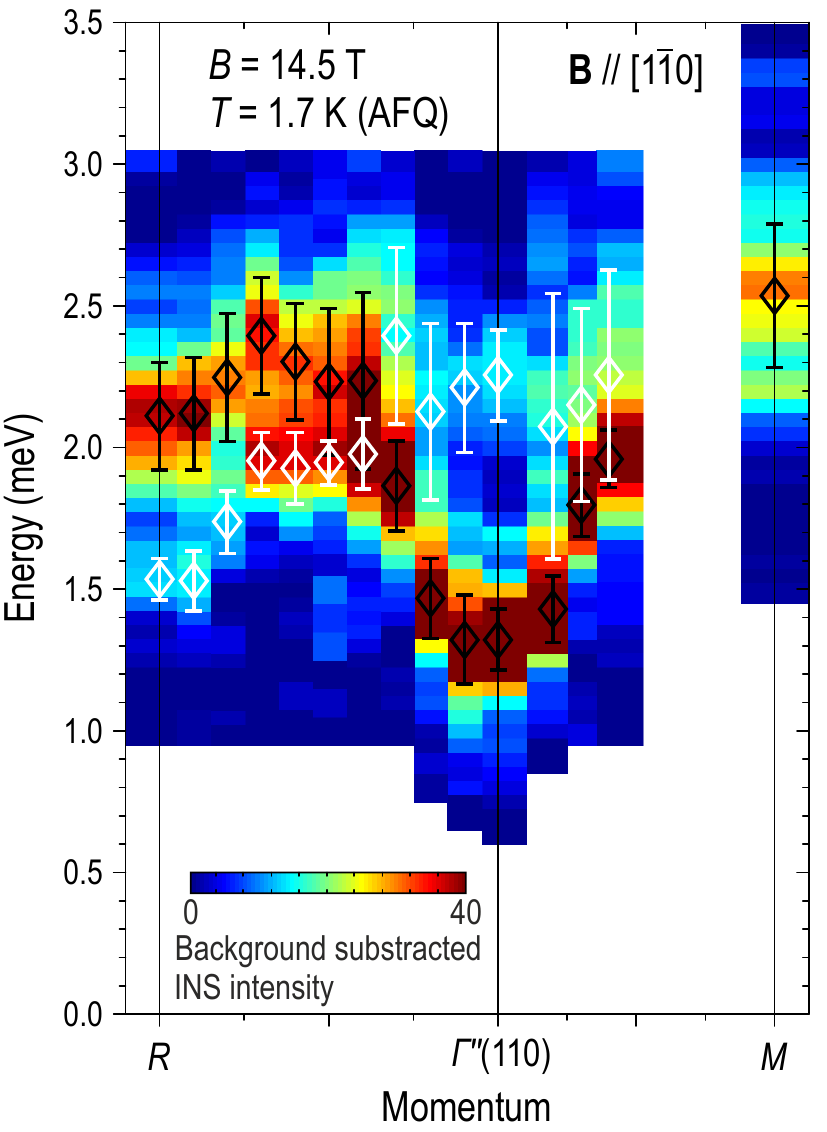}
\vspace{-20pt}\caption{Energy-momentum profile, which shows magnon spectrum at magnetic field of 14.5\,T along straight segments connecting the $R(\frac{1}{2}\,\frac{1}{2}\,\frac{1}{2})$, $\Gamma''(110)$ and $M(\frac{1}{2}\,\frac{1}{2}\,0)$ points. Black and white diamond markers represent peak positions, that resulted from fitting the energy profiles. Each energy scan was background corrected and smoothed with a one-dimensional Gaussian filter characterized by  a FWHM of 0.1\,meV.}
\label{Fig:CeB6cmap14p5T}
\end{wrapfigure}

In order to clarify this question, and argue that the deviation of the $g$-factor from the expected value cannot be interpreted as a mixing between two resonances, we compare a typical high field energy scan with the expected values of the A and B resonances. At 13\,T, the expected position of the A and B resonances are 0.96\,meV and 1.28\,meV. The respective positions are shown in fig.\,\ref{Fig:CeB6HighFieldFLEXXResolution}  with dashed and dash-dotted orange lines. It was not possible to determine the energy resolution in this experiment. Due to the thick aluminium walls of an extremely massive 15\,T magnet used for the experiment, the elastic line was largely broadened{\footnotemark}.
\footnotetext{See app.\,\ref{Appendix:ScatteringTAS} for details.}
As a resolution estimation we use the calculated{\footnotemark} \footnotetext{The question of the correct resolution calculation is very ambiguous, as in order to increase neutron flux we used the focusing mode on both monochromator and analyzer. Besides a $60^{\prime}$ radial collimator  was installed between the sample and the analyzer. We used TAKIN \cite{WeberGeorgii16} to estimate the energy resolution as a projection of the resolution ellipsoid, calculated with Eckold-Sobolev algorithm, on the energy-momentum plane for the PANDA spectrometer. According to our experience, we find a good agreement between the calculated and actually measured resolution for the PANDA spectrometer when using a comparable with FLEXX configuration, as well as our observation that energy scans from different instruments, measured with a same $k_{\text f}$, can be ploted together after  appropriate background and monitor corrections, as for example shown in fig.\,\ref{Fig:CeB6cmap14p5T}, where data measured at FLEXX and \textsc{ThALES} are shown together. Therefore we believe that the resolution estimation of the FLEXX spectrometer can be done with the described procedure.}
value of $\sim0.15$\,meV, shown with a horizontal black bar inside the peak in fig.\,\ref{Fig:CeB6HighFieldFLEXXResolution}. As one can see, the observed single resonance under no circumstances can be seen as a linear combination of the  `A' and `B' resonances. Exactly the same conclusion is valid for the maximum available field: 14.5\,T, where 1.07\,meV and 1.43\,meV are the expected energies of the resonances, previously determined with ESR.

In addition to the fact that we do not observe the second resonance, a very important conclusion can be drawn. At higher fields, the excitation that followed the energy dependence of the A resonance starts to deviate towards lower energies. To avoid the possibility that the observed deviation is a consequence of an incorrect magnetic field calibration, and the second resonance located at a lower energy simply cannot be distinguished from the background  due to insufficient statistics, we conducted an additional experiment.

\vspace{+5pt}

As a first step  we repeated the same 14.5\,T scan at the \textsc{ThALES} spectrometer, as we did at FLEXX, in order to check if the excitation's energy can be reproduced. The main result of our  measurements is shown in fig.\,\ref{Fig:CeB6SecondModeIN14}(a), where in addition to the previously observed resonance, at the energy that is the same within the error bars in both experiments, we discovered a new weak resonance at higher energies at fields above 12\,T. Its strong field dependence, as shown in fig.\,\ref{Fig:CeB6SecondModeIN14}(b), indicates that it is magnetic and likely corresponds to one of the upper modes, calculated in refs.\, \citenum{ThalmeierShiina03,ShiinaShiba03}. A further increase of the magnetic field, in order to verify the change in the resonance position, faces serious technical complications. An additional 2.5\,T can be implemented with the so-called Dy-booster. However, in order to reach 17\,T, the sample volume has to be reduced severely \cite{HabichtQuintero-Castro15}, thus sacrificing the signal intensity. The possibility of such an experiment will be addressed in future studies.

According to our previous experiments, within the AFM phase a strong FM mode at the zone center $\Gamma$ is hybridized with the maximum of intensity at the $R$ point and the spin-wave modes emanating from the AFM wave vectors. Together they form a continuous dispersive magnon band in a narrow energy
range between 0.2 meV and 0.7 meV \cite{JangFriemel14}. A magnetic field of 2.5\,T does not change the excitation energy at the zone center significantly but increases the magnon bandwidth twofold, as the dispersion now reaches 1.4\,meV at the $M(\frac{1}{2}~\frac{1}{2}~0)$ point in contrast to 0.7 meV in zero field [\ref{PortnichenkoDemishev16}]. Assuming a linear dependence for the $M$-point excitation, we can expect that the bandwidth will also increase significantly for  high field. Another important issue, that requires an answer, is the dispersion of the higher-energy resonance at the $\Gamma$ point, and whether it has a connection to the previously reported splitting of the excitation at the $R$ point \cite{FriemelJang15} [\ref{PortnichenkoDemishev16}].

To check this we did detailed mapping of the magnon dispersions  at the maximum avaliable field of 14.5\,T. Individual scans, measured at the FLEXX and \textsc{ThALES} spectrometers, at different $\mathbf{Q}$ points along high-symmetry directions, after appropriate background subtraction and counting monitor correction, were combined into a color-map. As expected, at the $R$ point we found both resonances that follow previously observed energy scaling upon field increase, but the peak at 2.53\,meV observed at the $M$ point suggests that the bandwidth of the strongest excitation branch, shown with black diamond markers in fig.\,\ref{Fig:CeB6cmap14p5T}, remains more or less constant within the AFQ phase. Dispersion of the newly found resonance points
\begin{wrapfigure}[17]{l}{0.50\textwidth}\vspace{-5pt}
\includegraphics[width=0.50\textwidth]{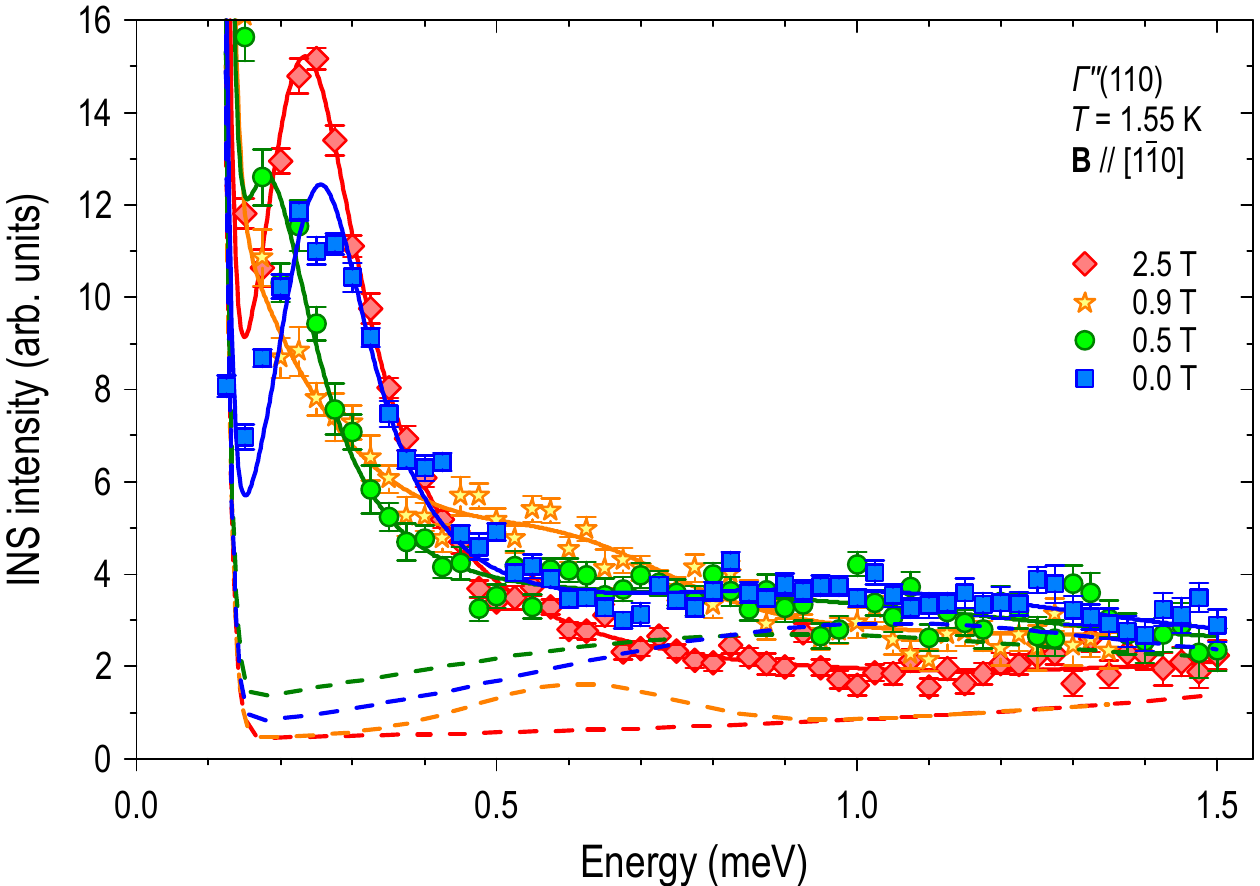}
\vspace{-25pt}\caption{INS spectra at $T=1.55$\,K obtained after integration of the four-dimensional TOF data set within $\pm0.10$\,r.l.u. along the in plane momentum direction with respect to the $\Gamma''(1\;1\;0)$ point and $\pm0.04$\,r.l.u. in the out-of-plane (vertical) direction, parallel to the magnet axis. Solid lines are fits to the data, where low-energy (mode I) and high-energy (mode II) excitations are described with Lorentzian and Gaussian lines, respectively. Dashed lines show individual contribution from the mode II  excitation.}
\label{Fig:CeB6IN5ModeII}
\end{wrapfigure}
towards the fact that the second resonance which arises at the $R$ point upon crossing through the phase III--III$^\prime$ transition and the one observed at the $\Gamma$ point above 12\,T belong to the same excitation branch, shown with white diamond markers in fig.\,\ref{Fig:CeB6cmap14p5T}. The question of dispersion propagation along the segment that connect $\Gamma$ and $M$ points can not be clearly determined at present, because mixing between the almost flat upper mode and relatively stiff lower branch exclude the possibility of accurate identification of the peak position.

At present we have successfully studied the evolution of the resonant mode at the $\Gamma$ point upon application of the magnetic field along the $[1\;\bar{1}\;0]$ crystallographic direction. Above 12\,T we managed to observe the onset of the second branch. From this moment we identify lower and higher energy excitations  as mode I and mode II. Another important observation is that the intensity of mode I after reappearing within the AFQ phase is strongly enhanced, as can be seen in fig.\,\ref{Fig:CeB6BdepGammaRPoint}(a). Because we expect that the total spectral weight of all excitations should be conserved, we need to verify the possibility of an additional resonance at the $\Gamma$ point being overlooked  at the  previous experiment [\ref{PortnichenkoDemishev16}].

\enlargethispage{10pt}
We present a newly analyzed field dependence for the $\Gamma(1\;1\;0)$ point in fig.\,\ref{Fig:CeB6IN5ModeII}, obtained from the previously published TOF data set [\ref{PortnichenkoDemishev16}]. Comparison of the 2.5\,T scan with the 0\,T scan shows a clear presence of an additional broad contribution, centered at $\hbar\omega=0.97\pm0.06$\,meV, in the case of the latter. Upon field increase, the second excitation, which we designate as mode II, mimics the behavior of mode I, as it is also suppressed to zero together with the antiferromagnetic order parameter. Based on the fact that the second resonance is absent upon entering AFQ phase, we conclude that its spectral weight contributed to the enhanced intensity of the mode I excitation. Detailed field dependence of the uncovered resonance, as well as updated values of the mode I energy in higher field that shows slight deviation towards lower than expected energies, will be discussed later, and the corresponding results are summarized in fig.\,\ref{Fig:CeB6Hdep110and001}(a).

\begin{figure}[!t]\vspace{-5pt}
\includegraphics[width=\textwidth]{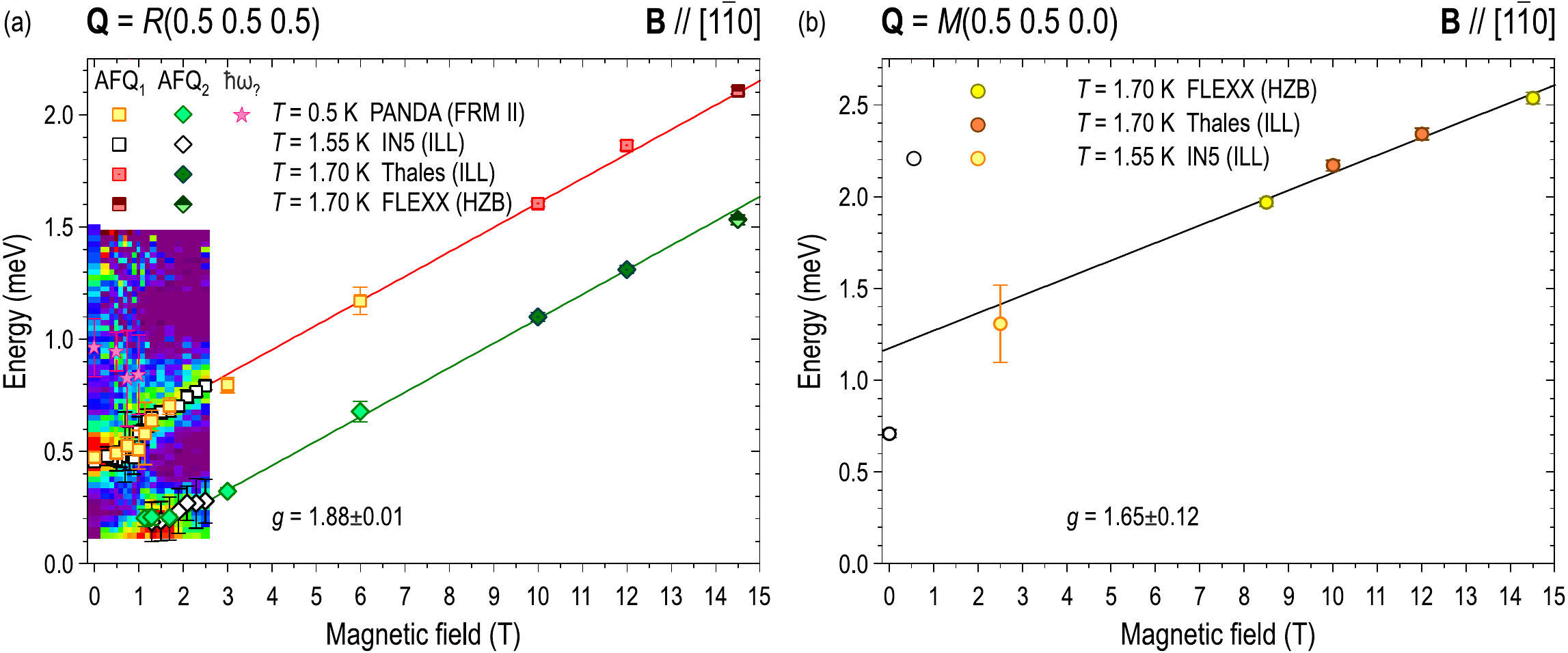}
\vspace{-25pt}\caption{Summary of the magnetic field dependence of excitations determined from fitting the energy profiles measured with INS for: (a) $R(\frac{1}{2}\,\frac{1}{2}\,\frac{1}{2})$ and (b) $M(\frac{1}{2}\,\frac{1}{2}\,0)$ points.}\vspace{-30pt}\enlargethispage{15pt}
\label{Fig:CeB6gfactor}
\end{figure}

Momentum  dependence of the $g$-factor at the $R$ and $M$ points, that combines our last experiments at the FLEXX, \textsc{ThALES}, IN5 and PANDA spectrometers, is summarized in fig.\,\ref{Fig:CeB6gfactor}. These measurements are of particular importance, as they allow proper model parameter adjustment. Previously reported measurements{\footnotemark}
\footnotetext{Field dependence at PANDA was measured by G.~Friemel \textit{et al.} and published in refs.\,\citenum{Friemel14,FriemelJang15}.}
of  both resonance peaks  at the $R$ point showed that increasing the field within the AFQ phase results in a monotonic growth of the resonances energies. We extended the observed dependence up to 14.5\,T, and we can confirm that the linear dependence persists, as shown in fig.\,\ref{Fig:CeB6gfactor}(a),  with the refined $g$-factor of $1.88\pm0.01$. An additional resonance, observed in the AFM phase labeled as $\hbar\omega_{?}$ corresponds to the broad peak observed in energy scans in refs.\,\citenum{FriemelLi12,Friemel14,FriemelJang15} and in fig.\,\ref{Fig:CeB6gfactor}(a). It was shown that the observed excitation exhibits a weak dispersion along the $(H~\frac{1}{2}~0)$ between 0.95\,meV and 1.05\,meV. This value is actually in a good agreement with the $\mathbf{Q}$-averaged value of $\Updelta=1.2$\,meV for the charge gap, that was determined by point-contact spectroscopy. At the same time because of its weak dependence upon temperature increase within the AFM phase, it could also  originate from the low-energy crystal-field excitation of Ce$^{3+}$ ion, and the observed $\mathbf{Q}$-dependence could be due to the presence of the intersite interaction. From the fact that the mentioned above excitation is absent at zero field within the AFQ phase at 2.5\,K, as well as it is strongly suppressed with field together with the AFM phase, we can conclude that it is clearly assosiated with the charge gap, that opens upon entering the AFM state \cite{Friemel14}. Local intensity maximum around $\sim1.25$\,meV, shown in fig.\,\ref{Fig:CeB6gfactor}(a),  originates from the spurious ``flat'' mode{\footnotemark}.
\footnotetext{The true nature of this spurion remains an open question. The mentioned ``flat'' mode can be found in energy-momentum profiles shown in fig.\,\ref{Fig:CeB6Map2p5Tand5T}, fig.\,\ref{Fig:CeLaB6TOF}(a,b), which corresponds to the 2.5\,T and zero field scans within the AFQ phase for parent and La-doped samples. From the fact that this excitation is absent in TAS measurements on  the same single crystal we conclude that the signal is a spurion.}
At the $M$ point  the intensity  of the single excitation{\footnotemark}
\footnotetext{Zero field excitation, observed below $T_{\text{N}}$, was reported by H.~Jang \textit{et al.} and published in ref.\,\citenum{JangFriemel14}.}
is almost zero within the AFM phase. Upon increasing field  up to 2.5\,T, an unambiguously inelastic excitation can be observed. Further increase of the magnetic field within the AFQ phase monotonically shifts excitation's energy  and the corresponding $g$-factor is $1.65\pm0.12$, as shown in fig.\,\ref{Fig:CeB6gfactor}(b). These results clearly demonstrate significant changes in the $g$-factor as a function of momentum, which is  not predicted with the theory described in sec.\,\ref{SubSec:MagneticFieldDependenceSpinExcitation} for the simple case of canonical anti- and  ferromagnet. The measured excitation spectrum dispersion, and in particular its magnetic field  dependence is of particular importance, as it allows accurate determination of the described in sec.\,\ref{SubSec:InteractionBetweenMultipoles} and in ref.\,\citenum{ThalmeierShiina03} quadrupolar and octupolar symmetry breaking terms as well as interaction matrix $D_{\Upgamma}^{}(\mathbf{q})$, thus allowing the final determination of the interaction Hamiltonian.

\subsection{\textit{g}-factor anisotropy in CeB$_{6}$}\label{SubSec:AnisotropyCeB6}

A not completely clear question in the physics of cerium hexaboride is the presence of magnetic anisotropy. Presumably one necessary condition for the observation of the ESR resonance -- the ferromagnetic correlations among the rare earth spins \cite{KrellnerFoerster08} is fulfilled in CeB$_6$, since a strong ferromagnetic mode was observed with neutron scattering \cite{JangFriemel14}. But in addition to ferromagnetic correlations, observed in case of tetragonal CeFePO \cite{KrellnerKini07,BrueningKrellner08,KrellnerFoerster08}, YbIr$_2$Si$_2$ \cite{KrellnerTaube12,SichelschmidtWykhoff07}  and YbRh$_2$Si$_2$ \cite{KrellnerTaube12,SichelschmidtWykhoff07a}, the presence of a strong magnetic anisotropy, that can reach a ratio of $g_{\text{ab}}/g_{\text{c}}=21$ for the latter, was assumed to play an important role. It is obvious that the simple cubic crystal structure \cite{Alper76} in CeB$_6$ will be an exception to the examples mentioned above, as no significant anisotropy can be expected \cite{Schlottmann13}.

Measurement of the angular dependence of the magnetization within the paramagnetic phase at 4.2\,K  and 0.94\,T has shown fully isotropic behavior for $[1~1~0]$ and $[1~0~0]$ crystallographic directions \cite{SluchankoBogach08a,SluchankoBogach10}. The same result can be found in ref.\,\citenum{SatoKunii84}, where inside the paramagnetic phase no difference between high symmetry directions $[1~1~1]$, $[1~1~0]$ and $[1~0~0]$ can be found. However, upon crossover into the ordered antiferroquadrupolar phase, the $[1~0~0]$ axis becomes hardest and the $[1~1~1]$ axis the easiest \cite{SatoKunii84}. This dependence is observed only up to $\sim15$\,T,{\footnotemark} \footnotetext{Data in ref.\,\citenum{HanzawaKasuya83} was measured at 1.3\,K, however, in this case it is out of importance, since at $\sim15$\,T the difference between the magnetization curves is minimal, in accordance with the data from \citenum{SatoKunii84}.} the field at which magnetizations for the $[1~1~1]$, $[1~1~0]$ and $[1~0~0]$ main crystal axes coincide, and with a further increase of the magnetic field,  anisotropy in the phase II changes sign, and up to $\sim50$\,T, $[1~0~0]$ and $[1~1~1]$  become the easy and hard axes respectively  \cite{HanzawaKasuya83}.

Similar measurements at $T=1.85$\,K, initially inside the antiferromagnetic phase, show that the angular dependence of the magnetization exhibits a significant change upon crossing the phase transition into the antiferroquadrupolar phase. Magnetization maxima and minima, which initially corresponded to magnetic field applied parallel to $[1~0~0]$ and $[1~1~0]$ crystallographic directions below $\sim1$\,T, swap themselves inside the antiferroquadrupolar phase. Besides, the amplitude of the anisotropic component also changes significantly. Its almost constant value, within the antiferromagnetic phase, rapidly drops to zero upon transition through phases III-II, followed by the gradual increase{\footnotemark} \footnotetext{ Not entirely true. In the mentioned experiment the magnetic field range was limited to 6\,T. Actual measurements of the  angular dependence were done at 0.37\,T, 0.94\,T, 1\,T, 3\,T and 5\,T. According to measurements in refs.\,\citenum{HanzawaKasuya83,SatoKunii84}, magnetization anisotropy depends nonmonotonically on the field. Its maximum value at $\sim5.5$\,T is surrounded by minima at $\sim2$\,T and $\sim15$\,T.}
as a function of the external magnetic field in the antiferroquadrupolar phase \cite{SluchankoBogach08a,SluchankoBogach10}.

To the best of our knowledge \cite{CameronFriemel16}, there were no systematic measurements of the anisotropy effects in CeB$_6$, either with neutron scattering or electron spin resonance, available until now. A relatively large number of neuron scattering measurements at various $\mathbf{Q}$ positions, but limited values of magnetic fields  applied along $[1~0~0]$ and $[1~1~0]$ crystallographic directions were published by A. Bouvet \cite{Bouvet93}. Since good agreement between the resonant mode observed in ESR and neutron scattering was confirmed in ref.\,\ref{PortnichenkoDemishev16}, recent ESR study of the $g$-factor angular dependence, performed at a fixed frequency of 60~GHz \cite{SemenoGilmanov16,SemenoGilmanov17} is of exceptional importance for us.

According to A.~V.~ Semeno \textit{et al}. \cite{SemenoGilmanov16,SemenoGilmanov17} ESR measurements revealed a significant anisotropy of the $g$-factor as a function of the applied field direction and temperature. Rotation of the sample around the $[1~\bar{1}~0]$ axis allowed them to measure the angular dependence of the $g$-factor for different crystallographic directions. The $g$-factor remains temperature independent and isotropic for generic field directions like $[1~1~0]$ or $[1~1~1]$, slightly varying around the average value $g\approx1.6$. Application of the magnetic field along the $[0~0~1]$ axis results in the anomalous temperature dependence of the $g$-factor. Its value  gradually increases from $g\sim1.4$ up to $g\sim1.7$, upon sample cooling from 3.2\,K below 2.2\,K. A strong broadening of the ESR line was also observed. Since simultaneous observation of the FM resonance by INS and ESR was previously confirmed, we expected to find a similar effect with INS.

Existing theory \cite{Schlottmann12,Schlottmann13}, which predicts angular dependence of the $g$-factor, fails to explain the results of the experiment. Assuming $\Upgamma_8$ ground state, theoretical expressions $g(\theta,\phi)$ have only one free parameter $\phi$, which defines spatial orientation of the 4$f$ orbitals. This parameter does not depend directly on temperature, however even if one assumes  its dependence  $\phi=\phi(T)$, calculated value of the effective $g$-factor for $\mathbf{B}\parallel[0~0~1]$ remains constant, since it is independent of the parameter $\phi$ value. Besides, theoretically calculated $g$-factor values lie between 2 and 2.23, much larger than actually measured in the experiment. Therefore the existing theory is unable to describe either the anomalous temperature dependence of the $g$-factor  nor its lower absolute values.

\enlargethispage{10pt}
Some of  the early attempts to calculate neutron scattering spectra within the antiferroquadrupolar phase of CeB$_6$ \cite{ThalmeierShiina03,ShiinaShiba03} are also of particular interest for this study. Transformation of the excitation energies and dispersions, calculated along high-symmetry directions of the Brillouin zone  for 7\,T and 14\,T, as well as their dependence on the magnetic field direction, suggest that two high-energy modes are almost insensitive to an increase of the magnetic field. However, the remaining four low-energy branches significantly shift to higher energies. Comparison of the calculations at certain points of reciprocal space with experiments done by A. Bouvet \cite{Bouvet93} show relatively good agreement, but the assumed quadrupolar and octupolar  order parameter degeneracy  of the model requires further clarification.  A limited number of measured data also does not allow us to identify existing modes completely. The author explicitly mentions that ``more accurate INS experiments as function of field and temperature than presently available are desirable'' \cite{ThalmeierShiina03}.

In order to fill gaps in the existing data, we carried out a systematic study of the $\Gamma$ point resonance  as a function of the magnetic field up to 15\,T for two different field configurations: $\mathbf{B}\parallel[0~0~1]$ and $\mathbf{B}\parallel[1~\bar{1}~0]$. An important issue, that requires clarification, is the temperature dependence of the zone-center resonance in high magnetic field applied along the $[0~0~1]$ direction. According to the recently published ESR data, the $g$-factor in this direction should be temperature dependent, resulting in a higher energy of the resonance (as compared to $\mathbf{B}\parallel[1~\bar{1}~0]$) at low temperatures and a gradual shift of its energy downwards upon warming. Here we verify this reported anisotropy of the $g$-factor with neutron scattering.

\begin{figure}[!t]
\includegraphics[width=\textwidth]{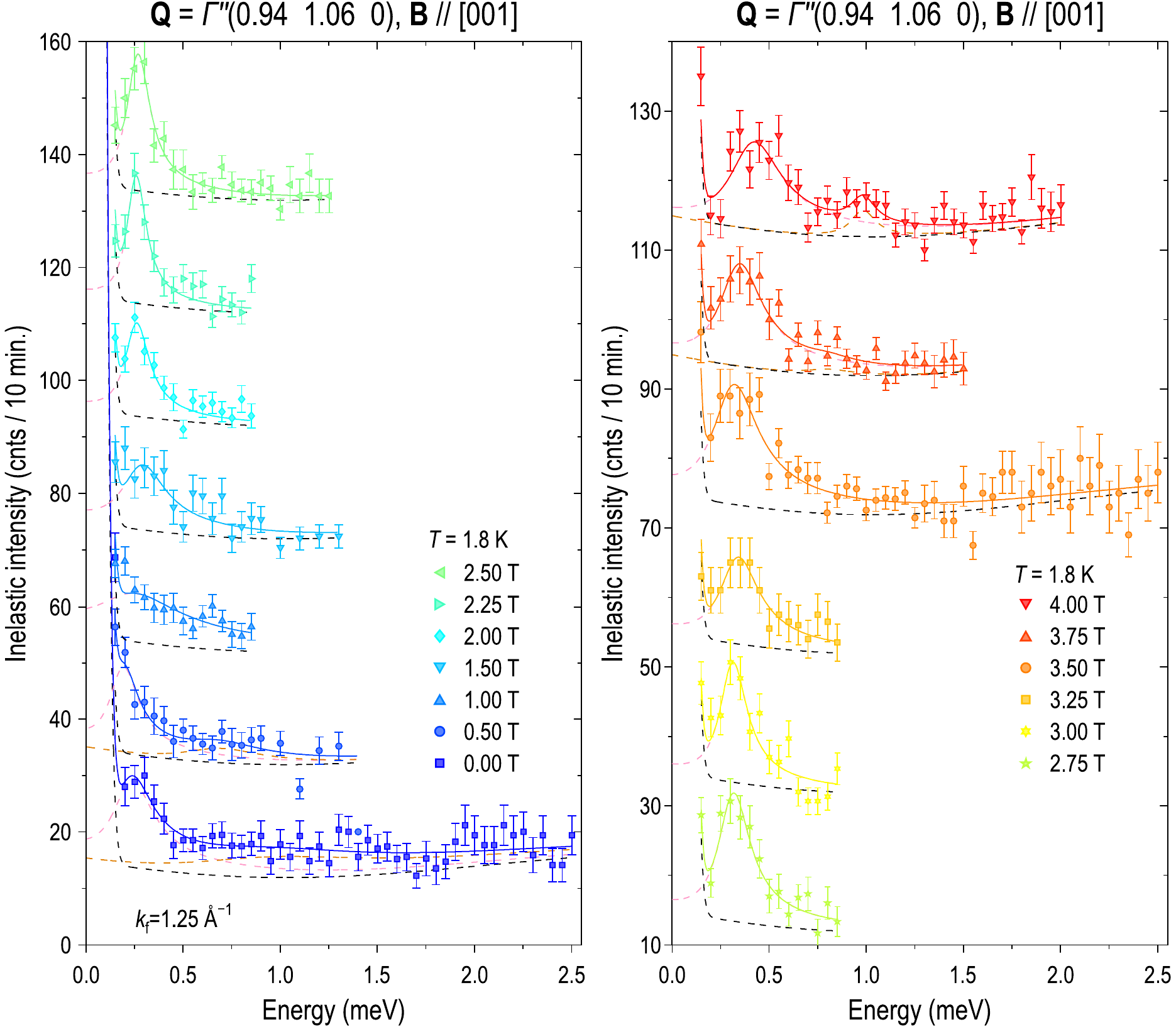}\vspace{-8pt}
\vspace{-15pt}\caption{INS spectra measured near the zone center at a slightly incommensurate wave vector. The spectra are shifted vertically for clarity with black dashed lines indicating the background baseline for each spectrum. Pink and orange dashed lines show individual contribution from both excitations. Solid lines represent Lorentzian fits.}\vspace{-25pt}
\label{Fig:CeB6GammaPANDA001}
\end{figure}

\begin{wrapfigure}[22]{l}{0.50\textwidth}\vspace{-15pt}
\includegraphics[width=0.50\textwidth]{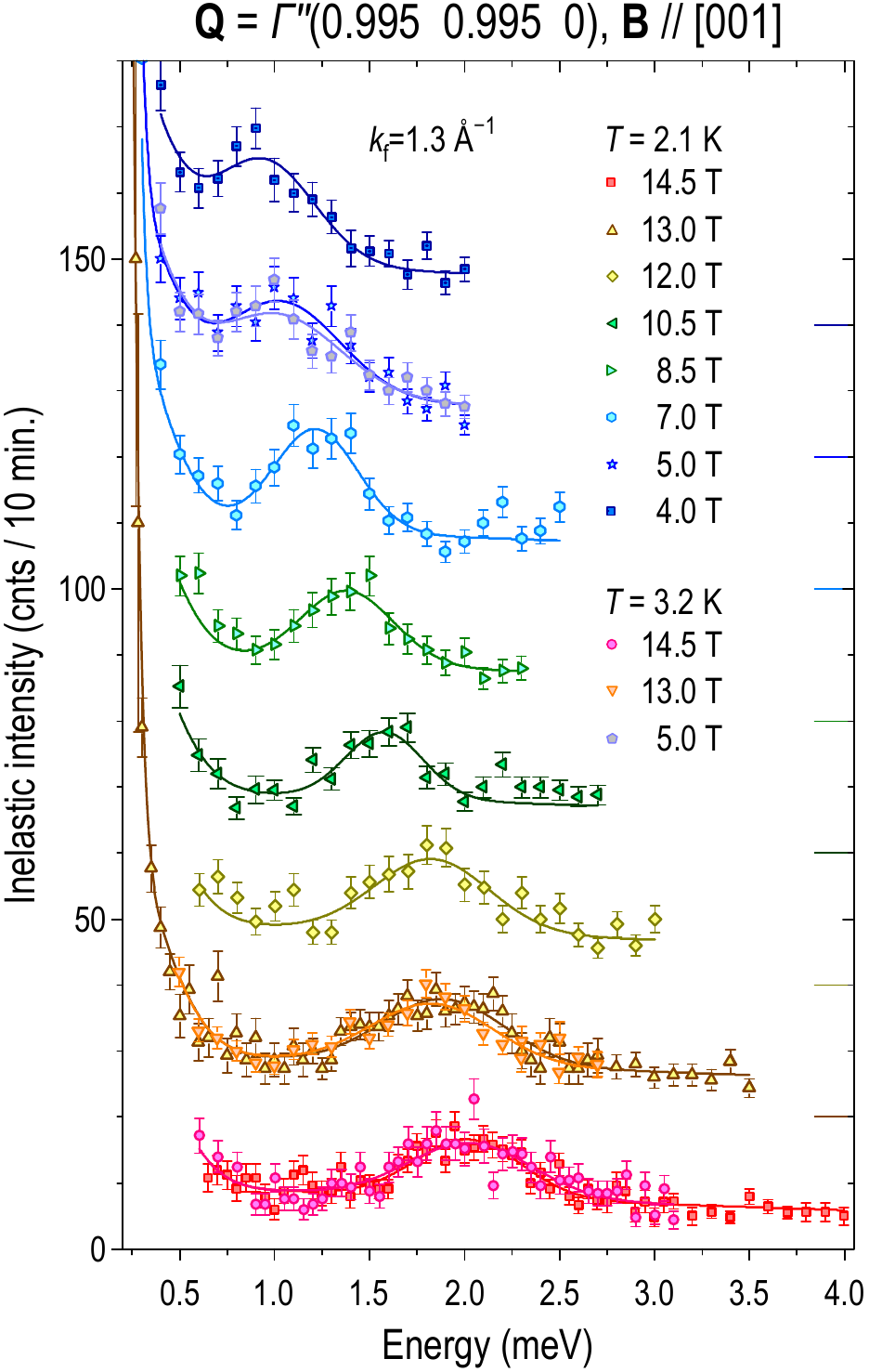}
\vspace{-25pt}\caption{INS spectra measured near the zone center at a slightly incommensurate wave vector. The spectra are shifted vertically for clarity. Solid lines represent Gaussian fits.}
\label{Fig:CeB6GammaFLEXXH001}
\end{wrapfigure}

INS experiments were performed at the three-axis spectrometer FLEXX \cite{HabichtQuintero-Castro15} at HZB, Berlin, and the cold three-axes spectrometer PANDA \cite{SchneidewindLink06} at MLZ, Garching. For all TAS experiments the final wave vector of the neutrons was fixed  to $k_{\text f}=1.25$ or $k_{\text f}=1.3$\,\AA$^{-1}$, as indicated in the legend in figs.\,\ref{Fig:CeB6GammaPANDA001}, and \ref{Fig:CeB6GammaFLEXXH001}. A cold Be filter was used in all TAS experiments to avoid higher-order neutron contamination. The sample environment for all experiments with the magnetic field aligned along the $[0~0~1]$ direction of the crystal comprised a 15\,T vertical-field VM-1B and  active shielded JVM1 5\,T  cryomagnets. A rod-shaped single crystal of CeB$_{6}$ with a mass of 4\,g was grown by the floating-zone method from a 99.6\,\% isotope-enriched $^{11}$B powder (to minimize neutron absorption), as described elsewhere \cite{FriemelLi12}.

In order to study anisotropy effects in CeB$_6$, we reoriented  the same single crystal with its crystallographic $[0\;0\;1]$ axis parallel to the magnetic field. This changes the resulting scattering plane to $(H\;K\;0)$. An anomalous temperature dependence of the $g$-factor was reported for this field direction, and therefore confirmation of such behavior was our first priority. Since we already knew, that due to significantly different background, the used magnets do not allow us to cover entire range of energies and fields simultaneously, the measurements were split into two parts. Measurements below 4 Tesla were performed at PANDA, and the high-field range 4--14.5\,T was complemented at FLEXX.

\vspace{-0pt}

\begin{wrapfigure}[16]{r}{0.50\textwidth}\vspace{-10pt}
\includegraphics[width=0.50\textwidth]{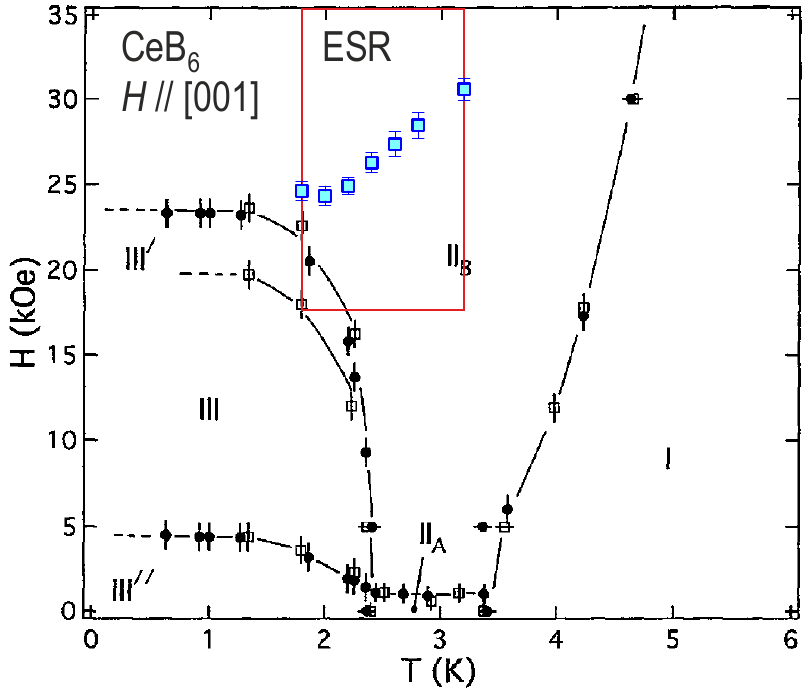}
\vspace{-15pt}\caption{The phase diagram of CeB$_6$ measured for the magnetic field parallel to $[001]$, reproduced from ref.\,\citenum{NakamuraGoto95}. The red rectangle  shows the part of the phase diagram where the temperature dependent $g$-factor was found with ESR. Blue squares mark the positions of the 60~GHz resonance at different temperatures \cite{SemenoGilmanov16,SemenoGilmanov17}.}
\label{Fig:CeB6PhaseDiagram}
\end{wrapfigure}

In order to  estimate the $g$-factor accurately, we repeated the same energy scans as a function of field. We first present results measured for the 0--4\,T field range, shown in fig.\,\ref{Fig:CeB6GammaPANDA001}. A reference 4\,T scan immediately reveals the existence of both resonances, which belong
to the modes I and II. The lower energy excitation is centered at $\hbar\omega_{1}=0.40\pm0.03\,\text{meV}$, whereas the higher energy mode at $\hbar\omega_{2}=0.98\pm0.04\,\text{meV}$. Individual contributions from both excitations are shown with the pink and orange dashed lines. We repeated the measurement at several values of  magnetic field, however, the intensity of the second resonance rapidly decreased with decreasing field. In addition,  the energy of the second resonance  decreased faster than energy of the first resonance, and for the 3.75\,T scan they were $\hbar\omega_{1}=0.33\pm0.02\,\text{meV}$ and $\hbar\omega_{2}=0.80\pm0.21\,\text{meV}$. It is almost impossible to resolve both modes below 3.75\,T. Convincingly, the second resonance reappeared  again only at 0.5\,T at $\hbar\omega_{2}=0.66\pm0.15\,\text{meV}$ with a much larger peak width. According to the phase diagram, shown in fig.\,\ref{Fig:CeB6PhaseDiagram}, transition from the  AFM phase III$^\prime$  to the  AFQ phase II is expected around $\sim2.5\,\text{T}$, substantially below 3.5\,T, the field at which we no longer observe the second resonance. The exact behavior of the second resonance in the gap between phases still remains unclear.


The behavior of the first resonance is significantly different from the experiment, where magnetic field was directed along $[1 \bar{1} 0]$. Magnetic field applied along $[001]$ cannot suppress its energy to zero together with the AFM order parameter, and the position of the resonance  does not change until the transition to the AFQ phase happens. Lack of change in resonance energy with increasing field prevents adequate definition of the $g$-factor.

Measurements, made within the higher field range are shown in fig.\,\ref{Fig:CeB6GammaFLEXXH001}. As it was already mentioned before, the magnet used in this experiment produces much higher background, therefore the reference 4\,T scan, measured  in order to have data overlap for both experiments, does not allow us to observe the mode I resonance, as it is masked by the elastic line.  The energy of the mode II excitation is the same within the error bars in both experiments. Upon further  increase of the magnetic field we observe a gradual shift of the second resonance. The corresponding $g$-factor obtained from the linear fit, shown with red solid line in fig.\,\ref{Fig:CeB6Hdep110and001}(b), now increases up to a value of $1.81\pm0.03$.

Because of time constrains we were not able to measure the full temperature dependence, however reference measurements for several values of magnetic field at 2.1\,K and 3.2\,K as indicated in the legend, showed  absolutely no effect on the $g$-factor. It is necessary to emphasize the absence of the excitation that belongs to mode I at high field. Although the elastic line is much broader, we expected to distinguish it at the maximum available field. Absence of a signal in the energy range 0.5--1.5\,meV for data measured at 14.5\,T suggests that either the mode I $g$-factor is significantly different for cases when magnetic field is parallel to $[1\;\bar{1}\;0]$ and $[0\;0\;1]$, or its intensity is strongly suppressed at high fields.

Relying on existing measurements, we can not clearly identify if the resonance observed above 12\,T  for the magnetic field aligned along $[1\;\bar{1}\;0]$ and excitation shown in fig.\,\ref{Fig:CeB6GammaFLEXXH001} for magnetic field parallel to $[0\;0\;1]$ have the same origin. At this point it is worthwhile to mention that the excitations have different width, with the FWHMs of $0.4\pm0.1$\,meV and $0.65\pm0.06$\,meV for the magnetic field along $[1\;\bar{1}\;0]$ and $[0\;0\;1]$ respectively. Broadening of the excitation upon field rotation between these high symmetry directions is actually predicted by theory in refs.\,\citenum{ThalmeierShiina03,ShiinaShiba03}, but a detailed analysis and comparison with the existing  theory will be discussed later.

Here we want to discuss the apparent contradiction regarding the absence of the temperature dependence of the $g$-factor,  derived from our results with INS and that of ESR. The observed discrepancy is further aggravated by the fact that earlier we declared perfect agreement between these methods. First, we need to clarify that we define the $g$-factor as a slope of the linear fit $\hslash\omega\kern-.2pt=\kern-.2pt\hslash\omega_{0}\kern-.3pt+{\kern-.3pt}g\mu_{\text B}B$ to the field dependence of the resonance energies. Because of  technical restrictions the ESR data are only available at relatively low fields in the range of 2.5--3\,T for the 60\,GHz (0.25\,meV) cavity. Accordingly, as a $g$-factor A.~V.~Semeno \textit{et al.} \cite{SemenoGilmanov16,SemenoGilmanov17} simply take the ratio $g=0.25\,\text{meV}/(B_\text{res}(T)\cdot\mu_{\text B})$. Thus, it is possible to define  resonance field unequivocally for each $g$-factor value. The corresponding resonance positions are overlayed on the phase diagram, in fig.\,\ref{Fig:CeB6PhaseDiagram}. Our measurements at FLEXX in high magnetic fields were performed fully in the AFQ phase, whereas the ESR measurements were done in the vicinity of the phase transition to the AFM phase, which could lead to the observed temperature dependence. It would be therefore incorrect to interpret this effect as a change in the $g$-factor.

\begin{wrapfigure}[14]{l}{0.50\textwidth}\vspace{-15pt}
\includegraphics[width=0.50\textwidth]{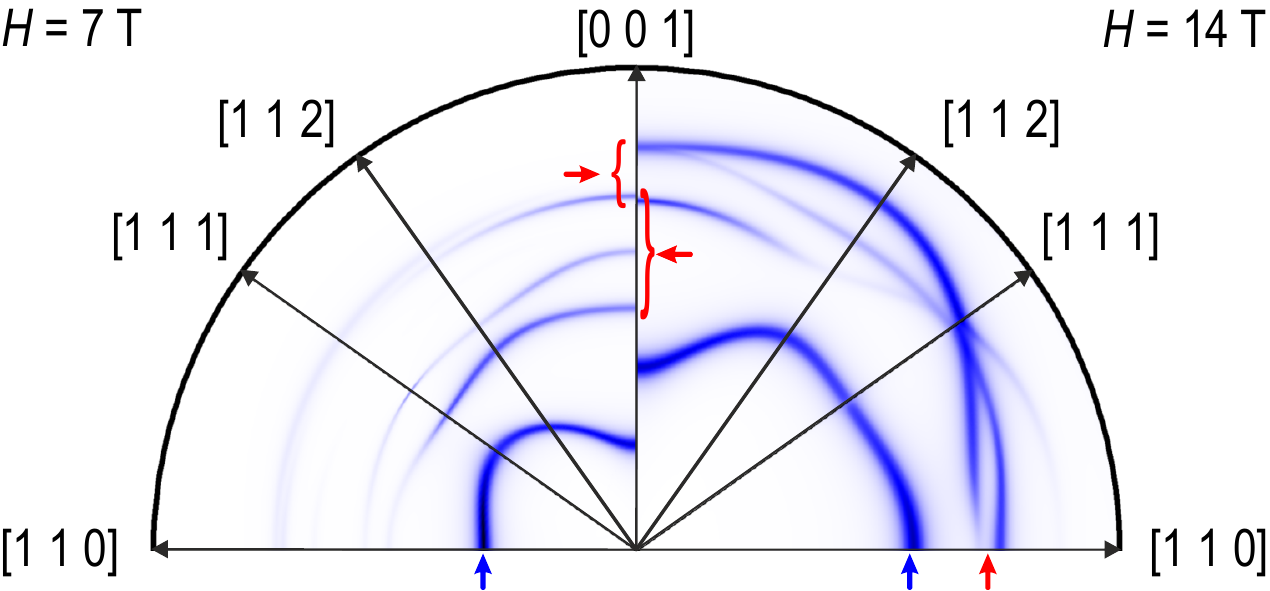}
\vspace{-15pt}\caption{Multipolar excitation branches for the $\Gamma$ point in polar representation, where the radial coordinates corresponds to the excitation energy within the limits of $0<\omega<6T_{0}$ (see ref.\,\citenum{ThalmeierShiina03} for details) and  azimuthal to the magnetic field direction. Left and right halves of the figure show results for 7\,T and  14\,T,  respectively. Small arrows indicate possible candidates for modes I and II, which are marked by the corresponding colors in \cref{Fig:CeB6Hdep110and001,Fig:ResonanceEnergyDep2}. \vspace{5pt}}
\label{Fig:ResonanceEnergyDep}
\end{wrapfigure}
It has been already shown that resonance found at the $R$ point within the AFM phase exhibits gradual broadening  and shifts towards lower energies upon warming \cite{FriemelLi12}. Such an order-parameter behavior should also be  expected for the zone-center resonance, as an intense quasielastic signal at zero field for the AFQ \cite{JangFriemel14} and paramagnetic (see sec.\,\ref{SubSec:KondoTemperatureInCeLaB6}) phases for $\Gamma$ and $R$ points was previously reported. In other words, the $0.31\pm0.01$\,meV resonance at 3\,T and $T=1.8$\,K, shown in fig.\,\ref{Fig:CeB6GammaPANDA001}, would correspond to $g=1.79\pm0.06$. Supposing 20\,\% suppression upon warming to 3.2\,K, the reported softening of the $g$-factor to 1.43 when magnetic field is parallel to $[0\;0\;1]${\footnotemark}
\footnotetext{As it was already mentioned, below 15\,T $[0\;0\;1]$ is the hardest axis, while $[1\;1\;0]$ and $[1\;1\;1]$ axes are the easiest.  AFQ phase stabilizations with magnetic field applied along the easy axis occurs at lower fields, therefore excitation energy is more robust upon heating. }
may be assumed{\footnotemark}
\footnotetext{Due to problems with the temperature stabilization we were not able   to heat up the sample and therefore check this assumption. See FRM-II experimental report \#\,13790 for details.}.

Before we start the discussion of the observed $g$-factor anisotropy, let us recall some of the theoretical predictions. Using the theory discussed in refs.\,\citenum{ThalmeierShiina03,ShiinaShiba03}, with corresponding calculations done by A. Akbari, we present angular dependence of the  multipolar excitation branches at the $\Gamma$ point, shown in fig.\,\ref{Fig:ResonanceEnergyDep}. The attempt to describe observed dependencies  with the existing theory faced some difficulties. First of all it is unclear to what extent we can trust the calculated intensities. Furthermore, the energy range of the calculated excitations also requires clarification.

\vspace{-2pt}
Despite these inaccuracies, existing theory already picks up the correct trend. For the 7\,T field applied along $[1\;\bar{1}\;0]$  one strong excitation is predicted, flagged with a blue arrow  on the left half in fig.\,\ref{Fig:ResonanceEnergyDep}. Its occurrence was previously reported [\ref{PortnichenkoDemishev16}]. At 14\,T the energy of the excitation is increased, as shown with a blue arrow for the right half of the figure.  The intensity of the higher energy branches, that used to be very weak, is strongly enhanced in this case, what we observe as an emergence of the second excitation in fig.\,\ref{Fig:CeB6SecondModeIN14}.

\vspace{-2pt}
Aligning the magnetic field along $[0\;0\;1]$  does not significantly change the energy of the higher energy branch, as shown with a red arrow. However, unlike the case with field applied along $[1\;\bar{1}\;0]$, it should be possible to observe two higher energy excitations, joined with a red bracket. The energy scan at 14.5\,T, shown in fig.\,\ref{Fig:CeB6GammaFLEXXH001}, unequivocally indicates the presence of only one higher energy resonance, which we assume might consist of two broadened modes that cannot be separated. Its width is larger, as compared to that for field applied along $[1\;\bar{1}\;0]$.  According to calculations, for the 7\,T scan with the magnetic field along $[0\;0\;1]$ we can expect even more pronounced line-width broadening, which we indeed  do not observe. The lower-energy excitation, predicted from theory, should have comparable intensity  regardless of the magnetic field orientations. However this resonance managed to avoid direct observation with INS for the case when magnetic field was along $[0\;0\;1]$ at 7 and 14.5\,T. The observed discrepancy  could be due to the already raised question of the calculated intensities accuracy, suggesting that this excitation has a strongly angular dependent spectral weight. In any case, direct observation of both resonances simultaneously, as well as theoretical predictions, suggests that the higher energy branch for both field orientations has the same nature, and its designation as mode II in both cases is justified.

\vspace{-0pt}\begin{figure}[!t]
\includegraphics[width=\textwidth]{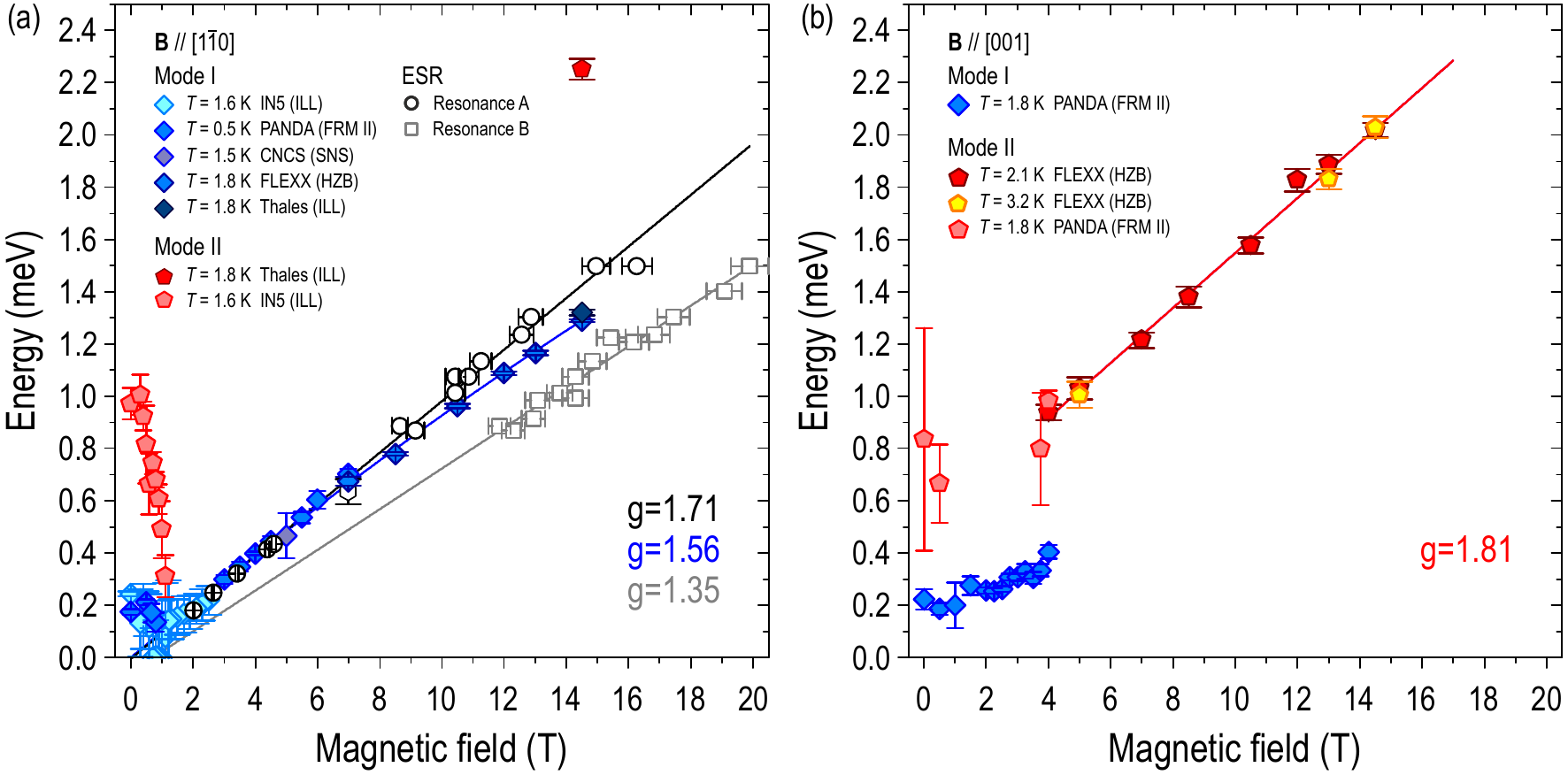}\vspace{-8pt}
\caption{Summary of the magnetic field dependence of zone-center excitations obtained from both INS and ESR spectra. Solid lines are linear fits of resonances A, B and excitations measured with neutrons for magnetic field parallel to (a) $[1\;\bar{1}\;0]$; (b) $[0\;0\;1]$. Corresponding $g$-factors have the same color. \vspace{-3pt}}
\label{Fig:CeB6Hdep110and001}
\includegraphics[width=\textwidth]{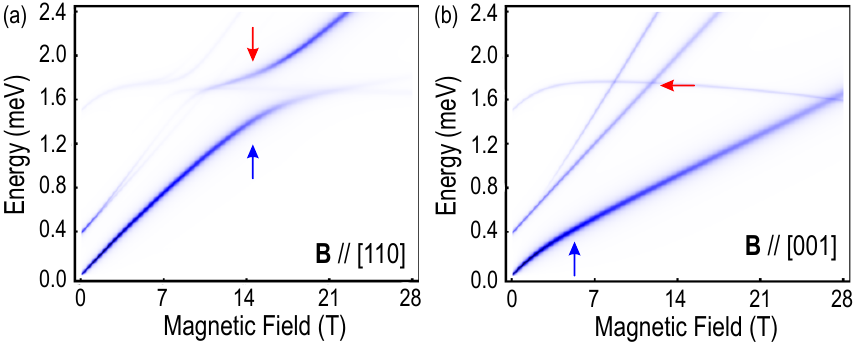}~\vspace{-8pt}
\vspace{-0pt}\caption{Multipolar excitation branches at the $\Gamma$ point as a function of magnetic field, applied parallel to (a) [110]; (b) [001]. Small arrows indicate possible candidates for modes I and II, which are marked by the corresponding arrows in fig.\,\ref{Fig:ResonanceEnergyDep}.}
\label{Fig:ResonanceEnergyDep2}\vspace{-30pt}
\end{figure}

\vspace{-2pt}
In \cref{Fig:CeB6Hdep110and001,Fig:ResonanceEnergyDep2} we compare results of the experiments with the calculated magnetic field dependence of zone-center excitations for the AFQ phase. Origin of the inelastic resonances in zero field inside the AFM phase remains unclear and is not discussed now.

\vspace{-2pt}
\enlargethispage{+8pt}
Within the AFQ phase we can clearly define two resonance modes. As one can see, the branches have a very clear anisotropy with respect to the field direction. The intensity of the modes is also strongly dependent on the field magnitude. For the field applied along $[1\;\bar{1}\;0]$, the field dependence of one strong excitation  is shown with blue solid line in fig.\,\ref{Fig:CeB6Hdep110and001}(a). As it was predicted by the theory, in fig.\,\ref{Fig:ResonanceEnergyDep2}(a), at higher fields above 8\,T it starts to deviate from the previously observed linear dependence towards lower energies. At higher fields the intensity of branches that used to be very weak is strongly enhanced (red arrow in fig.\,\ref{Fig:ResonanceEnergyDep2}(a)), what we observe as the appearance of mode II in fig.\,\ref{Fig:CeB6Hdep110and001}(a).

The situation with the field along $[0\;0\;1]$ is not yet fully understood, but there is a general trend of having stronger intensity at higher energies for mode II, shown in fig.\,\ref{Fig:ResonanceEnergyDep2}(b) with a red arrow and actually measured in fig.\,\ref{Fig:CeB6Hdep110and001}(b), unlike in the orthogonal field orientation. The fact that the stronger mode I was not observed during the experiment is not explained yet. In ref.\,\citenum{RegnaultErkelens88}, single resonance centered around $\sim0.4$\,meV was found in a 4\,T scan. This excitation is in  good agreement with the energy of  mode I, as shown in fig.\,\ref{Fig:CeB6Hdep110and001}(b). To the best of our knowledge, the only existing measurement in higher fields (except our measurements) is a 6\,T energy scan shown in fig.\,72 \cite{Bouvet93}. Due to the fact that the measurements were carried out with $k_{\text f}=1.55$\,\AA$^{-1}$, too coarse resolution does not allow a clear definition of the mode I position at $(0.95\;0.95\;0)$, but already at $(0.90\;0.90\;0)$ we can clearly resolve both resonances at $\hbar\omega_{1}\sim0.5\,\text{meV}$  and $\hbar\omega_{2}\sim1.0\,\text{meV}$. The energy of the second mode  is in  good agreement with that interpolated from our experiments $\sim1.1$\,meV, while the exact behavior of the first resonance is planned for the future experiment.

\vspace{-7pt}\enlargethispage{8pt}
\section{INS Measurements of Ce$_{\text{1}-x}$La$_x$B$_\text{6}$ Solid Solutions}{\label{Sec:INSCeLaB6}}

At the moment, all of our attention was focused on the strong magnon excitations at the zone-center and the propagation vector of the AFQ phases, as well as their field dependence. Common belief, on the one hand, indicates a mainly localized approach \cite{Ohkawa85,ShiinaShiba97,ShiinaSakai98}, that explains appearance of the field-induced Bragg peaks at the propagation vector of the AFQ phases,  anisotropy and field dependence of the $T_{\text{Q}}$ as well as NMR line shifts. On the other hand, a recently uncovered  resonance mode with a spin-gap $\sim0.5$\,meV at the $R$ point \cite{FriemelLi12}, like those observed in unconventional superconductors, seriously questioned the validity of this established picture. Taking into account the itinerant spin dynamics it became possible to reproduce the above mentioned resonance \cite{AkbariThalmeier12a}. However, further experiments \cite{JangFriemel14} showed that the existing theory overlooks an intense ferromagnetic collective mode,  that dominates the magnetic excitation spectrum of CeB$_6$.

In Ce$_{\text{1}-x}$La$_x$B$_\text{6}$ solid solutions, the Ce sublattice is randomly diluted by nonmagnetic La ions, adding an extra dimension to the parameter space of the phase diagram (fig.\,\ref{Fig:PhaseDiagram3D}). It was demonstrated that the Ce/La ratio has a notable influence both on the Fermi surface geometry \cite{GoodrichHarrison99} and on the Fermi-liquid properties \cite{NakamuraEndo06} of the material.

It is generally accepted that both the AFM and AFQ phases are suppressed by La doping. In a broad range of La concentrations between $\sim$\,20\% and $\sim$\,70\%, another low-temperature ``hidden order'' phase (so-called phase IV) is stabilized, whose origin until now remains poorly understood. Elastic neutron scattering  \cite{KuwaharaIwasa07} and polarized neutron diffraction \cite{KuwaharaIwasa09} have ascribed this mysterious phase to an ordering of higher-order magnetic multipoles (``octupolar order''). However, until recently the understanding of Ce$_{\text{1}-x}$La$_x$B$_\text{6}$ systems was restricted to only a few La concentrations, and the available data remained very limited. In particular, there were no systematic investigations of magnetic dynamics in phase IV, and it was not clear, how far this phase persists in the phase diagram. Neutron diffraction data are available for only a few La concentrations, namely for $x=0.25$ both on a single crystal in finite fields \cite{ErkelensRegnault87} and on a polycrystalline sample of the same composition in zero field \cite{FischerIwasa05}, as well as for $x=0.30$, where only the form factor of the $(\frac{1}{2} \frac{1}{2} \frac{1}{2})$ magnetic reflection has been investigated \cite{KuwaharaIwasa07}. No systematic refinement of the magnetic structure within phase~III has been performed for any La concentration. On the one hand, the results of powder neutron diffraction \cite{FischerIwasa05} suggest that  phase III persists in the same configuration as in the parent CeB$_\text{6}$ compound at least up to $x=0.25$, though with a somewhat reduced magnetic ordered moment. On the other hand, muon-spin relaxation ($\mu$SR) data on a sample with a similar composition \cite{SchenckGygax07} has indicated the presence of a short-range or spin-glass-like order with no evidence for long-range magnetic correlations typical of phase~III. Therefore, the persistence of the AFM order up to high La concentrations ($x\lesssim0.5$), as suggested by the established phase diagram as shown in fig.\,\ref{Fig:PhaseDiagram3D}, appears questionable. In other families of CeB$_6$-derived solid solutions, such as Ce$_{\text{1}-x}$Pr$_x$B$_\text{6}$ \cite{MignotAndre08} or Ce$_{\text{1}-x}$Nd$_x$B$_\text{6}$ \cite{MignotRobert09}, recent neutron diffraction measurements revealed several incommensurate magnetic phases, which cannot be excluded so far also in Ce$_{\text{1}-x}$La$_x$B$_\text{6}$.

\vspace{-7pt}\enlargethispage{8pt}
\subsection{Diffuse fluctuations in Ce$_{\text{1}-x}$La$_x$B$_\text{6}$}{\label{SubSec:DiffusefluctuationsCeLaB6}}
Within an itinerant description, the strength of magnetic interactions is mediated by the conduction electrons, and therefore depends on the low-energy electronic structure. It can be expressed within linear response theory by the Lindhard susceptibility~\cite{ChanHeine73}, which quantifies the propensity of a particular electronic structure towards nesting instabilities of the Fermi surface. This may often lead towards a magnetically ordered state \cite{InosovEvtushinsky09}, therefore the possibility that the AFQ state can be similarly linked to the electronic structure should not be overlooked. Previously, RKKY-type interactions between the quadrupolar Ce moments have been theoretically considered~\cite{ShibaSakai99}. However, until now a detailed 3D picture of the Fermi surface in CeB$_{6}$ has been notably absent from the literature. Studies to date~\cite{SoumaIida04,NeupaneAlidoust15,RamankuttyJong16} have been limited in scope owing to a variety of problems: CeB$_{6}$ is a fully 3D system, thereby requiring a detailed set of measurements acquired with different photon energies to measure the full 3D Fermi surface with sufficient accuracy. The material itself is hard to cleave, and is subject to surface reconstruction which can hide the true structure of the bulk~\cite{PatilAdhikary10,Trenary12}. In order to study  the Fermi surface with soft-x-ray ARPES, an innovative approach, analogous to a 3D tomographic imaging,  was implemented~[\ref{KoitzschHeming16}]. In order to probe different planes of $\mathbf{k}$-space the sample was cleaved along all high-symmetry crystallographic planes $[1~0~0]$, $[1~1~0]$ and $[1~1~1]$. This gave complete 3D information about the electronic structure. The Fermi surface images, taken for different cleavage planes, have either two- or fourfold symmetry for the $[1~1~0]$ and $[1~0~0]$ crystallographic planes respectively, or three- or sixfold symmetry, depending on the photon energy for the $[1~1~1]$ plane. These results agree with previous studies~\cite{SoumaIida04, NeupaneAlidoust15}, and the orbital character of the ellipsoids consists of extended Ce\,$5d$ states with admixtures of localized Ce\,$4f$ states near the Fermi energy, which is similar to other $4f$ systems which exhibit a neutron resonant mode~\cite{StockBroholm08}. Calculations from the first principles density functional theory shows even better agreement of the Fermi surface with the ARPES result \cite{Barman18}.

\begin{wrapfigure}[17]{r}{0.48\textwidth}\vspace{-5pt}
\includegraphics[width=\linewidth]{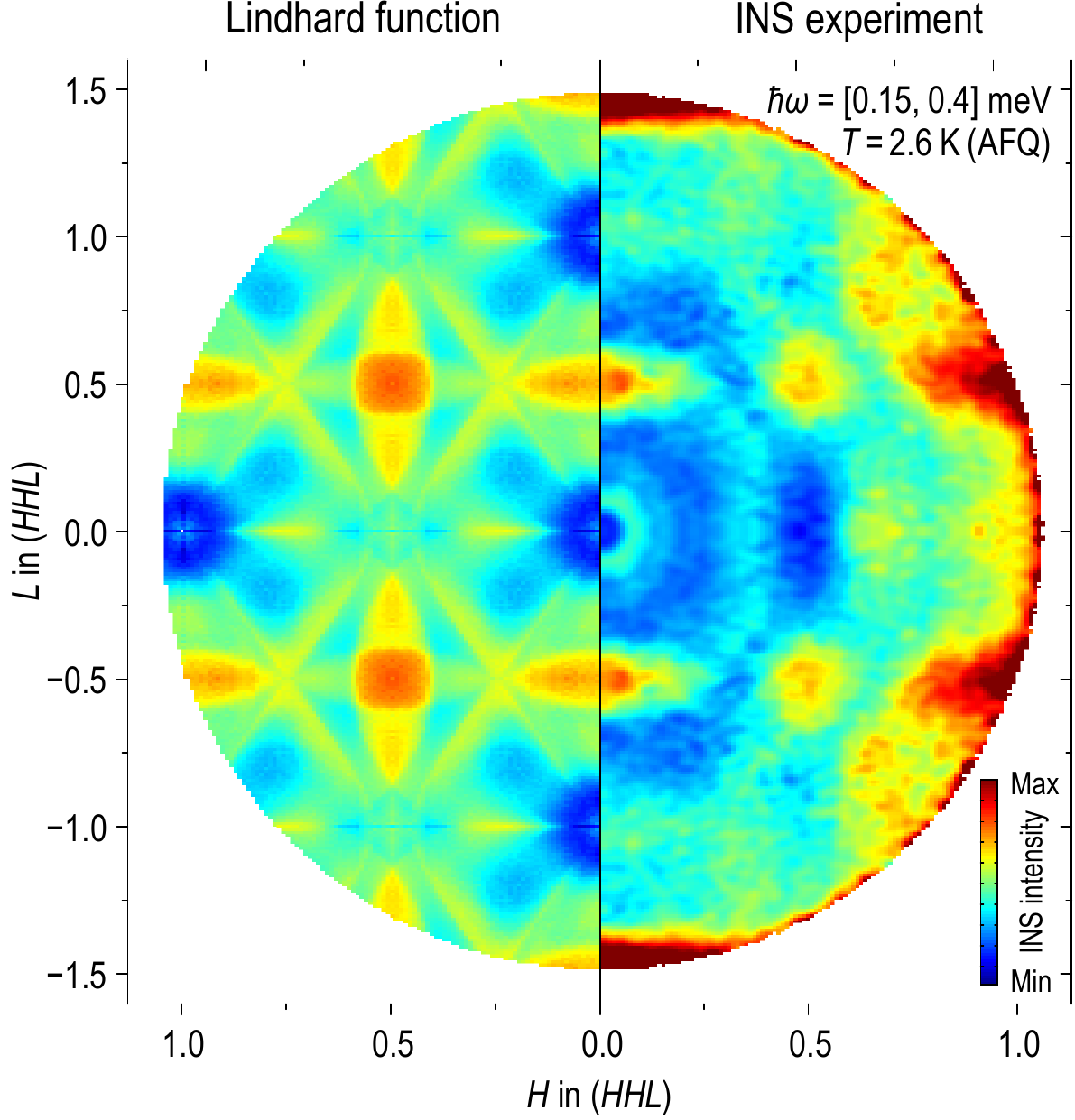}
\caption{A two-dimensional representation of the Lindhard function in the $(HHL)$ scattering plane (left) compared with the distribution of magnetic quasielastic intensity measured by INS at $T = 2.6$~K (right). Figure reproduced from ref.\,\ref{KoitzschHeming16}.}
\label{qxqyIN5Lindhard}
\end{wrapfigure}
Using a tight-binding fit to the experimentally obtained Fermi surface, it was possible to calculate the Lindhard function and compare it with INS data from our neutron investigation into CeB$_{6}$. This is shown in fig.\,\ref{qxqyIN5Lindhard}, where the left side of the figure portrays the calculated Lindhard function while the right side of the image is the distribution of magnetic quasielastic scattering at $T = 2.6$~K. It is evident that the two images are remarkably similar, as the calculation reproduces not only the maxima of intensity but also the qualitative shape of the scattering at the $R(\frac{1}{2} \frac{1}{2} \frac{1}{2})$ and $X(0 0 \frac{1}{2})$ points. The good agreement between the Lindhard susceptibility, calculated from the measured electronic structure, and the INS data illustrates the itinerant character of  magnetic excitations in CeB$_{6}$. Furthermore, it suggests that the propagation vector of the AFQ order is determined by the geometry of the Fermi surface. Here, itinerant electrons determine the RKKY interactions between Ce $4f$ multipolar moments, which can still be considered as local. In addition, the precursor signatures to the compex magnon spectrum of CeB$_{6}$, which are visible above $T_{\text N}$ and have been discussed in previous sections, appear to have their natural explanation in the low-energy electronic structure. The results summarized here indicate that the propensity towards a specific magnetic order has roots in favourable nesting conditions of the Fermi surface.

We know that the quasielastic intensity has several distinct local maxima in the following high-symmetry points within the Brillouin zone: (i) the ferromagnetic $\Gamma$ point, (ii) the AFQ propagation vector $R(\frac{1}{2}\,\frac{1}{2}\,\frac{1}{2})$ at the corner of the cubic Brillouin zone, and (iii) the $X(0\,0\,\frac{1}{2})$ point at the center of the Brillouin-zone face, where no static order is realized in pure CeB$_6$, but which becomes the propagation vector of the AFM order in samples above 40\% Nd doping as well as in pure NdB$_6$ \cite{MignotRobert09}. These results suggest that the concentration of magnetic spectral weight in these high-symmetry points could be influenced by chemical doping. Our recent measurement [\ref{NikitinPortnichenko18}], performed at the MACS spectrometer at NIST, showed a considerable effect of La and Nd doping on the spectral weight distribution between different high-symmetry points. Samples with up to 75\% La doping show monotonic increase in intensity at the $X$ point at the expense of the suppressed spectral weight at the $R$ point. This result seems to be consistent with the suppression of AFQ ordering by La substitution, and at the same time surprising, as no long range order can be observed at the $X$-point propagation vector. Both La and Nd substitution lead to an effective hole doping, therefore one should expect similar influence on the Fermi surface, and the corresponding strengthening of the intensity at the $X$ point should be even more pronounced in Nd-doped samples. As expected we observe strong suppression of the intensity at the $R$ point, consistent with AFQ phase suppression upon 30\% Nd doping. At the $X$ point we find a narrow diffuse peak, that connects the strong intensity maxima at the equivalent AFM wavevectors, $(\pm\frac{1}{4}\,\pm\frac{1}{4}\,\frac{1}{2})$, and an elongated broad peak extending along $(\frac{1}{3}\,\frac{1}{3}\,L)$. The presence of extended peaks in momentum space with multiple local maxima of QEMS intensity is a signature of itinerant frustration in this system.

\subsection{Kondo temperature of Ce$_{\text{1}-x}$La$_x$B$_\text{6}$ solid solutions}{\label{SubSec:KondoTemperatureInCeLaB6}}
It is well known that in CeB$_6$, as well as in many other heavy-fermion metals, the quasielastic line width $\Gamma$ has the following phenomenological temperature dependence: $\Gamma=k_{\text B}T_{\text K}+A \sqrt{T}$, where $T_{\text K}$ is the Kondo temperature. This dependence corresponds to the conventional spin relaxation rate typical for the paramagnetic state of most heavy-fermion compounds, such as CeCu$_6$, CeAl$_3$, CePb$_3$, CeRu$_2$Si$_2$, CeB$_6$, and many others. Despite a long history of examples where this dependence is realized \cite{Robinson00}, the physical meaning of the parameter $A$ is not yet clear. Also, there have been no systematic studies of the dependence of the neutron-derived Kondo temperature $T_{\text K}$ and the parameter $A$ on doping, on the details of the electronic structure, and on the position in reciprocal space.

The availability of the La-doped series of Ce$_{\text{1}-x}$La$_x$B$_\text{6}$ single crystals gives us an opportunity to address these long-standing open questions. From the theory point of view, a $\sqrt{T}$ scaling of the relaxation rate with temperature is expected for a single Kondo impurity in the high-temperature limit \cite{CoxBickers86, BickersCox87}. Here we have the possibility to follow its persistence separately at every high-symmetry point of the Brillouin zone ($\Gamma$, $R$, and $X$) in the whole range of Ce concentrations from a dense Kondo lattice into the dilute limit of low Ce concentrations. Therefore, here we employ INS to study the changes of the quasielastic line width and the neutron-derived Kondo energy scale, which can be extracted from the same data, by means of both triple-axis and time-of-flight spectrometers.

INS experiments on the quasielastic line width  temperature dependencies  were performed at the three-axis spectrometer FLEXX \cite{HabichtQuintero-Castro15} at HZB, Berlin, the cold neutron three-axis spectrometer IN12 \cite{SchmalzlSchmidt16} at ILL, Grenoble,  and cold three-axis spectrometer PANDA \cite{SchneidewindLink06} at MLZ, Garching. For the IN12 and PANDA  TAS experiments, the final wave vector of the neutrons was fixed  to  $k_{\text f}=1.3$\,\AA$^{-1}$, while the FLEXX experiment was done with  $k_{\text f}=1.25$\,\AA$^{-1}$. In order to  avoid higher-order neutron contamination, a cold Be filter was used in PANDA and FLEXX experiments. Full mapping of the quasielastic intensity distribution in $\mathbf{Q}$-space was done with the cold neutron multi-disk chopper spectrometer (LET) \cite{BewleyTaylor11} at ISIS, Didcot and the disk chopper time-of-flight spectrometer IN5 \cite{OllivierMutk11} at ILL, Grenoble. The incident neutron wavelength was fixed at 5\,\AA\:(3.27\,meV). The sample environment for all experiments except PANDA comprised a similar ``orange'' type  helium cryostat. For the PANDA experiment, a liquid-free closed cycle (CC) cryostat (see app.\,\ref{Appendix:INSFit}) was used. Rod-shaped single crystals of Ce$_{1-x}$La$_{x}$B$_{6}$ ($x=$ 0, 0.23, 0.28, 0.5, 0.75) with a typical mass of $\sim4$\,g were grown by the floating-zone method from an isotope-enriched $^{11}$B powder to minimize neutron absorption, as described elsewhere \cite{FriemelLi12}.

\begin{figure}[!t]
\includegraphics[width=\textwidth]{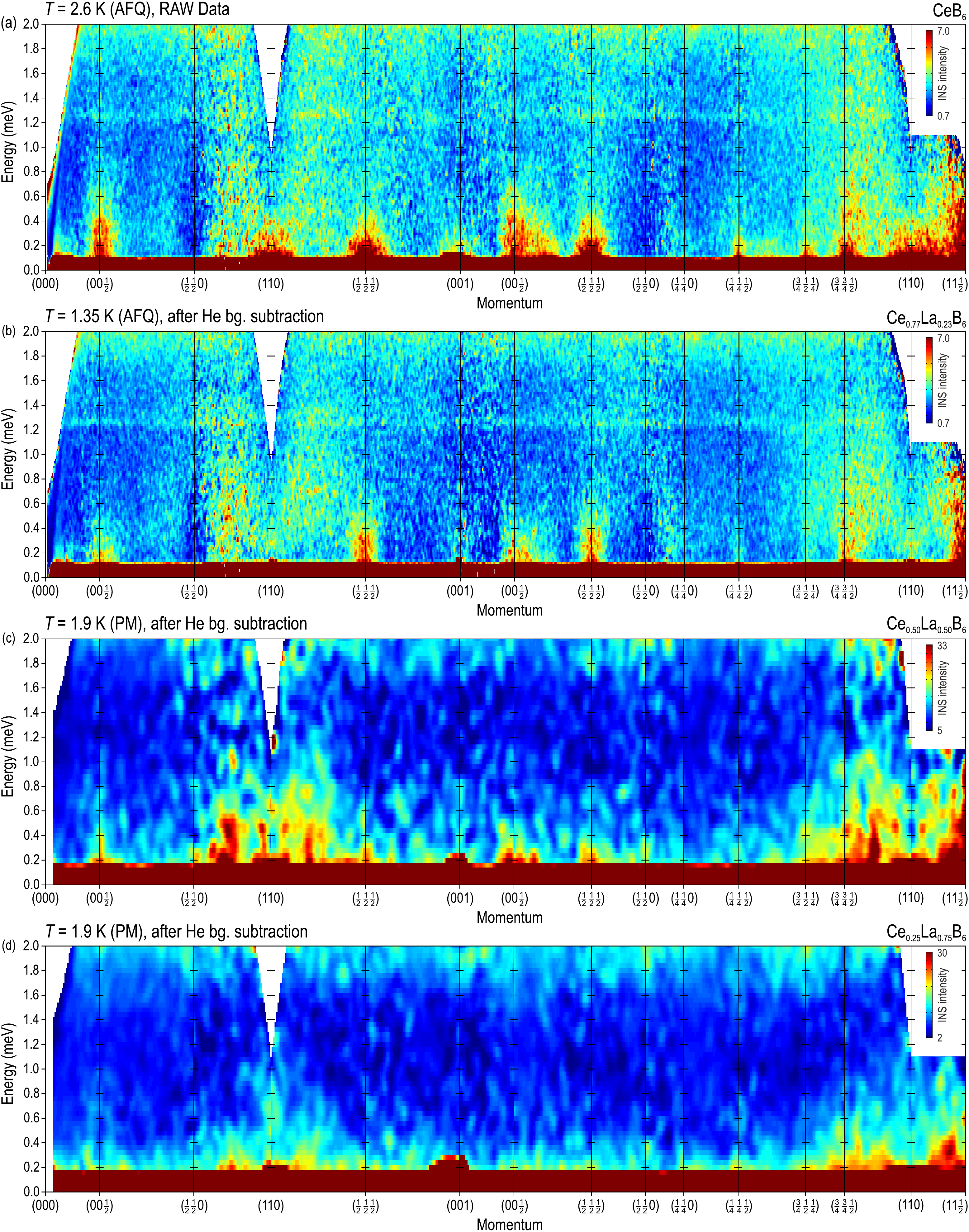}\vspace{-5pt}
\caption{Energy-momentum profiles along a polygonal path through high-symmetry directions in $\mathbf{Q}$-space for (a) CeB$_6$; (b) Ce$_{0.77}$La$_{0.23}$B$_6$; (c) Ce$_{0.50}$La$_{0.50}$B$_6$; (d) Ce$_{0.25}$La$_{0.75}$B$_6$. Background contamination from the He exchange gas was subtracted from the data in panels (b-d) as explained in  app.\,\ref{Appendix:HeExchangeGasBackgroundSubstraction}. Each straight segment in panels (a-b) represents integration of the TOF data within $\pm0.05$\,\AA$^{-1}$ along both perpendicular directions  to the desired in-plane momentum direction. Because of insufficient statistics, corresponding range in panels (c-d) was increased to  $\pm0.10$\,\AA$^{-1}$.}
\label{Fig:CeLaB6TOF}\vspace{-20pt}
\end{figure}

\vspace{-5pt}
However, a certain issue for the observed redistribution of intensity between $R$ and $X$ points exists. Unlike the reported measurement on the same 23\% doped sample [\ref{NikitinPortnichenko18}], where a more pronounced effect of the spectral weight redistribution  between these points was found, we hardly observe this effect in fig.\,\ref{Fig:CeLaB6TOF}(b). This comparison raises several technical concerns, as datasets were measured at different instruments and processed differently. One would expect that the final result should not depend on data processing. For sure, a similar reference measurement of pure CeB$_6$ at the MACS spectrometer and comparison with the data from panel (a) would allow to resolve  the observed discrepancy. However, the use of a separate experiment for this question is inappropriate.  We assume that this inconsistency can be overcome if we recall the fact that the results from ref.\,\ref{NikitinPortnichenko18} actually show momentum-space distributions of the magnetic intensity after subtracting same area in momentum space at an elevated temperature, that acts as an empty-can background. In both cases, in TOF and TAS measurements, we usually observe higher background at large momentum transfer (see fig.\,\ref{Fig:HeSubtractionManual}), which we assume originates from multiple scattering. This effect could lead to the observed inconsistency. Measurements performed on highly doped 50\% and 75\% La doped samples, as shown in panels (c) and (d), are in  good agreement with experiments from [\ref{NikitinPortnichenko18}] and confirm rapid suppression of the diffuse peak at the $R$ point.

\vspace{-0pt}

\vspace{-5pt}\enlargethispage{5pt}
Before discussing the temperature dependence of the quasielastic linewidth, we present a set of energy-momentum profiles for all available single crystals. Measurements were carried out either within the AFQ or PM phases, as indicated in the caption for each panel in fig.\,\ref{Fig:CeLaB6TOF}. Upon 23\% La doping, the uniformly distributed intensity previously observed in CeB$_6$ \cite{JangFriemel14}, shown in panel (a), undergoes a minor redistribution, as shown in panel (b). Weak intensity observed near the AFM ordering vectors $(\frac{1}{4}\;\frac{1}{4}\;\frac{1}{2})$ in the parent compound almost completely disappears,  in accordance with a rapid suppression of the AFM~phase.

\begin{wrapfigure}[25]{r}{0.50\textwidth}\vspace{-0pt}
\includegraphics[width=0.50\textwidth]{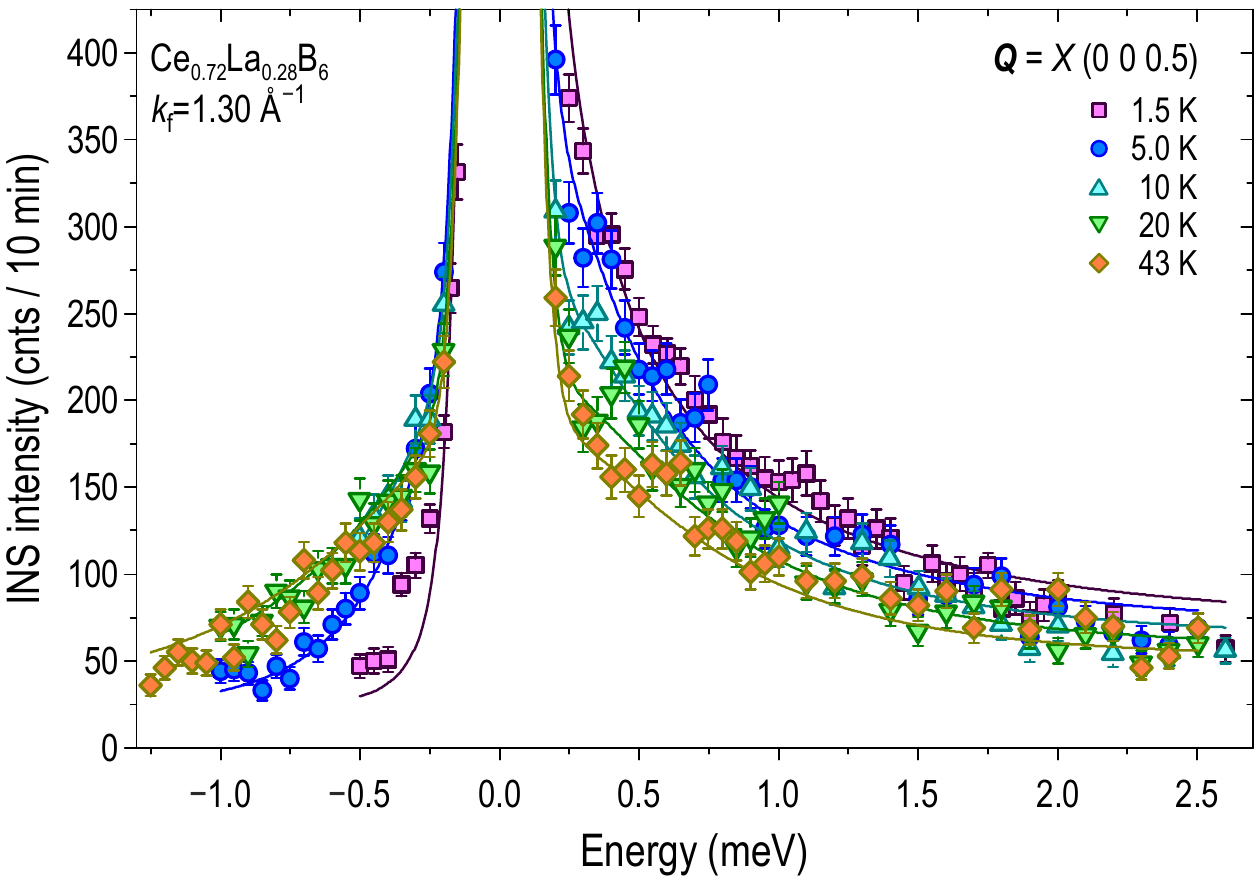}
\vspace{-25pt}\caption{Temperature dependence of the low-energy magnetic scattering for the 28\% doped sample. The solid lines are fits with quasielastic Lorentzian profiles.}
\label{Fig:CeLa0p28B6TdepIN14}
\vspace{+8pt}
\includegraphics[width=0.50\textwidth]{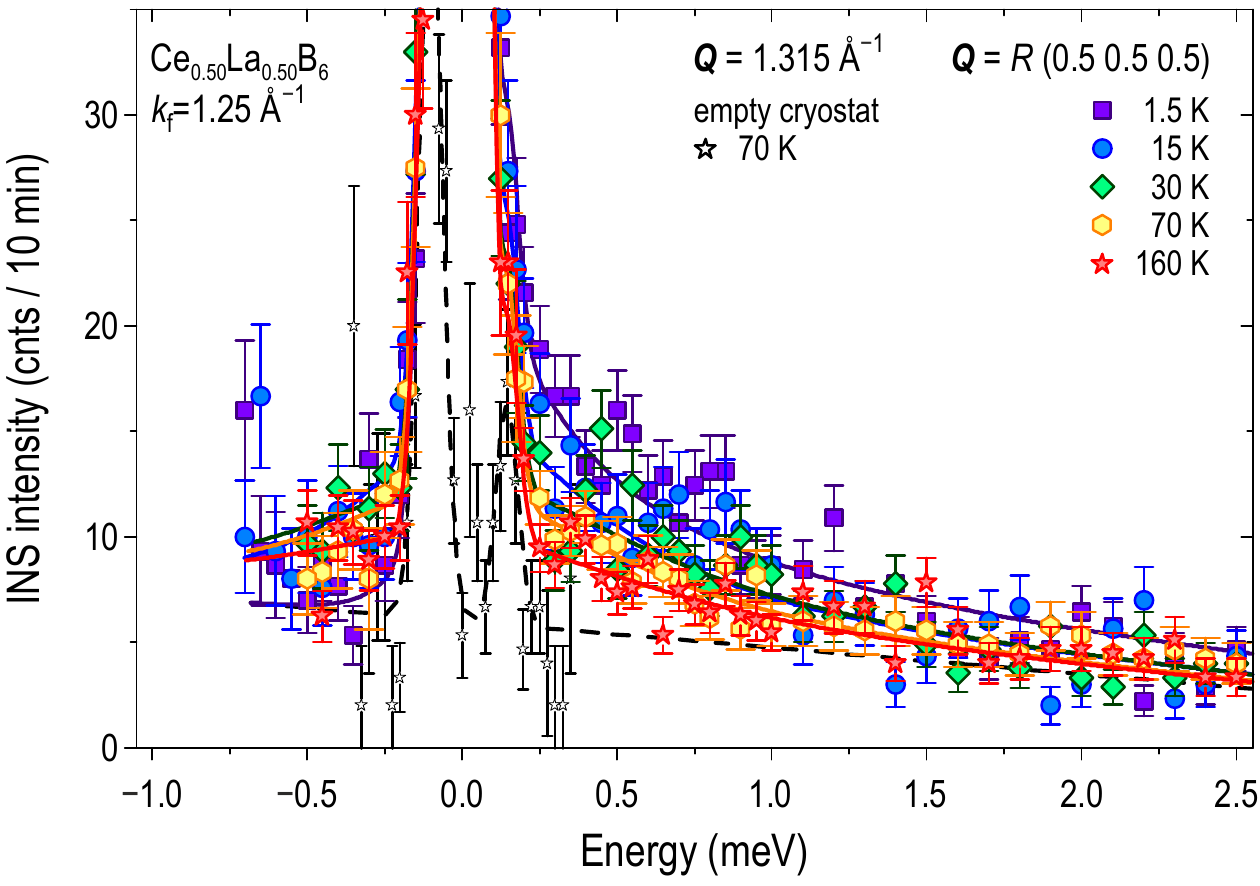}
\vspace{-25pt}\caption{Temperature dependence of the low-energy magnetic scattering for the 50\% doped sample. Background signal, from the empty cryostat, shown with black stars, has been added to all fits. The solid lines are fits with quasielastic Lorentzian profiles. }
\label{Fig:CeLa0p50B6TdepFLEXX}
\end{wrapfigure}
\begin{figure}[!t]
\centering
\includegraphics[width=0.8\textwidth]{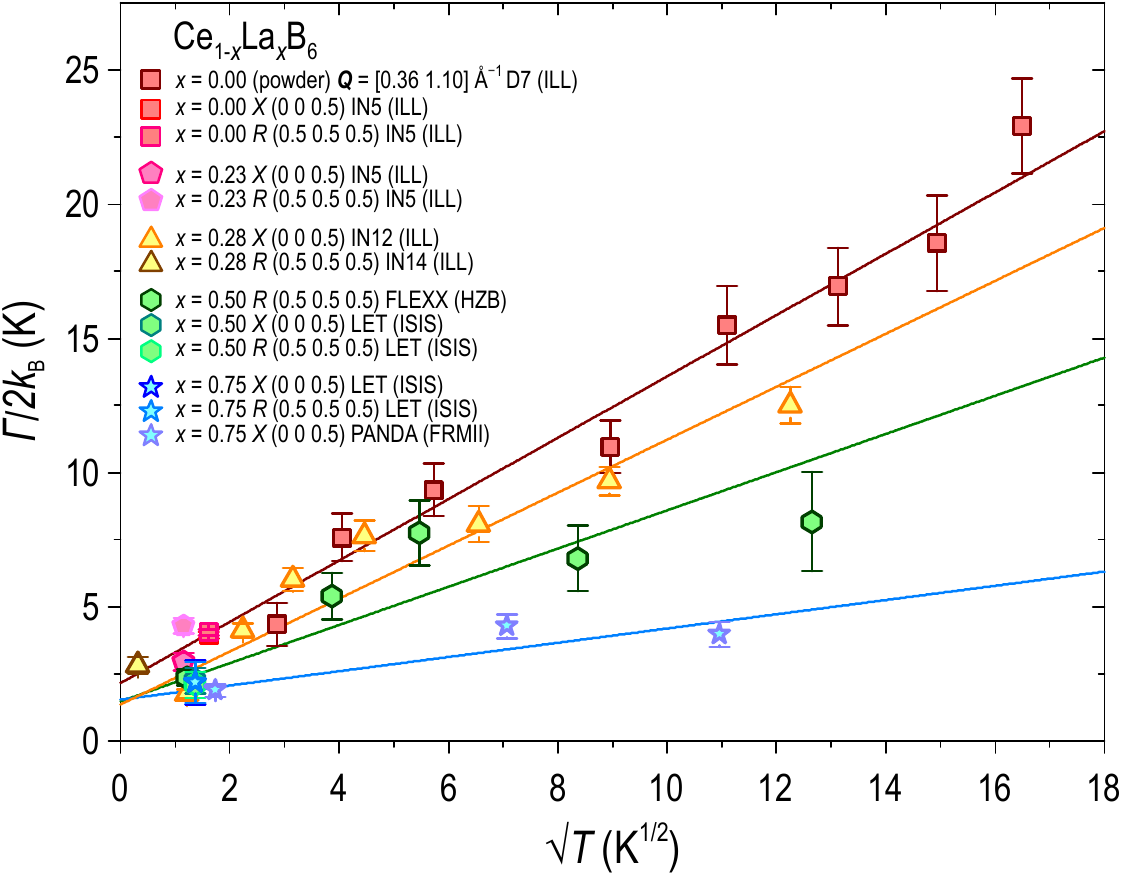}\vspace{-5pt}
\vspace{-0pt}\caption{Summary, of all measurements on samples with different La-doping concentrations, of the temperature dependence of the quasielastic line width, $\Gamma/2$, fitted to a $T^{1/2}$ law. Powder measurements for the parent compound, that correspond to the $X$ point within the momentum transfer range $\mathbf{Q}=[0.36~1.10]$\,\AA$^{-1}$, were plotted from ref.\,\citenum{HornSteglich81}.}\vspace{-20pt}
\label{Fig:SqrtGamma}
\end{figure}

Now we turn to the temperature dependence of the quasielastic linewidth. In fig.\,\ref{Fig:CeLa0p28B6TdepIN14} I show the evolution of the quasielastic magnetic scattering signal with temperature for the sample with the 28\% La doping measured at the $X$ point. Experimental results can be well described by a quasielastic Lorentzian profile of the form given by eq.\,\ref{Eq:Quasielastic} \cite{GoremychkinOsborn00}. Upon temperature increase we observe a monotonic suppression of the signal and linewidth exhibits a gradual broadening{\footnotemark}.
\footnotetext{This experiment was affected by a cryostat leak, as seen by the increased background above 77\,K (not shown in fig.\,\ref{Fig:CeLa0p28B6TdepIN14}). See ILL Exp. Rep. \#4-01-1339 for details, thus seriously questioning the accuracy of the high temperature data. However, since a good agreement of the data affected  by the leak  with the $T^{1/2}$ scaling was confirmed, we assume that all extrinsic effects discussed in app.\,\ref{Appendix:INSFit} possess less influence on the linewidth compared to the sample with higher La concentration.}
The temperature dependence of the quasielastic line width, shown in  fig.\,\ref{Fig:SqrtGamma}, follows the conventional $\sqrt{T}$ law, and as will be discussed later is not very different from that for the parent compound.

As the next step we investigate the temperature dependence of the quasielastic signal, measured at the $R$ point{\footnotemark}
\footnotetext{The inconsistency, in measuring the $R$ point instead of the $X$ point, unlike in all other cases, was due to the technical limitations on the minimum $2\theta$ angle. According to fig.\,\ref{Fig:CeLaB6TOF}(c), a certain fraction of the spectral weight is  present in the $R$ point, as well as residual quasielastic linewidth, which within the error, is the same for both points. Therefore, we expected that defining the linewidth from the $R$ point will only require longer counting time.}
for the 50\% La doping. When trying to fit data with the quasielastic Lorentzian profile we noticed a minor deviation of signal shape, which at the same time broadened the linewidth at high temperatures significantly. This effect was due to the scattering from the cryostat walls, as explained in app.\,\ref{Appendix:ScatteringTAS}, and was successfully overcome after including signal from the empty cryostat to the fit. We assume that we faced this problem due to the low signal intensity. Compared with the parent compound we have half as many Ce atoms, that is further aggravated by the  strong suppression of spectral weight at the $R$ point reported above. Except for the low signal intensity, there is no difference with respect to the lower doping levels. In consistency with the previous sample, the intensity of the signal is strongly suppressed  with doping and accompanied by a corresponding broadening of the linewidth.

\enlargethispage{10pt}
\begin{wrapfigure}[13]{r}{0.50\textwidth}\vspace{-10pt}
\includegraphics[width=0.50\textwidth]{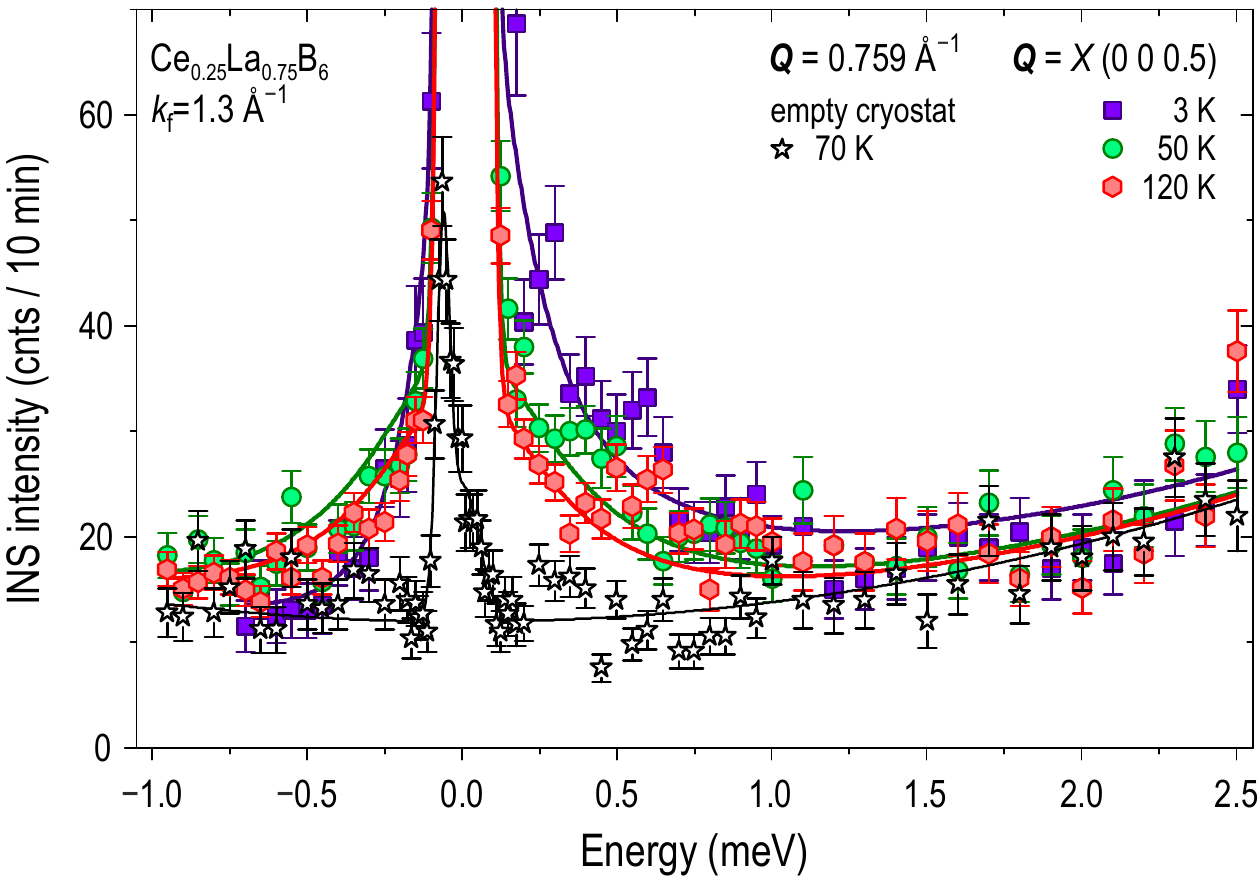}
\vspace{-25pt}\caption{Temperature dependence of the low-energy magnetic scattering for the 75\% doped sample. Background signal from the empty cryostat, shown with black stars has been added to all fits. The solid lines are fits with quasielastic Lorentzian profiles.}
\label{Fig:CeLa0p75B6TdepPANDA}
\end{wrapfigure}
Measurement on a highly doped compound Ce$_{0.25}$La$_{0.75}$B$_{6}$, which we performed on the cold-neutron triple-axis spectrometer PANDA at FRM-II, is shown in fig.\,\ref{Fig:CeLa0p75B6TdepPANDA}. Since we already knew that scattering from the sample environment could significantly influence the fit results of the quasielastic line, for a convincing determination of the linewidth upon heating we decided to measure the background from an empty cryostat completely for the full energy range, as indicated in the legend. Besides, we carried out measurement in a CC cryostat. Because the sample is a metal and has good thermal conductivity, it was directly screwed to the cold-head. This allowed us to pump the sample space continuously  and operate the cryostat without He exchange gas. Therefore, all extrinsic effects due to the change of the  exchange gas pressure inside the cryostat upon warming can be completely excluded. Despite the fact that the measured temperature dependence suffers from a lack of the experimental data for different temperatures, thanks to  all the technical efforts it gives a reliable evidence for the linear temperature dependence of the quasielastic lines width on $\sqrt{T}$.


\vspace{-0pt}

\begin{wrapfigure}[13]{r}{0.50\textwidth}\vspace{-0pt}
\includegraphics[width=0.50\textwidth]{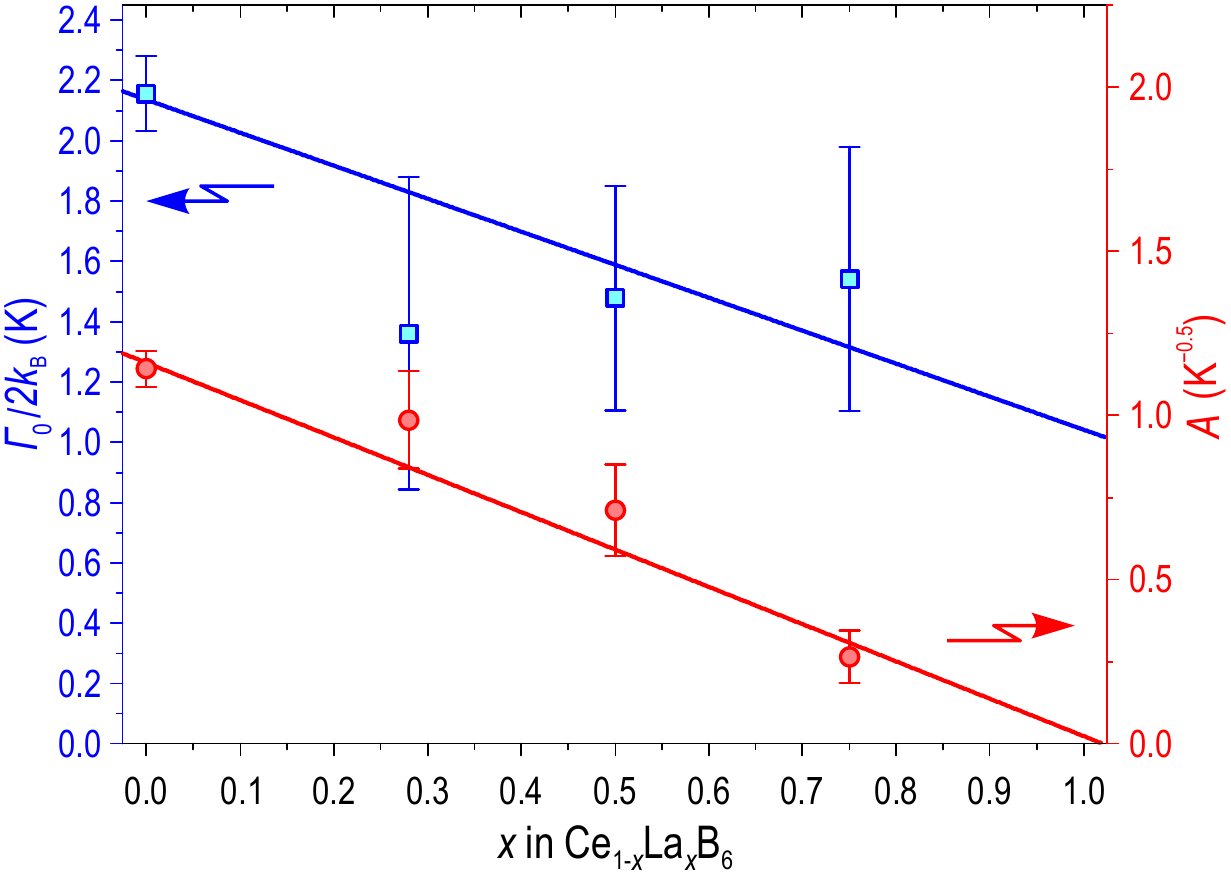}
\vspace{-15pt}\caption{Doping dependence of the: (left) Kondo temperature, defined as $T_{\text K}=\Gamma_{0}/(2 k_{\text B})$; (right) parameter $A$. Values are obtained from the linear fits $\Gamma=k_{\text B}T_{\text K}+A \sqrt{T}$ to fig.\,\ref{Fig:SqrtGamma}.}
\label{Fig:KondoTemperature}
\end{wrapfigure}

The summary of all our measurements on samples with different La-doping concentrations on different instruments and wave vectors is shown in fig.\,\ref{Fig:SqrtGamma}. As a reference measurement on the parent CeB$_6$ sample,  we used results published in ref.\,\citenum{HornSteglich81}. As one can see, all samples  comply with the above mentioned law. Upon 28\% La dilution a slight decrease of the quasielastic linewidth  at higher temperatures, compared with the parent compound, can be found. Results for the 50\% doped sample show a small deviation{\footnotemark}
\footnotetext{As will be shown in app.\,\ref{Appendix:INSFit}, it is not trivial to  obtain quasielastic linewidth from the fit of the INS spectrum, particularly in case of high La concentrations.  At the time of measurements, we did not fully understand the extremely important role of accurate background estimation, therefore as shown in fig.\,\ref{Fig:CeLa0p50B6TdepFLEXX}, an empty cryostat was measured only in the narrow energy range in the vicinity of the elastic line, since scattering from cryostat walls in this experiment (see app.\,\ref{Appendix:ScatteringTAS}) complicated the correct determination of the linewidth for low temperatures.}
from $T^{1/2}$ law scaling, however one can still fit them with the same law, and further decrease of the linewidth persists. The same result was obtained when we increased La concentration to 75\% and changed the Kondo-lattice limit of  pure CeB${_6}$ to the dilute limit of low Ce concentrations, where Ce atoms can be treated as isolated Kondo impurities.  All temperature dependencies were fitted with the  $\Gamma=k_{\text B}T_{\text K}+A \sqrt{T}$ law, and corresponding  Kondo temperature, defined as $T_{\text K}=\Gamma_{0}/(2 k_{\text B})$ and parameter $A$ are shown in fig.\,\ref{Fig:KondoTemperature}. Before considering the change as a function of La doping, let us recall the prior knowledge of the Kondo temperature expected value, obtained from other techniques.

The parent compound CeB$_6$ shows a typical Kondo-type behavior, with a resistivity minimum, that starts to increase upon cooling as expected (see sec.\,\ref{Sec:KondoScreening}). Although CeB$_6$ represents a classical Kondo lattice, its  temperature dependence of resistivity is well described with a model valid for the dilute Kondo state, and the corresponding $T_{\text K}=5\sim10$\,K \cite{TakaseKojima80}. This result significantly differs from the Kondo temperature $\sim3$\,K, determined from the lowest temperature quasielastic linewidth, measured with INS \cite{HornSteglich81}. Measurements of the other extreme case, with just a few percent of Ce left, thus representing genuine impurity model, gives a lower  $T_{\text K}=1\sim3$\,K value \cite{Felsch78,SamwerWinzer76,Winzer75}.

\begin{table}\vspace{-25pt}
\begin{tabular}{c@{~~~~~~~~~~~}c@{~~~~~~~~}c@{~~~~~~~~~~~~}c@{~~~~~~~}}\\
\toprule
$x$ in Ce$_{1-x}$La$_{x}$B$_{6}$~(at--\%)       & $T_{\text K}$~(K)        & Method                        & reference \\
\toprule
        0                               &$5\sim10$\tablefootnote{\label{first} Here $T_{\text K}$ was determined by fitting the temperature dependence of resistivity with a theory of dilute Kondo system, which could lead to exaggerated Kondo temperature determination. }
                                                                            & resistivity                   & \cite{TakaseKojima80}   \\
        0                               &  3\tablefootnote{\label{second} Low-temperature $T=5$\,K width of the quasielastic neutron line.}
                                                                            & INS                           & \cite{HornSteglich81}   \\
\midrule
$0 \leq x \leq75$                     &  2                    & resistivity                              & \cite{SatoSumiyama85}   \\
\midrule
97                                    &  1                    & resistivity                     & \cite{SatoSumiyama85}   \\
$97.10 \leq x \leq99.39$                  & $2.8$               &  magnetic susceptibility      & \cite{Felsch78}\\
$97.10 \leq x \leq99.39$                  & $1.05$              & resistivity                   & \cite{SamwerWinzer76}   \\
$99.89 \leq x \leq99.93$                  & $1.40\pm0.05$       &  magnetic susceptibility      & \cite{Felsch78}\\
$98.80 \leq x \leq99.39$                   & $1.1\pm0.2$         & resistivity                   & \cite{Winzer75}   \\
\midrule
\bottomrule
\end{tabular}
\caption{Comparison of the Kondo temperatures, determined with various methods for  two limiting cases:  the isolated impurity model with extremely low Ce atoms concentration as well as for a dense Kondo lattice.}
\label{Tab:KondoTemperature}\vspace{-15pt}
\end{table}
It is clear that in case of a dense arrangement of impurities, one cannot neglect the interaction between them. This effect is observed as a doping dependence of the   residual resistivity. According to  discussion in ref.\,\citenum{SatoKunii84}, disregarding coherence effects might lead to the exaggerated value of the Kondo temperature, and might be the case in \cite{TakaseKojima80}. Therefore, authors  suggest that in order to accurately estimate the Kondo temperature, one should fit data at higher temperatures, at which coherence effects are less pronounced. By a lucky coincidence, crystal field splitting in CeB$_6$ is large enough, therefore it will not affect the fit result at temperatures below the characteristic temperature of the $\Upgamma_{8}$--$\Upgamma_{7}$ splitting. Determined in this way the Kondo temperature is claimed to be weakly dependent on La concentration up to 75\% doping level, with a corresponding value $T_{\text K}=2$\,K. Further increase of the La concentration gradually lowers the Kondo temperature to 1\,K \cite{SatoKunii84,SatoSumiyama85}. Determined Kondo temperatures for various La concentrations,  summarized in  table\,\ref{Tab:KondoTemperature}, should be compared with our results.

In fig.\,\ref{Fig:KondoTemperature} we present the doping dependence of the Kondo temperature, extracted from the residual quasielastic linewidth at $T\rightarrow0$ and parameter $A$. Despite all the difficulties that accompanied our attempts to determine the quasielastic linewidth at high temperatures and La doping, we managed to get unambiguous confirmation of the $T^{1/2}$ law scaling. Unfortunately, clearly observed dependence of parameter $A$, shown with a red solid line,
has no explanation, since  its physical meaning  is not yet clear. Our result suggests that it is proportional to the concentration of the Ce moments in Ce$_{1-x}$La$_{x}$B$_{6}$. On the other hand, the  Kondo temperature determined with INS for the parent compound is in good agreement with the one determined from resistivity. Besides, weak dependence of the Kondo temperature \cite{SatoKunii84,SatoSumiyama85} upon La doping is also confirmed. Shown with  blue markers, the Kondo temperature has a certain tendency to a marginal decrease at high doping. For certain one could claim that the large value of error bars  does not allow us to distinguish between a possible decrease or a constant $T_{\text K}$ value. However, since for the measurement of the 75\% doped sample we took into account all requirements for accurate linewidth estimation, we claim the reported reduction of the Kondo temperature to be real.

\chapter{Ce$_{3}$Pd$_{20}$Si$_{6}$}\label{Chap:ResultsCe3Pd20Si6}

\section{Introduction to Ordering Phenomena in Ce$_{3}$Pd$_{20}$Si$_{6}$}

\begin{wrapfigure}[12]{r}{0.50\textwidth}\vspace{-12pt}
\noindent\includegraphics[width=0.50\textwidth]{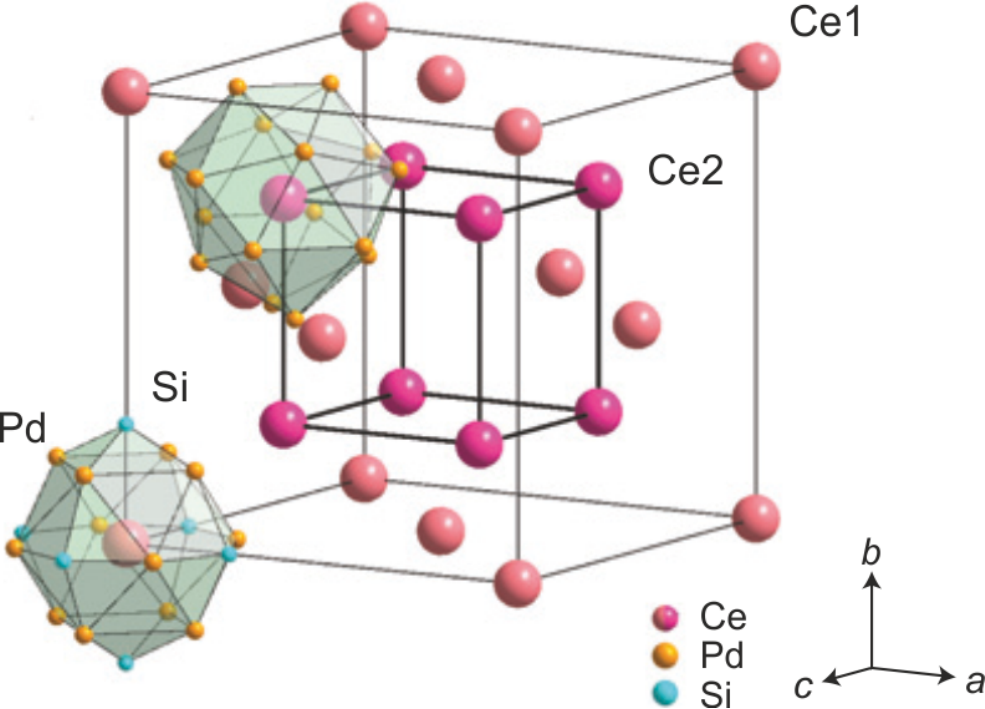}\vspace{-0pt}
\caption{The crystal structure of Ce$_{3}$Pd$_{20}$Si$_{6}$, showing two interpenetrating sublattices of cerium ions. Figure from ref.\,\citenum{CustersLorenzer12}.}
\label{Fig:Ce3Pd20Si6CrystalStructure}
\end{wrapfigure}

In general, the magnetic dynamics of heavy-fermion systems is governed by the interplay of local-moment fluctuations and the dynamic spin response of itinerant heavy quasiparticles~\cite{Stewart84, FuldeLoewenhaupt85}. Because the multipolar ordering is often coupled to another magnetically ordered state, the instances where the orbital ordering appears as a separate solitary phase are rare. Several Ce-based compounds, such as CeB$_6$, Ce$_{3}$Pd$_{20}$Si$_{6}$, and Ce$_{3}$Pd$_{20}$Ge$_{6}$, are of particular interest as they display such standalone multipolar ordered phases. The so-called clathrate materials, $R_{3} \mathrm{Pd}_{20} X_{6}$ ($R$\,=\,rare earth, $X$\,=\,Si, Ge), are far more complex than CeB$_{6}$, because they consist of two interpenetrating sublattices of cerium, as indicated in fig.~\ref{Fig:Ce3Pd20Si6CrystalStructure}. One of them (8c) possesses a simple-cubic structure like in CeB$_6$, whereas the other one (4a) has a geometrically frustrated face-centered-cubic (fcc) structure, with each cerium ion being surrounded by either palladium/germanium or silicon-palladium/silicon-germanium nonmagnetic ``cages''. This unprecedented coexistence in the same material of two inequivalent Kondo lattices with different symmetry has been considered theoretically \cite{BenlagraFritz11}, and the two sublattices were predicted to exhibit two drastically different Kondo temperatures due to the competitive Kondo-screening effects.

\begin{wrapfigure}[21]{r}{0.50\textwidth}\vspace{0pt}
\includegraphics[width=0.5\textwidth]{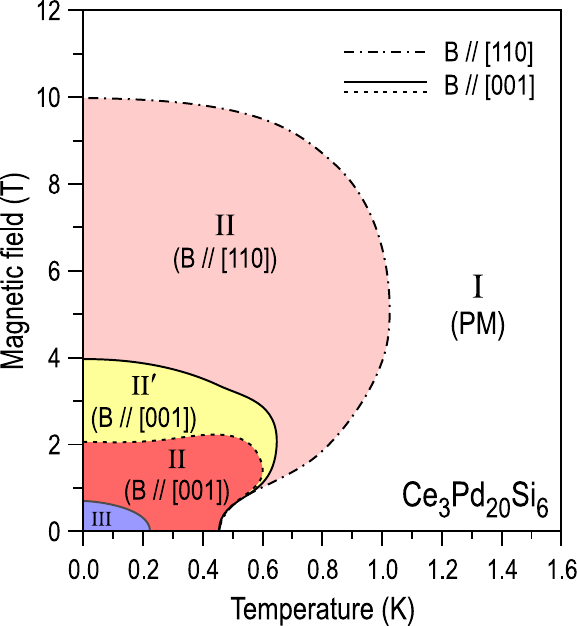}\vspace{-4pt}
\caption{Schematic phase diagram of Ce$_3$Pd$_{20}$Si$_6$ after refs.\,\citenum{GotoWatanabe09, MitamuraTayama10, OnoNakano13}, showing strong anisotropy of the hidden-order phase~II with respect to the magnetic field direction. Phase I represents the paramagnetic state, phase~II is associated with antiferroquadrupolar order, phase~III is the incommensurate dipolar magnetic phase~\cite{CustersLorenzer12, Lorenzer12, DeenUnpublished}, whereas the order parameter of phase~II$^\prime$ that appears only for $\mathbf{B}\parallel[001]$ still remains unclear. Figure from ref.\,\ref{PortnichenkoPaschen16}\vspace{-3pt}.}
\label{Fig:Ce3Pd20Si6PhaseDiagram}
\end{wrapfigure}
The magnetic phase diagram of Ce$_3$Pd$_{20}$Si$_6$ \cite{MitamuraTayama10, OnoNakano13, PaschenLarre14} nearly replicates that of CeB$_6$, though with reduced characteristic temperature and magnetic-field scales. The ground state, labeled phase~III, appears to be AFM in nature with a maximal reported $T_{\text{N}}$ of 0.31~K~in the highest-quality stoichiometric samples, according to a corpus of available studies performed on both single crystals \cite{ProkofievCusters09, MitamuraTayama10, OnoNakano13} and powders \cite{TakedaKitagawa95, DuginovGritsaj00, StrydomPikul06, PaschenMueller07, GotoWatanabe09, CustersLorenzer12}. Evidence of the AFM order comes from neutron scattering, which shows the presence of static AFM order~\cite{Lorenzer12}, and magnetization measurements which indicate either AFM order or an ordered octupolar phase~\cite{MitamuraTayama10}. The AFM phase can also be suppressed by a magnetic field of only 0.7\,T applied along the $[100]$ crystallographic direction \cite{MitamuraTayama10, OnoNakano13}. This places Ce$_3$Pd$_{20}$Si$_6$ very close to a quantum critical point (QCP) \cite{StrydomPikul06, PaschenMueller07}, which has been reached in polycrystals by the application of a magnetic field of less than 1~T \cite{CustersLorenzer12}, and which is likely accessible also under a small hydrostatic or chemical pressure \cite{ProkofievCusters09}. Such a proximity leads to non-Fermi-liquid behavior and, in particular, to very high values of the electronic specific-heat coefficient, $\gamma=\lim_{T\rightarrow0}{\scriptsize\Delta}C(T)/T$. Reportedly, this can reach up to $\sim$\,8\,J/(mol$_\text{Ce}\cdot$K$^2$) near the QCP, making Ce$_3$Pd$_{20}$Si$_6$ one of the heaviest-electron systems known to date \cite{TakedaKitagawa95, CustersLorenzer12}.

Like in CeB$_6$, the AFM phase in Ce$_3$Pd$_{20}$Si$_6$ is surrounded by the so-called phase~II, which is also attributed to an AFQ order \cite{MitamuraTayama10, OnoNakano13} yet remains much less studied. In zero field, this phase persists only in a narrow temperature window between $T_\text{N}$ and \mbox{$T_\text{Q}\approx0.45$\,--\,0.5\,K}, according to thermodynamic measurements \cite{StrydomPikul06, MitamuraTayama10, OnoNakano13}. It is initially stabilized by the application of small magnetic fields of a few teslas, but is eventually suppressed at even higher fields, leading to another qualitative similarity to the AFQ phase of CeB$_6$ \cite{GoodrichYoung04}, albeit in a much more accessible field range. To date, the propagation vector of phase~II in Ce$_{3}$Pd$_{20}$Si$_{6}$ remains unknown, although we have found strong evidence in our neutron scattering measurements that, for the 8c sublattice, it matches the $( \frac{1}{2} \frac{1}{2} \frac{1}{2})$ of the AFQ phase in CeB$_{6}$~\cite{PortnichenkoCameron15}.

\section{Momentum-Space Structure of Quasielastic Spin Fluctuations in Ce$_3$Pd$_{20}$Si$_6$}

Magnetic dynamics in heavy-fermion metals usually represents an intricate tangle of the local-moment fluctuations and the spin-dynamical response of itinerant heavy quasiparticles \cite{Stewart84, FuldeLoewenhaupt85}. The strong hybridization of the localized 4$f$ electron states with the conduction band makes these two contributions difficult to decouple. For instance, in the most classical heavy-fermion compounds like CeCu$_6$, CeRu$_2$Si$_2$ and CeAl$_3$, the low-temperature inelastic neutron scattering (INS) signal consists of a momentum-independent single-site quasielastic magnetic scattering (QEMS) attributed to localized Kondo-type excitations and an inelastic contribution from inter-site magnetic correlations due to RKKY interactions, which merge together at higher temperatures \cite{AeppliYoshizawa86, RossatMignod88, LazukovAlekseev02, TidenAlekseev07}.

Despite these qualitative similarities, the momentum-space distribution of the QEMS response in the paramagnetic state is strongly material dependent. Thus, CeRu$_2$Si$_2$ exhibits incommensurate magnetic fluctuations peaked at $(0.3~0~0)$ and $(0.3~0.3~0)$ wave vectors \cite{RossatMignod88}, while short-range antiferromagnetic (AFM) correlations were found near $(1\,0\,0)$ in CeCu$_6$ \cite{AeppliYoshizawa86, RossatMignod88}. The resulting momentum-space structure of the zero-frequency susceptibility, $\chi_0(\mathbf{Q},\omega\!=\!0)$, carries essential information about the material's electronic properties and its tendency to magnetic instabilities driven by the RKKY coupling. From more recent examples, a direct relationship between the Fermi-surface nesting properties and the short-range magnetic correlations, resulting in a diffuse neutron-scattering signal, was demonstrated for Tb$_2$PdSi$_3$~\cite{InosovEvtushinsky09}. Of particular relevance for the present work are also recent results on CeB$_6$ \cite{FriemelLi12, JangFriemel14}, where a maximum in the normal-state QEMS intensity was found to coincide in momentum space with the $(\frac{1}{2} \frac{1}{2} \frac{1}{2} )$ propagation vector of the magnetically hidden order that sets in below $T_\text{Q}=3.2$\,K and is usually attributed to the antiferroquadrupolar (AFQ) ordering of the localized Ce\,4$f$ quadrupolar moments \cite{EffantinRossat-Mignod85, SantiniCarretta09, KuramotoKusunose09}. At even lower temperatures below $T_\text{N}=2.3$\,K, it succumbs to a multi-$\mathbf{k}$ commensurate AFM order \cite{EffantinRossat-Mignod85, ZaharkoFischer03}, which stabilizes a narrow band of dispersive magnetic excitations in the INS response \cite{FriemelLi12, JangFriemel14} that can be explained by the formation of a low-energy resonant spin-exciton mode \cite{AkbariThalmeier12}.

The very low temperature scales and complications due to the intricate crystal structure of Ce$_3$Pd$_{20}$Si$_6$ have so far precluded any direct observations of the AFQ order by diffraction methods, either with or without the application of magnetic field. As a result, the propagation vector of phase~II remains unknown, whereas in the structurally much simpler CeB$_6$ it has been determined as $\mathbf{q}_{\text{AFQ}}=(\frac{1}{2} \frac{1}{2} \frac{1}{2})$ both by neutron scattering \cite{EffantinRossat-Mignod85, FriemelLi12} and resonant x-ray diffraction \cite{NakaoMagishi01, TanakaStaub04, MatsumuraYonemura09, MatsumuraYonemura12}. One essential complication that hinders similar measurements on Ce$_3$Pd$_{20}$Si$_6$ is that its crystal structure includes two interpenetrating sublattices of Ce ions on crystallographically inequivalent 4a (Ce1) and 8c (Ce2) Wyckoff sites in a cubic unit cell with the $Fm\bar{3}m$ space group. This results in the unit cell parameter of the simple cubic Ce2 sublattice being half that of the fcc Ce1 sublattice, $a=12.28$\,\AA. As a consequence, additional magnetic Bragg reflections due to an AFQ order of the same kind as in CeB$_6$, residing on the Ce2 sublattice, would coincide with the much stronger $(111)$ structural Bragg reflections and be therefore much more difficult, if not impossible, to observe directly.

Previous INS measurements on Ce$_3$Pd$_{20}$Si$_6$ have been performed, to the best of our knowledge, only on polycrystalline samples \cite{PaschenLaumann08, DeenStrydom10}. They revealed a clear crystalline electric field (CEF) line at 3.9\,meV \cite{PaschenLaumann08} and suggested the presence of an additional unresolved low-energy peak centered at 0.31\,meV, which reportedly persisted up to ambient temperature and was attributed to another CEF excitation \cite{DeenStrydom10}. However, follow-up measurements performed on the same powder sample with a better energy resolution, which we will present further on, show a clear magnetic signal centered at much lower energies, consistent with a quasielastic response.

In the present study, we investigated the low-energy spin dynamics of Ce$_3$Pd$_{20}$Si$_6$ using single-crystal INS spectroscopy. For TAS measurements, we used one large single crystal with a mass of 1.89~g, whereas for the TOF experiment, it was coaligned with an additional larger crystal, resulting in the total sample mass of $\sim$\,5.9\,g. Both crystals were characterized by resistivity measurements, indicating a sharp magnetic transition at $T_{\text N}=0.23$\,K. The crystals were mounted on a copper sample holder in the $(HHL)$ scattering plane to allow access to all high-symmetry directions of the cubic Brillouin zone (BZ). The mosaic spread of the sample, determined from the full width at half maximum (FWHM) of the rocking curves measured on structural Bragg reflections during sample alignment, was better than 0.5$^\circ\!$. For the low-temperature TAS measurements, the sample was first mounted in a $^3$He/$^4$He dilution refrigerator{\footnotemark}.
\footnotetext{In this configuration, the sample was fixed to a thermometer block with no temperature sensor attached directly to the sample. Consequently, due to the absence of an exchange gas, the real sample temperature could be somewhat higher than the set-point value. From the absence of magnetic Bragg scattering \cite{Lorenzer12} we conclude that the actual sample temperature was above $T_\text{N}$, hence the base-temperature data are representative of the paramagnetic state. We additionally confirmed this by checking that no detectable change in the inelastic-scattering intensity could be seen upon increasing the temperature to 0.9\,K.}
Afterwards, for $T$-dependent measurements, the sample was remounted in a conventional $^4$He closed-cycle refrigerator with an exchange gas. The TAS measurements were taken using the 4F2 cold-neutron spectrometer at the Laboratoire L\'eon Brillouin (LLB), Saclay, France, operated with a fixed final neutron wave vector $k_\text{f}=1.3$\,\AA$^{-1}$ that corresponds to an energy resolution of 0.11\,meV, defined as the FWHM of the elastic line.

\begin{figure}[t]\vspace{-10pt}
\includegraphics[width=\textwidth]{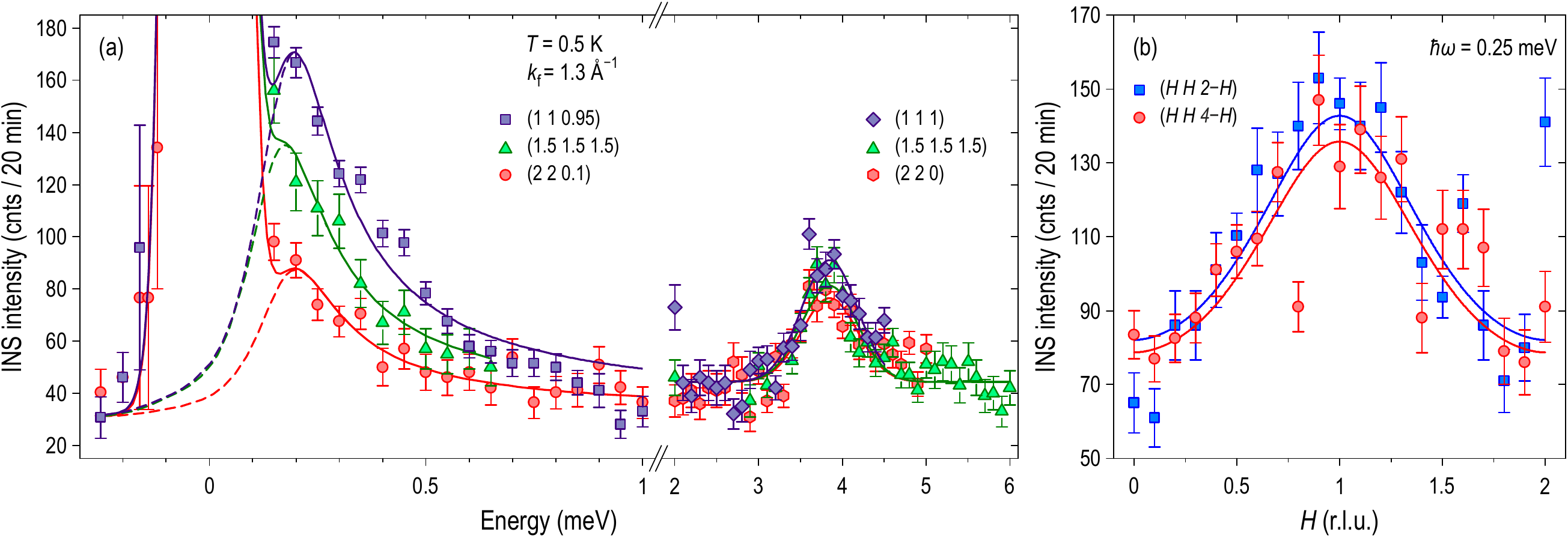}\vspace{-5pt}
\caption{(a)~Energy dependence of the magnetic scattering at base temperature ($T\approx0.5$\,K), measured at several wave vectors, as indicated in the legend. The left side of the panel shows the $\mathbf{Q}$-dependent QEMS response, fitted to the quasielastic Lorentzian line shape as given by eq.\,(\ref{Eq:Quasielastic}), whereas the right-hand-side presents data measured across the $\mathbf{Q}$-independent CEF line centered at 3.9\,meV, fitted to Gaussian profiles. The dashed lines show the magnetic signal without the elastic incoherent scattering contribution. (b)~Momentum dependence of the signal at $\hslash\omega=0.25$\,meV along two equivalent trajectories: ($H$\,$H$\,2\,--\,$H$) and ($H$\,$H$\,4\,--\,$H$), fitted with Gaussian profiles. Figure from ref.\,\ref{PortnichenkoCameron15}\vspace{-25pt}.}
\label{Fig:Qdep}\vspace{-0pt}
\end{figure}
The unprocessed low-temperature energy scans at several representative wave vectors are shown in fig.\,\ref{Fig:Qdep}\,(a). At low energies, below 1\,meV, we observe a QEMS signal that can be described by a quasielastic Lorentzian line shape, given by eq.\,\ref{Eq:Quasielastic}. This signal has nearly identical shape in energy at different wave vectors, but its intensity varies strongly in $\mathbf{Q}$-space, as evidenced by a twofold difference in the Lorentzian amplitude between the $(111)$ and $(220)$ points. Note that the measurements were done at slightly incommensurate wave vectors to avoid the contamination from phonons and the Bragg tail.

At higher energies, one can see a CEF line centered at 3.9\,meV, already known from previous INS measurements on powder samples \cite{PaschenLaumann08, DeenStrydom10}. As expected, it exhibits no dispersion and its intensity is nearly constant in momentum space apart from a minor form-factor suppression towards higher $|\mathbf{Q}|$. This qualitatively different behavior distinguishes it from the QEMS signal, characterized by a strongly $\mathbf{Q}$-dependent dynamical structure factor. At the same time, the low-energy CEF line at $\sim$\,0.31~meV suggested in ref.\,\citenum{DeenStrydom10} from data with a lower energy resolution can be clearly excluded by our present measurements.

Two constant-energy scans measured at 0.25\,meV along equivalent Brillouin-zone diagonals, as presented in fig.\,\ref{Fig:Qdep}\,(b), demonstrate a broad $\mathbf{Q}$-space distribution of the quasielastic intensity, peaked at the $(111)$ wave vector. From here on we show momentum in reciprocal lattice units ($1\,\text{r.l.u.}~= 2\piup/a$ with $a=12.28$\,\AA). The peak width of $\,\sim\kern.5pt$1\,r.l.u. is suggestive of short-range dynamical AFM correlations over distances of the order of one lattice constant, or two interatomic distances of the Ce\,8c sublattice. The momentum scans along ($H$\,$H$\,2\,--\,$H$) and ($H$\,$H$\,4\,--\,$H$) are essentially identical apart from a small form-factor suppression of intensity at higher $|\mathbf{Q}|$, which confirms that the periodicity of the signal in momentum space matches with that of the Ce\,8c sublattice, i.e. a translation by a vector with all even Miller indices results in an equivalent $\mathbf{Q}$ vector. On the other hand, the $(111)$ and $(002)$ points that are expected to be equivalent for the Ce~4a sublattice show different intensity, indicating that the QEMS response breaks the symmetry of the face-centered cubic BZ and should therefore originate predominantly from magnetic correlations on the Ce\,8c sites. This situation is reminiscent of that in iron pnictides, where spin fluctuations inherit the symmetry of the unfolded BZ because of the higher symmetry of the magnetic Fe sublattice with respect to the crystal itself \cite{ParkInosov10}.

\begin{figure}[!t]\vspace{-10pt}
\includegraphics[width=\textwidth]{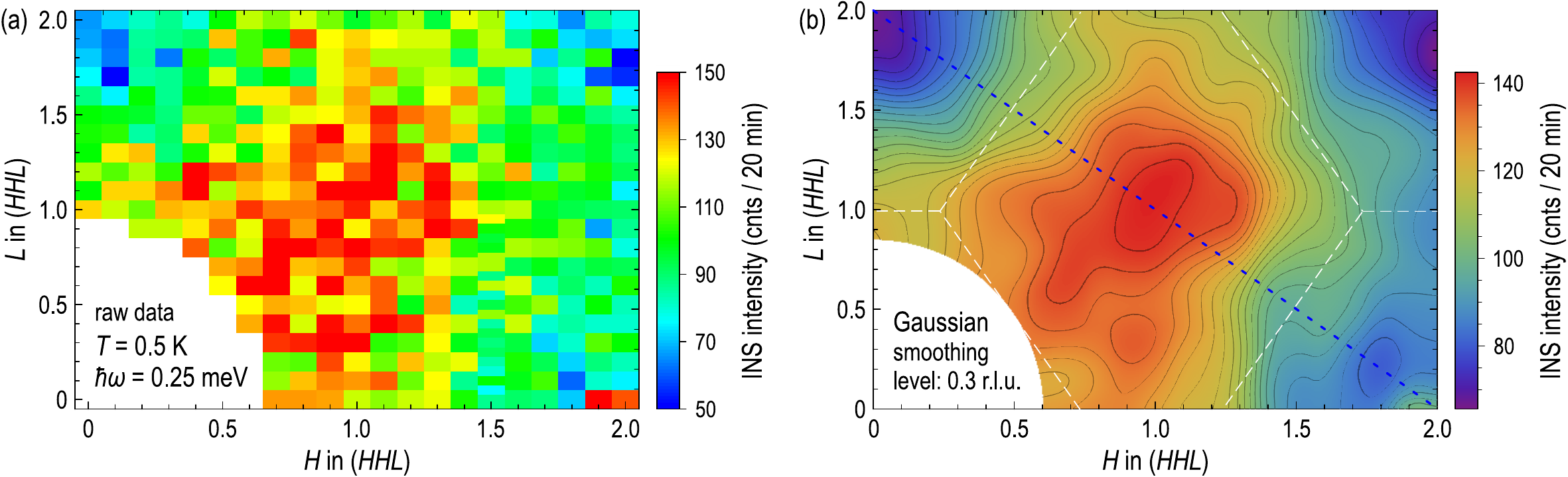}\vspace{-5pt}
\caption{Momentum dependence of the low-$T$ QEMS intensity in the $(HHL)$ scattering plane, measured at an energy transfer of 0.25\,meV: (a)~raw INS data; (b)~a contour map showing the same data smoothed with a two-dimensional Gaussian filter characterized by the FWHM of 0.3\,r.l.u. White dashed lines mark BZ boundaries of the face-centered cubic Ce$_3$Pd$_{20}$Si$_6$ lattice. Blue dashed line shows the  scan trajectory from fig.\,\ref{Fig:Qdep}(b).  Figure from ref.\,\ref{PortnichenkoCameron15}}\vspace{-15pt}
\label{Fig:ColorMaps}\vspace{-0pt}
\end{figure}
A more complete picture of the quasielastic intensity distribution in $\mathbf{Q}$-space is given by fig.\,\ref{Fig:ColorMaps}, showing a constant-energy map of the low-temperature QEMS response at 0.25\,meV over the entire $(HHL)$ scattering plane. From the fact that the quasielastic line shape remains essentially unchanged with $\mathbf{Q}$, as follows from fig.\,\ref{Fig:Qdep}, we can conclude that such an intensity map is also representative of the total energy-integrated spectral weight distribution in momentum space. It shows a broad anisotropic hump of intensity centered at $(111)$, with weaker side lobes extending along the $(110)$ and $(001)$ directions. In the large BZ corresponding to the Ce\,8c sublattice, this wave vector would coincide with the zone corner ($R$ point), matching with the AFQ propagation vector of CeB$_6$, where a similarly broad local maximum of the QEMS intensity was also found above $T_\text{N}$ \cite{FriemelLi12}. Yet, the lowest intensity in our dataset is observed in the vicinity of the $(002)$, $(220)$ and $(222)$ wave vectors, i.e. near the center of the large BZ. This is in remarkable contrast to CeB$_6$, which hosts strong ferromagnetic fluctuations at these points \cite{JangFriemel14}.

\begin{wrapfigure}[15]{r}{0.5\textwidth}\vspace{-15pt}
\includegraphics[width=0.5\textwidth]{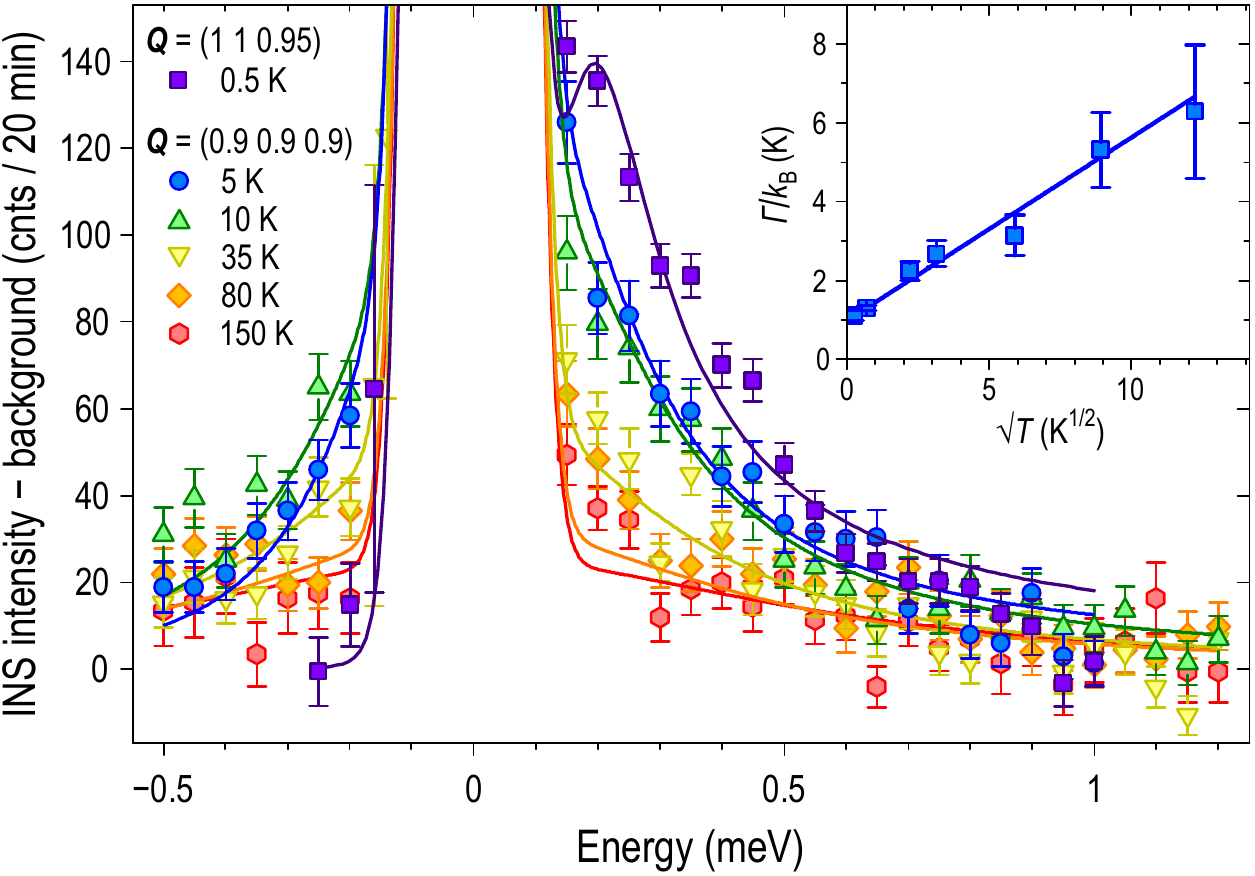}\vspace{-0pt}
\caption{Temperature evolution of the background-subtracted QEMS intensity, fitted to the sum of a quasielastic Lorentzian (eq.\,\ref{Eq:Quasielastic}) and the incoherent elastic line. The inset shows the $T$-dependence of the normalized quasielastic line width, $\Gamma(T)/k_\text{B}$, plotted vs. $T^{1/2}$ to emphasize the $\Gamma_0+A\sqrt{T}$ dependence. Figure from ref.\,\ref{PortnichenkoCameron15}.\vspace{-3pt}}
\label{Fig:Tdep}\vspace{-4pt}
\end{wrapfigure}
Next, we consider the temperature dependence of the QEMS response near its maximum at the $(111)$ wave vector, as shown in fig.\,\ref{Fig:Tdep}. Here, all but the lowest-temperature datasets were measured in the closed-cycle $^4$He refrigerator. As this cryostat produced a different background from the one observed with the $^3$He/$^4$He dilution fridge, the data in fig.\,\ref{Fig:Tdep} are plotted after subtraction of the corresponding constant background levels (shared for all data measured under the same conditions). Upon warming, we can observe a monotonic suppression and broadening of the quasielastic signal. The temperature dependence of the quasielastic line width, $\Gamma(T)$, presented in the inset to fig.\,\ref{Fig:Tdep}, follows the conventional $T^{1/2}$ law \cite{Robinson00}, $\Gamma(T)/k_\text{B}=\Gamma_0/k_\text{B}+A\kern.5pt\sqrt{T}$. From the residual width at \mbox{$T=0$}, the characteristic neutron-deduced Kondo temperature, \mbox{$T_\text{K}=\Gamma_0/k_\text{B}=(0.97\pm0.07)$\,K}, can be inferred.

\begin{wrapfigure}[31]{l}{0.5\textwidth}\vspace{-5pt}
\includegraphics[width=0.5\textwidth]{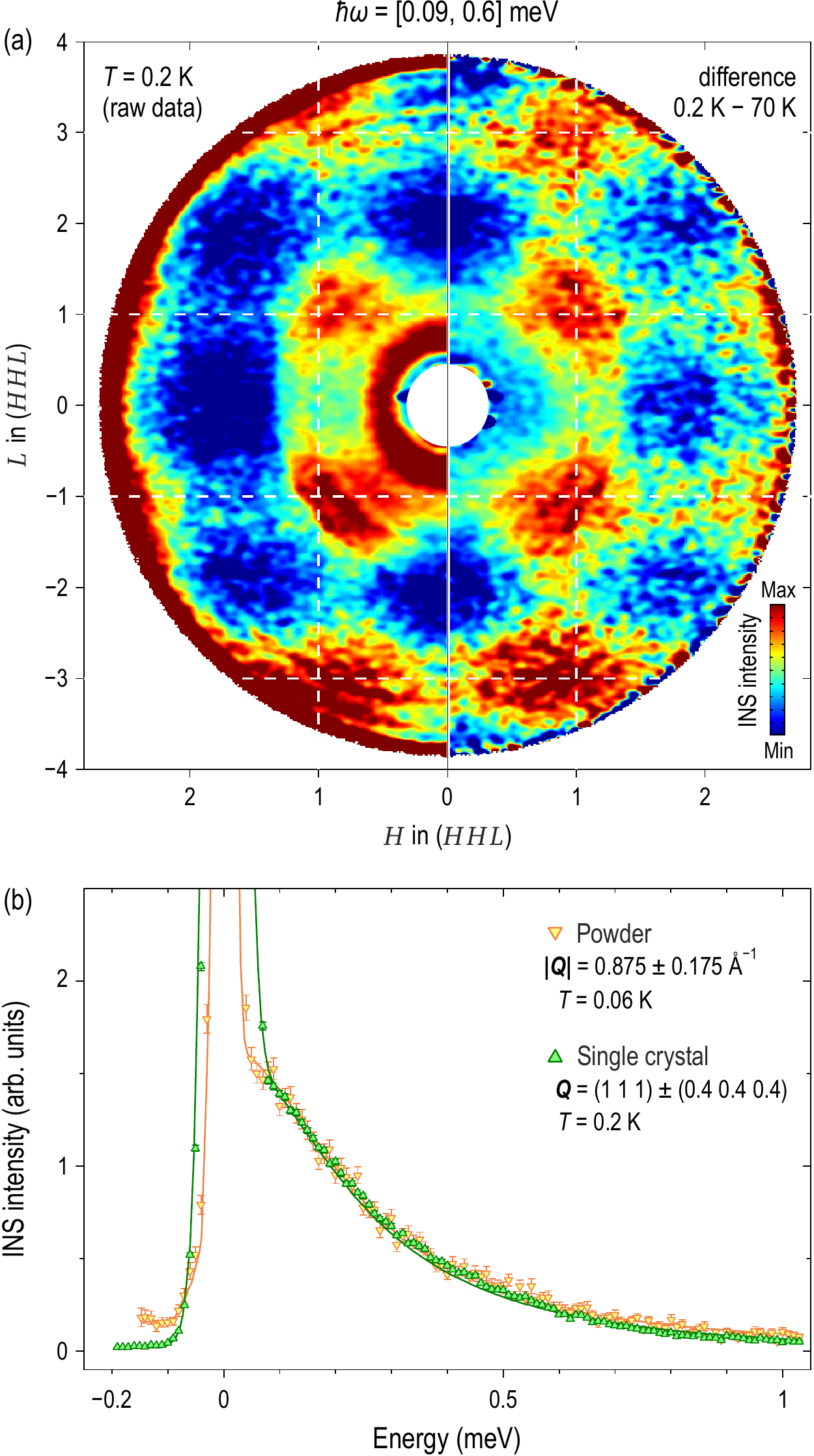}\vspace{-10pt}
\caption{(a)~Momentum dependence of the QEMS intensity in the $(HHL)$ scattering plane, obtained by integrating the TOF data in the energy window $0.09\,\text{meV} \leq \hslash\omega \leq 0.6\,\text{meV}$. The left panel shows raw data collected at $T\approx0.2$\,K, whereas the right panel presents the same dataset after subtraction of the high-temperature nonmagnetic background measured at $T=70$\,K. White dashed lines mark BZ boundaries corresponding to the simple cubic Ce\,8c sublattice. (b)~High-resolution energy profiles of the QEMS intensity from powder and single-crystal samples, integrated around the $(111)$ wave vector, as indicated in the legend. Figure from ref.\,\ref{PortnichenkoCameron15}.}
\label{Fig:TOF-Edep}\vspace{-0pt}
\end{wrapfigure}

To get a broader overview of the $\mathbf{Q}$-space and to ensure that no additional magnetic contributions are present at lower energy transfers, we also performed TOF measurements on a larger sample using the cold-neutron chopper spectrometer (CNCS) \cite{EhlersPodlesnyak11} at the Spallation Neutron Source (Oak Ridge National Laboratory, USA). The energy of the incident neutrons was set to $E_{\text i}=2.49$\,meV, providing an energy resolution with a FWHM of 0.051\,meV, that is twice better than in the TAS experiment. The dataset collected at $T=0.2$\,K (setpoint value) was complemented by a higher-temperature background measurement at 70\,K, where the low-energy magnetic spectral weight is considerably reduced, according to fig.\,\ref{Fig:Tdep}. The corresponding intensity maps of the $(HHL)$ plane before and after subtraction of the high-temperature background, integrated immediately above the elastic line in the energy range $0.09\,\text{meV} \leq \hslash\omega \leq 0.6\,\text{meV}$, are presented in fig.\,\ref{Fig:TOF-Edep}\,(a). In agreement with fig.\,\ref{Fig:ColorMaps}, we see a broad intensity maxima at wave vectors with all odd Miller indices, i.e. at the corners of the large BZ corresponding to the Ce\,8c sublattice. Down to the lowest accessible energy (0.09\,meV), we find neither any additional magnetic contributions that could be reconciled
by symmetry with the face-centered Ce\,4a sublattice, nor any ferromagnetic fluctuations like those found in CeB$_6$ at the zone center \cite{JangFriemel14}.

In fig.\,\ref{Fig:TOF-Edep}\,(b), we also compare energy profiles of the QEMS signal, obtained from our TOF data by integration within $\pm0.5$~r.l.u.\hspace{-0.6pt} on either side of the $(111)$ wave vector, to the corresponding energy dependence measured earlier on a powder sample \cite{DeenStrydom10} at the IRIS spectrometer (ISIS, UK) with an even better energy resolution of 0.025\,meV. The perfect agreement between the two curves confirms that the signal is sample-independent and only originates from the $(111)$ fluctuations. The exact form of the signal deviates from the perfect Lorentzian line shape, but could be reconciled with a generalized Voigt profile, shown as solid lines.

To summarize, our results demonstrated the presence of low-energy dynamical magnetic correlations in the paramagnetic state of Ce$_3$Pd$_{20}$Si$_6$, which could be responsible for the excess magnetic entropy in specific heat \cite{TakedaKitagawa95}. According to their $\mathbf{Q}$-space symmetry, they are associated with the same simple-cubic Ce\,8c sublattice that was shown earlier to host static AFM order below $T_{\text N}$ \cite{Lorenzer12}. This suggests that the remaining Ce\,4a ions are magnetically inactive, which could be either due to the frustration on the face-centered cubic sublattice, strong Kondo screening of their magnetic moments, or both. The possibly large difference in the Kondo scales on different sublattices would be in line with the theoretically suggested regime of competing Kondo effects \cite{BenlagraFritz11}. Despite the strikingly similar magnetic phase diagrams of Ce$_3$Pd$_{20}$Si$_6$ and CeB$_6$, both exhibiting an AFQ phase, their spin-fluctuation spectra are markedly different: Ferromagnetic correlations that dominate the spectrum of CeB$_6$ are absent in Ce$_3$Pd$_{20}$Si$_6$, while the dynamical AFM correlations in Ce$_3$Pd$_{20}$Si$_6$ are much more short-range and extend over distances of only about one lattice constant. Nevertheless, from the presence of strong quasielastic scattering at the BZ corner in both compounds, which coincides in CeB$_6$ with the propagation vector of the AFQ phase, we may tentatively surmise that the AFQ phase in Ce$_3$Pd$_{20}$Si$_6$ may also reside at the same wave vector in the large BZ, which is $\mathbf{q}_{\text{AFQ}}=(111)$.

\section{Incommensurate Short-Range Multipolar Order Parameter of Phase~II in Ce$_3$Pd$_{20}$Si$_6$}

Some heavy-fermion materials show so-called hidden-order phases, which are invisible to conventional diffraction techniques, and whose microscopic origin remained controversial for
decades. Such hidden-order phases have been observed in a variety of compounds containing $4f$ and $5f$ elements, for example, URu$_2$Si$_2$ \cite{MydoshOppeneer11, MydoshOppeneer14}, NpO$_2$ \cite{PaixaoDetlefs02}, skutterudites~\cite{YogiNiki09,AokiSanada08}, YbRu$_2$Ge$_2$ \cite{JeevanGeibel06}, or CeB$_6$ \cite{EffantinRossat-Mignod85, RossatMignod87, CameronFriemel16}. It is often assumed that the multipolar moments of the $f\!$ electrons in their specific crystal-field environment play a decisive role in the formation of these phases \cite{ThalmeierTakimoto08, KuramotoKusunose09, SantiniCarretta09}. The competition or coexistence of multipolar ordering (MPO) with more conventional magnetic order parameters, such as ferro- or antiferromagnetism, gives rise to complex magnetic-field\,--\,temperature phase diagrams in these compounds, often with multiple quantum critical points \cite{WengSmidman16, PaschenLarre14}, that provide a rich playground for experimental and theoretical investigations.

Up to now, the true order parameter and the propagation vector of phase~II in Ce$_{3}$Pd$_{20}$Si$_{6}$ have remained elusive and have not been observed directly by any diffraction method. Magnetization and specific-heat measurements revealed strong anisotropy of this phase with respect to the direction of applied magnetic field: For $\mathbf{B}\parallel[001]$ it is suppressed by as little as 2~T, for $\mathbf{B}\parallel[110]$ this critical field, $B_\text{II}$, dramatically increases to 10~T, whereas for $\mathbf{B}\parallel[111]$ this phase persists to even higher fields ($\sim$\,16\,--\,18~T by extrapolation) \cite{MitamuraTayama10, OnoNakano13}. Moreover, only for $\mathbf{B}\parallel[001]$ the system exhibits another transition to the phase~II$^\prime$ that is stabilized between 2 and 4~T (see fig.\,\ref{Fig:Ce3Pd20Si6PhaseDiagram}), whose microscopic origin remains unknown.

We employed elastic neutron scattering to reveal the order parameter of phase~II in Ce$_3$Pd$_{20}$Si$_6$ for the first time. All measurements were taken using the cold-neutron triple-axis spectrometers \textsc{4F2} (Laboratoire L\'eon Brillouin, Saclay, France) and \textsc{Thales} (Institut Laue-Langevin, Grenoble, France). The energy analysis was used to separate the weak elastic scattering signal from inelastic contributions. Both spectrometers were operated with the fixed final neutron wave vector $k_\text{f}$ set to 1.3\,\AA$^{-1}$ and a cold beryllium filter installed between the sample and the analyzer to suppress higher-order contamination from the monochromator. In all the reported experiments, we used the same sample as in ref.\,\citenum{PortnichenkoCameron15}, consisting of two coaligned single crystals with a total mass of $\sim$\,5.9\,g and a mosaic spread better than 0.5$^\circ\!$. The crystals were mounted on a copper sample holder in the $(HHL)$ scattering plane in a $^3$He/$^4$He dilution refrigerator inside a cryomagnet. All measurements with magnetic field along the $[1\overline{1}0]$ direction were done at \textsc{4F2} using a vertical-field 9~T magnet, whereas the measurements with magnetic field along $[001]$ were performed at \textsc{Thales} using a horizontal-field 3.8~T magnet available at ILL.

\begin{figure}[t]\vspace{-1em}
\includegraphics[width=\textwidth]{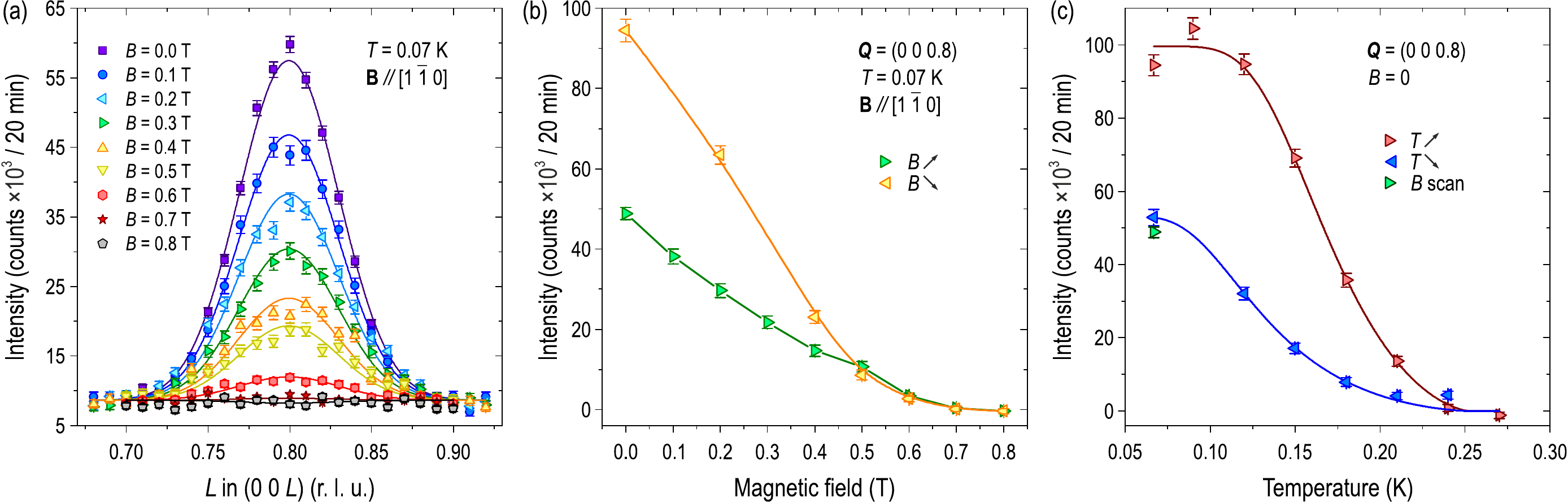}\vspace{-5pt}
\caption{Magnetic-field and temperature dependence of the incommensurate $(0~0~0.8)$ magnetic Bragg peak that represents the order parameter of phase~III. (a)~Unprocessed elastic neutron scattering data measured at $T=0.07$\,K for various fields applied along $[1\overline{1}0]$, fitted to a Gaussian model to extract the peak intensity. (b)~Field dependence of the background-subtracted magnetic Bragg peak amplitude for increasing~($\triangleright$) and decreasing~($\triangleleft$) field, $\mathbf{B}\parallel[1\overline{1}0]$. The observed hysteresis below 0.5~T is a result of domain selection. (c)~Temperature dependence of the same amplitude for increasing~($\triangleright$) and decreasing~($\triangleleft$) temperature that was measured in zero field after the application of a 0.8~T field. The point labeled ``$B$ scan'' is the first point of the ``$\kern.7pt\triangleright$'' dataset in panel (b), measured before the field was applied. All data points in panels (b) and (c) were obtained from Gaussian fits similar to those shown in panel~(a). Figure from ref.\,\ref{PortnichenkoPaschen16}.}\vspace{-15pt}
\label{Fig:Ce3Pd20Si6AFM110}
\end{figure}

\vspace{-5pt}\enlargethispage{5pt}
To verify that our sample is in the low-temperature phase~III, we first measured the $(0\,0\,0.8)$ magnetic Bragg reflection at the base temperature, $T=0.07$\,K, and investigated its magnetic-field and temperature dependencies, which are presented in fig.\,\ref{Fig:Ce3Pd20Si6AFM110}. According to the longitudinal $(0\,0\,L)$ scans in panel (a), the peak is centered at $L\approx1/5$, which is very close to the earlier result by Lorenzer~\textit{et~al.} \cite{Lorenzer12, DeenUnpublished}, who first observed this peak at an incommensurate wave vector with $L=0.792$. With the application of a magnetic field $\mathbf{B}\parallel[1\overline{1}0]$, the magnetic Bragg intensity is suppressed as shown in fig.\,\ref{Fig:Ce3Pd20Si6AFM110}\,(b). The full suppression is observed at $B_\text{III}=0.7$\,T, which coincides with the transition to phase~II. Before that, a domain-selection transition occurs at $B_\text{ds}=0.5$\,T, evidenced by a hysteresis of magnetic intensity measured in increasing and decreasing field. We note that the magnetic domains with the propagation vector $\mathbf{q}_\text{III}\parallel(001)\perp\mathbf{B}$ are favored at the expense of two other domains with $\mathbf{q}_\text{III}\parallel(100)$ and $\mathbf{q}_\text{III}\parallel(010)$ that both form a 45$^\circ$ angle to the field direction, which results in a nearly twofold increase of the $(0~0~0.8)$ Bragg intensity after field cycling. This situation is qualitatively different to the domain selection in the antiferromagnetic (AFM) phase of CeB$_6$, where the application of magnetic field along $[1\overline{1}0]$ suppresses Bragg intensity in the horizontal scattering plane and favors out-of-plane magnetic domains~\cite{EffantinBurlet82, CameronFriemel16}. As a function of temperature, the magnetic Bragg intensity, plotted in fig.\,\ref{Fig:Ce3Pd20Si6AFM110}\,(c), follows an order-parameter-like behavior with an onset temperature $T_\text{N}\approx0.23$\,K, in good agreement with transport and thermodynamic measurements \cite{ProkofievCusters09}.

\begin{figure}[t]\vspace{-1em}
\includegraphics[width=\textwidth]{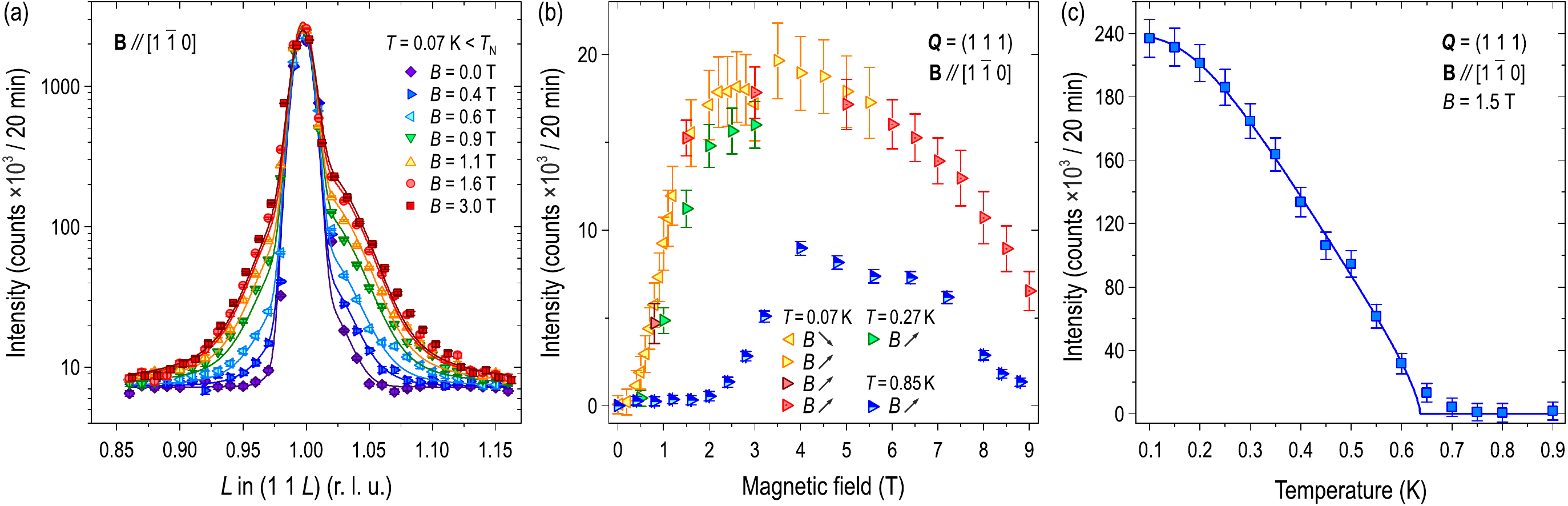}\vspace{-5pt}
\caption{Magnetic-field and temperature dependence of the diffuse magnetic signal that we associate with the order parameter of phase~II, surrounding the $(111)$ structural Bragg reflection. (a)~Typical unprocessed elastic neutron scattering scans measured at $T=0.07$\,K for various fields applied along $[1\overline{1}0]$, fitted with a sum of a Gaussian and Voigt profiles to account for the structural and magnetic contributions on top of a constant background (note the logarithmic intensity scale). (b)~Field dependence of the background-subtracted diffuse magnetic intensity for increasing ($\triangleright$) and decreasing ($\triangleleft$) field, $\mathbf{B}\parallel[1\overline{1}0]$, at three different temperatures. (c)~Temperature dependence of the same intensity, measured in a constant field of 1.5~T. All data points in panels (b) and (c) were obtained from fits similar to those shown in panel (a). Figure from ref.\,\ref{PortnichenkoPaschen16}.}\vspace{-15pt}
\label{Fig:Ce3Pd20Si6AFQ110}
\end{figure}

\enlargethispage{5pt}
Now we turn to the discussion of our main result: the observation of an elastic magnetic signal in the vicinity of the $(111)$ wave vector. Because of the peculiar crystal structure of the $R_{3}\text{Pd}_{20} X_{6}$ compounds with interpenetrating simple-cubic and fcc sublattices, this wave vector simultaneously corresponds to the zone corner ($R$ point) for the simple-cubic sublattice and to the zone center ($\Gamma$ point) for the fcc sublattice. As we noted earlier \cite{PortnichenkoCameron15}, the propagation vector of a G-type antiferroquadrupolar (AFQ) order analogous to that of CeB$_6$, residing on the 8$c$~Ce sublattice, would thus overlap with the $(111)$ structural reflection. The corresponding weak magnetic Bragg peak \cite{FriemelLi12} would be thus exceedingly difficult to detect. In fig.\,\ref{Fig:Ce3Pd20Si6AFQ110}\,(a) we present the magnetic-field dependence of the elastic scattering intensity along the $(11L)$ direction, plotted on the logarithmic intensity scale. In zero magnetic field, only the sharp $(111)$ structural Bragg reflection is observed, whereas magnetic field induces an additional diffuse contribution seen as a much broader peak of nearly field-independent width. This magnetic peak reaches its maximal intensity around 3\,--\,4~T and then starts to decrease again, as shown in fig.\,\ref{Fig:Ce3Pd20Si6AFQ110}\,(b). The signal persists both below and above $T_\text{N}$, yet in a narrower field range towards higher temperatures. A comparison with the phase diagram in fig.\,\ref{Fig:Ce3Pd20Si6PhaseDiagram} clearly establishes that it corresponds to the stability range of phase~II, which also gets narrower upon warming. The temperature dependence of the diffuse intensity, measured in the field of 1.5~T, is shown in fig.\,\ref{Fig:Ce3Pd20Si6AFQ110}\,(c). It shows an order-parameter-like suppression, evidencing a phase transition at $T_\text{Q}(1.5\,\text{T})\approx0.65$\,K, as expected for phase~II in this field.

\vspace{-5pt}
We note that the energy analysis allows us to suppress most of  the diffuse elastic contribution from the much broader quasielastic fluctuations reported earlier in ref.\,\citenum{PortnichenkoCameron15}, which were centered at the same wave vector. This ensures that the observed signal corresponds to truly static magnetic correlations that represent the order parameter of phase~II. Its broad width in momentum (as compared to the structural Bragg reflection) indicates that this is a short-range order with a correlation length of about 120\,\AA\ or $\sim$10 lattice constants. The elastic intensity that is absent in zero field and then starts increasing with field after entering phase~II is analogous to the behavior of the AFQ Bragg intensity in CeB$_6$ \cite{RossatMignod87, CameronFriemel16}. However, because of the much lower value of the critical field, $B_\text{II}$, in Ce$_{3}$Pd$_{20}$Si$_{6}$ the signal saturates and exhibits a maximum at moderate field values ($\sim$\,3\,--\,5~T in fig.\,\ref{Fig:Ce3Pd20Si6AFQ110}\,(b) for $\mathbf{B}\parallel[110]$), which in the case of CeB$_6$ would be shifted to much higher fields. This distinctive field dependence results from the field-induced dipolar moments modulated by the underlying orbital order \cite{EffantinRossat-Mignod85, TakigawaYasuoka83, CustersLorenzer12}\,---\,a renowned signature of the AFQ state. Hence, our results demonstrate that the order parameter of phase~II in Ce$_3$Pd$_{20}$Si$_6$ represents a short-range version of the AFQ ordering with the same propagation vector and possibly a similar structure as in CeB$_6$, residing on the simple-cubic 8$c$~Ce sublattice.

\begin{wrapfigure}[12]{l}{0.5\textwidth}\vspace{-15pt}
\includegraphics[width=0.5\textwidth]{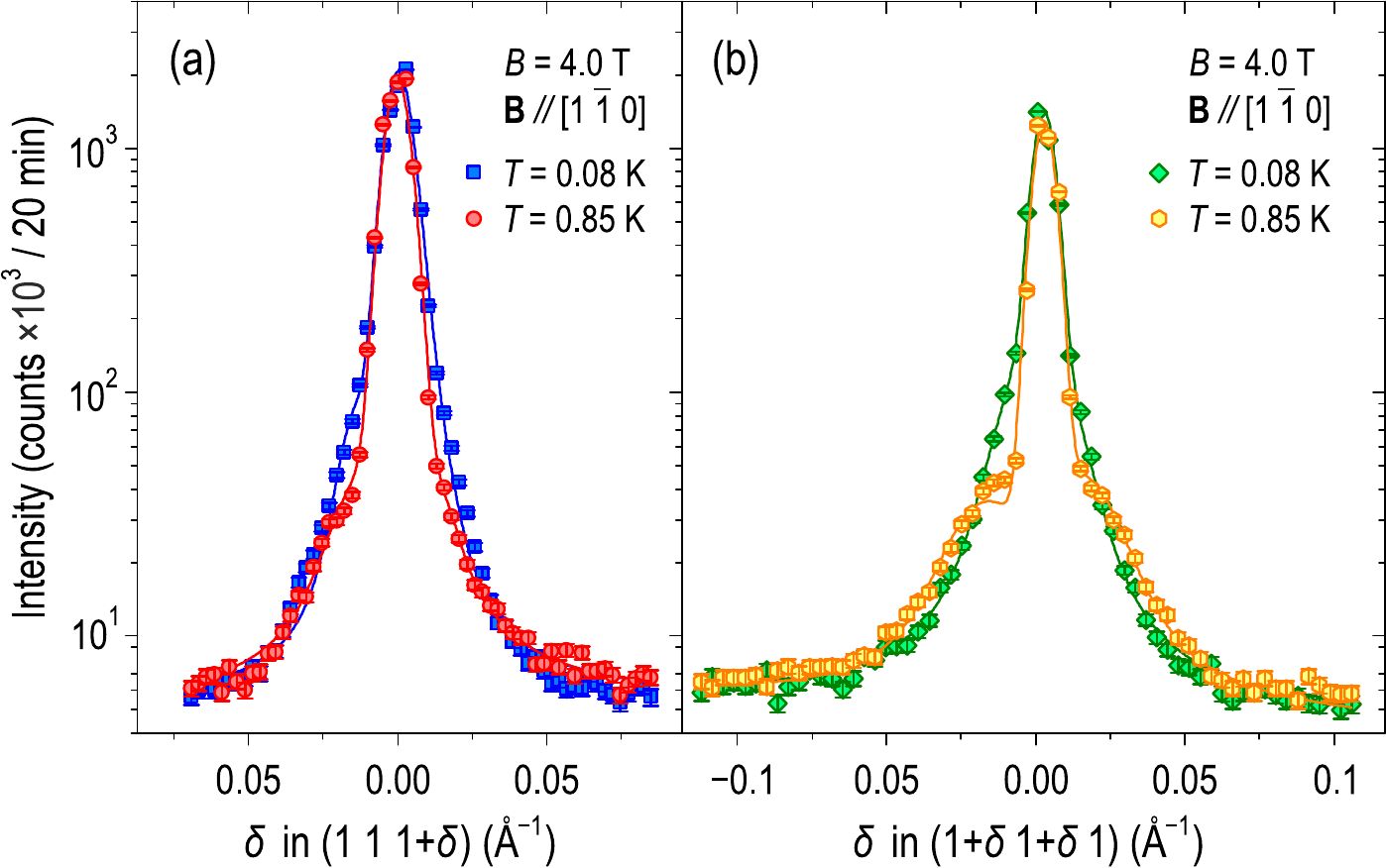}\vspace{-5pt}
\caption{Variation of the diffuse peak shape with temperature: (a)~along the \mbox{$(1~1~1\!+\!\delta)$} direction; (b)~along the \mbox{$(1\!+\!\delta~1\!+\!\delta~1)$} direction. The magnetic field of 4\,T was applied vertically along $[1\overline{1}0]$, perpendicular to both scan directions. Figure from ref.\,[\ref{PortnichenkoPaschen16}].\vspace{-0pt}}
\label{Fig:Ce3Pd20Si6AFQTdep}
\end{wrapfigure}
A closer inspection of the diffuse peak shape reveals its variation with temperature, which is illustrated in fig.\,\ref{Fig:Ce3Pd20Si6AFQTdep}. The measurements are done in a constant field of 4.0\,T, applied along the same $[1\overline{1}0]$ vertical direction perpendicular to the scattering plane. Along the \mbox{$(1~1~1\!+\!\delta)$} scan direction, shown in panel~(a), the magnetic intensity is suppressed near the center of the peak upon raising the temperature to 0.85\,K, immediately before the suppression of phase~II. The effect is even more pronounced along the \mbox{$(1\!+\!\delta~1\!+\!\delta~1)$} scan direction, shown in panel~(b), where we observe a considerable broadening of the peak manifested in the transfer of the magnetic spectral weight away from the central reflection (note the enhanced intensity in the ``tails'' of the peak). These data suggest a tendency towards the formation of an incommensurate modulated quadrupolar structure, as we demonstrate in the following with a dedicated experiment using an orthogonal field orientation.

\begin{sidewaysfigure}[ht]
\centering
\includegraphics[width=0.90\textwidth]{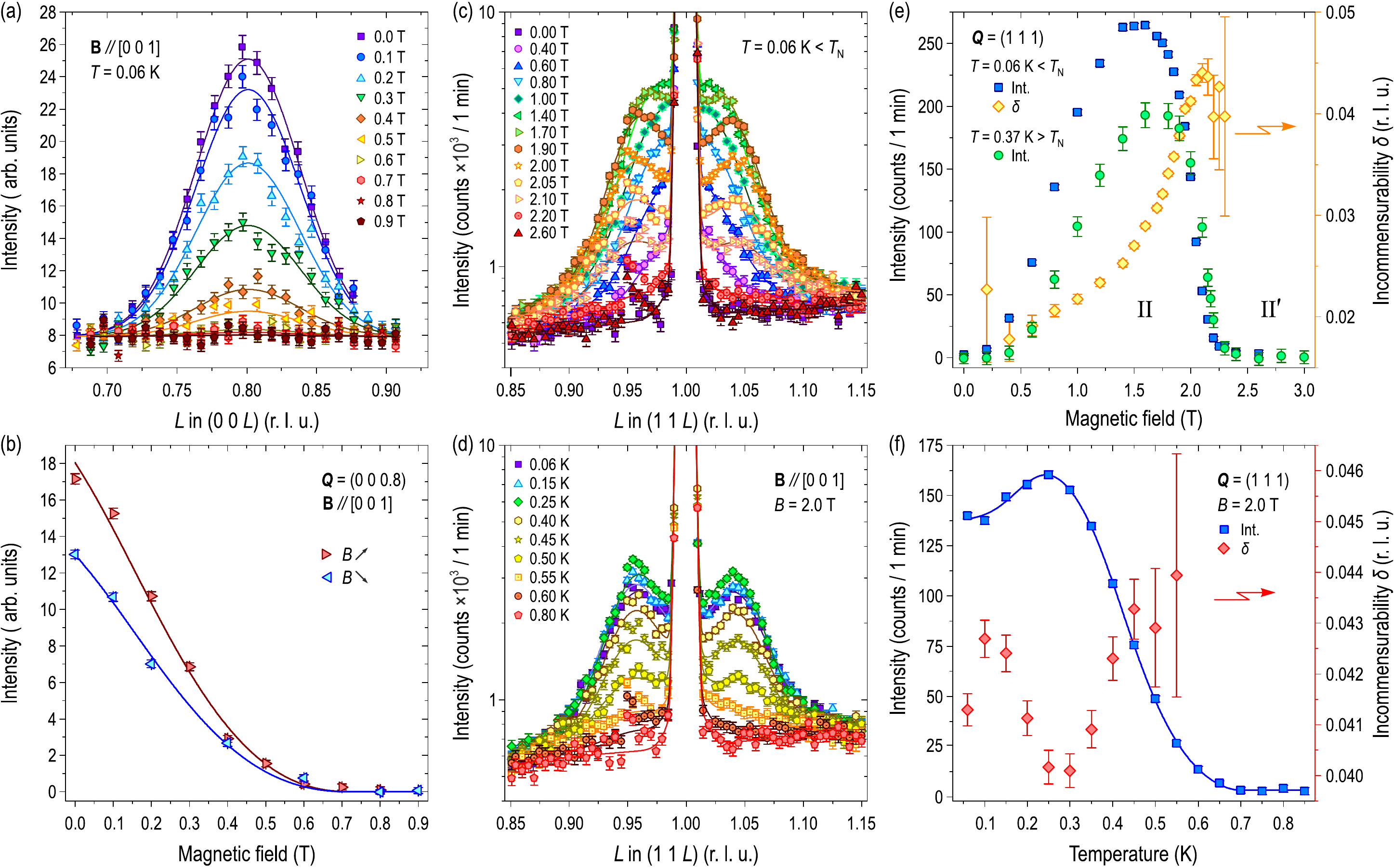}\vspace{-0pt}
\caption{(a)~Magnetic-field dependence of the $(0~0~0.8)$ magnetic Bragg peak for $\mathbf{B}\parallel[001]$, measured at $T=0.06$\,K. (b)~The corresponding field dependence of the Bragg peak amplitude for increasing~($\triangleright$) and decreasing~($\triangleleft$) field, extracted from the Gaussian fits. (c)~Unprocessed elastic-scattering scans along $(11L)$, measured at $T=0.06$\,K for various fields applied along $[001]$. (d)~Temperature dependence of the elastic magnetic intensity at a constant field of 2.0\,T, where the incommensurability of the signal is maximized. (e)~Magnetic-field dependence of the diffuse magnetic intensity at $T=0.06\,\text{K}<T_{\mathrm N}$ and $0.37\,\text{K}>T_{\mathrm N}$ (squares and circles, left axis) and of the incommensurability parameter $\delta$ (diamond symbols, right axis). (f)~Temperature dependence of the diffuse magnetic intensity (squares, left axis) and of the incommensurability parameter $\delta$ (diamond symbols, right axis) at a constant magnetic field of 2.0\,T. Figure from ref.\,\ref{PortnichenkoPaschen16}.}
\label{Fig:Ce3Pd20Si6AFMAFQ001}\vspace{-0pt}
\end{sidewaysfigure}

\vspace{-2.5pt}
Next, we present similar measurements performed in a horizontal magnetic field, $\mathbf{B}\parallel[001]$. This field orientation corresponds to the lowest critical field for phase~II, $B_\text{II}^{[001]}\approx2$\,T, and in addition stabilizes the enigmatic phase~II$^\prime$ that is not found for any other high-symmetry field direction. Hence, a natural question that served as an initial motivation for these measurements is whether the diffuse magnetic signal near $(111)$ persists into phase~II$^\prime$ or gets suppressed already at $B_\text{II}^{[001]}$. Again, we start our presentation with the analysis of the magnetic Bragg peak at $\mathbf{q}_\text{III}=(0~0~0.8)$, which is shown in fig.\,\ref{Fig:Ce3Pd20Si6AFMAFQ001}\,(a,b). In contrast to phase~II, the low-temperature phase~III is much more isotropic and nearly insensitive to the direction of the magnetic field. This can be seen from fig.\,\ref{Fig:Ce3Pd20Si6AFMAFQ001}\,(b), where the full suppression of magnetic intensity occurs at the same field of 0.7~T as for $\mathbf{B}\parallel[110]$. However, in contrast to fig.\,\ref{Fig:Ce3Pd20Si6AFM110}\,(c), the peak intensity in this configuration decreases after cycling the field, which can be explained by a partial suppression of the unfavorable magnetic domain with $\mathbf{q}_\text{III}\parallel(001)\parallel\mathbf{B}$ as a result of the domain selection.

\vspace{-2.5pt}
Figures \ref{Fig:Ce3Pd20Si6AFMAFQ001}\,(c) and (d) show the development of the diffuse magnetic signal near $(111)$ in the magnetic field $\mathbf{B}\parallel[001]$ at the base temperature of 0.06\,K, and its temperature dependence in constant magnetic field of 2.0\,T, respectively. Here we observe an increasingly incommensurate response with two broad magnetic satellites centered at $(1~1~1\!\pm\!\delta)$, which surround the structural $(111)$ reflection and move further away from the commensurate position as the magnetic field is increased. From the fitting results shown in fig.\,\ref{Fig:Ce3Pd20Si6AFMAFQ001}\,(e), it is evident that the magnetic intensity starts to increase upon entering phase~II, reaches a saturation around 1.5\,T, and then rapidly drops to zero across the transition to phase~II$^\prime$, proving that the order parameter of II$^\prime$ is qualitatively different from that of phase~II and is not represented by the observed diffuse signal. In fig.\,\ref{Fig:Ce3Pd20Si6AFMAFQ001}\,(e), we also present a similar field dependence of the intensity measured at an elevated temperature of 0.37\,K~$>$~$T_{\mathrm N}$, with a qualitatively similar behavior.

\vspace{-2.5pt}\enlargethispage{5pt}
The field dependence of the incommensurability parameter $\delta$, resulting from the fits of the diffuse magnetic intensity, is also shown in fig.\,\ref{Fig:Ce3Pd20Si6AFMAFQ001}\,(e) with diamond symbols. It demonstrates a clear monotonic increase and gets maximized near the transition between phases II and II$^\prime$, beyond which it can no longer be determined because of the vanishing peak intensity. We have extracted the incommensurability parameter by imposing a constraint on the correlation length at low magnetic fields, i.e. the width of the peak in momentum space at low magnetic fields was extrapolated from its values at higher fields. This assumption results in a finite incommensurability even in the low-field limit, which is however smaller than the peak width and therefore cannot be clearly resolved in the raw data. The low-field datasets in fig.\,\ref{Fig:Ce3Pd20Si6AFMAFQ001}\,(c) can be equally well described with a broader commensurate peak.

Further, we followed the temperature dependence of the incommensurate magnetic response at the field of 2.0\,T, where the incommensurability parameter is maximized. The corresponding data are shown in fig.\,\ref{Fig:Ce3Pd20Si6AFMAFQ001}\,(d), and the resulting fitting parameters are plotted in fig.\,\ref{Fig:Ce3Pd20Si6AFMAFQ001}\,(f) vs. temperature. We observe a non-monotonic behavior of the peak intensity with a local maximum around 0.25\,K, which is consistent with the upturn in the transition field $B_{\text{II\,--\,II}^\prime}(T)$ seen in the phase diagram (fig.\,\ref{Fig:Ce3Pd20Si6PhaseDiagram}, dotted line). The incommensurability parameter remains nearly constant as a function of temperature with only minor variations of the order of 10\%, as shown in fig.\,\ref{Fig:Ce3Pd20Si6AFMAFQ001}\,(f) with diamond symbols.

Our results demonstrate the existence of static short-range AFQ correlations propagating along $(111)$, which represent the order parameter of phase~II in Ce$_{3}$Pd$_{20}$Si$_{6}$, seen here for the first time directly in a scattering experiment. With the application of magnetic field, these correlations become increasingly incommensurate and finally vanish across the transitions to either phase~II$^\prime$ or phase~I (for $\mathbf{B}\parallel[001]$ and $\mathbf{B}\parallel[1\overline{1}0]$, respectively). Under the assumption that the field-induced dipolar magnetic correlations are modulated by the underlying orbital order \cite{CustersLorenzer12}, this implies the existence of a rather unusual incommensurate orbitally ordered state whose propagation vector can be continuously tuned by the external magnetic field.

We now discuss possible mechanisms which may lead to this field-tuned incommensurate multipolar order. First we recall that quadrupolar structures with incommensurate modulations have been previously observed, for instance, in PrPb$_3$ \cite{OnimaruSakakibara05, OnimaruKusunose16} and in the so-called ``phase IV'' (IC1) of the solid solution Ce$_{0.7}$Pr$_{0.3}$B$_6$ \cite{KishimotoKondo05, TanakaSera06}. Incommensurate octupolar order was also considered as a candidate for the hidden order parameter in URu$_2$Si$_2$ \cite{Hanzawa07}. In the case of Ce$_{1-x}$Pr$_x$B$_6$, it has been suggested that the incommensurability of the $O_{xy}$-type quadrupolar order results from the frustration imposed by the competition among the AFQ  and AFM exchange interactions in combination with RKKY interactions between the Ce and Pr multipoles, and thermal fluctuations are necessary to stabilize the incommensurate MPO phase. Compared to these cases, the present situation in Ce$_3$Pd$_{20}$Si$_6$ is very unusual in two respects:
\begin{enumerate}\vspace{-5pt}
\item[(i)] The incommensurability varies continuously with field, with no apparent lock-in of the wavevector as opposed to PrPb$_3$ \cite{OnimaruSakakibara05}.\vspace{-5pt}
\item[(ii)] The order is rather short-ranged despite the fact that the compound is stoichiometric, without obvious sources of strong quenched disorder.\vspace{-5pt}
\end{enumerate}

\clearpage
\clearpage

The rather localized nature of the Ce orbitals suggests that the incommensurate ordering wave vector has itinerant origin, determined by the Fermi-surface geometry. On the one hand, in the itinerant approach similar structures can be obtained as exotic types of density-wave phases, one prominent example being the incommensurate orbital antiferromagnetism associated with circulating orbital currents \cite{ChandraColeman02} or different kinds of multipolar density waves \cite{RauKee12, IkedaSuzuki12, Das12, ThalmeierTakimoto14}, which were proposed among other scenarios as possible explanations for the hidden-order state in URu$_2$Si$_2$ \cite{MydoshOppeneer11, MydoshOppeneer14}. On the other hand, an alternative scenario, which is more conceivable for our system with strongly localized $f$ orbitals, would involve long-ranged indirect RKKY-type interactions between multipolar moments, which are mediated by the heavy conduction electrons \cite{RossatMignod87, OnimaruSakakibara05, OnimaruKusunose16}. The experimentally established phase diagram also supports this scenario, as it compares remarkably well with earlier theoretical predictions derived from an effective pseudospin model for RKKY-coupled multipoles of the $\Gamma_8$ quartet at the $8c$~Ce site [see Supplementary Information of ref.\,\citenum{CustersLorenzer12}]. Using a microscopic study and a Ginzburg-Landau analysis, it was shown that, in a finite magnetic field, the AFQ order can induce dipolar AFM order with the same symmetry. Depending of the field direction, this dipolar order is either stable in the entire AFQ phase (generic field direction away from $[001]$) or only in part of it (field along $[001]$). This is in striking agreement with our experimental findings. A stabilization of $T_{\mathrm{Q}}$ with field, as observed in experiments, is expected for the quadrupolar moment $O^0_2$, which induces a dipolar moment $J^z$ for field along $[001]$. On the other hand, the order in phase II$^\prime$, which remains elusive in the present study, could be of $O_{xy}$ type as theoretically suggested. This hypothesis remains to be clarified by future experiments.

While incommensurability was not considered initially in the framework of this theoretical model, it is possible that the momentum-space structure of the RKKY interaction, as expressed by the Lindhard function, displays a rather weak momentum dependence near its maximum: Such a situation, arising from a complex underlying band structure, would reflect itinerant frustration. A weak momentum dependence over a range of momenta implies that the position of the maximum can acquire sizeable shifts as function of an applied Zeeman field. Hence, we propose that the RKKY interaction displays a shallow peak at the ordering wave vector with small incommensurability $\delta$ in weak fields, and this peak is continuously shifted to larger $\delta$ with the application of an external magnetic field.

Given the discrete character of the orbital degrees of freedom, a plausible picture for a multipolar state with small incommensurability $\delta$ is that of antiphase domain walls of density $\propto 1/\delta$ in a commensurate background. A periodic arrangement of domain walls yields a sharp Bragg peak. However, these domain walls are naturally susceptible to pinning by defects, which would destroy long-range order and result in a state with short-ranged correlations. The susceptibility to quenched disorder is greatly enhanced by the postulated weak momentum dependence of the RKKY interaction, as this also implies a weak selection of an ordering wavevector. We propose this scenario as a possible explanation of the observed small correlation length.

In summary, we provided direct evidence for field-induced dipolar magnetic correlations in Ce$_3$Pd$_{20}$Si$_6$, experimentally confirming the previously suggested AFQ order parameter of the hidden-order phase II. We suggest that itinerant frustration, reflected in a particularly weak momentum dependence of the RKKY interaction near its maximum wave vector, is responsible for the experimental findings and can explain both the field-dependent incommensurability and the short-range nature of the multipolar order. To verify the scenario of itinerant frustration, detailed band-structure calculations for Ce$_3$Pd$_{20}$Si$_6$ would be required; those are not available to date. Alternatively, photoemission tomography might be used to experimentally determine the low-energy bands which can be used to parameterize the band structure and calculate the Lindhard function, as recently done for CeB$_6$~\cite{KoitzschHeming16}.
\clearpage
\appendix

\vspace{+25pt}
\chapter{Spurions}\label{Appendix:Spurions}

\vspace{+25pt}
\setlength{\epigraphwidth}{0.67\textwidth}
\epigraph{An expert is a person who has found out by his own painful experience all the mistakes that one can make in a very narrow field.}{\vspace{0.3em} Niels Bohr}
\vspace{+0pt}

This appendix contains supplementary material, which may be helpful in providing a more comprehensive understanding of the difficulties that a researcher may face while conducting a neutron scattering experiment or analyzing the measured data. Unfortunately, I noticed a tendency, that no one is really interested and motivated to mention the problems, that may arise during an experiment. Although in my opinion,  the knowledge of technical nuances can help significantly to interpret the results of the experiment.

It is extremely important for the experimentalist not to use the spectrometer as a ``blackbox'', since convenient programs have recently appeared for processing neutron data which seduce the user to get the result in just several clicks. It is simply impossible to estimate how much time such toolsets saved for me. However, no program is capable of determining the presence of the attenuator or a closed slit, which  incorrectly reacted  to the applied   magnetic field. The program is also incapable of determining the dark angles of the magnet and to check if they block the incident beam. Problems which seem trivial at a first glance, can sometimes lead to very unexpected results after averaging the data.  I am even afraid to discuss more complex questions, such as how to average the data measured at different sides of the multi-analyzer with respect to the incoming beam. There are lots of other questions which still puzzle me.

In one of the TOF experiments I have spent many hours  in an attempt to determine that the rotation of the sample occurs in the wrong direction. The problem appeared as a complete mess of the structural Bragg peak positions. After all, in the case of the TOF experiment it is not so important in which direction the sample rotates. Thus, without exact knowledge where in reciprocal  space  measurements were done, I continued to collect data over a broad A3 range. My mistake was that the actual sample rotation, and the one defined in the script were confused. However, I hardly imagine how much time I would have spent if I had focused my attention on a small region where I expected the magnetic signal, without doing an overview A3 scan, that immediately showed an incorrect position of the structural Bragg reflections. Ironically, my colleague's next experiment was associated with same difficulties. The opportunity to learn about similar failures from other people within neutron scattering community would be very useful to me.

Below I will try to give a detailed description of every  spurious signal that appeared as an artifact  during some of the experiments in which I participated. As it was already shown by other experimentalists, critical and unbiased attitude to the enigmatic peaks, which were found in a variety  of inelastic neutron scattering experiments, allows us to explain their nature. A mysterious 20\,meV signal, observed in TOF neutron data on pnictides \cite{ZbiriSchober09,MittalSu08}, was found to be due to multiple scattering events involving the walls of the sample environment \cite{PintschoviusReznik14}. Thus, the examples from my experience should not by no means be interpreted as  criticism{\footnotemark}
    \footnotetext{According to ``Pauli-Effect'', there is a higher probability  of technical equipment to encounter critical failure in the presence of certain people. Another facetious interpretation sounds like: ``a functioning device and Wolfgang Pauli may not occupy the same room''.}
of specific spectrometers, but as a guide for experimentalists to distinguish real signals from background and spurious artefacts. ``If you don't look for a mistake, you can't find one''.{\footnotemark}
    \footnotetext{Television series \textit{Scrubs}, season 1, episode 22.}

\vspace{-10pt}\enlargethispage{5pt}
\section{Aluminium Powder Lines}

The signal in INS experiments can be contaminated by artifacts originating from the sample environment, higher-order neutrons from the monochromator, or accidental Bragg scattering. Sample holders and cryogenic sample equipment are typically made of polycrystalline aluminum \cite{ShiraneShapiro02} due to its low absorption and incoherent scattering cross-sections, which helps to minimize the background. However, the coherent cross-section according to table\,\ref{Tab:SummaryCohIncAbs} is not negligible. As a result, contamination from the powder spheres, that originate from the strongest $(1 1 1)$, $(2 0 0)$, $(2 2 0)$, and $(1 1 3)$ Bragg reflections, allowed for the face-centered cubic  crystal structure of Al, might appear.

\begin{wrapfigure}[14]{r}{0.55\textwidth}\vspace{-10pt}
\includegraphics[width=0.55\textwidth]{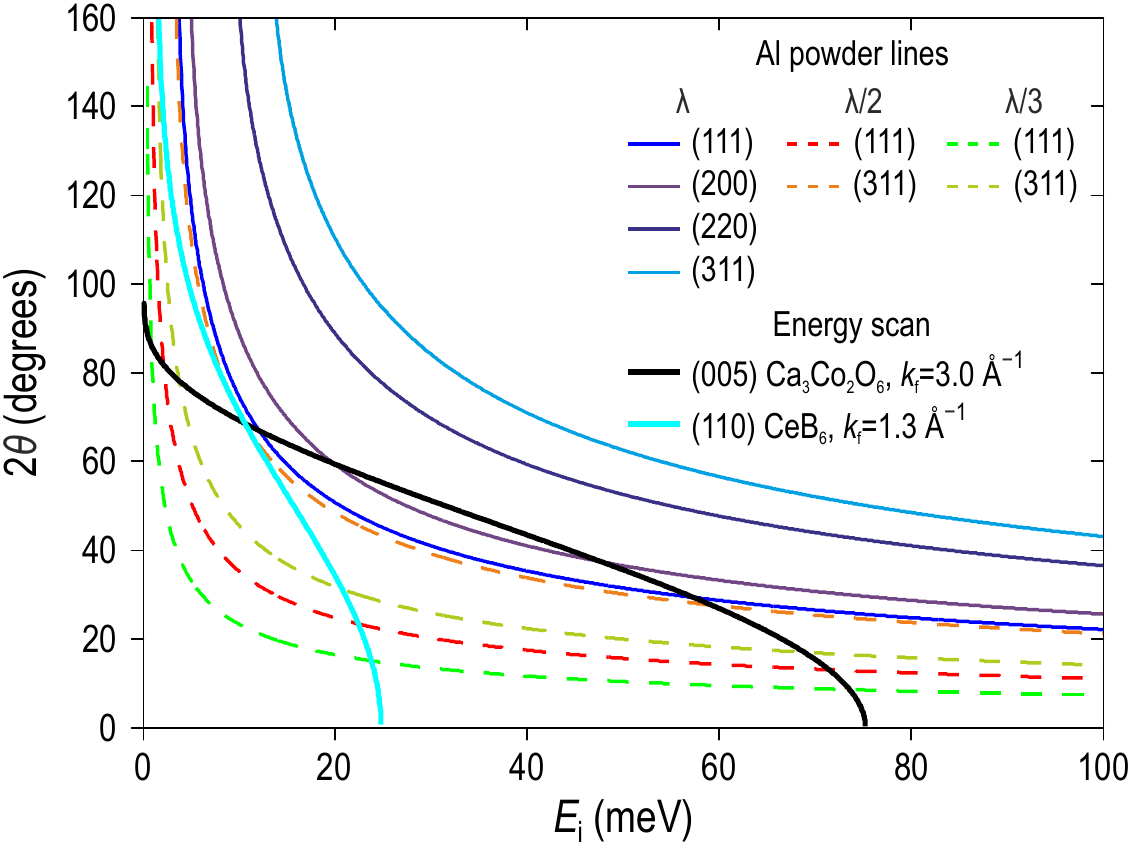}\vspace{-0pt}
\caption{Scattering angles of the strongest aluminium Bragg peaks as a function of a neutron energy. Black and cyan solid lines show a typical energy scans at a constant $|\mathbf{Q}|$ with different  values of $k_{\text{f}}$, as was mentioned in sec.\,\ref{SubSec:TAS}.}
\label{Fig:PorderLinesA4Ei}\vspace{-0pt}
\end{wrapfigure}

Let us calculate at which scattering angle aluminium powder lines will appear. According to Bragg's law, $2d\text{sin}\theta=n\lambda$, where $\lambda\,[\text{\AA}]=9.044/\sqrt{E\,[\text{meV}]}$ is the neutron wavelength, and $d=a/\sqrt{h^2+k^2+l^2}$ is an interplanar distance  with the lattice parameter $a=4.0862\,\text{\AA}$.  Besides, scattering will also occur for neutrons with the  wavelengths  $\lambda/2$ and $\lambda/3$. The corresponding scattering angle $2\theta$ as a function of incident neutron energy is shown{\footnotemark} in fig.\,\ref{Fig:PorderLinesA4Ei}. So finally we got a trivial result: when doing a neutron scattering experiment one should be careful and try to avoid $2\theta$ and $E_{\text{i}}$ values, which fulfill Bragg's law for Al. \footnotetext{$(200)$ and $(220)$ Bragg reflections also present for $\lambda/2$ and $\lambda/3$ wavelength. They are not shown to keep the figure understandable and not too messy.}

\enlargethispage{0pt}
A typical energy scan with a fixed-$k_{\text{f}}$, which is often performed during the experiment, also has a certain dependence between the scattering angle and neutron energy (see sec.\,\ref{SubSec:TAS} for details). As one can see, the energy scan around $(005)$ for Ca$_{3}$Co$_{2}$O$_{6}$ with a $k_{\text{f}}=3\,\text{\AA}^{-1}$ intersects with several Al powder lines, while the energy scan with shorter $k_{\text{f}}$ in case of CeB$_{6}$ intersects only with powder lines that originate from higher order neutrons. At a first glance it is not clear how  the signal from the aluminum lines can be observed in the experiment. In the first case, the scan is done with $E_{\text{f}}=18.65\,\text{meV}$. Powder lines $(111)$ and $(200)$ intersect with an energy scan when $E_{\text{i}}$ equals $12.07\,\text{meV}$, $19.92\,\text{meV}$, $47.81\,\text{meV}$ or $57.37\,\text{meV}$. None of these values will satisfy Bragg's law for Al, since $E_{\text{i}} \neq E_{\text{f}}$, therefore one would expect that no signal from Al will be detected, however this is not the case. In addition to the inelastic position on which a spectrometer nominally operated, elastic process with  $k_{\text{f}}^{\prime}=k_{\text{i}}$  occurs, since the $2\theta$ spectrometer angle coincides with   Bragg's law for Al. Despite the fact that the analyzer is tuned to a different wavelength, because of incoherent scattering at the analyzer, a certain number of elastically scattered neutrons will hit the detector. As a result, a spurious signal, which originates from the Bragg scattering on Al, can be found in the inelastic scan. In the case of hexaboride only $\lambda/2$ and $\lambda/3$ powder lines can be observed, however this possibility is completely excluded by high-order neutron filters.

\begin{figure}[h]
\includegraphics[width=\textwidth]{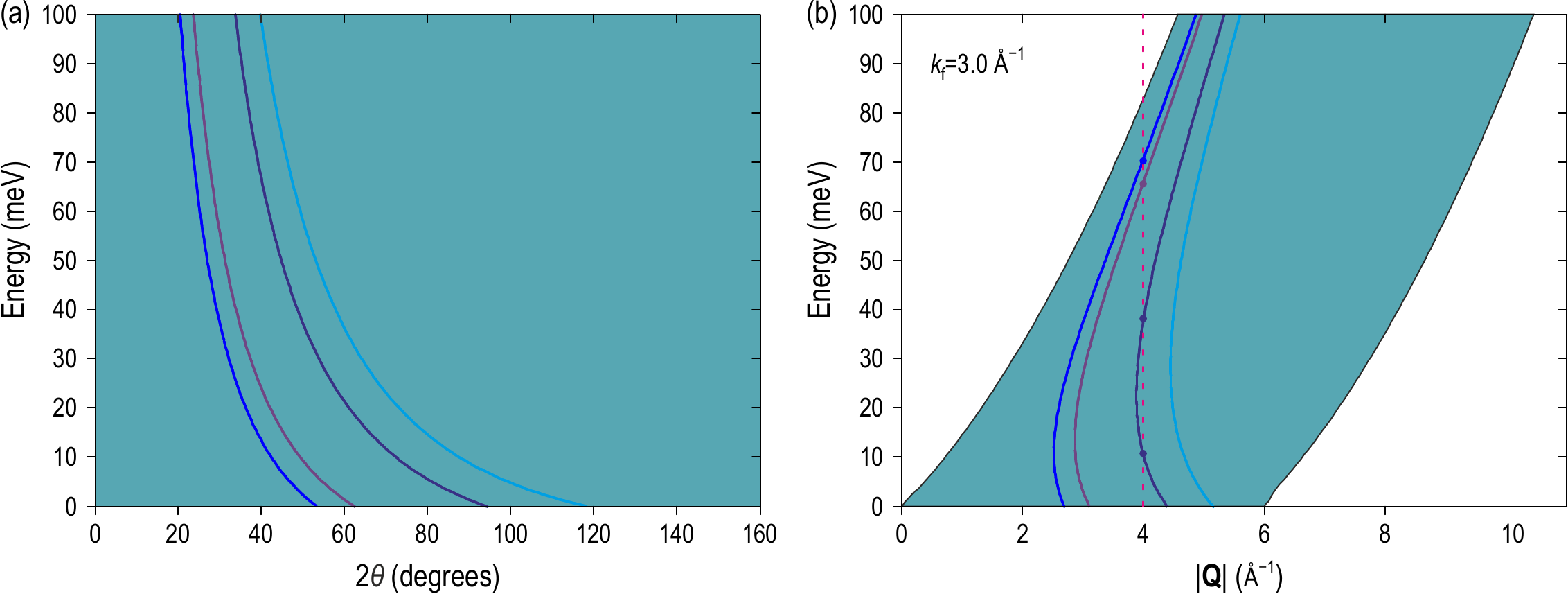}~\vspace{-5pt}
\vspace{-0pt}\caption{ Energy vs. (a) $2\theta$; (b) $|\mathbf{Q}|$ for a fixed $k_{\text{f}}=3\,\text{\AA}^{-1}$. Shaded area shows the appropriate transformation from angle to momentum space. Solid lines are aluminum powder lines.}
\label{Fig:PorderLinesA4EtoQE}\vspace{-15pt}\enlargethispage{15pt}
\end{figure}

It is not really convenient to predict  powder lines with fig.\,\ref{Fig:PorderLinesA4Ei}, since no one plots energy spectra as a function of incident neutron energy. Dependence, calculated as a function of  $|\mathbf{Q}|$ and $\hbar\omega$, will be much more representative. Let us  imagine that with a fixed $k_{\text{f}}$ we have covered a $2\theta=0...160^{\circ}$ range and measured an energy transfer range $\hbar\omega=0...100\,\text{meV}$. This situation would occur, for example, when using a multianalyzer on a triple-axis neutron spectrometer. Using eq.\,\ref{eq:ScatteringVector} and eq.\,\ref{eq:EnergyConservation} we can recalculate the specified $2\theta$ into momentum transfer $|\mathbf{Q}|$. These calculations are shown in fig.\,\ref{Fig:PorderLinesA4EtoQE}, where panel (a) is shown in angular coordinates and panel (b) after appropriate transformation. The aluminum powder lines are shown with solid lines, with colors that match the ones from fig.\,\ref{Fig:PorderLinesA4Ei}.

According to fig.\,\ref{Fig:PorderLinesA4EtoQE}(b), important conclusions can be drawn. The most obvious and simple one is that it is very important to do  detailed analysis of the expected spurion peaks positions prior to the experiment, and try to adjust future scans in a way that they avoid intersections with powder lines. In the worst case, neglecting this simple rule, and for example doing broad energy scan in the vicinity  of $|\mathbf{Q}|=4\,\text{\AA}^{-1}$ with a $k_{\text{f}}=3\,\text{\AA}^{-1}$, shown with a magenta dashed line in fig.\,\ref{Fig:PorderLinesA4EtoQE}(b),  will lead to the intersection of the powder lines four times. Because of the nonmonotonic dependence  of the spurions, one would intersect (220) powder line twice, followed by additional contribution from (200) and  (111) lines at higher energies. Another important conclusion is that in a constant-energy  map, powder lines will appear as concentric rings, which radius nonmonotonically depends  on the chosen energy transfer value. Such measurements are shown in fig.\,1 in  ref.\,[\ref{JainPortnichenko13}].

\vspace{-7pt}\vspace{-7pt}

\begin{figure}[!t]
\includegraphics[width=\textwidth]{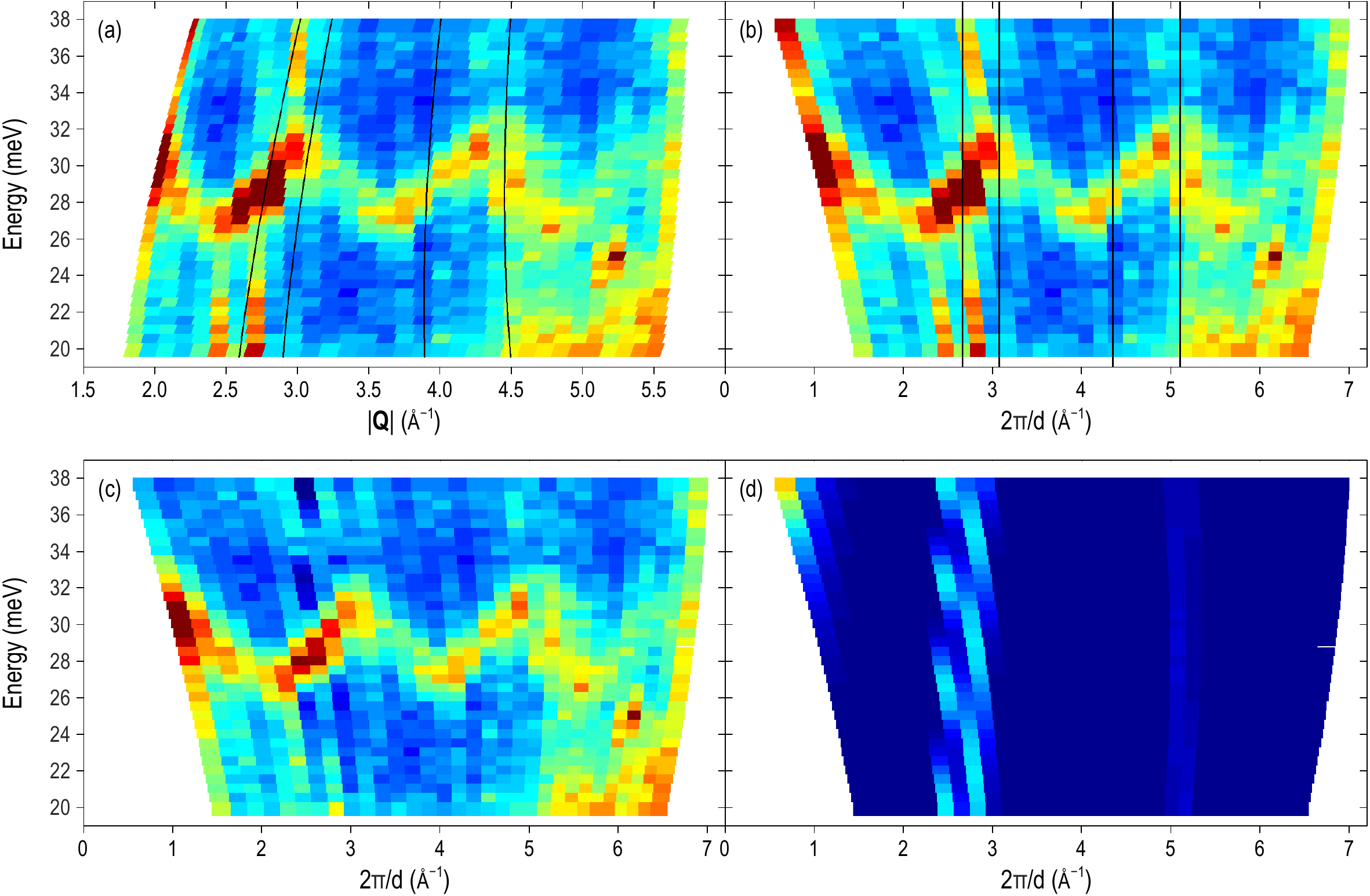}~\vspace{-5pt}
\vspace{-5pt}\caption{(a) Energy-momentum map as measured on Ca$_3$Co$_2$O$_6$ using FlatCone multianalyzer, presented in (E,$|\mathbf{Q}|$) coordinates.(b) The same data after coordinate transformation to $2\pi/d$; (c) The same data after subtraction of the Al powder lines and contribution from the tail of the direct beam. (d) Background model that was used for subtraction.  }\vspace{-25pt}
\label{Fig:PorderLinesCa3Co2O6}
\end{figure}

We figured out how powder lines could appear during the experiment, and now it becomes possible to subtract them. This was successfully done in one of our experiments (see  ref.\,[\ref{JainPortnichenko13}]). As shown in fig.\,\ref{Fig:PorderLinesCa3Co2O6}(a), an energy scan  performed using the FlatCone multianalyzer at the IN8 spectrometer at ILL shows Al powder lines as well as additional contamination from the tail of the direct beam, which appears at small $|\mathbf{Q}|$ and large energy transfer values. As a first step we have to transform the data as if they were measured for elastic scattering of aluminum. In order not to confuse the meaning of $|\mathbf{Q}|$ for the sample and aluminium, we can introduce $|\mathbf{Q}_{\text{Al}}|=2\pi/d=2k_{\text{i}}\text{sin}\theta$. Transforming coordinates in this way makes it clear that the Al lines become vertical irrespective of the energy transfer, as shown in fig.\,\ref{Fig:PorderLinesCa3Co2O6}(b). Now we can select energy windows, where there is no  dispersive  magnon band, and integrate over them. After fitting  of every line and the tail of the direct neutron beam centered at $2\theta=0$ with a Gaussian function, the corresponding background was subtracted from the data, as shown in fig.\,\ref{Fig:PorderLinesCa3Co2O6}(c). The subtracted that consists of powder lines and direct neutron beam is shown in fig.\,\ref{Fig:PorderLinesCa3Co2O6}(d).

\vspace{-0pt}

\vspace{-5pt}\enlargethispage{10pt}
\section{Time-Independent Background}\label{Appendix:TIB}

As it was already mentioned in sec.\,\ref{SubSec:NeutronSources}, for the TOF experiments it is possible to get rid of a certain part of the background: time-independent background (TIB), which for example can originate from  neutrons scattered from other experiments and natural radioactivity.  In addition, radioactive samples can also contribute to the background level \cite{ReifarthBredeweg04}. One should not forget the fact that neutron detectors are also sensitive to $\gamma$-rays (see sec.\,\ref{SubSec:NeutronDetector} for details), and  is easy to understand a high level of $\gamma$-radiation  if we recall sec.\,\ref{SubSec:NeutronAbsorption}, where it was mentioned that some materials emit $\gamma$-rays after  absorbing a neutron. Described $\gamma$-ray background could also give a rise to background counts \cite{Kornilov15}.

\begin{wrapfigure}[24]{r}{0.49\textwidth}\vspace{-0pt}
\includegraphics[width=0.49\textwidth]{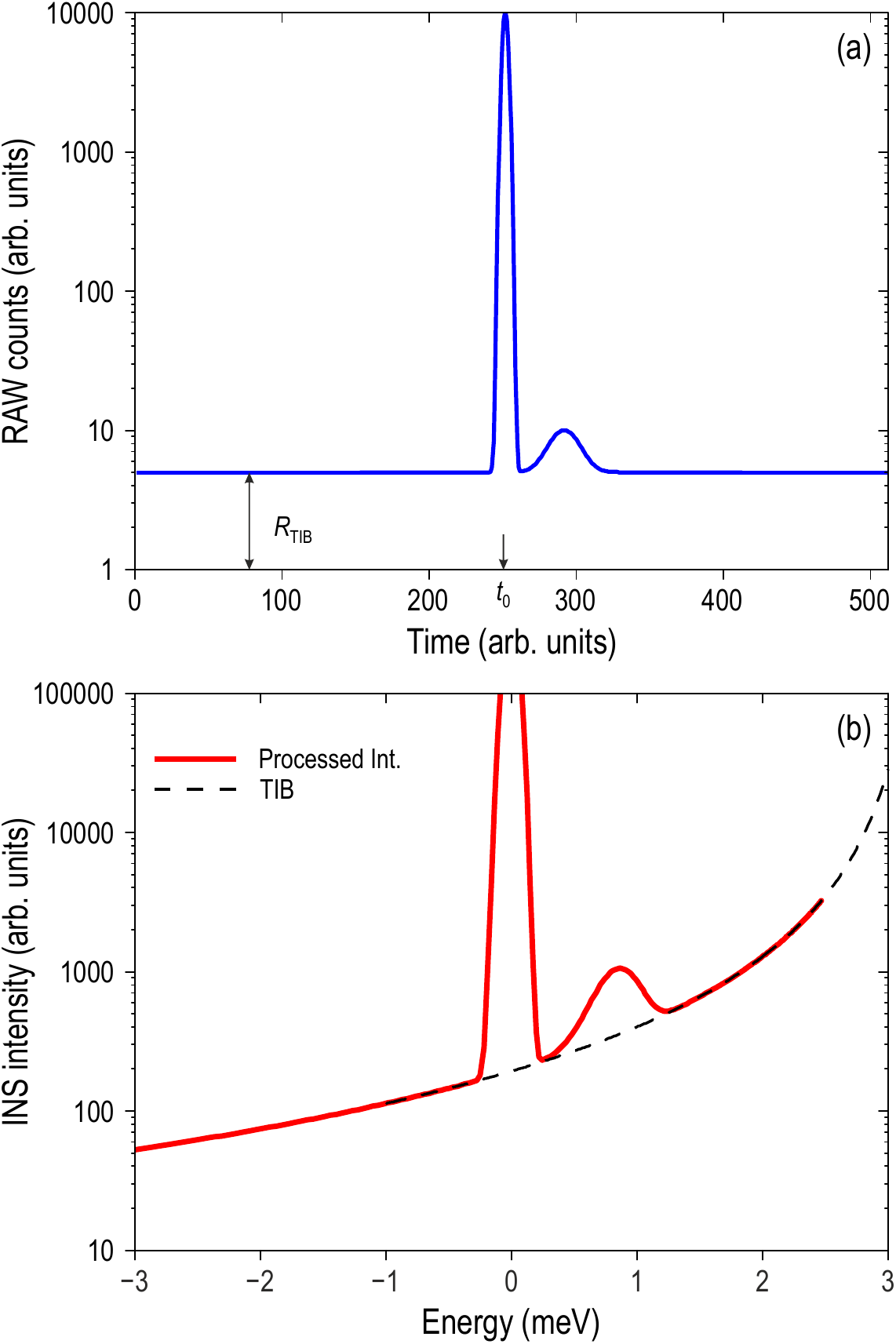}\vspace{-0pt}
\caption{Simulation of a typical (a) TOF spectra, with the elastic line centered at $t_{0}$ and TIB $R_{\text{TIB}}$; (b) energy spectrum, after conversion using eq.\,\ref{eq:Re}. Separately converted value $R_{\text{TIB}}$ is shown as a background with a dashed line.}
\label{Fig:TIB}\vspace{-0pt}
\end{wrapfigure}\enlargethispage{0pt}

First of all let me give a brief explanation how time of flight spectra  are converted to energy spectra. After the neutron pulse reaches the sample, neutrons scatter either elastically or inelastically. As a result, elastically scattered neutrons  will reach the detector bank at the moment of time $t_0$, given by:
\vspace{-0pt}\begin{equation}
t_{0}=\sqrt{\frac{m_{\text{n}}}{2E_{\text{i}}}}\cdot s,
\label{eq:TIBt0}
\vspace{0pt}\end{equation}
where $s$ is the sample-to-detector distance. Thus, the energy of inelastically scattered neutrons can be written as:
\vspace{-0pt}\begin{equation}
E_{\text{f}}=E_{\text{i}} \left( \frac{t_0}{t}\right)^{2},
\label{eq:TIBEf}
\vspace{0pt}\end{equation}
where $t$ is the time of flight for any neutron. A typical count rate in time space, schematically shown in fig.\,\ref{Fig:TIB}(a). It consists of the most intense elastic peak  centered at the time $t_{0}$, the intensity of which away from any Bragg peaks is mainly determined by incoherent scattering, a weak inelastic signal and a constant background $R_{\text{TIB}}$. Conversion from the TOF spectra $R(t)$  to an energy spectrum $R(\hbar\omega)$ is given by
\vspace{-0pt}\begin{equation}
R(\hbar\omega)=R(t)\frac{\text{d}t}{\text{d}\omega}\cdot\frac{k_{\text{i}}}{k_{\text{f}}}.
\label{eq:RtTORe}
\vspace{0pt}\end{equation}
Substituting the expression for $E_{\text{f}}$, given by eq.\,\ref{eq:TIBEf}, in eq.\,\ref{eq:EnergyConservation} we obtain:
\vspace{-0pt}\begin{equation}
t=\frac{t_{0}\sqrt{E_{\text{i}}}}{\sqrt{E_{\text{i}}-\hbar\omega}},
\hspace{30pt}
\frac{\text{d}t}{\text{d}\omega}=\frac{t_{0}\sqrt{E_{\text{i}}}}{2}\frac{1}{(E_{\text{i}}-\hbar\omega)^{3/2}}.
\label{eq:TOFtime}
\vspace{0pt}\end{equation}
Finally, keeping in mind that $k_{\text{i}}/k_{\text{f}}=\sqrt{E_{\text{i}}/E_{\text{f}}}=\sqrt{E_{\text{i}}/(E_{\text{i}}-\hbar\omega)}$,  we can rewrite eq.\,\ref{eq:RtTORe} as:
\vspace{-0pt}\begin{equation}
R(\hbar\omega)=R(t)\frac{t_{0}{E_{\text{i}}}}{2}\frac{1}{(E_{\text{i}}-\hbar\omega)^{2}}=
R(t)\sqrt{\frac{{s^2}m_{\text{n}}{E_{\text{i}}}}{8}}\frac{1}{(E_{\text{i}}-\hbar\omega)^{2}},
\label{eq:Re}
\vspace{0pt}\end{equation}
as shown in fig.\,\ref{Fig:TIB}(b). Omitting the constant expressed by the first term,  correct fitting  of the TOF data should be done with the background given by $\sim R_{\text{TIB}}(E_{\text{i}}-\hbar\omega)^{-2}$, where $R_{\text{TIB}}$ is a free parameter and has to be adjusted to provide the best fit to the measured data.

At some facilities TIB subtraction is done automatically. Measuring the background level during the certain time between the pulses when no ``useful'' neutrons are expected and subtracting it from a TOF data allows you to get rid of TIB completely. In cases where such kind of algorithm is not performed automatically, and  as a result data contains background that monotonically{\footnotemark} increases with energy, it is extremely important to take it into account, especially when fitting the data at various $\mathbf{Q}$ points (see app.\,\ref{Appendix:INSFit} for details).

\footnotetext{TIB is not the only possible contribution to background that increases with energy. Multiple scattering also show energy dependence at large momentum transfer values. See ref.\,\citenum{KitagawaMiyazawa68,Copley74,Berk93,DawidowskiBermejo98,Schober14} for details.}

\vspace{-10pt}\enlargethispage{8pt}
\section{Scattering from Cryostat Walls in a TOF Experiment}\label{Appendix:ScatteringTOF}
\vspace{-8pt}

\begin{wrapfigure}[11]{r}{0.5\textwidth}\vspace{-15pt}
\includegraphics[width=0.5\textwidth]{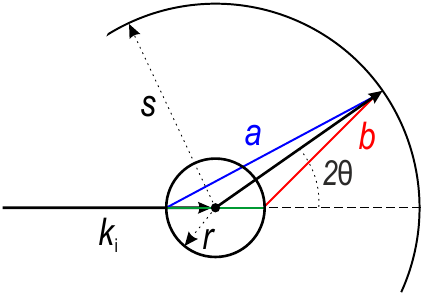}
\caption{Top view of a typical TOF spectrometer, which illustrates scattering from  cryostat walls.}
\label{Fig:TOFemptycryotop}\vspace{-4pt}
\end{wrapfigure}
Incoherent scattering of the incident neutron beam on the Al walls of a cryostat or cryomagnet can significantly broaden the elastic line, as it typically appears in the spectrum as a pair of peaks slightly shifted to positive and negative energy transfer values with respect to the elastic position. Their intensities depend strongly on the thickness of the inner walls of the cryogenic device, the purity of the used aluminum, and the degree of collimation of the incident and scattered neutron beams. This type of background contamination was observed both in the TAS and TOF data. In order to accurately describe experimental results, corresponding peaks have to be included in the fitting model for the shape of the elastic line.  In case of a  TOF experiment, the apparent shape of the elastic line also becomes momentum-dependent after the data are transformed into energy-momentum space. We can observe this effect in fig.~\ref{Fig:CeB6Map2p5Tand5T}(b), where thicker Al walls of the cryomagnet at CNCS  produce higher background, and at larger momentum transfer values the elastic line splits into two ``branches''. Let us try to figure out where these elastic ``branches'' come from.

In the TOF experimental geometry, as shown in fig.\,\ref{Fig:TOFemptycryotop}, the opposite segments of the cryostat's inner wall  are illuminated by the direct neutron beam. Despite the fact that according to table\,\ref{Tab:SummaryCohIncAbs}, the incoherent scattering cross-section of aluminum is relatively low, because of high neutron flux, the two illuminated spots that are offset from the nominal sample position serve as incoherent scattering centers, and it becomes possible to observe the signal from an empty cryostat.

Let the sample-to-detector distance be $s$, cryostat radius $r$, and the energy of the incident neutron beam $E_{\text{i}}$. Neutrons that are scattered elastically from the sample will reach the detector bank in time $t_{0}={s}/{v_{\text{i}}}$, where $v_{\text{i}}$ is the neutron velocity. Now we have to calculate the time when neutrons  which are scattered from the  cryostat walls will reach the detector. These two possible paths are shown in fig.\,\ref{Fig:TOFemptycryotop} with blue and red color. For a neutron that travel along these routes it is necessary to take into account the time which it would spend when flying between the cryostat wall and the sample (shown with green color in fig.\,\ref{Fig:TOFemptycryotop}). Thus  we can write down both times as:
\vspace{-3pt}\begin{equation}
t_a=\frac{a}{v_{\text{i}}}-\frac{r}{v_{\text{i}}},
\hspace{30pt}
t_b=\frac{b}{v_{\text{i}}}+\frac{r}{v_{\text{i}}}.
\label{eq:TOFtimeAandB}
\vspace{-3pt}\end{equation}
According to eq.\,\ref{eq:TIBEf} we can now define two different final energies $E_{\text{f}}^{a}$ and $E_{\text{f}}^{b}$, which will only depend on the ratio ${t_0}/{t_x}$, written as:
\vspace{--3pt}\begin{equation}
\frac{t_0}{t_a}=\frac{s}{a-r},
\hspace{30pt}
\frac{t_0}{t_b}=\frac{s}{b+r}.
\label{eq:TOFemptycryoTimeDif}
\vspace{-3pt}\end{equation}
Using the geometric relationships for both paths, we can write them as
\vspace{-3pt}\begin{equation}
a=\sqrt{s^2+r^2+2sr\text{cos}2\theta},
\hspace{30pt}
b=\sqrt{s^2+r^2-2sr\text{cos}2\theta}.
\label{eq:TOFpathAandB}
\vspace{-3pt}\end{equation}
It is very easy to evaluate $\text{cos}2\theta$ if one recalls the Bragg condition for elastic scattering $|\mathbf{Q}|=2 |\mathbf{k}_{\text{i}}|\text{sin}\theta$, the relation between energy and wave vector $E_{\text{i}}\,[\text{meV}]=2.072 k_{\text{i}}^2\,[\text{\AA}^{-1}]$ and the list of trigonometric identities:\vspace{-0pt}\begin{equation}
\text{cos}2\theta=1-2\text{sin}^2\theta=1-2\left( \frac{|\mathbf{Q}|}{2|\mathbf{k}_{\text{i}}|}\right)^2=1-2\left( \frac{2.072|\mathbf{Q}|^2 }{4E_{\text{i}}}\right)=1-\frac{1.036|\mathbf{Q}|^2 }{E_{\text{i}}}
\label{eq:TOFcostwotheta}
\vspace{0pt}\end{equation}

Substituting the expression for $E_{\text{f}}$, given by eq.\,\ref{eq:TIBEf}, in eq.\,\ref{eq:EnergyConservation}, and taking into account expressions for both paths given by eq.\,\ref{eq:TOFpathAandB}  together with eq.\,\ref{eq:TOFcostwotheta}, we obtain:
\vspace{-3pt}\begin{equation}
\hbar\omega_{a}=E_{\text{i}}\left( 1-\frac{s^2}{(a-r)^2}\right)=
E_{\text{i}}\left( 1-\frac{s^2}{\left(\sqrt{s^2+r^2+2sr\left(1-\frac{1.036|\mathbf{Q}|^2 }{E_{\text{i}}}\right)}-r\right)^2}\right)
\label{eq:EmptyCryoModeI}
\vspace{-3pt}\end{equation}
\vspace{-3pt}\begin{equation}
\hbar\omega_{b}=E_{\text{i}}\left( 1-\frac{s^2}{(b+r)^2}\right)=
E_{\text{i}}\left( 1-\frac{s^2}{\left(\sqrt{s^2+r^2-2sr\left(1-\frac{1.036|\mathbf{Q}|^2 }{E_{\text{i}}}\right)}+r\right)^2}\right)
\label{eq:EmptyCryoModeII}
\vspace{-3pt}\end{equation}

Finally we derived the $|\mathbf{Q}|$ dependence of the incoherent scattering from the cryostat walls.  If we consider the limiting case $|\mathbf{Q}|\rightarrow0$, both $\hbar\omega_{a}$ and $\hbar\omega_{b}$ become zero. In the case of a finite $|\mathbf{Q}|$ value, $\hbar\omega_{a}<0$ while $\hbar\omega_{b}>0$, therefore one ``branch'' will appear on the energy gain side and the other one on the energy loss respectively. Using above mentioned formulas, it becomes possible to determine the inner radius of the cryostat by fitting shifted position of the elastic line during data analysis. The intensity ratio between the two ``branches'' is determined by the neutron flux before and after the sample, which is influenced by the total beam attenuation by the sample itself. Moreover, any accumulation of ice (e.g. due to a lead in the cryomagnet) or hydrogen-containing grease on the cryostat walls will dramatically enhance the spurious lines. In particular, a persistent leak in the cryostat would result in their intensity increase with time (see app.\,\ref{Appendix:Water}).

The easiest way to suppress spurious scattering from the cryostat walls  is to use radial collimators, which would prohibit the passage of neutrons scattered from any point except the sample, located in the center of the cryostat. The use of such collimators significantly weakens the  described effect and allows us to uncover low-energy  excitations even at large momentum transfers.

\vspace{-10pt}\enlargethispage{5pt}
\section{Scattering from Cryostat Walls in a TAS Experiment}\label{Appendix:ScatteringTAS}

\begin{wrapfigure}[12]{r}{0.5\textwidth}\vspace{-12pt}
\includegraphics[width=0.5\textwidth]{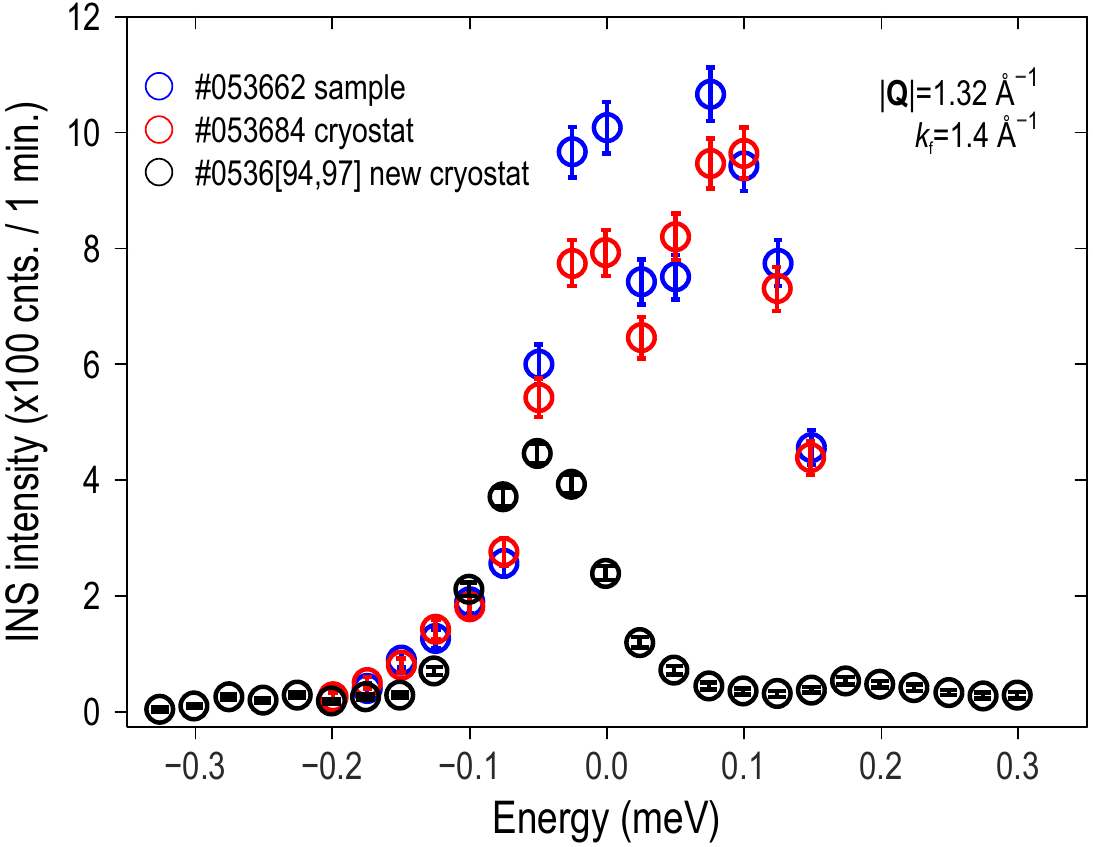}\vspace{-0pt}
\caption{Energy spectrum which shows spurions for different cryostats, with and without sample.}
\label{Fig:WaterInCryoHZB1}\vspace{-0pt}
\end{wrapfigure}
In one of our TAS experiments, an energy scan through the elastic line showed an unusual shape with two almost equally strong peaks, shown in fig.\,\ref{Fig:WaterInCryoHZB1} with blue and red  markers. What is even more surprising,  in the first case the sample was inside the cryostat, and in the second case the same empty cryostat was measured. Thus, comparing these two scans one can conclude that incoherent scattering from the cryostat is much stronger than the incoherent line of the sample. We assume this happens due to the presence of moisture inside the cryostat, since we already know (see table\,\ref{Tab:SummaryCohIncAbs}) that, among the simplest and most possible elements, hydrogen has the largest incoherent scattering cross-section. To test this, we replaced the cryostat with an absolutely identical one. Note, that the ``new'' cryostat was warm, and thus we could  exclude the possibility  of ice being present inside. The corresponding scan of the new cryostat is shown with black markers in fig.\,\ref{Fig:WaterInCryoHZB1},  and as can be seen it has a much lower background. After we cooled down the cryostat, it showed an absolutely identical background level, which was averaged. Apart from the most intense elastic line, it is important to mention two weak satellites at $\sim-0.24$\,meV and $\sim0.18$\,meV.

\clearpage
\vspace{-0pt}\begin{figure}[!t]
\includegraphics[width=1\textwidth]{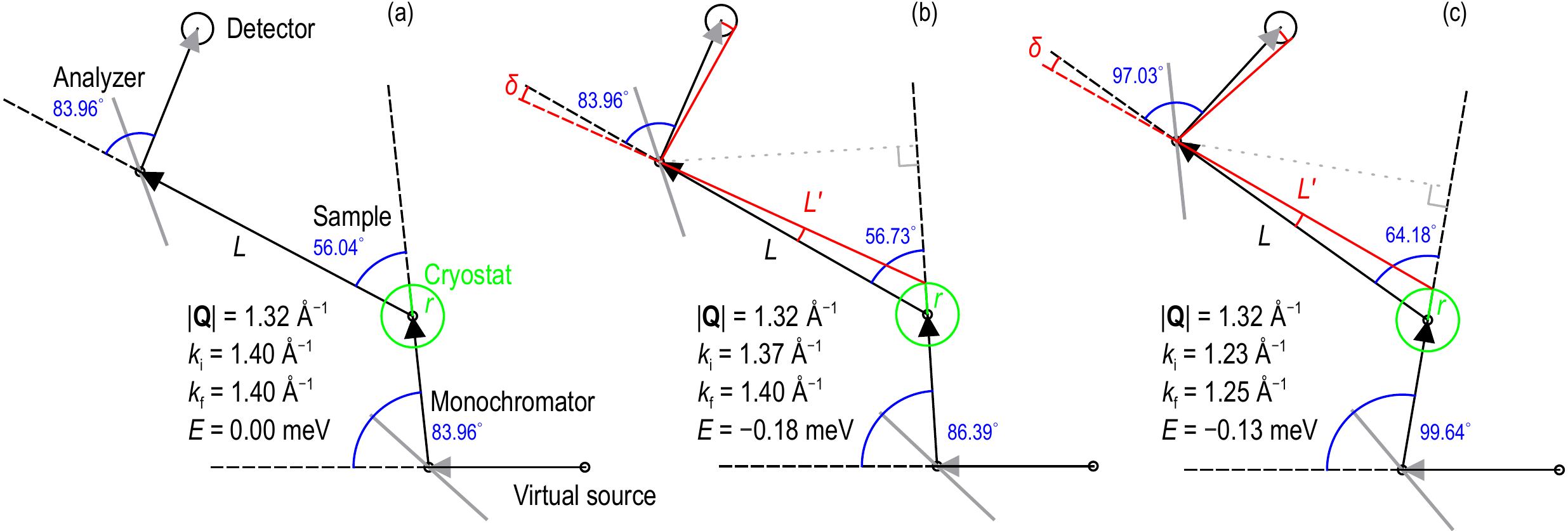}~\vspace{-5pt}
\vspace{-5pt}\caption{Top view of a typical triple axis spectrometer, which illustrates neutron path for: (a) elastic scattering from a sample; (b) incoherent scattering from a cryostat wall with $k_{\text{i}}$, and then how neutrons can get into the detector, although the analyzer is tuned to a different wavelength; (c) same process with a different $k_{\text{f}}$. The image was created using TAKIN \cite{WeberGeorgii16}.}
\label{Fig:EmptyCryoTASviewkf}\vspace{-25pt}
\end{figure}

Below, I argue that observed peaks are due to elastic scattering from the cryostat walls. First of all, one has to recall basic knowledge about TAS geometry (see. sec.\,\ref{SubSec:TAS}). Vectors $\mathbf{Q}$, $\mathbf{k}_{\text{i}}$ and $\mathbf{k}_{\text{f}}$ form a triangle in the scattering plane, which satisfies the condition $|\mathbf{Q}|^2={|\mathbf{k}_{\text{i}}|}^2+{|\mathbf{k}_{\text{f}}|}^2-2{|\mathbf{k}_{\text{i}}|}{|\mathbf{k}_{\text{f}}|}\text{cos}2\theta_{\text{S}}$, where $2\theta_{\text{S}}$ is the sample scattering angle. Corresponding scattering angles of the analyzer and monochromator are $2\theta_{\text{A}}$ and $2\theta_{\text{M}}$. Let the spectrometer be in the elastic position, with the corresponding  scattering angle of the monochromator, sample and analyzer as shown{\footnotemark}
\footnotetext{Layout of the PANDA spectrometer is shown. Distance from the virtual source to monochromator, as well as to the sample was shortened. Size of the cryostat and detector are exaggerated, as well as $\delta$ angle.}
in fig.\,\ref{Fig:EmptyCryoTASviewkf}(a).
When we change the energy transfer value in the $k_{\text{f}}$-fixed mode at a fixed $|\mathbf{Q}|$, $k_{\text{i}}$ and $2\theta_{\text{S}}$ values are automatically adjusted, in order to fulfill the condition mentioned above, as shown in fig.\,\ref{Fig:EmptyCryoTASviewkf}(b). In both configurations, opposite segments of the cryostat's inner wall are illuminated by the direct neutron beam, as was mentioned in app.\,\ref{Appendix:ScatteringTOF}. The analyzer is aligned in such a way that Bragg condition $2d_{\text{A}}\text{sin}\theta_{\text{A}}=\lambda_{\text{f}}$ is fulfilled. In the elastic configuration, neutrons which are scattered incoherently from the cryostat walls can not fulfill the condition above, unless the wavelength acceptance $\Updelta\lambda/\lambda$ is too large. However in case of non-zero energy transfer, scattering from the cryostat wall can fulfill the condition $2d_{\text{A}}\text{sin}(\theta_{\text{A}}+\delta)=\lambda_{\text{i}}$. This path is shown{\footnotemark}
\footnotetext{Scattering from the second wall is simply not shown. This case will appear in the solution to eq.\ref{eq:EmptyCryoSollutionTAS}.}
with red in fig.\,\ref{Fig:EmptyCryoTASviewkf}. If the detector size is large enough{\footnotemark},
\footnotetext{A typical $^{3}_{2}\text{He}$ detector is 1 inch in diameter.}
and $\delta$ is small, neutrons scattered from the cryostat in this way can reach the detector and might appear as a spurious peak at a non-zero energy transfer. It is clear that after changing $k_{\text{f}}$,  the $2\theta$ angle at every axis is adjusted appropriately, as shown in fig.\,\ref{Fig:EmptyCryoTASviewkf}(c). As a result, scattering from the cryostat with a different wavelength $\lambda_{\text{i}}$ can fulfill  Bragg's law $2d_{\text{A}}\text{sin}(\theta_{\text{A}}+\delta)=\lambda_{\text{i}}$ for the analyzer.

\begin{wrapfigure}[20]{r}{0.5\textwidth}\vspace{-0pt}
\includegraphics[width=0.5\textwidth]{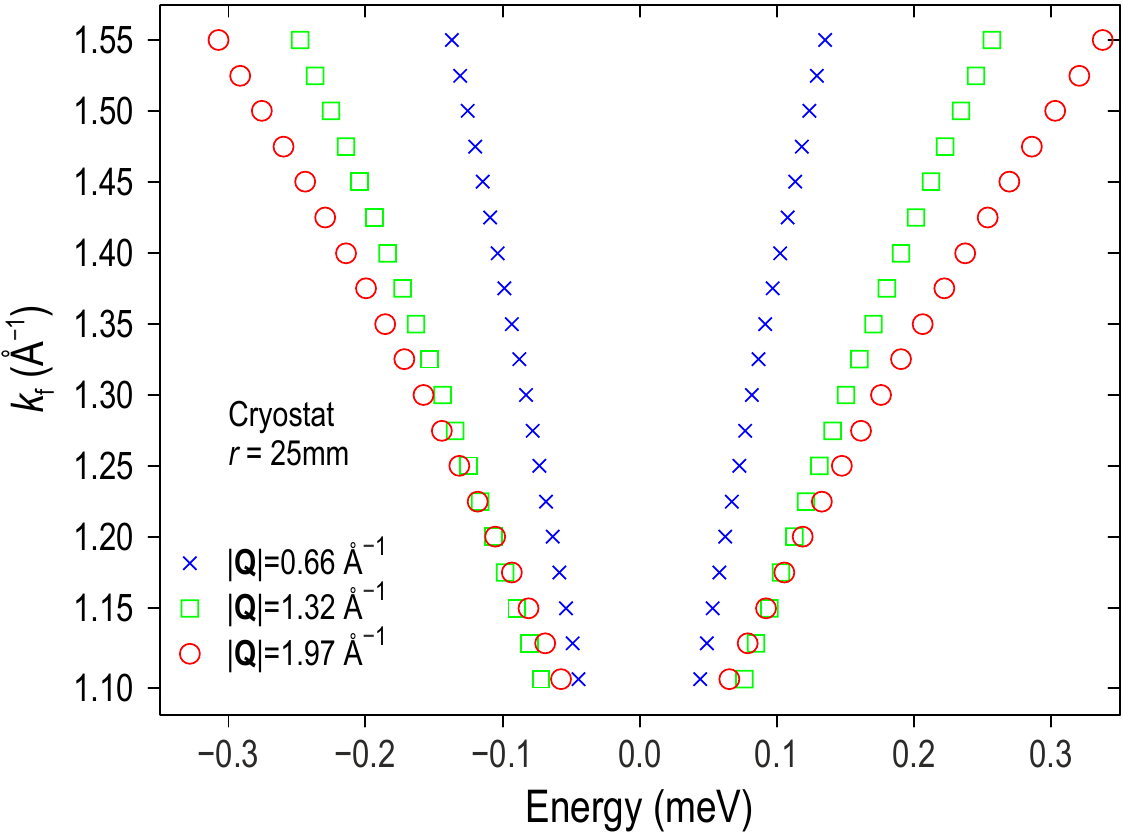}\vspace{-2pt}
\caption{Solution of the  eq.\,\ref{eq:EmptyCryoSollutionTAS} for different ${k}_{\text{f}}$ and $|\mathbf{Q}|$ values, calculated for PG  analyzer ($d_{\text{A}}=3.355\,\text{\AA}^{}$), which describes scattering from the cryostat walls.}
\label{Fig:EmptyCryoTASkf}\vspace{-0pt}
\includegraphics[width=0.5\textwidth]{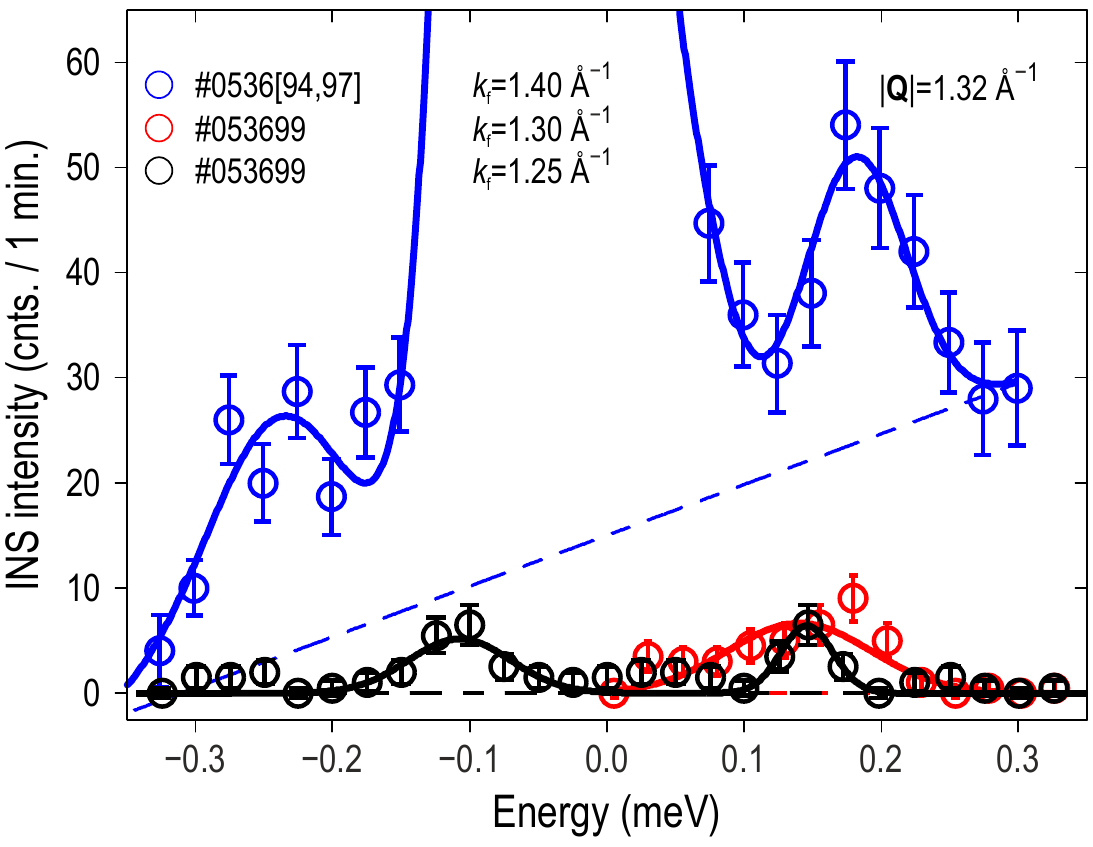}\vspace{-2pt}
\caption{An energy spectrum, which shows spurious signal measured for different $k_{\text{f}}$ values.}
\label{Fig:WaterInCryoHZB2}\vspace{-0pt}
\end{wrapfigure}\enlargethispage{15pt}
It is relatively difficult to obtain  an analytical expression for the  energy dependence $\hbar\omega(|\mathbf{Q}|,{k}_{\text{f}})$ of the spurious peak. In order to understand the general behavior we can limit ourselves to calculating only  specific ${k}_{\text{f}}$ values. In order to allow neutrons with the wavelength $\lambda_{\text{i}}$ to reach the detector, one should fulfill condition $\text{sin}(\theta_{\text{A}}+\delta)/\text{sin}(\theta_{\text{A}})={k}_{\text{f}}/{k}_{\text{i}}$
for any fixed ${k}_{\text{f}}$ value. Let the distance between the sample and analyzer be $L$, cryostat radius $r$, and the distance between the cryostat wall and the analyzer $L^{\prime}$. For the scattering described above, the condition $L\text{cos}(90-2\theta_{\text{S}})=L^{\prime}\text{cos}(90-2\theta_{\text{S}}-\delta)$
is also met. Finally using the $2\theta_{\text{S}}$ angle obtained from the scattering triangle, we can write:
\vspace{-8pt}\begin{align}
\begin{cases}
\delta &=\text{arcsin}\left(\frac{k_{\text{f}}}{k_{\text{i}}}\text{sin}(\theta_{\text{A}})\right)-\theta_{\text{A}}, \\
2\theta_{\text{S}} &=\text{arccos}\left(\frac{k_{\text{i}}^2+k_{\text{f}}^2-|\mathbf{Q}|^2}{2k_{\text{i}}k_{\text{f}}}\right),\\
L^{\prime}&=L\frac{\text{cos}(90-2\theta_{\text{S}})}{\text{cos}(90-2\theta_{\text{S}}-\delta)},\\
r^2&=L^2+{L^{\prime}}^2-2LL^{\prime}\text{cos}(\delta).
\end{cases}
\label{eq:EmptyCryoSollutionTAS}
\vspace{0pt}\end{align}\vspace{-8pt}

The numerical solution for most commonly used ${k}_{\text{f}}$ values  is shown in fig.\,\ref{Fig:EmptyCryoTASkf}, where red, green and blue markers show energy transfer for different momentum transfer $|\mathbf{Q}|$, at which the spurious peak from the sample environment is expected. Analyzing the equation eq.\,\ref{eq:EmptyCryoSollutionTAS}, one can come to the evident conclusion that, due to the finite size of the detector, scattering  from large $\delta$ angles will not reach the detector. Therefore the simplest way to suppress spurious signal is either to increase{\footnotemark}
\footnotetext{Changing cryostat radius from 2.5\,cm to 10\,cm, when measuring at $|\mathbf{Q}|=1.32\,\text{\AA}^{-1}$ with ${k}_{\text{f}}=1.25\,\text{\AA}^{-1}$, will increase $\delta$ angles from  $1.31^{\circ}$ and $-1.26^{\circ}$ to  $5.45^{\circ}$ and $-4.74^{\circ}$ respectively. It is obvious that in the second case the angle is too large and such scattering will not be observed.}
the diameter of the cryostat or to use collimation. The calculated spurious peak positions are in good  agreement with the measured positions. Two satellites, which were assumed to be the signal from the walls of an empty cryostat, are also shifted closer to the elastic line at smaller ${k}_{\text{f}}$ values, as  shown in fig.\,\ref{Fig:WaterInCryoHZB2}.

\section{Water Inside the Cryostat}\label{Appendix:Water}

\begin{wrapfigure}[27]{r}{0.5\textwidth}\vspace{-10pt}
\includegraphics[width=0.5\textwidth]{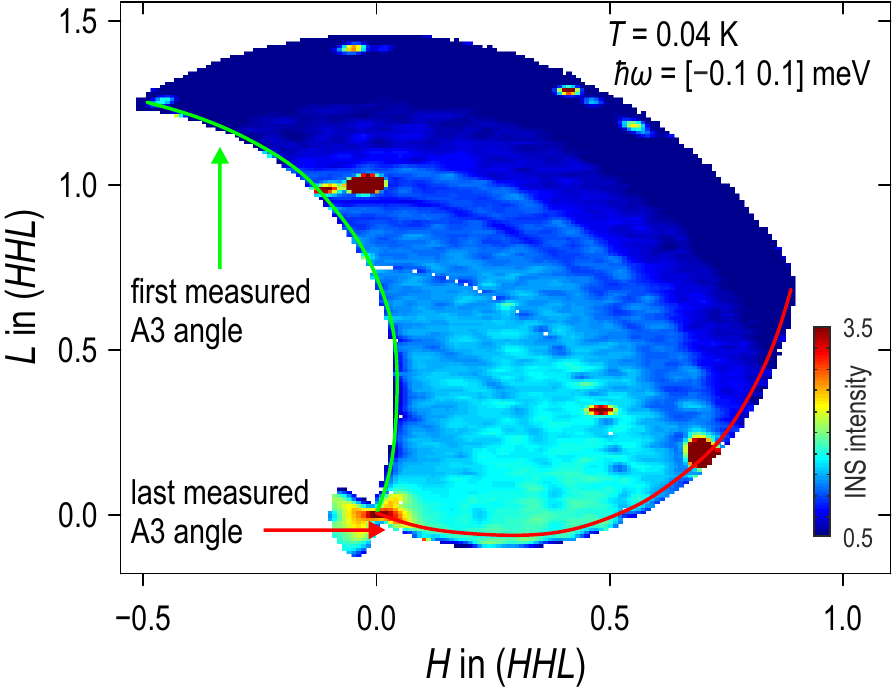}\vspace{-5pt}
\caption{Constant-energy map of the INS intensity, obtained by integrating the TOF data measured at LET in the energy window $[-0.1\,\,0.1]$\,meV. Time-dependent background, which increases between the first and the last measured A3 angle is assumed to be due to the water, which freezes on a cryostat wall.}
\label{Fig:WaterInCryoISISqxqy}\vspace{-0pt}
\includegraphics[width=0.5\textwidth]{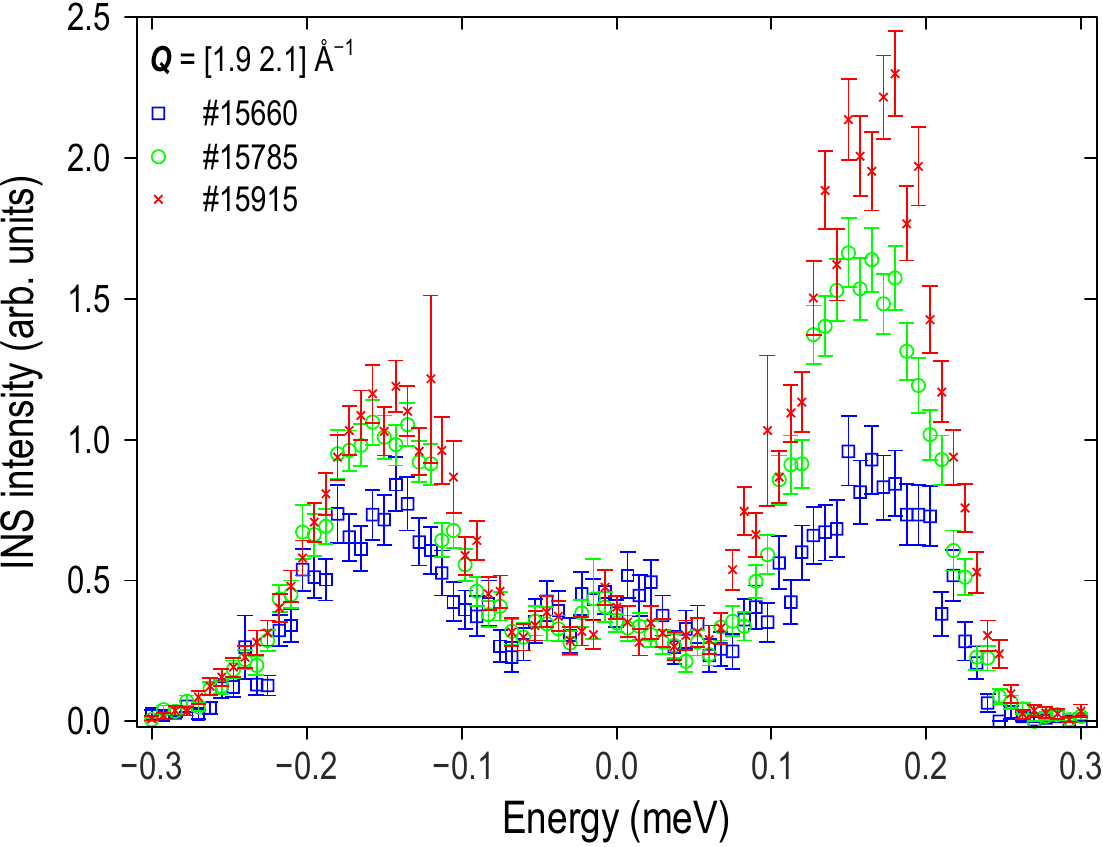}
\caption{Energy spectrum which shows time development of the spurios signal, which originates from the frozen water on  the cryostat wall. Calculations of the scattering from an empty cryostat is done in  app.\,\ref{Appendix:ScatteringTOF}.}
\label{Fig:WaterRAWCounts}\vspace{-4pt}
\end{wrapfigure}
As we have already seen in previous sections, there is a high chance to observe spurios signals which originate from the sample environment. Here I will present one of the extreme  examples, where the scattering from an empty cryostat, already described in  app.\,\ref{Appendix:ScatteringTOF}, was largely enhanced due to the freezing of water on the outer cryostat walls. This happened because the outer heat shielding from the cryostat was erroneously removed, in order to reduce the amount of aluminum in the beam. The whole TOF detector bank is evacuated down to a cryogenic vacuum, however as it is very difficult to pump out water from a vacuum system,  moisture can freeze on the outer cryostat wall, which has approximately the temperature of the liquid nitrogen vessel. In case of even a small leak, one should observe significant increase of the background with time. This behavior is shown in fig.\,\ref{Fig:WaterInCryoISISqxqy}, where each individual A3 angle has a higher background. Note, that at higher momentum transfer values, the elastic background  is even lower. It looks so because the spurious signal from the cryostat wall appears on the energy gain and loss side at higher $|\mathbf{Q}|$ values (see eq.\,\ref{eq:EmptyCryoModeI}, eq.\,\ref{eq:EmptyCryoModeII}).

In order to confirm that background increases with a time, the same A3 angle range was measured multiple times. One-dimensional energy profiles obtained from the same individual files by integration within $|\mathbf{Q}|=[1.9\,\,2.1]\,\text{\AA}^{-1}$ are shown in fig.\,\ref{Fig:WaterRAWCounts}. The intensity of two peaks, which appear at $\sim\pm0.15\,\text{meV}$, increases with time, while the elastic line remains constant. This is fully consistent with the suggested explanation that the appearance of the background is due to water freezing on the wall of the cryostat.

\begin{figure}[h]
\includegraphics[width=\textwidth]{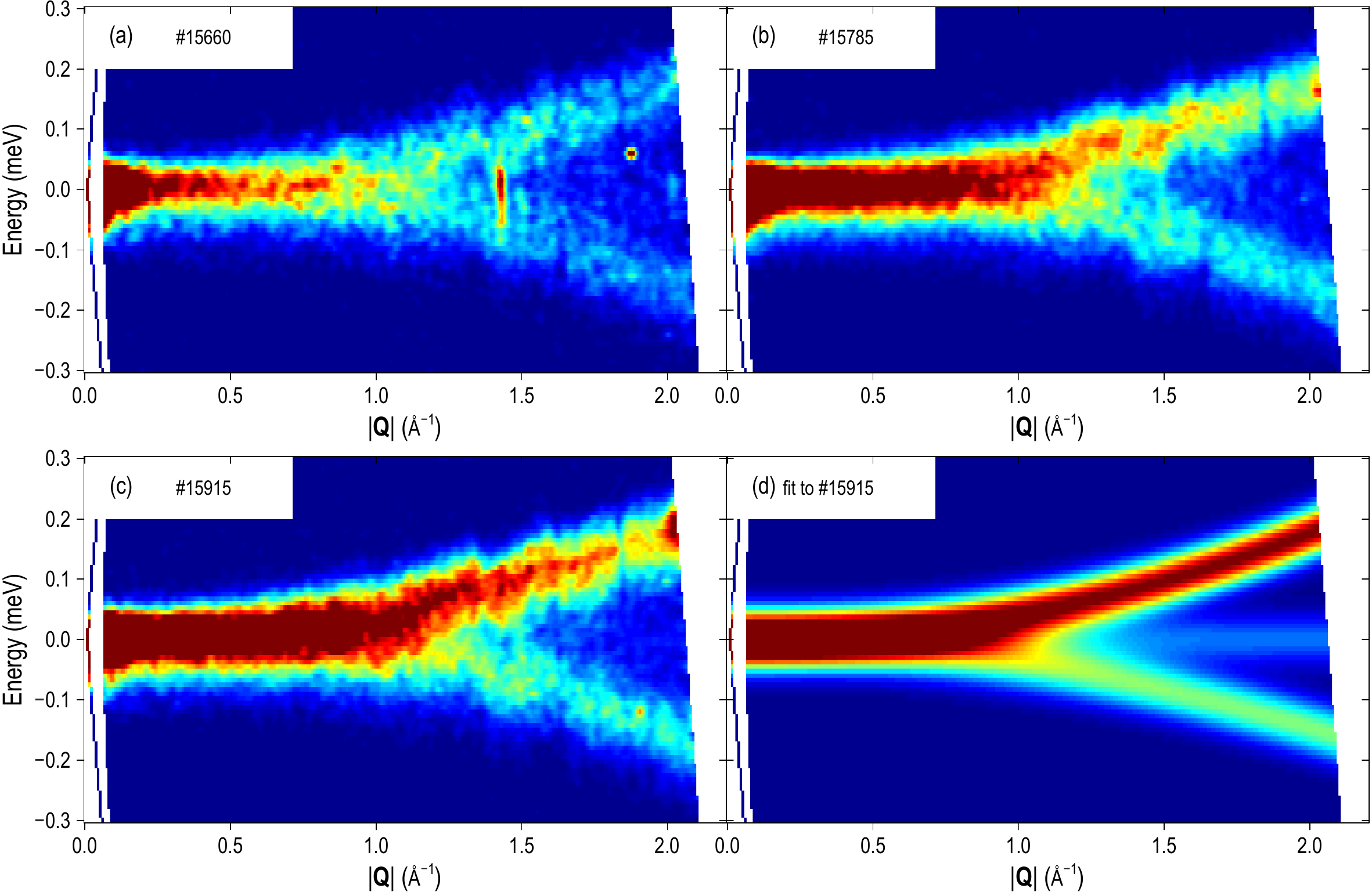}~\vspace{-5pt}
\vspace{-0pt}\caption{Energy-momentum map of the INS intensity, obtained after powder-averaging the TOF data measured at LET. Same A3 angle was measured three times as shown in panels (a), (b) and (c). Panel (d) shows fitting of the file \#15915 with eq.\,\ref{eq:EmptyCryoModeI}, eq.\,\ref{eq:EmptyCryoModeII}, and elastic line.}
\label{Fig:WaterInCryoISIS}\vspace{-0pt}
\end{figure}

Powder averaged energy-momentum maps of each scan are shown in fig.\,\ref{Fig:WaterInCryoISIS}\,(a-c). According to app.\,\ref{Appendix:ScatteringTOF}, the elastic line is split into two branches, which appear above and below incoherent scattering from a sample. This immediately explains the lower background at higher momentum transfers in fig.\,\ref{Fig:WaterInCryoISISqxqy}, as integrations within a narrow energy range $[-0.1\,\,0.1]$\,meV simply prevents the spurious signal from getting into the elastic channel.

Using a script, which takes into account the presence of incoherent scattering by the sample, it becomes possible to determine the inner radius of the cryostat.  As shown in fig.\,\ref{Fig:WaterInCryoISIS}\,(d), where powder-averaged TOF data of one individual scan were fitted with eq.\,\ref{eq:EmptyCryoModeI} and eq.\,\ref{eq:EmptyCryoModeII}. Additionally, adding the possibility of having a different radius for each ``branch'', corresponding radii of the cryostat are $r_a=6.15\pm0.04$\,cm and $r_b=7.59\pm0.03$\,cm. The reason why radii should be different  is not obvious to me, and I assume this is an artifact of the model, which does not take into account the change in distance to the out of plain detectors, which also contribute when doing powder averaging of the data. Unfortunately, such  kind of simulations can not be used to subtract the elastic line from the data, where low lying inelastic excitations are masked with the elastic line, and it was done mainly as an exercise in order to understand the origin of the spurious signal.

\vfill

\vspace{-0pt}

\clearpage
\section{Helium Exchange-Gas Background}\label{Appendix:HeExchangeGasBackgroundSubstraction}

\begin{wrapfigure}[14]{r}{0.5\textwidth}\vspace{-1em}
\includegraphics[width=0.5\textwidth]{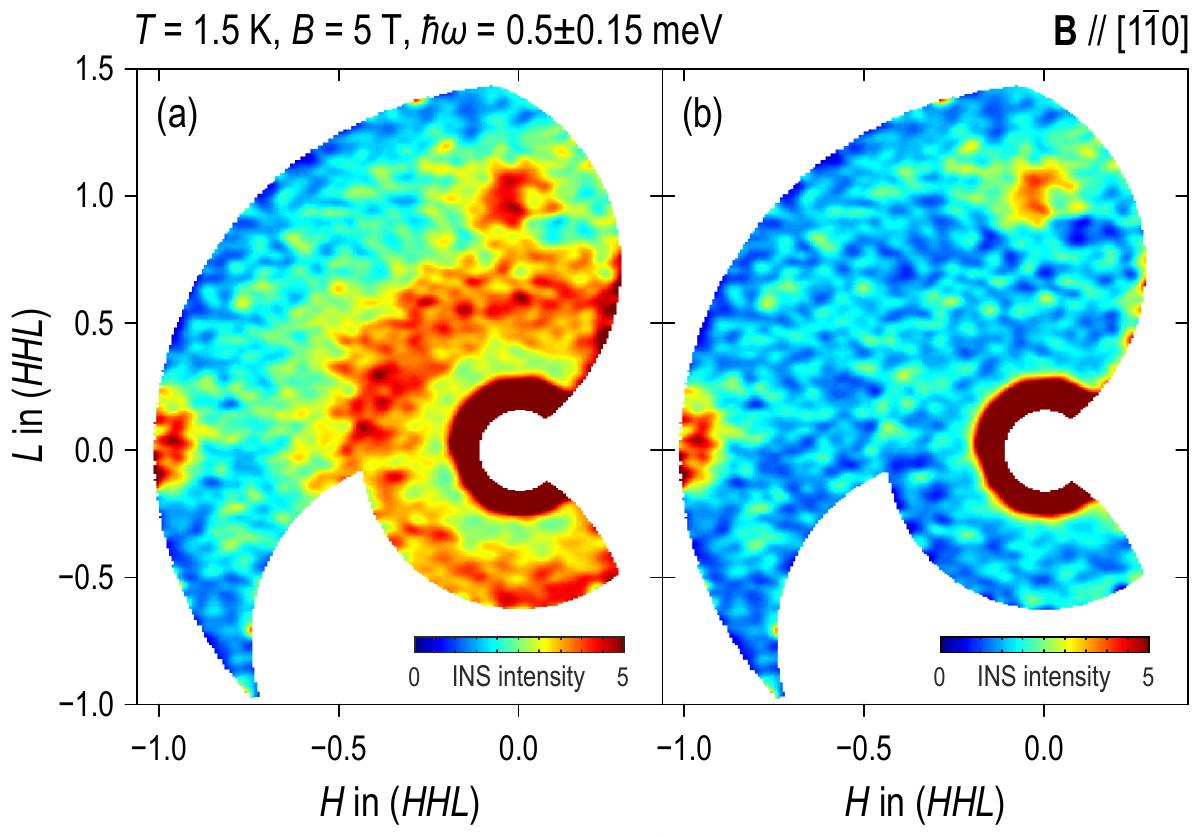}
\caption{Constant-energy map of the INS intensity, obtained by integrating the TOF data measured at CNCS in the energy window [0.35 0.65]\,meV. The data show resonant FM excitations centered at $\Gamma''(110)$ and $\Gamma'(001)$: (a) as initially measured, (b) after substraction of the He background according to eq.\,(\ref{Eq:He_gas}).}
\label{Fig:He_sub}\vspace{-4pt}
\end{wrapfigure}

Additional background contamination may originate from the He gas used for heat exchange between the sample and the cryostat in some types of the cryogenic sample environment \cite{ShiraneShapiro02}. It originates from nonmagnetic scattering on free He nuclei and is both momentum and temperature dependent, which can lead to a misinterpretation of experimental results \cite{BuyersKjems85, BuyersKjems86}. This type of background contamination was observed in our TOF data at differen facilities. It appears as a ring of intensity within every constant-energy cut, as can be seen in fig.~\ref{Fig:He_sub}, where the contamination is observed in the momentum range 0.25~r.l.u.~$\leq |\mathbf{Q}| \leq$~0.5~r.l.u. Scattering from single free nuclei can be analytically described as \cite{Squires96}
\begin{equation}\label{Eq:He_gas}
S(\mathbf{Q},\omega)\propto\,\biggl(\frac{\beta}{4\pi E_{\text r}}\biggl)^{1/2}\times\exp\biggl[-\frac{\beta}{4E_{\text r}}(\hbar\omega-E_{\text r})\biggr],
\end{equation}
where $\beta=(k_{\text B}T)^{-1}$ and $E_{\text r}=\hslash^2|\mathbf{Q}|^2/(2M)$ is the recoil energy, $M$ being the mass of the nucleus. The CNCS data presented in fig.~\ref{Fig:CeB6Map2p5Tand5T}(b) have been background-corrected by subtracting the analytical form of the He signal given by eq.\,(\ref{Eq:He_gas}).

\begin{wrapfigure}[9]{r}{0.5\textwidth}\vspace{-12pt}
\includegraphics[width=0.5\textwidth]{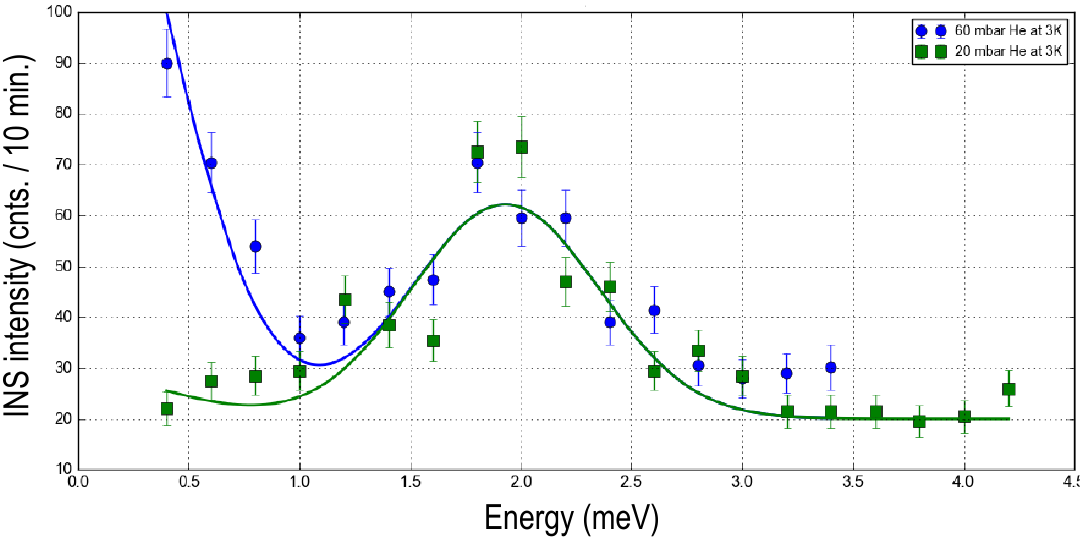}\vspace{-5pt}
\caption{Energy scan, which shows  contamination from the He exchange gas. Figure was taken from the \href{http://wiki.mlz-garching.de/panda:measurement:helium}{wiki:PANDA} web-page.}
\label{Fig:HePanda}\vspace{-4pt}
\end{wrapfigure}
Besides, exchange gas produce strongly enhanced elastic background, thus significantly broadening the elastic line. When cooling a top loading cryostat the ambient exchange gas pressure decreases to roughly 60\,mbar. Such high exchange gas pressure produces high background, as shown in fig.\,\ref{Fig:HePanda}. Since it is not possible to pump out exchange gas completely at low temperatures. One should pay attention to this technical nuance and pump the exchange gas to roughly 100\,mbar at room temperature.

For the purpose of this fitting, first of all one has to restrict the TOF dataset to the volume of energy-momentum space that contains  no magnetic signal. In particular case of CeB$_{6}$ this
\begin{wrapfigure}[34]{r}{0.5\textwidth}\vspace{-0pt}
\includegraphics[width=0.5\textwidth]{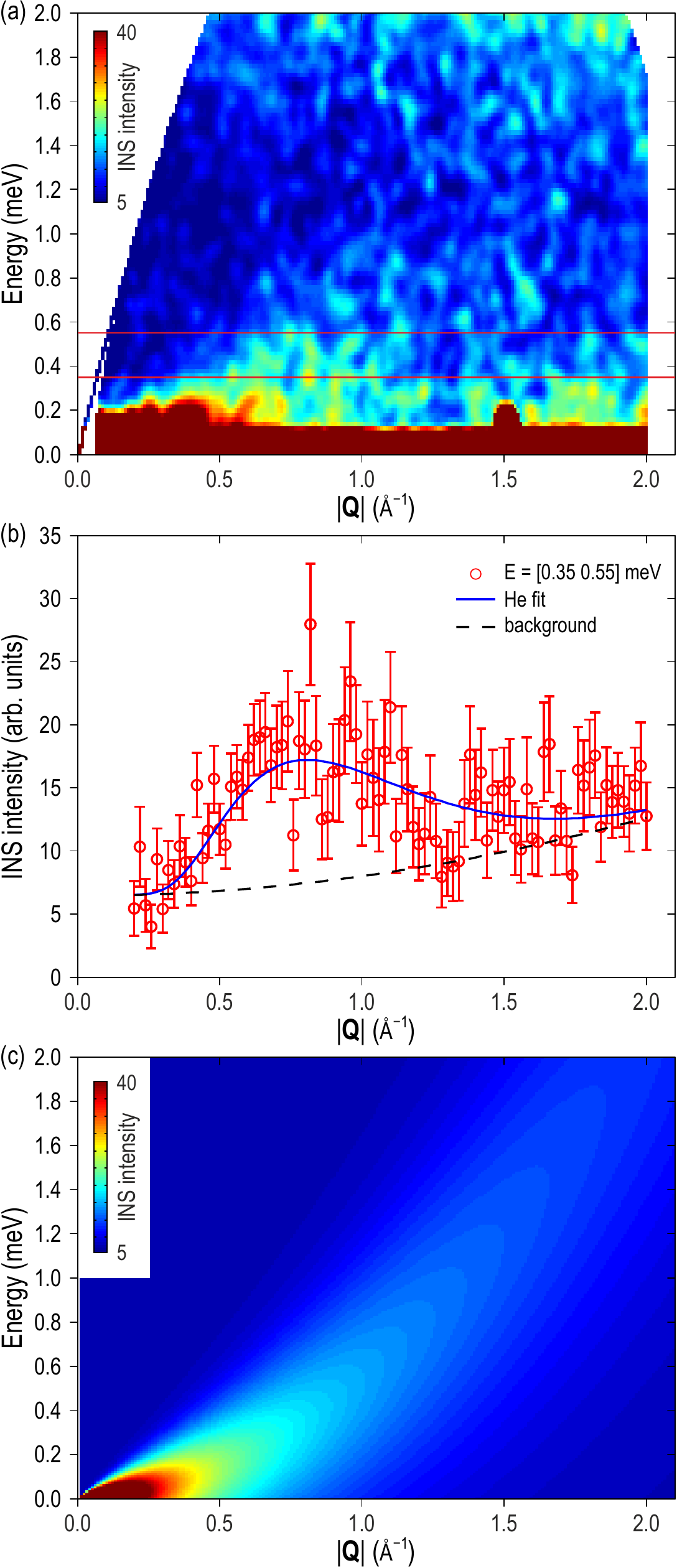}\vspace{-5pt}
\caption{Energy-momentum map of the INS intensity, obtained after powder-averaging the TOF data measured at LET. Red lines shows the energy integration window. (b) $\mathbf{Q}$-dependence, obtained by integrating the data shown in panel (a) in the energy window [0.35\,0.55]\,meV. Solid line represents fit, given by eq.\,(\ref{Eq:He_gas}) with a quadratic background, shown with a dashed line. (c) Simulation of the He recoil signal.}
\label{Fig:HeSubtractionManual}\vspace{-0pt}
\end{wrapfigure}
is not so trivial task, as it is important to avoid individual files which contain information about $\Gamma$, $R$, $X$ and $M$ points, because intense magnetic signal can be observed there. Thus after selecting appropriate files, they need to be averaged and treated as a powder, as shown{\footnotemark}
\footnotetext{Pay attention to the fact the the data in fig.~\ref{Fig:HeSubtractionManual} represents experiment with the Ce$_{0.25}$La$_{0.75}$B$_{6}$ sample and is shown to emphasize how important it is to take into account weak He sinal. For details see app.\,\ref{Appendix:INSFit}.}
in fig.~\ref{Fig:HeSubtractionManual}(a). As a next step it is important to choose an appropriate energy integration range. Although the intensity of the signal decreases with increasing values of both momentum and energy transfer, one should not be jealous and choose a range that is not too close to the elastic line. Fig.~\ref{Fig:HeSubtractionManual}(b) presents the $\mathbf{Q}$-dependence of INS intensity within an energy window $E = [0.35\,0.55]$\,meV, fitted to the He signal given by eq.\,\ref{Eq:He_gas} on top of  a quadratic background. The amplitude of this contribution was kept as a free parameter and adjusted to provide the best fit to the measured data. A good validation of the fit result is to repeat the entire procedure for different energy ranges, that do not overlap at best. This is a good sign if the value of the free parameter is the same within the error bar for several energy rages, however if it is not the case one should consider to verify selected individual files and check the presence of inelastic or quasielastic  excitations  within the energy integration ranges.  As the last step it is necessary to calculate a complete energy-momentum dependence of the spurious signal, as shown in fig.~\ref{Fig:HeSubtractionManual}(c).

\clearpage
\vspace{-0pt}\begin{figure}[!t]
\includegraphics[width=\textwidth]{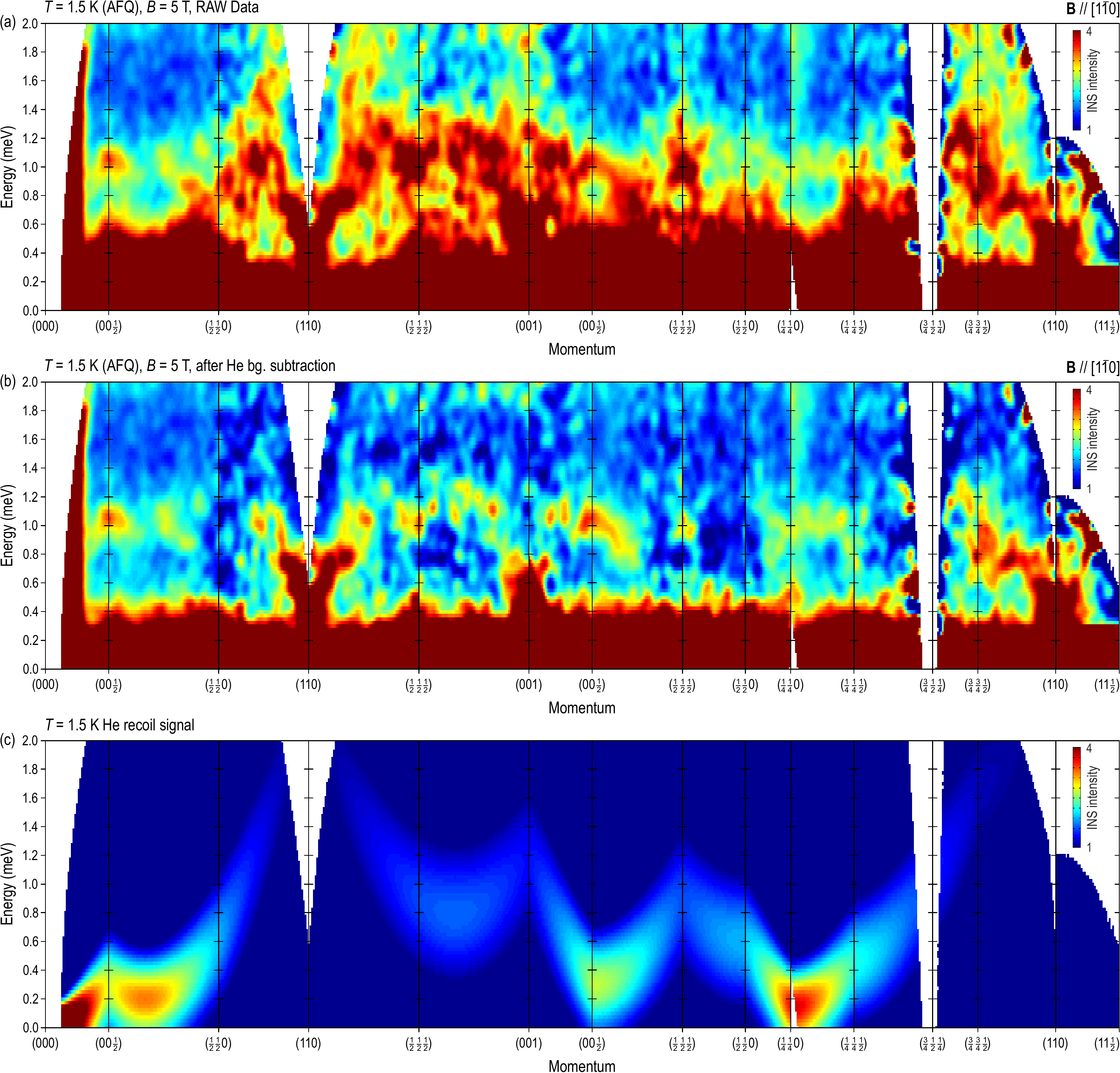}~\vspace{-5pt}
\vspace{-0pt}\caption{Energy-momentum profiles along high-symmetry directions measured at CNCS for CeB${_6}$: (a) in the AFQ state $B = 5$\,T,  (b) after subtracting He background contamination, (c) simulation of the He ``excitation spectrum'' along the same high-symmetry directions, that was subtracted from. }
\label{Fig:CeB6HeSignalSummary}\vspace{-25pt}
\end{figure}

The quality of the resulting subtraction is demonstrated by fig.~\ref{Fig:CeB6HeSignalSummary}, where energy-momentum profiles through the 5\,T dataset measured for CeB${_6}$ are shown. Panel (a) presents  initially measured data,  and panel (b)  shows the described above background-correction procedure, which helps to reveal resonances at various high-symmetry points. Panel (c) is shown to get a general impression on how strong the He contamination could be. As the amplitude of the signal, the value of the free parameter obtained with the procedure described above was used. Since every individual energy-momentum cut  shown in fig.~\ref{Fig:CeB6HeSignalSummary} is obtained by averaging the signal within a certain range along the momentum direction perpendicular to the specified trajectories, simulation of the He ``excitation spectrum'' was done by averaging  the signal within the same volume of the reciprocal space.

The appearance of such a  background is not always \textit{a priori} predictable. We all know that gas pressure depends strongly on temperature. Cryostats or cryomagnets have a very uneven distribution of the temperature inside. Slightly above the neutron beam level, the VTI regulator is located, and thus the temperature in this area is almost equal to the set-point temperature. Closer to the loading flange, on top of the cryostat, the exchange gas temperature is approximately equal to room temperature. In accordance with these simple facts, it follows that the amount of gas at the beam level is  actually strongly dependent on the set-point temperature of the sample and will change with heating or cooling. This is very easy to verify, by simply comparing the fit results for different temperatures measured in one experiment, when the chance having different volume of the exchange gas is excluded.

%

\vspace{-15pt}
\section{Fitting the Data}\label{Appendix:INSFit}
\vspace{-5pt}

During a typical neutron scattering experiment one has to deal  mainly with an  energy scan at constant-$\mathbf{Q}$ position or  fix energy transfer $\hbar\omega$ and run a $\mathbf{Q}$-scan. Here I would like to draw your attention to the seemingly routine procedure of processing the data. As one can see, because of various inherent properties of the instruments as well as non-ideal sample environment, spurious signal may appear in both elastic and inelastic spectra. In some cases such artifacts might appear as sharp, well defined peaks,  which quickly disappear when the measuring point or instrument alignment is changed, thus being easily identified as a spurious signal. However undefined and unexpected signals which mimic intrinsic scattering features can easily be misinterpreted. Since all the intensive features  have already  been measured on ``previous generations'' of instruments, each new piece of research pushes modern equipment to its limit in terms of sensitivity and count rate. In some cases I had to deal with intensities that lie a few counts above the background level of the instrument. In such cases, the counting time could be as high as 60 minutes per point.  Here I want to show several examples where presence of the He background  as well as scattering from the sample environment could significantly influence  the interpretation of the very weak magnetic signal. This problem was most acute in  sec.\,\ref{SubSec:KondoTemperatureInCeLaB6}, where it was necessary to fit many quasielastic lines as a function of temperature.

Let me start with the case where the effect of helium background can be excluded. The Ce$_{0.25}$La$_{0.75}$B$_6$ sample was mounted in a liquid-free bottom-loading closed cycle cryostat. The metallic sample was screwed directly to the cold head, and therefore it became possible to operate the cryostat without exchange gas, continuously pumping the vacuum chamber.  The results of this experiment, shown in fig.\,\ref{Fig:CeLa0p75B6TdepPANDA}, were already discussed in sec.\,\ref{SubSec:KondoTemperatureInCeLaB6}, therefore here I would like to make an  emphasis on the necessity  to determine correct shape of the background. In order to determine the width of the quasielastic line, we performed an energy scan. At higher temperatures one should  expand a scan to the energy gain side (negative energy transfer values), as in this case the intensity of the signal at negative energies is strongly enhanced  because of the Bose population factor.

\enlargethispage{10pt}
\begin{figure}[!t]
\vspace{-10pt}\includegraphics[width=\textwidth]{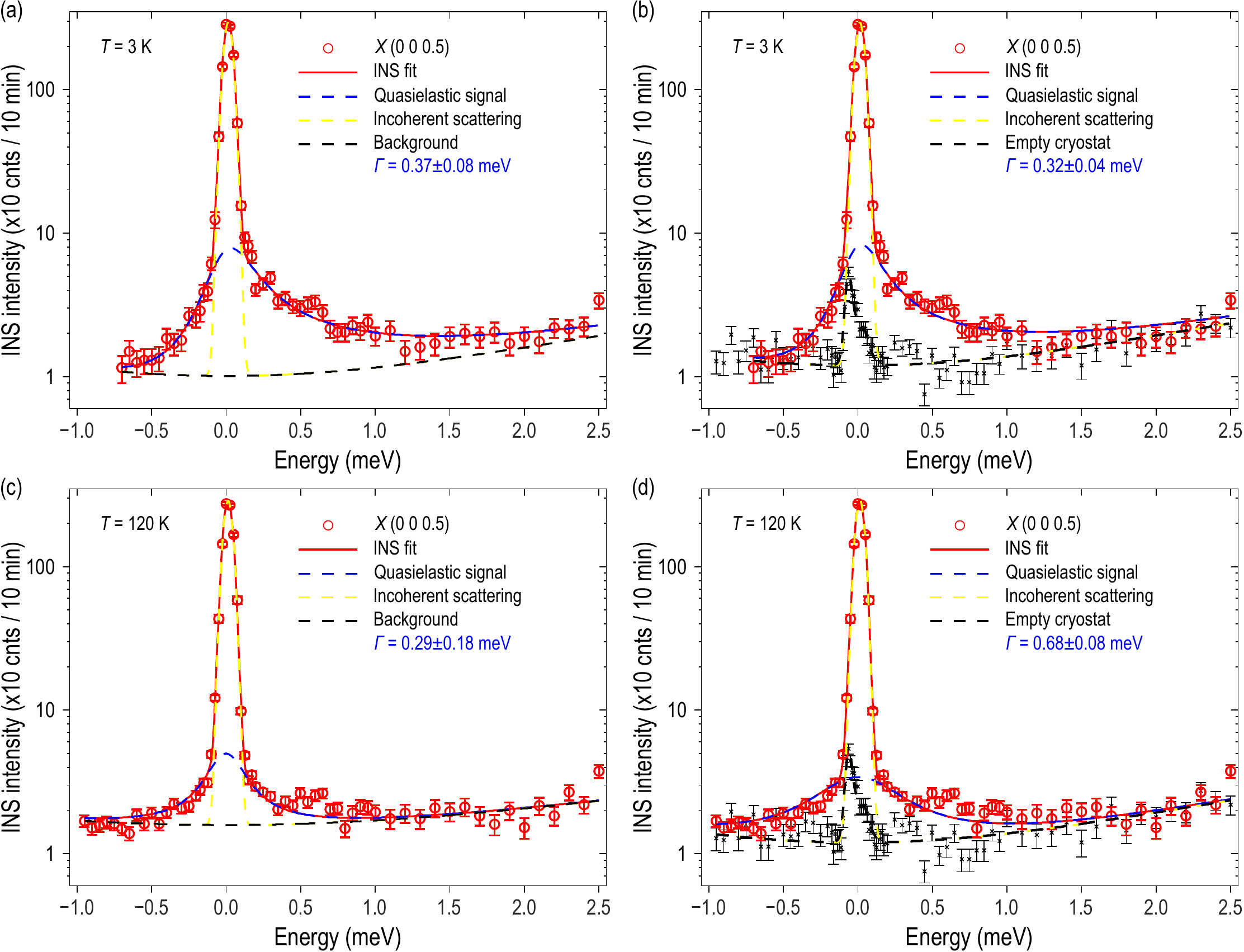}~\vspace{-10pt}
\vspace{-0pt}\caption{Energy scan measured at the $X(0\,\,0\,\,0.5)$ point, fitted to the sum of gaussian peaks, quadratic background and quasielastic line at: (a) $T=3\,\text{K}$; (b) same temperature, but as the background signal, empty cryostat scan is used. Same fit model at a higher temperature $T=120\,\text{K}$ is (c) and (d) respectively.}\vspace{-25pt}
\label{Fig:XpointEmptyCryo}
\end{figure}

When one tries to determine the peak width, the strongest signal at zero energy transfer, which is usually attributed to the incoherent scattering from the sample, is fitted to the gaussian line, unless convolution with the spectrometers resolution function is done. Linear, quadratic, or a sum of both functions is used to describe a non monotonic background. After adjusting all parameters to provide the best fit to the measured data, the width of the quasielastic line is plotted as a function of temperature (see fig.\,\ref{Fig:SqrtGamma}). Such a fit at $T=3\,\text{K}$ is shown in fig.\,\ref{Fig:XpointEmptyCryo}\,(a), and the corresponding line width value $\Gamma=0.37\pm0.08\,\text{meV}$. If we recall app.\,\ref{Appendix:ScatteringTAS}, it becomes clear that the fit described above  lacks an important part: additional scattering from an empty cryostat. An empty cryostat scan shows that the background of the instrument is perfectly described by a quadratic function and scattering from the cryostat wall, in the vicinity of the elastic line, with two additional gaussian functions. An empty cryostat scan and fit are shown in fig.\,\ref{Fig:XpointEmptyCryo}\,(b) with black markers and a dashed line, respectively. After including emptycan measurement as a background into the fit model and running a global fit of both data sets,  the line width value reduces to  $\Gamma=0.32\pm0.04\,\text{meV}$. As one can see both values are the same within the errorbars. Thus, the exact shape of the background hardly{\footnotemark}
\footnotetext{In the case of an ``orange'' type cryostat this is not entirely true, since the amount of aluminum in the beam is larger and it becomes difficult to describe the elastic line with only one gaussian peak.}
affects the width of the quasielastic peak at low temperatures.

\vspace{-5pt}
However, in the case of high temperatures  the situation changes fundamentally. As shown in fig.\,\ref{Fig:XpointEmptyCryo}\,(c), an attempt to fit a $T=120\,\text{K}$ data without a pre-defined background, using same simple model as in panel\,(a), fails. Despite the fact that the data are well described by the fit function, a too narrow peak width $\Gamma=0.29\pm0.18\,\text{meV}$ as well as an incorrectly described background, of course in comparison with the known one from the empty cryostat scan,  indicate that  peak width could be misinterpreted. This problem automatically disappears and the peak width is correctly determined when the background of an empty cryostat is included in the model, as shown in fig.\,\ref{Fig:XpointEmptyCryo}\,(d). Of course one can oppose, that by using a shared background between all data sets it is easy to overcome this problem. In some cases it is true, and to some extent this improves the results, as this trick give peaks width $\Gamma=0.28\pm0.05\,\text{meV}$ and $\Gamma=0.50\pm0.08\,\text{meV}$ for low and high temperatures. But similarly  one can use a linear function as a background, and obtain a different  set of the peak width values. Therefore here I wanted to make an  emphasis not on the absolute value of the peak width  and how accurate it is, but on the fact that  in case of weak signals, the measured width of the quasielastic signal should be seriously questioned without proper investigation of the instrument's background.

\enlargethispage{5pt}
\vspace{-5pt}
As the next step, I want to discuss the case in which it is quite easy to make a mistake in determining the signal width because of the spurious signal from the He exchange gas. TOF measurements of  Ce$_{0.50}$La$_{0.50}$B$_6$ and Ce$_{0.25}$La$_{0.75}$B$_6$  at LET shown in fig.\,\ref{Fig:CeLaB6TOF}\,(c-d), turned out to be slightly contaminated. If one does detailed analysis of the He scattering ``spectrum'', shown in fig.\,\ref{Fig:CeB6HeSignalSummary}\,(c), it becomes clear that for both $X(0\,\,0\,\,0.5)$ and $R(0.5\,\,0.5\,\,0.5)$ points, spurious signal appears at different energies  with different intensities, and in cases when the magnetic signal is very weak it can substantially change the width of the peak. In app.\,\ref{Appendix:HeExchangeGasBackgroundSubstraction}, I have already shown that in case of TOF{\footnotemark}
\footnotetext{The same trick should be possible for the TAS experiment, by measuring the signal at various $|\mathbf{Q}|$ points at the same temperature, and then include it in the fit model. Because of exceptionally high time efforts, as well as the difference of the exchange gas amount  with a temperature change, this method has never been used by me. Pumping exchange gas to the lowest possible limit, and thus sacrificing the time of temperature stabilization works well.}
measurements, it becomes possible to subtract the spurious signal. Below I would like to demonstrate unprocessed data, where amplitude of the He spurious signal, defined as described in  app.\,\ref{Appendix:HeExchangeGasBackgroundSubstraction}, has been included in the fit as a spurious peak.  After comparing fit results with the model where the helium signal was not taken into account, it becomes obvious that it is very easy to misinterpret the quasielastic  peakwidth.

\vspace{-5pt}\begin{figure}[!t]
\includegraphics[width=1\textwidth]{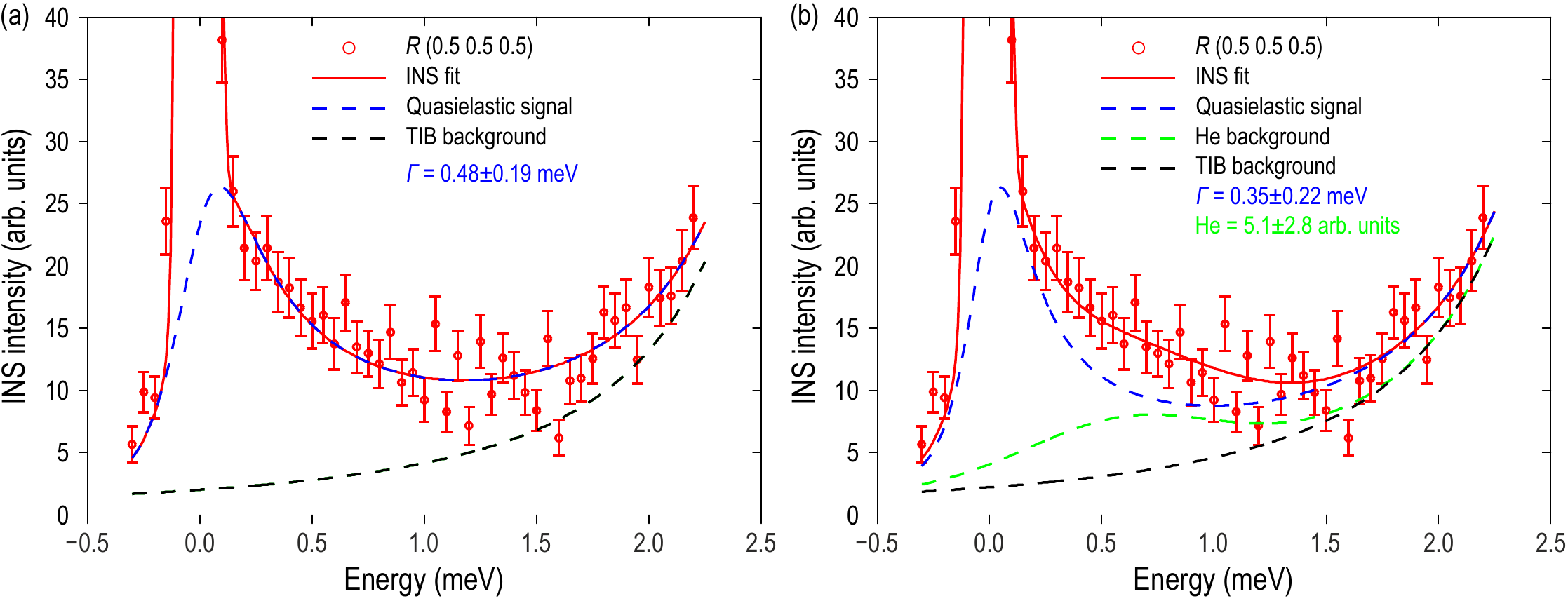}\vspace{-5pt}
\caption{Energy scan measured at the $R(0.5\,\,0.5\,\,0.5)$ point, fitted to the sum of gaussian peaks, TIB background and quasielastic line,  (a) without He contribution; (b) with He spurious signal.}\vspace{-10pt}
\label{Fig:RpointHe}
\end{figure}

In fig.\,\ref{Fig:RpointHe} an energy scan measured at the $R(0.5\,\,0.5\,\,0.5)$ point is shown. It is important to note that the above-raised question of the  background shape naturally disappearing in the case of TOF data. This profile in most cases is perfectly described by just one number $R_{\text{TIB}}$, as was shown in app.\,\ref{Appendix:TIB}. Incoherent scattering from the sample  was again fitted with a gaussian profile. In both cases, whether He contamination is taken into account or not, apparent quality of the fit is perfect. However, a comparison of the quasielastic peak width  points out  the fact that the sample environment has a large influence  on the width of a weak magnetic signal.  Moreover, if one compare compares $\mathbf{Q}$ vectors of different length, like $R$, $X$ and $\Gamma(110)$  points, the spurious helium signal will  affect the width of the peak in different ways, resulting in a spurious momentum-dependence of the peak width. The fit result at the $\Gamma(110)$ point is sufficiently stable against spurious He signal since it appears at high energy transfer values. On the other hand, one needs to be extremely careful when fitting $R$ and $X$ points.

\vspace{-5pt}\begin{figure}[!t]
\includegraphics[width=1\textwidth]{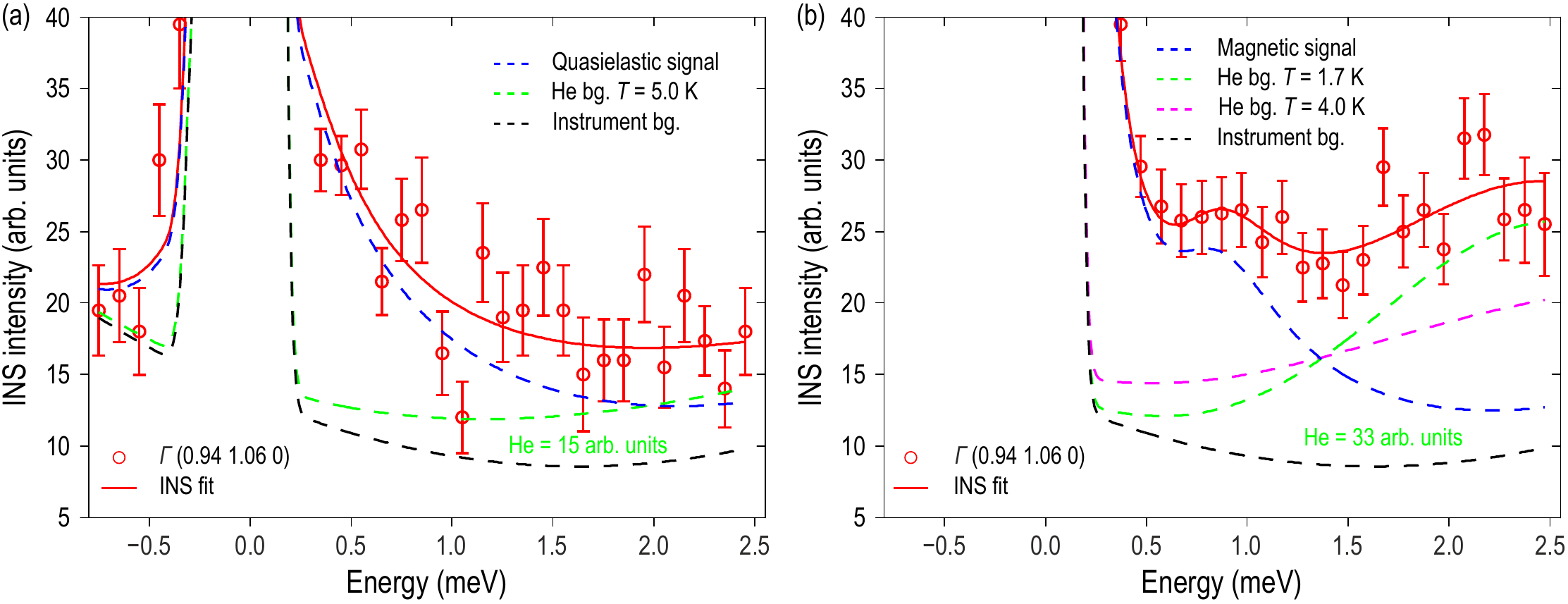}\vspace{-2pt}
\caption{Energy scan  at the $\Gamma''(110)$ point, at a slightly incommensurate wave vector  to avoid the contamination from the Bragg tail, measured (a) above $T_{\text{N}}$, and (b) below $T_{\text{N}}$. Difference in  spurious He signal for different temperatures is additionally shown in panel (b). Negative slope of the background was confirmed during empty cryostat measurements.}\vspace{-10pt}
\label{Fig:GpointHe}
\end{figure}\vspace{-0pt}

As the last example I want to show measurements where He exchange gas  produced a  spurious peak after cooling the sample below $T_{\text{N}}$, with an intensity comparable to that of the magnetic signal. Without prior knowledge about the absence of a magnetic signal  at this energy, as well as of the He scattering ``spectrum'', I would have spent a lot of time  determining that the peak is nonmagnetic. In a previous experiment,  below $T_{\text{N}}$, we found two excitations  at the $\Gamma''(110)$ point in CeB$_6$. For quite a long time we could not cool down the sample. Therefore to rationally use the time, we decided to measure the sample in the paramagnetic state at $T=5\,\text{K}$, where quasielastic magnetic signal with a known width is   present, and use it as a background. The corresponding scan is shown in fig.\,\ref{Fig:GpointHe}\,(a), however at that point He background was not included in the fit model. Finally, after reaching base temperature, I realized, that in addition to the two expected signals, a broad contribution  at $\sim2.5\,\text{meV}$ is present, as shown in fig.\,\ref{Fig:GpointHe}\,(b). In order to reach base temperature, He exchange gas pressure was significantly increased by the sample environment technician. After including it in the model I was able to fit data for both temperatures, with a different exchange gas amount, as indicated in each panel. In addition, He contribution at higher temperature is show in panel(b) with a magenta dashed line. As it can be seen, the signal becomes much broader at higher temperatures, thus an attempt to measure a high-temperature background in some cases may fail.

As a conclusion I would like to give several, in my opinion useful, recommendations. Triple-axis spectrometers are more effective for parametric studies, where the magnetic scattering intensity has to be measured as a function of external parameters (temperature, magnetic field, or pressure) for a few given positions in the energy-momentum space. Therefore, for an accurate interpretation of the  obtained data when measuring temperature dependencies of various excitations, it is desirable to avoid a cryostat with exchange gas. This is especially important when the position of the investigated peak intersects with the position where a signal from the exchange gas is expected. In the case, when you need to determine quasielastic peak width, I would strongly suggest to use a cryostat with larger inner diameter. For example at PANDA,  to reduce background, a vacuum chamber with diameter 540\,mm is available. If the thermal shielding is removed, only a thin ring of the aluminium from the vacuum chamber will be illuminated with neutrons, which according to app.\,\ref{Appendix:ScatteringTAS} will not produce a spurious signal in the vicinity of the elastic line. Measuring an empty cryostat in some cases can also help.

\vspace{-30pt}
\chapter{Animation (online only)}\label{Appendix:Animation}

\enlargethispage{20pt}
\vspace{-55pt}\begin{figure}[h]
\centering
\animategraphics[controls, loop, width=0.75\textwidth, timeline=timeline.txt]{2}{figureS}{1}{18}
\caption{Animation illustrating magnetic field dependence of the magnon spectrum along high-symmetry directions in CeB$_6$. Each frame consists of three color maps integrated along straight segments connecting the $X(00\frac{1}{2})$, $R(\frac{1}{2}\frac{1}{2}\frac{1}{2})$, $\Gamma''(110)$, and $X''(11\frac{1}{2})$ points, which we combined to form a continuous polygonal path in reciprocal space. The controls at the bottom of the figure can be used to modify the frame rate, pause the animation, or browse through individual frames.
The animation can be viewed in the online supplementary materials of ref.\,\ref{PortnichenkoDemishev16}.}
\label{Fig:Animation}\vspace{0pt}
\end{figure}\vspace{-10pt}

\clearpage
\section{Detailed Magnetic Field Dependence}\label{Appendix:DetailedMagneticFieldDependence}

Here in fig.~\ref{Fig:Animation} I present an animation showing the full set of INS data on CeB$_6$ measured at the cold-neutron TOF spectrometer IN5 as a function of magnetic field. These data were obtained as two-dimensional cuts along the $(H\,H\,\frac{1}{2})$, $(H\,H\,1\!-\!H)$, and $(1\,1\,L)$ high-symmetry directions from our four-dimensional TOF dataset by integrating within $\pm$\,0.15~r.l.u. along the momentum direction perpendicular to the plane of the figure in the $(HHL)$ scattering plane, and $\pm$\,0.04~r.l.u. in the out-of-plane (vertical) direction parallel to the magnet axis. Evolution of both resonant modes at the zone center ($\Gamma$) and zone corner ($R$) can be observed.

\prefacesection{Publication List} \fancyhead[LO]{Publication List} \fancyhead[RE]{\quad}

\begin{enumerate}[{[E1]},ref={E\arabic*},leftmargin=20.5pt,labelsep=8pt,itemsep=4pt]

\item\label{JainPortnichenko13} A.~Jain, \underline{P.~Y.~Portnichenko}, H.~Jang, G.~Jackeli, G.~Friemel, A.~Ivanov, A.~Piovano, S.~M.~Yusuf, B.~Keimer, and D.~S.~Inosov, \href{http://link.aps.org/doi/10.1103/PhysRevB.88.224403}{Phys. Rev. B {\bf 88}, 224403 (2013)}.

\item\label{PortnichenkoCameron15} \underline{P.~Y.~Portnichenko}, A.~S.~Cameron, M.~A.~Surmach, P.~P.~Deen, S.~Paschen, A.~Prokofiev, J.-M.~Mignot, A.~M.~Strydom, M.~T.~F.~Telling, A.~Podlesnyak, and D.~S.~Inosov,  \href{http://link.aps.org/doi/10.1103/PhysRevB.91.094412}{Phys. Rev. B {\bf 91}, 094412 (2015)}.

\item\label{SurmachBrueckner15} M.~A.~Surmach, F.~Br\"{u}ckner, S.~Kamusella, R.~Sarkar, \underline{P.~Y.~Portnichenko}, J.~T.~Park, G.~Ghambashidze, H.~Luetkens, P.~K.~Biswas, W.~J.~Choi, Y.~I.~Seo, Y.~S.~Kwon, H.-H.~Klauss, and D.~S.~Inosov, \href{http://link.aps.org/doi/10.1103/PhysRevB.91.104515}{Phys. Rev. B {\bf 91}, 104515 (2015)}.

\item\label{PortnichenkoRomhanyi16} \underline{P.~Y.~Portnichenko}, J.~Romhanyi, Y.~A.~Onykiienko, A.~Henschel, M.~Schmidt, A.~S.~Cameron, M.~A.~Surmach, J.~A.~Lim, J.~T.~Park, A.~Schneidewind, D.~L.~Abernathy, H.~Rosner, J.~van~den~Brink, and D.~S.~Inosov, \href{http://dx.doi.org/10.1038/ncomms10725}{Nature Commun. {\bf 7}, 10725 (2016)}.

\item\label{KoitzschHeming16} A.~Koitzsch, N.~Heming, M.~Knupfer, B.~B\"{u}chner, \underline{P.~Y.~Portnichenko}, A.~Dukhnenko, N.~Shitsevalova, V.~Filipov, L.~Lev, V.~Strocov, J.~Ollivier, and D.~S.~Inosov, \href{http://dx.doi.org/10.1038/ncomms10876}{Nature Commun. {\bf 7}, 10876 (2016)}.

\item\label{CameronTymoshenko16} A.~S.~Cameron, Y.~V.~Tymoshenko, \underline{P.~Y.~Portnichenko}, J.~Gavilano, V.~Tsurkan, V.~Felea, A.~Loidl, S.~Zherlitsyn, J.~Wosnitza, D.~S.~Inosov, \href{http://dx.doi.org/10.1088/0953-8984/28/14/146001}{ J. Phys.: Cond. Matter {\bf 28}, 146001 (2016)}.

\item\label{PortnichenkoDemishev16} \underline{P.~Y.~Portnichenko}, S.~V.~Demishev, A.~V.~Semeno, H.~Ohta, A.~S.~Cameron, M.~A.~Surmach, H.~Jang, G.~Friemel, A.~V.~Dukhnenko, N.~Yu.~Shitsevalova, V.~B.~Filipov, A.~Schneidewind, J.~Ollivier, A.~Podlesnyak, D.~S.~Inosov,  \href{http://link.aps.org/doi/10.1103/PhysRevB.94.035114}{Phys. Rev. B {\bf 94}, 035114 (2016)}.

\item\label{PortnichenkoPaschen16} \underline{P.~Y.~Portnichenko}, S.~Paschen, A.~Prokofiev, M.~Vojta, A.~S.~Cameron, J.-M.~Mignot, A.~Ivanov, D.~S.~Inosov, \href{https://link.aps.org/doi/10.1103/PhysRevB.94.245132}{Phys. Rev. B {\bf 94}, 245132 (2016)}.

\item\label{SurmachPortnichenko16} M.~A.~Surmach, \underline{P.~Y.~Portnichenko}, J.~T.~Park, J.~A.~Rodriguez-Rivera, D.~L.~Sun, Y.~Liu, C.~T.~Lin, D.~S.~Inosov, \href{http://dx.doi.org/10.1002/pssb.201600162}{Phys. Stat. Sol. B {\bf 254}, 1600162 (2017)}.

\item\label{TymoshenkoOnykiienko17} Y.~V.~Tymoshenko, Y.~A.~Onykiienko, T.~Mueller, R.~Thomale, S.~Rachel, A.~S.~Cameron, \underline{P.~Y.~Portnichenko}, D.~V.~Efremov, V.~Tsurkan, D.~L.~Abernathy, J.~Ollivier, A.~Schneidewind, A.~Piovano, V.~Felea, A.~Loidl, D.~S.~Inosov, \href{https://link.aps.org/doi/10.1103/PhysRevX.7.041049}{Phys. Rev. X {\bf 7}, 041049 (2017)}.

\item\label{JangPortnichenko17} D.~J.~Jang, \underline{P.~Y.~Portnichenko}, A.~S.~Cameron, G.~Friemel, A.~V.~Dukhnenko, N.~Y.~Shitsevalova, V.~B.~Filipov, A.~Schneidewind, A.~Ivanov, D.~S.~Inosov, M.~Brando,
    \href{https://doi.org/10.1038/s41535-017-0068-x}{npj Quantum Materials {\bf 2}, 62 (2017)}.

\item\label{NikitinPortnichenko18} S.~E.~Nikitin, \underline{P.~Y.~Portnichenko}, A.~V.~Dukhnenko, N.~Yu.~Shitsevalova, V.~B.~Filipov, Y.~Qiu, J.~A.~Rodriguez-Rivera, J.~Ollivier, D.~S.~Inosov,
    \href{https://link.aps.org/doi/10.1103/PhysRevB.97.075116}{Phys. Rev. B {\bf 97}, 075116 (2018)}.

\end{enumerate}

\vspace{35pt}
{\Huge \textbf{Preprints}}

\begin{enumerate}[{[E1]},ref={E\arabic*},leftmargin=20.5pt,labelsep=8pt,itemsep=4pt]
\setcounter{enumi}{12}
\item\label{CameronYerin18} A.~S.~Cameron, Y.~S.~Yerin, Y.~V.~Tymoshenko, \underline{P.~Y.~Portnichenko}, A.~S.~Sukhanov, M.~Ciomaga Hatnean, D.~McK.~Paul, G.~Balakrishnan, R.~Cubitt, D.~S.~Inosov, \href{https://arxiv.org/abs/1810.03876}{arXiv:1810.03876}.

\item\label{PortnichenkoNikitin18} \underline{P.~Y.~Portnichenko}, S.~E.~Nikitin, A.~Prokofiev, S.~Paschen, J.-M.~Mignot, J.~Ollivier, A.~Podlesnyak, S.~Meng, Z.~Lu, D.~S.~Inosov, \href{https://arxiv.org/abs/1810.12740}{arXiv:1810.12740}.

\end{enumerate}


\bibliography{PhD}\bibliographystyle{my-apsrev}


\prefacesection{Acknowledgements} \fancyhead[LO]{Acknowledgements} \fancyhead[RE]{\quad}

First of all I want to thank my supervisor Jun.-Prof. Dmytro~S.~Inosov for giving me the opportunity to perform my research at TU Dresden and for his help in my work. His inborn intuition allowed us to revive interest to the seemingly long-forgotten topic of underlying physics in CeB$_6$, as well as the related Ce$_{3}$Pd$_{20}$Si$_{6}$ compound. His meticulousness instilled in me a love for accurate design of figures, and the ability to properly highlight important items in the proposal allowed us to obtain numerous days of beamtime for neutron measurements.

It would be impossible to do my research without excellent single-crystal samples, therefore I want to thank N.~Yu.~Shitsevalova, A.\,V.~Dukhnenko and V.~B.~Filipov for providing us with a whole family of pure, La, and Nd doped CeB$_6$ crystals, and also S.~Paschen and A.~Prokofiev for  Ce$_{3}$Pd$_{20}$Si$_{6}$ and Ce$_{3}$Pd$_{20}$Ge$_{6}$ crystals.

Since my work was related with experiments, which were conducted at large-scale neutron facilities, it becomes impossible to fully appreciate the assistance provided to me from the beam-line scientists. I would like to thank J.-M. Mignot, Y.~Sidis, S.~Petit, and P.~Bourges from Laboratoire L\'{e}on Brillouin not only for brilliant instrument support, but also for their help in the ``fight'' with the administration to gain unaccompanied facility access. In my opinion, P.~Boutrouille deserves to be named ``cryostat wizard'', for his extraordinary ability to handle the cryogenic equipment. P.~\v{C}erm\'{a}k and I.~Radelytskyi from  the J\"{u}lich Center for Neutron Science helped enormously in my attempts to force the PANDA spectrometer to work properly, and I hope A.~Schneidewind and her family were not very annoyed by the number of night calls from me. The speed, at which J.~T.~Park  managed to start the experiment and  to process the data always amazed me. Many hours were spent together with Z.~H\"{u}sges, Z.~Lu and D.~L.~Quintero-Castro from the Helmholtz-Zentrum Berlin, trying to determine the best configuration of the spectrometer and identify the origin of spurious signals. Every experiment with A.~Ivanov and J.~Ollivier  at the  Institute Laue-Langevin gave me a fruitful learning experience on various technical details about the spectrometer and sample environment. I acknowledge the help of A.~Podlesnyak, D.~L.~Abernathy and G.~Ehlers from the Oak Ridge National Laboratory for their support during the experiments and advice to visit the historically important X-10 Graphite Reactor, K-25 facility for uranium enrichment, HFIR, and other interesting, as well as scary buildings, like the Experimental Gas-Cooled Reactor at which an operational decontamination shower can still be found outside the reactor building. I was pleased to conduct experiments with J.~A.~Rodriguez-Rivera and Y.~Qiu at the NIST Center for Neutron Research. I also thank all my collaborators  who helped me in carrying out a lot of experiments: B.~Frick (ILL), J.~White (PSI), R.~Bewley, T.~Guidi, D.~Adroja (ISIS), R.~Mole (ANSTO), K.~Proke\v{s} and I.~Glavatskyy (HZB). I would like to thank A.~Akbari from APCTP in Pohang, South Korea, for his theoretical support with  CeB$_6$ project, and in particular for the calculations shown in\,\cref{Fig:ResonanceEnergyDep,Fig:ResonanceEnergyDep2}.

I also would like to thank the whole staff at the Research/Instructional Support Workshops at the Department of Physics, and especially  M.~Siegel, D.~Andris, J.~Kretschmer, and U.~Dorn. For the purpose of numerous experiments, special sample holders, adaptors and very small nuts and bolts were made by them. They were  always sympathetic to my demands and tried to do their best in  making all necessary parts even on short notice.

Our neutron scattering experiments required careful alignment of the available single crystals, therefore I would like to acknowledge B.~B\"{u}chner, S.~Seiro, S.~Wurmehl, C.~Blum at IFW Dresden, and L.~H.~Tjeng, A.~Komarek, D.~Sokolov at  MPI-CPfS, for providing access to the Laue machines  to orient crystals by means of the backscattering method.

Over the past five years I took part in a number of experiments, together with my colleagues from the neutron scattering group M.~A.~Surmach, A.~S.~Cameron, Y.~A.~Onykiienko, Y.~V.~Tymoshenko from TU Dresden, S.~E.~Nikitin, A.~Sukhanov, O.~Stockert, D.~J.~Jang, M.~Brando from MPI-CPFS, G.~Friemel, H.~Jang, A.~Jain from MPI Stuttgart, A.~Chikina, D.~V.~Vyalikh from TU Dresden, and A.~Fedorov from IFW Dresden.

I credit D.~S.~Inosov and A.~S.~Cameron for a careful proofreading of this thesis.

My work on this thesis was supported by the Research Training Group GRK\,1621, the German Research Foundation (DFG) under grants No.~IN\,\mbox{209/1-2} within the priority programme SPP\,1458, individual grants IN\,\mbox{209/4-1}, IN\,\mbox{209/3-1}, and the Collaborative Research Center SFB\,1143 (project C03).


\prefacesection{Versicherung} \fancyhead[LO]{Versicherung} \fancyhead[RE]{\quad}

Hiermit versichere ich, dass ich die vorliegende Arbeit ohne unzul\"{a}ssige Hilfe Dritter und ohne Benutzung anderer als der angegebenen Hilfsmittel angefertigt habe; die aus fremden Quellen direkt oder indirekt \"{u}bernommenen Gedanken sind als solche kenntlich gemacht. Die Arbeit wurde bisher weder im Inland noch im Ausland in gleicher oder \"{a}hnlicher Form einer anderen Pr\"{u}fungsbeh\"{o}rde vorgelegt.
Bis zum jetzigen Zeitpunkt habe ich keinerlei erfolglose Promotionsverfahren absolviert.

Die vorliegende Dissertation wurde am Institut f\"{u}r Festk\"{o}rper- und Materialphysik der Fakult\"{a}t Physik der Technischen Universit\"{a}t Dresden unter Leitung von Jun.-Prof. Dr. Dmytro~S.~Inosov angefertigt.

Ich erkenne hiermit die Promotionsordnung der Fakult\"{a}t Physik  der Technischen Universit\"{a}t Dresden vom 23.02.2011 an.

\vspace{40pt}

\makebox[1.5in]{\hrulefill}\hfill\makebox[1.5in]{\hrulefill}

\vspace{-7pt}
\hspace*{0mm}\phantom{~~~~~~}(Ort, Datum)\hfill(Unterschrift)\phantom{~~~~~~}

\end{document}